# Nonlinear response of wrinkled premixed flames to time- and space-dependent forcing and stretch.

Guy Joulin [1*] and Bruno Denet [2#].

[1] Institut P-prime, UPR 3346 CNRS, ENSMA, Université de Poitiers,
1 rue Clément Ader, B.P. 40109, 86961 Futuroscope Cedex, Poitiers, France.

[2] Aix-Marseille Univ., IRPHE UMR 7342 CNRS, Centrale-Marseille,
Technopole de Château-Gombert, 49 rue Joliot-Curie, 13384 Marseille Cedex 13, France.

---

[*] guy.joulin@lcd.ensma.fr . Corresponding author.          Sept. 15, 2019

[#] bruno.denet@irphe.univ-mrs.fr .

**Abstract**

Premixed-flame wrinkling is studied via a Michelson-Sivashinsky [MS] type of evolution equation retaining the Darrieus-Landau [DL] instability, a curvature effect and a geometric nonlinearity. Here it also keeps *forcing* by longitudinal shearflow and wrinkle *stretch* by transverse flow; both imposed stimuli vary in time [$t$] and space as to make the front slope comprise a given fluctuating spatial harmonics and unknown pole-decomposed pieces. A DL-free Burgers version is examined in parallel, also with Neumann conditions and symmetry. As is shown for both models, solving $N_{tot}$ equations of motion for the poles in principle yields the front dynamics, the arclength increment $V(t)$ and its time-average $V_{av}$. Yet this could be worked out analytically [or nearly so] only in high-frequency [HF] or low-frequency [LF] limits. These tackle $N_{tot}=1$ and $N_{tot}=2$ pairs of poles per cell, then $N_{tot} \gg 1$ pairs forming two piles viewed as continua, one per crest. All such analyses handle various wrinkle wavelengths.

HF stimuli dominate the flame evolutions. Corrections, obtained explicitly if $N_{tot}=1$ or reduced to a root search if $N_{tot} = 2$, are explicit if $N_{tot} \gg 1$ thanks to exact resolutions of Tricomi-type integral equations for unsteady pole densities. Despite ample pole motions that make some commute between crests, $V_{av}$ grows in a nearly parabolic way with the combined intensity $u'$ of forcing and stretch. The qualitative influences of DL instability are: (i) an offset of $V_{av}$, known from the MS flame fronts at $u' \to 0$ and maximal if $N_{tot}$ reaches a definite $N_{opt}$ that grows in proportion to the wrinkle wavelength, and (ii) a nonzero initial $dV_{av}/du'$ if [and only if] $N_{tot} > N_{opt} \gg 1$.

LF stimuli and DL instability can induce multiple branches and relaxation phenomena. Numerical $t$-averages are needed even if $V(t)$ is analytically known. For $N_{tot}=1,2$ and short wrinkles, or Burgers fronts, $V_{av}$ transitions from $\sim u'^2$ to sublinear as $u'$ grows; for longer wrinkles $V_{av}$ keeps its MS value at moderate $u'$'s, then bifurcates to an ultimately sublinear growth that depends on the stimulus phases. For very long wrinkles [$N_{tot} \sim N_{opt} \gg 1$] coupled integral equations give analytical slope and pole-density profiles, but pile heights/contents need a $t$-dependent numerical search of up to two roots to get $V(t)$. If $N_{tot} \geq N_{opt}$ a unique solution, whose crests cyclically exchange poles, exists for all $u' \neq 0$; if $N_{tot} > N_{opt}$, $V_{av}$ transitions from offset $(u')^{2/3}$ to affine as $u'$ grows, whereas $N_{tot} = N_{opt}$ yield $\sim u'^{1/2}$ variations if no wrinkle stretch acts and $\sim u'$ if some does. $N_{tot} < N_{opt}$ and low $u'$'s yield a continuum of $t$-periodic shapes whose crest piles have constant pole contents, whilst larger $u'$'s restore uniqueness and pole tranfers; all $V_{av}$s nearly keep their MS value for $u'$ below a $N_{tot}/N_{opt}$-dependent threshold, then branch off to a linear growth that depends on the stimulus phases. For LF Burgers fronts, a single crest survives at a time in each cell and $V_{av}$ roughly is $\sim u'^2$. Stimuli of fixed spatial shapes also yields a final $V_{av} \sim u'^2$, which may turn linear in both models if the $t$-independent $N_{tot}$ is allowed to linearly vary from run to run with the stretch intensity.

A summary, a discussion and hints of generalizations are provided, and open problems are evoked.

# I. INTRODUCTION

The theory of wrinkled premixed flames is uneasy: these are unstable free boundaries, hydrodynamically coupled with adjacent gases of unlike densities, with ever present vorticity on the burnt side. And shearflows, eddies or turbulence in the fresh gas further feed front wrinkling.

Sivashinsky [1] derived the first evolution equation for unforced flames, *i.e.* propagating in uniform flows of far-upstream gases. Besides curvature effects and an eikonal nonlinearity it accounts for the Darrieus [2]-Landau [3] [DL] hydrodynamic instability caused by a nonunity fresh-to-burnt density ratio $E > 1$, provided the Atwood number $\mathcal{A} \equiv \frac{E-1}{E+1}$ is small. If rescaled, the equation so obtained in [1], and taken up numerically in [4], takes on the canonical form

$$\varphi_t + \tfrac{1}{2}(\varphi_x)^2 - \nu \varphi_{xx} - \mathcal{I}[\varphi] = 0, \qquad (1.1)$$

usually termed [5] the Michelson-Sivashinsky [MS] equation. The unknown $\varphi(t,x)$ in (1.1) stands for the flame location/shape if observed in a frame $(x,y)$ that drifts at flat-flame speed $u_L$ towards $y < 0$, with $y > \varphi(t,x)$ in burnt gas. The subscripts denote partial derivatives in scaled time $t$ or abscissa $x$. The constant $\nu > 0$ tells how the local burning speed $u_n$ reacts to curvature [7], $1 - u_n / u_L \sim \mathcal{A}^2 \nu \varphi_{xx}$; if the actual reciprocal wrinkle wavenumber is used as $x$-wise reference length for spatially periodic fronts, $\nu$ also is the neutral-to-actual wavelength ratio. The spatially nonlocal, linear operator $\mathcal{I}[.]$ encodes the DL wrinkling instability: as $\mathcal{I}[e^{i\kappa x}] = |\kappa| e^{i\kappa x}$ the growth/decay rate of Fourier modes $\varphi_\kappa \sim f_\kappa(t) e^{i\kappa x}$ of the linearized (1.1) reads $\frac{d}{dt} \ln(|f_\kappa|) = |\kappa| - \nu \kappa^2$. The nonlinear contribution to $\varphi_t$ in (1.1) mainly [8] is geometric [1]: tilted flame elements with $u_n \approx u_L$ propagate along their local normal at an angle $\alpha(t,x) \sim \mathcal{A}\varphi_x(t,x)$ to the mean direction of propagation [$y$-axis], and $u_L - u_L / \cos(\alpha) \approx -\tfrac{1}{2} u_L \alpha^2$.

Originally derived in [1] as a leading order for $\mathcal{A} \to 0$, (1.1) still rules spontaneous flame wrinkling once another two orders in $\mathcal{A}$ are kept [8,9], because the contributions of front-generated $x$-wise gas flow [$\sim \mathcal{A}^2 u_L \varphi_x$] and vorticity merely improve coefficients that were all rescaled to unity in (1.1). Hence (1.1) captures [4] free flame evolutions revealed by direct simulations that assume $\mathcal{A} = O(1)$ [10, 11] and by experiments [12], up to $\mathcal{A}$-dependent rescaling. It also governs incipient nuclear flames in Ia-Supernovae [13], reactive infiltration fronts [14], interfaces in doped semiconductors [15], and it modeled plasma instability [16].

Closed-form solutions to (1.1) are accessed via pole-decompositions [17, 18] that convert the search for $\varphi(t,x)$ to a $2N$-body problem for poles of the analytically continued front slope $\varphi_x(t,z)$, $z \equiv x+iB$ [[19] and Refs. therein]. Pairwise pole interactions from $\varphi_x^2$ explain how cusped crests [local maxima of $\varphi(t,x)$] form and how their coalescence builds up the longest wrinkle that fits in a channel if $2\pi$-periodicity in $x$ is assumed. Then $N \leq N_{opt}(\nu) \equiv \lfloor \frac{1}{2\nu} + \frac{1}{2} \rfloor$ pairs of poles, $\lfloor . \rfloor \equiv$ IntegerPart(.), may settle at $z = iB_{k\infty} (\mathrm{mod}\, 2\pi)$ in a cell, with $B_{-k\infty} = -B_{k\infty}$ ensuring that $\varphi(t,x)$ will be real. Such steady shapes centered on $x = 0$ have $\varphi(t \to \infty, x) = -tV_\infty + \varphi_\infty(x)$, where $\varphi_\infty(-x) = \varphi_\infty(x) = \varphi_\infty(x+2\pi)$ and $V_\infty = (2\pi)^{-1} \int_{-\pi}^{+\pi} \frac{1}{2}(\varphi_{\infty x})^2 dx$ satisfy [18]

$$\varphi_{\infty x}(x) = -\nu \sum_{0 \neq k = -N}^{N} \cot(\tfrac{1}{2}(x - iB_{k\infty})),$$
$$V_\infty = \mathcal{V}(\nu N), \quad \mathcal{V}(n) \equiv 2(1-n)n \,.$$
(1.2)

The pole altitudes $B_{k\infty}$ obey coupled algebraic equations, to be displayed later. In case $N > N_{opt}(\nu)$ pairs of poles are present at $t=0$, $N - N_{opt}(\nu)$ of them ultimately drift to $\pm i\infty$ and cease to count for $z$ real: $\varphi_t(t \to \infty, x) = -\mathcal{V}(\nu N_{opt}(\nu))$. The long steady wrinkles with $2N \sim 1/\nu \gg 1$ have smoothly distributed poles whose density obeys a singular integral equation, analytically solved if $2\nu N$ is $2\nu N_{opt} \approx 1$ [18] or less [20]; the front slopes and shapes ensue and the un-scaled, dimensioned effective burning speed $u_e$ then has $u_e/u_L - 1 \sim \mathcal{A}^2 \mathcal{V}(\nu N)$.

There also exist bi-coalesced steady solutions to (1.1) whose slopes have $2n$ extra poles $\pi + ib_{j\infty} (\mathrm{mod}\, 2\pi)$, $b_{-j\infty} = -b_{j\infty}$. Their front-length increment $V = \mathcal{V}(\nu N_{tot})$ only depends on the total number $N_{tot} = N + n \leq N_{opt}(\nu)$ of involved pairs of poles; pole densities, and cell shapes with unequal crests, were analytically determined in [20] for $N_{tot} \sim 1/\nu \gg 1$, $0 \leq n/N \leq \infty$.

Such exact results only concern propagations into an initially quiescent gas though. It would be of interest to extend them to situations where the fresh gas involves $x$-periodic shear-flows or vortices that act as extra sources of flame wrinkling. Some [21, 22] stressed the need for including the interplay of incoming velocity modulations and DL instability to accurately model weakly or moderately turbulent flames. But no analytical work could handle nonlinear instability-forcing interactions directly, *i.e.*, without using *a priori* averages [and closures] or simulations [23, 24]; even the unsteadily forced version of (1.1) needed those [26-28].

Reference [25] partly filled the gap, by considering steady fronts subject to a steady forcing term $u(x)$ added to the right of (1.1) to account for an imposed extra shear-flow of fresh gas in the $y$-direction. Flame shapes and speeds, and their changes with forcing intensity, were accessed in closed form on selecting such a $u(x)$ that the front slope still comprises a pole decomposed piece, just like in (1.2) yet superimposed to a nonzero base slope chosen in a class.

The present work adapts the latter strategy to wrinkled fronts forced by an unsteady $u(t,x)$. As the approach can also encompass an imposed extra $x$-wise flow $w(t,x)$ of fresh gas, the nonlinear response of DL-unstable flames to both stimuli can be taken up analytically. A DL-free, Burgers type of dynamics is studied in parallel for comparison, by the same methods.

This text is organized as follows. Section II introduces the front dynamics, and Sec. III tells how to access front shape or length via poles. Next to simple solutions [Sec. IV] the control parameters are inventoried, and a linear response is studied [Sec. V]. The long Sections VI & VII develop high- or low-frequency nonlinear analyses, respectively, each with a few pairs of poles or dense piles thereof; in any case the front-shape dynamics, the arclength evolution and time-average are accessed as analytically as one could. The concluding Sec.VIII offers a summary and a discussion, suggests extensions, and evokes open problems. Appendix A sketches how to get to the front evolution, and the Appendices B to E compile supplemental material.

## II. FRONT SHAPE DYNAMICS

The augmented, MS-type evolution equation adopted here for the scaled flame shape reads

$$\varphi_t + w(t,x)\varphi_x + \tfrac{1}{2}(\varphi_x)^2 - \nu\varphi_{xx} - \mathcal{I}[\varphi] = u(t,x), \quad (2.1)$$

$$\mathcal{I}[\varphi] \equiv (2\pi)^{-1} \fint_{-\pi}^{+\pi} \cot(\tfrac{1}{2}(x-x'))\varphi_x(t,x')dx', \quad (2.2)$$

where $\fint(.)dx'$ is a Cauchy principal-value integral and $\varphi(t,x) = \varphi(t,x+2\pi)$. Neumann conditions are imposed at $x = 0 \pmod{\pi}$, so $\varphi(t,x) = \varphi(t,-x)$ will also hold. The forcing, $u(t,x)$, and stretch, $w(t,x)$, functions have kinematic origins [Appendix A]. Both stem from fresh-gas motions ahead of the flame that do *not* result from the front distortions, contrary to the DL-instability term $\mathcal{I}[\varphi]$ and to the stretch effect $\propto \varphi_x\varphi_x$ [again lumped in with $\tfrac{1}{2}(\varphi_x)^2$] of the $x$-wise flow [$\propto \varphi_x$] induced by wrinkling. While $u(t,x)$ gives the front extra pulls or pushes in the mean direction of propagation [ $y$- axis] and reinforces or opposes $\mathcal{I}[\varphi]$ depending on the relative phases, $w(t,x)\varphi_x$ further stretches [or squeezes] laterally the front wrinkles generated by $u(t,x) + \mathcal{I}[\varphi]$. $u(t, x+2\pi) = u(t,x) = u(t,-x)$ and $w(t, x+2\pi) = w(t,x) = -w(t,-x)$ will undulate in time as well to mimic $y$-wise flame propagations across stacked line-arrays of counter-rotating eddies, with each array taken $2\pi$-periodic and mirror symmetric in $x$.

The goal is to solve (2.1) for $\varphi(t,x)$ and the scaled front-length increment [Appendix A]

$$V(t) = \langle \tfrac{1}{2}(\varphi_x)^2 \rangle, \quad \langle . \rangle \equiv (2\pi)^{-1} \int_{-\pi}^{+\pi} (.) dx, \quad (2.3)$$

as to access the wrinkling-induced increase $u_e(t)/u_L - 1 \sim \mathcal{A}^2 V(t)$ in effective burning speed $u_e(t)$ and its *a posteriori* time-average. All this should ideally be done analytically for general $w(t,x)$ and $u(t,x)$ with the good symmetries [see above] and $\langle u(t,x) \rangle = 0$ [Appendix A].

A Burgers-type of passive dynamics will also be studied for comparison with (2.1), *viz.*,

$$\hat{\varphi}_t + \hat{w}(t,x)\hat{\varphi}_x + \tfrac{1}{2}(\hat{\varphi}_x)^2 - \hat{\nu}\hat{\varphi}_{xx} = \hat{u}(t,x), \quad (2.4)$$

to get $\hat{\varphi}(t,x)$ and $\hat{V}(t) \equiv \langle \tfrac{1}{2}(\hat{\varphi}_x)^2 \rangle$ from the counterparts $(\hat{u}(t,x), \hat{w}(t,x))$ of $(u(t,x), w(t,x))$.

## III. SHAPES, POLES & VELOCITIES.

The tasks look hopeless as stated above: the disguised linear equation (2.4) has no known general solution, and the nonlocal $\mathcal{I}[.]$ in (2.1) does not really help. A way out is to try and adapt the strategy of Ref. [25]. To this end, the unknown flame shape in (2.1) is first split as:

$$\varphi(t,x) = h(t) + \Phi(t,x) + \phi(t,x),$$
$$\phi(t,x) \equiv -2\nu \sum_{\substack{k=-N_{tot} \\ 0 \neq k=-N_{tot}}}^{k=+N_{tot}} \ln[\sin(\tfrac{1}{2}(x - z_k(t)))] . \quad (3.1)$$

$\Phi(t,x) = \Phi(t, x+2\pi) = \Phi(t,-x)$, a base-shape with $\langle \Phi(t,x) \rangle = 0$, is assumed given whereas $h(t)$ and the $z_k(t)$, $1 \leq |k| \leq N_{tot}$, are unknown. The complex $z_k(t) = \Re[z_k] + i\Im[z_k]$ denote the movable [or solution dependent] poles of the slope once it is analytically continued as $\varphi_x(t,z)$ in $z = x + iB$ plane; by convention $z_{-k}$ is the complex conjugate $\bar{z}_k$ of $z_{+k}$, and $\Im[z_{k \geq 1}] > 0$.

For $z \to z_k(t)$, $\varphi_x(t,z) \approx -2\nu/(z - z_k)$ blows up. As the double poles of $\tfrac{1}{2}(\varphi_x)^2$ and $-\nu\varphi_{xx}$ neutralize one another, the left-hand side $e(t,z)$ of (2.1) generically has $z_k(t)$ as simple pole, hence $u(t,z)$ also does; that, however, is not allowed if $u(t,x)$ is to represent a *solution in-dependent* imposed stimulus. Requiring $e(t, z_k(t)) < \infty$ for each $k$ yields $2N_{tot}$ pole equations [now differential ones] to determine the $z_k(t)$. Not to clutter with duplicated formulae, it is appropriate now to specify the stretch $w(t,x)$ in (2.1) and the base shape $\Phi(t,x)$ in (3.1),

$$\Phi(t,x) = \Phi(t)\cos(x) , \qquad (3.2)$$
$$w(t,x) = w(t)\sin(x) . \qquad (3.3)$$

These have the desired parities in $x$ and involve arbitrary $\Phi(t)$, $w(t)$. Using the trig identities $\cot(\frac{1}{2}(x-z_k))\cot(\frac{1}{2}(x-z_j)) + 1 \equiv \cot(\frac{1}{2}(z_k - z_j))[\cot(\frac{1}{2}(x-z_k)) - \cot(\frac{1}{2}(x-z_j))]$ in $(\varphi_x)^2$ and $(\sin(x) - \sin(z_k))\cot(\frac{1}{2}(x-z_k)) \equiv \cos(x) + \cos(z_k)$ in $w\varphi_x$ the pole equations so obtained read:

$$\frac{dz_k}{dt} = \sum_{j \neq k, j = -N_{tot}}^{j = +N_{tot}} \nu \cot[\tfrac{1}{2}(z_j - z_k)] - i\,\mathrm{sgn}[\Im(z_k)] + [w(t) - \Phi(t)]\sin(z_k), \quad 1 \leq |k| \leq N_{tot} . \qquad (3.4)$$

The piece with the signum function sgn[.] in (3.4) tends to drive $z_{\pm k}(t)$ to the real axis and to render the front singular at $x = \Re(z_{\pm k})$, but the $z_{-k} - z_k$ interaction prevents $z_{\pm k}$ from getting real. Such sgn[.] terms stem from the DL instability and are absent from the Burgers version of (3.4) in which all symbols, except $t$ and dummy indices, will be equipped with carets.

$\langle u(t,x) \rangle = 0$ is next invoked to get $\frac{d}{dt}h$ from $\langle e(t,x) \rangle = 0$, and what is leftover from $e(t,x)$ *defines* the forcing function $u(t,x)$. In the present case where (3.2)(3.3) hold this produces:

$$\frac{dh}{dt} = -2\nu N_{tot}(1 - \nu N_{tot}) - \tfrac{1}{2}\Phi(t)(\tfrac{1}{2}\Phi(t) - w(t)) - 2\nu(\Phi(t) - w(t))\sum_{k=1}^{N_{tot}} \Re[\cos(z_k)], \qquad (3.5)$$

$$u(t,x) = u_1(t)\cos(x) + u_2(t)\cos(2x) , \qquad (3.6)$$

$$\begin{aligned} u_1(t) &= \tfrac{d}{dt}\Phi(t) + 2\nu N_{tot}(\Phi(t) - w(t)) + (\nu - 1)\Phi(t) , \\ u_2(t) &= \tfrac{1}{2}\Phi(t)(w(t) - \tfrac{1}{2}\Phi(t)) \end{aligned} \qquad (3.7)$$

Different choices than (3.2)(3.3) replace $[w(t) - \Phi(t)]\sin(z_k)$ by $[w(t, z_k) + \Phi_x(t, z_k)]$ in (3.4), since both result from the convection of each pole by the analytically continued $w(t,z)$ and base slope $\Phi_x(t,z)$. But only on assuming (3.2) and (3.3) could the Ansatz (3.1) single out a class of forcing functions $u(t,x)$ that are *independent of the pole trajectories*, see (3.6)(3.7).

To be sure, adding $2\Phi_1(t)\ln[\cos(\frac{1}{2}x)]$ to (3.2) and $w_1(t)\tan(\frac{1}{2}x)$ to (3.3) would not spoil the latter crucial property, since $(\tan(\frac{1}{2}x) - \tan(\frac{1}{2}z_k))\cot(\frac{1}{2}(x-z_k)) - 1 \equiv \tan(\frac{1}{2}x)\tan(\frac{1}{2}z_k)$ vanishes if summed over $1 \leq |k| \leq N_{tot}$ [if $z_k$ is a pole of $\varphi_x(t,z) = -\varphi_x(t,-z)$, so is $-z_k$]; this would add $[w_1(t) - \Phi_1(t)]\tan(\frac{1}{2}z_k)$ to (3.4) and new $w_1(t)$- and $\Phi_1(t)$-dependent functions to (3.5)(3.7). Yet no physical interpretation of an unbounded $w(t,x)$, or of a $\Phi(t,x)$ with infinite length and effective speed $u_e$, could be found [with $u_L$ constant vs. $x$]: $\Phi_1(t)$ and $w_1(t)$ are set to zero.

Hatted symbols [except $t$] aside, the Burgers analog of (3.5) has $2\hat{v}\hat{N}_{tot}(0 - \hat{v}\hat{N}_{tot})$ in lieu of $2\nu N_{tot}(1 - \nu N_{tot})$, and that of (3.7) has $(\hat{v} - 0)\hat{\Phi}(t)$ as the last term of $\hat{u}_1(t)$.

As for the arclength increment Eq. (3.1), along with $\langle \Phi_t(t,x) \rangle = 0$ and the $x$-wise Fourier series of $\ln[\sin(\frac{1}{2}(x - z_k))]$, imply $\langle \varphi_t(t,x) \rangle = \frac{d}{dt}h - 2\nu\Sigma_{k=1}^{N_{tot}}\Im[\frac{d}{dt}z_k]$; from $\langle u(t,x) \rangle = 0$ and (2.1) (2.3) one also gets $\langle \varphi_t \rangle + \langle \varphi_x w(t,x) \rangle + V(t) = 0$. Jointly with (3.2)(3.3)(3.5) this leads to:

$$V(t) = 2\nu\sum_{k=1}^{N_{tot}}\Im[\frac{dz_k}{dt}] + \mathcal{V}(\nu N_{tot}) + \tfrac{1}{4}\Phi(t)^2 + \\ 2\nu(\Phi(t) - w(t))\sum_{k=1}^{N_{tot}}\Re[\cos(z_k)] + 2\nu w(t)\sum_{k=1}^{N_{tot}}\Re[e^{iz_k}] \;. \quad (3.8)$$

The Burgers analog of (3.8) has $2\hat{v}\hat{N}_{tot}(0 - \hat{v}\hat{N}_{tot})$ instead of $\mathcal{V}(\nu N_{tot}) = 2\nu N_{tot}(1 - \nu N_{tot})$.

The findings of [18-20, 25] are all based on special cases of (3.1)-(3.8). In particular if $\Phi(t) \equiv 0 \equiv w(t)$ and for steady poles $z_k = iB_{k\infty}(\mathrm{mod}.2\pi)$, (3.4) yields the [then algebraic] pole equations that belong to (1.2) in Sec. I, and $V(t)$ in (3.8) resumes the constant $V_\infty = \mathcal{V}(\nu N_{tot})$.

Whenever the forcing function has the form (3.6) – with $u_1(t)$ and $u_2(t)$ expressed via a real $\nu > 0$, an integer $N_{tot} \geq 0$ and two functions $\Phi(t), w(t)$ as is specified in (3.7) – then (3.1) (3.2) yield the front-shape solution to (2.1) in terms of pole trajectories obeying (3.4), provided the stretch $w(t,x)$ conforms to (3.3); the corresponding scaled arclength increment $V(t)$ follows from (3.8). Although the very structure (3.6)(3.7) of $u(t,x)$ was deduced and $w(t,x)$ is constrained by (3.3) for overall consistency, the present semi-inverse approach still needs to solve (3.4) for the $z_k(t)$s; that is far from easy a task, not to mention the use of (3.1) (3.8) in practice so as to effectively determine the instantaneous front shape and its length.

## IV. SIMPLE SOLUTIONS

### A. Poleless fronts

The $N_{tot} = 0$ shapes are the simplest ones and generalize the unstable flat fronts $\varphi_x(t,x)$ $\equiv 0$: the pole equations (3.4) get void, all summations disappear, the front shape $\varphi(t,x)$ reads $h(t) + \Phi(t)\cos(x)$ and $u_1(t)$ in (3.7) simplifies to $\frac{d}{dt}\Phi(t) + (\nu - 1)\Phi(t)$. Despite the still active DL mechanism, the wrinkling-induced length increment $V(t) = \frac{1}{4}\Phi(t)^2$ from (3.8) reduces to the scaled base-shape arclength $\langle \frac{1}{2}(\Phi_x)^2 \rangle$ whatever the stretch intensity $w(t)$ is; $w(t)$ matters for internal consistency though, because $u_2(t) = \frac{1}{2}\Phi(t)(w(t) - \frac{1}{2}\Phi(t))$ by (3.6).

Next, choosing the base-shape amplitude to be $\Phi(t) = 2w(t)$ and $w(t) = w_n \sin(\omega_n t - \sigma_n)$ for some $(w_n > 0, \omega_n > 0, \sigma_n)$, yields $u_2(t) \equiv 0$ and a phase-shifted sine for $u_1(t)$: as can be checked $\varphi(t,x) = h(t) + 2w_n \sin(\omega_n t - \sigma_n)\cos(x)$ is an exact, not necessarily stable, solution to (2.1)(2.2) if $u(t,x) = -2w_n \operatorname{sgn}(1-\nu)[\omega_n^2 + (1-\nu)^2]^{1/2} \sin(\omega_n t - \sigma_n - \tan^{-1}(\frac{\omega_n}{1-\nu}))\cos(x)$. Because $u_1(t)$ in (3.7) depends linearly on $\Phi(t)$ one can also superimpose several $w_n \sin(\omega_n t - \sigma_n)$ with various $(w_n, \omega_n, \sigma_n)$, even a continuum thereof, in $\Phi(t) = 2w(t)$; here they combine linearly in $\varphi(t,x)$ and $u(t,x)$ as well. In any such instance the time-averaged $V(t)$ will be

$$V_{av} = \frac{1}{4}[\Phi^2]_{av} = [w^2]_{av}, \quad [.]_{av} \equiv (2\pi/\omega)^{-1} \int_{-\pi/\omega}^{+\pi/\omega} (.)dt, \qquad (4.1)$$

with $2\pi/\omega > 0$ the smallest period of $V(t)$. This may hold if $w(t)$ is aperiodic [$\omega \to 0$], *e.g.*, if it obeys $\frac{d}{dt}w = -w/\theta + D^{1/2}\beta(t)$ for prescribed constants $(\theta > 0, D > 0)$ and a white-in-time centered Gaussian noise $\beta(t)$. The obtained Gaussian $w(t) = \frac{1}{2}\Phi(t)$ has $[w]_{av} = 0$ and $[w^2]_{av}$ $\sim D\theta < \infty$: (4.1) still holds despite $u(t,x) = u_1(t)\cos(x)$ having $[u_1^2]_{av} \sim [\beta^2]_{av} = \infty$.

The Burgers counterparts behave similarly: $\hat{u}_1(t) = \frac{d}{dt}\hat{\Phi}(t) + (\hat{\nu} - 0)\hat{\Phi}(t)$, $\hat{V}(t) = \frac{1}{4}\hat{\Phi}(t)^2$ and $\hat{u}_2(t)$ has the same structure as $u_2(t)$.

### B. Reusing MS shapes

The results from Sec. III also drastically simplify when the base-shape amplitude $\Phi(t)$ is chosen to coincide with the stretch intensity $w(t)$, in which case $u(t,x)$ in (3.6) has $u_2(t) =$ $\frac{1}{4}\Phi(t)^2$ and, again, $u_1(t) = \frac{d}{dt}\Phi(t) + (v-1)\Phi(t)$. Now the pole dynamics does play a role yet (3.4) reads exactly the same as in the MS case. This leads to two outcomes for $t \to \infty$.

(i) $0 \leq N_{tot} \leq N_{opt}(v)$: the $z_k(t)$ all settle at the same steady locations $z_{k\infty}$ as if $w(t) \equiv 0 \equiv \Phi(t)$, and the polar piece superimposed to the unsteady base shape in (3.1) gets steady; note that $N_{opt}(v) \geq 1$ needs $v \leq 1$. The speed increment $V(t \to \infty)$ follows from (3.8):

$$V(t) = \mathcal{V}(vN_{tot}) + \tfrac{1}{4}w(t)^2 + 2vw(t)\sum_{k=1}^{N_{tot}}\Re[e^{iz_{k\infty}}] \ . \tag{4.2}$$

If $w(t)$ fluctuates about zero and has $[w(t)]_{av} = 0$, $V_{av}$ reads:

$$V_{av} = \mathcal{V}(vN_{tot}) + \tfrac{1}{4}[w(t)^2]_{av} \ . \tag{4.3}$$

(ii) $N_{tot} > N_{opt}(v) \geq 0$: $N_{tot} - N_{opt}(v)$ upmost poles $z_{k>N_{opt}(v)}$ escape to $+i\infty$ [hence $e^{iz_k} \to 0$], while only $N_{opt}(v)$ of them settle at steady locations $z_{k\infty}$. For the runaway poles (3.4) gives:

$$\frac{dz_k}{dt} \underset{t\to\infty}{\to} \sum_{j\neq k, j=N_{opt}+1}^{j=+N_{tot}} v\cot[\tfrac{1}{2}(z_j - z_k)] + i(vN_{tot} + vN_{opt}(v) - 1), \quad N_{opt}(v)+1 \leq k \leq N_{tot} \ . \tag{4.4}$$

Because the above $z_j - z_k$ terms disappear by anti-symmetry once (4.4) is summed over the $k$ s, and the poles $z_{1\leq k \leq N_{opt}}$ get steady, one obtains $2v\sum_{k=1}^{N_{tot}}\Im[\frac{d}{dt}z_k] \underset{t\to\infty}{\to} \mathcal{V}(vN_{opt}) - \mathcal{V}(vN_{tot})$ to use in (3.8): as a net result $N_{opt}(v)$ replaces $N_{tot}$ in equations (4.2)(4.3).

In either case $V_{av}$ is a sum of the free-front steady value $V_\infty$ and of the time averaged base-front arclength. Interestingly, (4.2) contains a DL affected, linear-in-$w(t)$ component which when small controls the fluctuations of $V(t)$ but altogether disappears from $V_{av}$.

At this stage one must stress that specifying $v < \tfrac{1}{2}$ and $1 < N_{tot} \leq N_{opt}(v)$ does not uniquely determine the front shape: several 'steady' solutions to (1.1) exist with the same $N_{tot}$ and $v$ [29]. Among those the bi-coalesced patterns evoked in Sec. I, each with its own set of steady poles $z_{k\infty}$ yet a common $\mathcal{V}(vN_{tot})$, are stable when Neumann conditions are imposed at $x = 0$ (mod.$\pi$). If $\Phi(t) = w(t)$ these ultimately provide several initial-condition-dependent shapes $\varphi(t,x)$ from (2.1) and speed increments $V(t)$ from (4.2), despite the stretch and forcing functions in (2.1) being unchanged, see (3.6)(3.7) and Fig.1; all share the same $V_{av}$ if $[w]_{av} = 0$, by (4.3). Just like previously, (4.3) may hold for random $w(t)$s, even for some that lead to $[u_1^2]_{av} = \infty$ as with the Langevin equation evoked below (4.1), now used with $\Phi(t) = w(t)$.

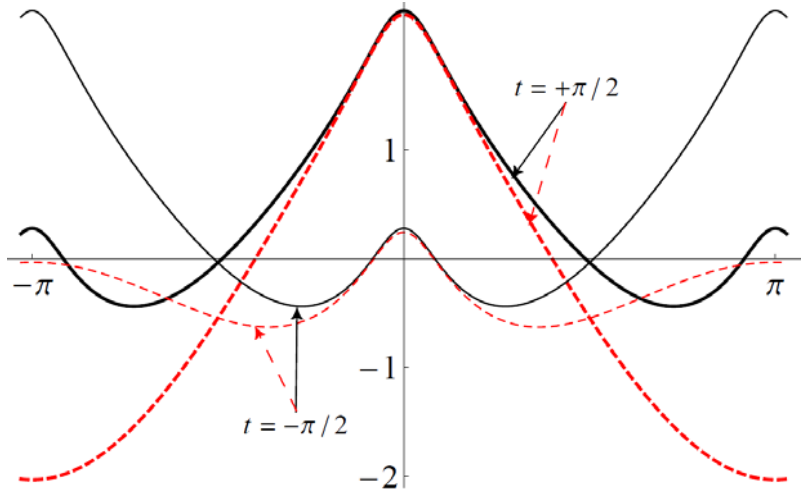

FIG.1. (Color online) Two solutions $\varphi(t,x)$ to (2.1)(2.2), both for $v = \frac{1}{4}$, $N_{tot} = 2 = N_{opt}(\frac{1}{4})$ and $\Phi(t) = w(t) = \sin(t)$, at $t = \pm\frac{\pi}{2}$. The solid-black or dashed-red shapes comprise a steady polar MS piece with $z_{1\infty} = z_{2\infty} - \pi = i\frac{1}{2}\tanh(2v)$, or $(z_{1\infty}, z_{2\infty}) = i[\tanh^{-1}(3v) \pm \frac{1}{2}\cosh^{-1}(1/(1-9v^2))]$ from Ref.[19], respectively. The two $\varphi(t,x)$s have different shape and $V(t)$ evolutions though share the same forcing $u(t,x) = (\cos(t) + \frac{3}{4}\sin(t))\cos(x) + \frac{1}{4}\sin^2(t)\cos(2x)$, stretch $w(t,x) = \sin(t)\sin(x)$, and $V_{av} = \frac{1}{2} + \frac{1}{8}$.

The picture is simpler for the Burgers analogs because the $\hat{z}_{k\geq 1}(t)$ all escape to $+i\infty$ if $\hat{\Phi}(t) = \hat{w}(t)$. Formally $\hat{N}_{opt}(\hat{v}) = 0$, and the polar part $\hat{\phi}_x(t,x)$ of $\hat{\varphi}_x(t,x)$ ultimately disappears: $\hat{V}(t) = \frac{1}{4}\hat{w}(t)^2$ and its $t$-average respectively replace (4.2) and (4.3) in the long-time limit, with the understanding that $\hat{u}(t,x)$ now has $\hat{u}_1(t) = \frac{d}{dt}\hat{w}(t) + (\hat{v} - 0)\hat{w}(t)$ and $\hat{u}_2(t) = \frac{1}{4}\hat{w}(t)^2$.

Unfortunately for models of flame propagations across line-arrays of eddies, $u_2(t) \sim \Phi(t)^2$ does not vanish on time-average even if $\Phi(t) = w(t)$ do: *handling $[u(t,x)]_{av} = 0$ needs $\Phi(t) \neq w(t)$ in (3.4) and a nontrivial pole dynamics*, e.g. if $\Phi(t) = 2w(t)$ and hence $u_2(t) = 0$. The special case where $\Phi(t) \equiv 0$ [flat base front], and where $u(t,x)$ from (3.6)(3.7) also has $u_2(t) = 0$ and vanishes on average if $w(t)$ does, is no exception; neither are the Burgers fronts.

# V. MEETING THE POLE DYNAMICS

Tackling the pole equations (3.4) in full generality needs several items to be specified: $\nu$ [wrinkle wavenumber], $2N_{tot}$ [total pole number], initial pole distribution, and the functional forms, intensities and frequencies of two independent, periodic functions $w(t)$ and $\Phi(t)$.

Numerical integrations of (3.4) using various $2\pi/\omega$-periodic $\Phi(t) - w(t)$ with zero average indicate that the poles ultimately acquire the form(s) $iB_k(t)(\mathrm{mod}.2\pi)$ or $\pi + ib_\alpha(t)(\mathrm{mod}.2\pi)$, or both: only such configurations are considered from now on.

Next, one must choose $N_{tot}$ and the numbers $N$ and $n = N_{tot} - N$ of $B$- and $b$- type poles: for technical reasons [see next Sections] the cases $n = 0$, $N \sim n = O(1)$, $N \sim n \gg 1$ will be analyzed separately, each time with wrinkle wavenumbers $\nu$ such that $N_{tot} \sim N_{opt}(\nu)$.

As to the choice of $\Phi(t)$ and $w(t)$ most of the analyses will focus on simple harmonics:

$$\Phi(t) = \Phi \sin(\omega t), \quad w(t) = w \sin(\omega t + \gamma), \tag{5.1}$$

with constant $\Phi$ and $w$, whereby $\Phi(t) - w(t)$ is a weighted sine. Adjusting the phase $\gamma = \pm \cos^{-1}(\Phi/2w)$ will guarantee that $[u_2(t)]_{av} = 0$ in (3.7), $u_2(t) \equiv 0$ inclusive, and comply with $[u(t,x)]_{av} = 0$. Satisfying a single constraint $[u_2(t)]_{av} \equiv \frac{1}{4}[\Phi(t)w(t)]_{av} - \frac{1}{2}[\Phi(t)^2]_{av} = 0$ on two functions becomes easier when more flexible $\Phi(t)$ and $w(t)$ with zero means are considered; how or whether results from (5.1) can be adapted will be indicated in due course.

The shortest scaled period $2\pi/\omega > 0$ of $w(t) - \Phi(t)$ is yet another free parameter. As seen below, even the simplest version of the pole equations (3.4) cannot be solved analytically if $\omega = O(1)$ and $\Phi(t) \neq w(t)$: Beside numerical integrations, high-frequency [HF] or low-frequency [LF] asymptotic analyses generically need be invoked, see Sections VI and VII.

To wit, the simplest configuration where $N_{tot} = 1$, and $z_1(t) = iB(t)$ with $\tanh(\frac{1}{2}B) > 0$,

$$\frac{dB}{dt} = \nu \coth(B) - 1 + [w(t) - \Phi(t)]\sinh(B), \tag{5.2}$$

already leads to a nontrivial dynamics. Before taking up its analysis we rewrite (5.2) as

$$\frac{d\lambda}{dt} = \sinh(\lambda) - \frac{1}{2}\nu \sinh(2\lambda) + \Phi(t) - w(t), \tag{5.3}$$

with $\tanh(\frac{1}{2}B) = e^{-\lambda}$; this converts the multiplicative $\Phi - w$ of (5.2) to an additive one, but has still another virtue. The identity $\tanh^{-1}(e^{-\lambda}) - \tanh^{-1}(e^{\lambda}) = i\frac{\pi}{2}\text{sgn}(\lambda)$, valid for any real $\lambda$, indeed shows that $iB(t)$ jumps by $\pm\pi$ when $\lambda(t)$ changes sign. Since a pair of remote poles [$\Re[B] \gg 1$] contributes $4\nu e^{-B}\cos(x) + ...$ to $\varphi(t, x)$, the jump reflects the change $\cos(x) \to -\cos(x) = \cos(x \pm \pi)$ and a sudden move of front-crest locations by half a wavelength.

Although $B(t) = +\infty$ ceases to contribute to the front shape when $\lambda(t) = 0^+$, $z_{\pm 1}(t)$ may not simply leave the stage and lower the number of poles by 2: the integer $2N_{tot}$ explicitly enters (3.7) and has to stay fixed if the structure and evolution of $u(t, x)$ are to be kept unchanged. $B(t)$ must therefore come back into the play at $\lambda(t) = 0^-$: it does so in finite time, endowed with an additive $-i\pi$ that makes $\frac{d}{dt}B \approx [w(t) - \Phi(t)]\sinh(B)$ reverse its sign but leaves $\tanh(B)$ as it is. If $N_{tot} > 1$ pairs of poles $iB_k(t)$ are involved, successive jumps may occur. Facing divergent $\Im[z_k(t)]$ at times that are not known in advance renders crude numerical integrations of (3.4) impractical; this can be overcome if one writes $\tan(\frac{1}{2}z_{k\geq 1})$ as $ie^{-\lambda_k}$.

The only easy situation about (5.2) is when $\Phi(t) - w(t) \ll 1$, with $\Phi(t) \sim w(t)$ not to duplicate what subsection IV.B covered. Provided $\nu < 1$, $B(t)$ can be linearized around the steady pole altitude $B_\infty = \tanh^{-1}(\nu)$; equivalently (5.3) is linearized around $\lambda_\infty = \cosh^{-1}(1/\nu)$. Owing to linearity one can use $\Phi(t) = \Phi\sin(\omega t)$ and $w(t) = w\sin(\omega t + \gamma)$ as in (5.1). After a transient that decays like $\exp(-\omega_\infty t)$, to first order in $\Phi \sim w \ll 1$ one gets:

$$\lambda(t) - \lambda_\infty = -\frac{\Phi}{\omega}\cos(\omega t) + \frac{\Phi}{\omega}\frac{\cos[\omega t - \tan^{-1}(\omega/\omega_\infty)]}{(1+\omega^2/\omega_\infty^2)^{1/2}}$$
$$+ \frac{w}{\omega}\cos(\omega t + \gamma) - \frac{w}{\omega}\frac{\cos[\omega t + \gamma - \tan^{-1}(\omega/\omega_\infty)]}{(1+\omega^2/\omega_\infty^2)^{1/2}} + ... \, , \quad (5.4)$$

where $\omega_\infty = 1/\nu - \nu$ is the reciprocal time of relaxation of $\lambda(t) - \lambda_\infty$ when $\Phi = 0 = w$. In the limit $\omega^2 \gg \omega_\infty^2$, $\lambda(t) - \lambda_\infty$ in (5.4) reduces to $(\cos(\omega t + \gamma)w - \cos(\omega t)\Phi)/\omega + O(\Phi\omega_\infty/\omega^2)$: only the forcing terms in (5.3) matter and $\lambda(t)$ is in quadrature of phase with them. In the LF limit $\omega^2 \ll \omega_\infty^2$ the pole response nearly cancels $(\cos(\omega t + \gamma)w - \cos(\omega t)\Phi)/\omega$, and $\lambda(t) - \lambda_\infty = (\sin(\omega t)\Phi - \sin(\omega t + \gamma)w)/\omega_\infty + o(1)$ quasi-steadily follows the *combined stretch*, i.e., the joint influence $w(t) - \Phi(t)$ of stretch and of its geometrical counterpart due to the base-slope.

Once time-averaged using $[\frac{d}{dt}B]_{av} = 0$, $\cos(iB) \equiv \coth(\lambda)$ and $e^{-B} \equiv \tanh(\frac{1}{2}\lambda)$, (3.8) leads to:

$$V_{av} - \mathcal{V}(\nu) - \tfrac{1}{8}\Phi^2 = \\ -\frac{\nu^2(\Phi^2 - 2\Phi w \cos(\gamma) + w^2)}{(\omega^2 + \omega_\infty^2)} - \frac{\nu w\,[\Phi\sin(\gamma)\omega + (w - \Phi\cos(\gamma))\omega_\infty]}{2\cosh^2(\tfrac{1}{2}\lambda_\infty)(\omega^2 + \omega_\infty^2)} + \dots, \quad (5.5)$$

with $\mathcal{V}(\nu) = 2\nu(1-\nu)$ and $\tfrac{1}{8}\Phi^2$ the averaged base-front length. Whereas $V(t) - \mathcal{V}(\nu)$ fluctuates by $O(\Phi, w)$, $V_{av}$ is invariant by time shifts and by the reversals $(\Phi, w) \leftrightarrow (-\Phi, -w)$, and $V_{av} - 2\nu(1-\nu)$ *is quadratic in* $(\Phi, w)$ *whatever* $\omega$ *is*; computing $\lambda(t)$ to next-to-linear order would be necessary to access $V(t) - \mathcal{V}(\nu)$ to the same accuracy. If $\Phi$ is chosen proportional to $w$, $V_{av} = 2\nu(1-\nu) + O(w^2)$: up to a known offset this bears a similarity with the Clavin-Williams [CW] formula [30, 31] for the speed increment of weakly turbulent, passive flames [no DL mechanism] subject to $w(t,x) = w(t)\sin(x)$ and $u(t,x) \approx u_1(t)\cos(x)$, $u_1(t) \sim w(t)$.

The Burgers analog of (5.3), without a DL-induced $\sinh(\hat{\lambda})$ on its right-hand side, can be handled similarly. The linear expansion around $\hat{\lambda}_\infty = 0$, *i.e.*, around $\Re[\hat{B}_\infty] = \infty$, yields the same formula [except for the carets] for $\hat{\lambda}(t)$ as (5.4), now with $\hat{\omega}_\infty = \hat{\nu}$. Being devoid of DL instability mechanism the Burgers fronts are enslaved by the forcing function, hence the varying $\operatorname{sgn}(\hat{\lambda}(t))$: the crests of $\hat{\varphi}(t,x)$ are located where $\hat{u}(t,x) \approx \hat{u}_1(t)\cos(x)$ is maximum.

If $\nu > 1$ the then DL-stable fronts from (5.3) behave likewise, with $\lambda_\infty = 0$ and $\omega_\infty = \nu - 1$.

## VI. HIGH FREQUENCIES

### A. One pair of poles

Nonlinear high-frequency studies of the pole dynamics are best taken up in the simplest case where $N_{tot} = 1$ and $\Phi(t) = \Phi\sin(\omega t)$, using $w(t) \equiv 0$ as a starter. Anticipating from (5.4) that $O(1)$ variations of $B(t)$ and of $\lambda(t)$ necessitate $\Phi = O(\omega)$ if $\omega \gg 1$, (5.3) is rewritten as:

$$\frac{d\lambda}{d\sigma} = \frac{\sinh(\lambda) - \tfrac{1}{2}\nu\sinh(2\lambda)}{\omega} - \delta\sin(\sigma), \\ \sigma \equiv \omega t, \quad \delta \equiv -\Phi/\omega. \quad (6.1)$$

Once $\lambda(\sigma) = \lambda(\sigma + 2\pi)$ is reached the $\sigma$-averaged (6.1) requires that the solvability condition

$$[\sinh(\lambda) - \tfrac{1}{2}\nu\sinh(2\lambda)]_{av} = 0 \qquad (6.2)$$

be met, whatever $\omega \neq 0$ is. Guided by (5.4) for $\omega^2 \gg \omega_\infty^2$ the solution to (6.1) is sought as

$$\lambda(\sigma) = \delta\cos(\sigma) + \lambda_0 + \lambda_1(\sigma)/\omega + ... , \qquad (6.3)$$

with $\lambda_0$ unknown yet shown from (6.1) to be $\sigma$-independent [compare with (5.4)]. Plugging (6.3) into (6.2) and only keeping the leading order in $\omega \gg 1$ determines $\lambda_0$ as a root of

$$\sinh(\lambda_0)I_0(\delta) - \tfrac{1}{2}\nu\sinh(2\lambda_0)I_0(2\delta) = 0, \qquad (6.4)$$

where $I_0(u) \equiv [e^{u\cos(\sigma)}]_{av} = 1 + \tfrac{1}{4}u^2 + \tfrac{1}{64}u^4 + ...$ is the usual $0^{\text{th}}$-order modified Bessel function.

Before proceeding to the pole trajectory via $\tanh(\tfrac{1}{2}B(\sigma)) \approx e^{-\delta\cos(\sigma)-\lambda_0}$ one notes that (6.4) is invariant by the combined reversals $(\lambda_0,\delta) \leftrightarrow (-\lambda_0,-\delta)$ since $(\lambda,\delta) \leftrightarrow (-\lambda,-\delta)$ leaves (6.1) invariant; hence one may select $\lambda_0 > 0$. But the separate invariance of (6.4) by $\delta \leftrightarrow -\delta$ alone stems from the extra symmetry of (6.1) by $(\delta,\sigma) \leftrightarrow (-\delta,\sigma-\pi)$.

One may also note that a two-time method, wherein $\tfrac{d}{d\sigma}$ in (6.1) becomes $\tfrac{\partial}{\partial\sigma} + \omega^{-1}\tfrac{\partial}{\partial t}$ as when analyzing the Kapitza pendulum [32], would replace the constant $\lambda_0$ with $\lambda_0(t)$, and the right-hand sides of (6.2) and (6.4) with the $\sigma$-averaged $\tfrac{\partial}{\partial t}\lambda(\sigma,t)$ and $\tfrac{d}{dt}\lambda_0(t)$, respectively. Stability over the $t = O(1)$, slower time-scale therefore requires the root $\lambda_0$ of (6.4) to satisfy $\tfrac{\partial}{\partial\lambda_0}[\sinh(\lambda_0)I_0(\delta) - \tfrac{\nu}{2}\sinh(2\lambda_0)I_0(2\delta)] < 0$ to ensure decay of the fast transient, which implies:

$$\lambda_0 = \Re[\cosh^{-1}(I_0(\delta)/\nu I_0(2\delta))] . \qquad (6.5)$$

If $\nu < 1$ this is $\lambda_0 = \cosh^{-1}(I_0(\delta)/\nu I_0(2\delta)) \geq 0$ provided $|\delta| \leq \delta_c$, where $\delta_c(\nu) \geq 0$ obeys $I_0(\delta_c) = \nu I_0(2\delta_c)$ [e.g., $\delta_c(1) = 0$, $\delta_c(\tfrac{1}{2}) \approx 1.109$, $\delta_c(\tfrac{1}{3}) \approx 1.512$, $\delta_c(0) = \infty$ ]; and $\lambda_0 = 0$ if $|\delta|$ lies at or beyond the bifurcation point $\delta_c(\nu)$. If $\nu \geq 1$, (6.5) yields $\lambda_0 = 0$ irrespective of $\delta$.

One is not yet fully done though, because $\lambda(\sigma) \approx \delta\cos(\sigma) + \lambda_0$ may change sign, which also matters, see subsection V.B . As long as $|\delta| \leq \delta_*(\nu)$, where $0 \leq \delta_*(\nu) \leq \delta_c(\nu)$ obeys $I_0(\delta_*) = \nu I_0(2\delta_*)\cosh(\delta_*)$ [e.g. $\delta_*(\tfrac{1}{3}) = 1.045$, $\delta_*(\tfrac{1}{2}) = 0.798$], $\lambda(\sigma)$ remains positive, $\tanh(\tfrac{1}{2}B(\sigma)) \approx e^{-\delta\cos(\sigma)-\lambda_0}$ stays below 1, the pole $iB(\sigma)$ oscillates along the imaginary axis in $z$ plane, and the front crests are located at $x = 0 \pmod{2\pi}$. If $|\delta| > \delta_*(\nu)$, however, the crest locations will jump by $\pi$ to the right and back, once during each full period of oscillation, see Fig.2. This always occurs for $|\delta| > \delta_c(\nu)$, in which case $\lambda_0 = 0$ is the relevant root of (6.4): $iB(\sigma)$ then spends half its life along the imaginary axis and is shifted $x$-wise by $\pi$ during the other half.

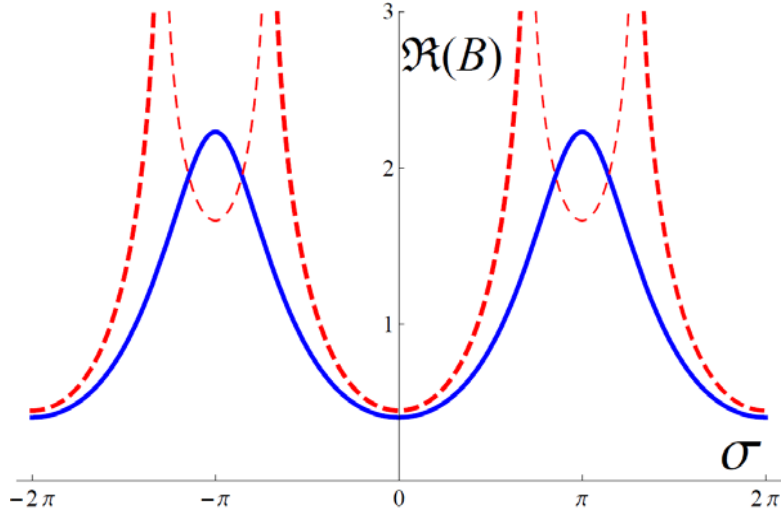

FIG.2. (Color online). Leading-order trajectories $\Re[B(\sigma)]$ for $\nu = \frac{1}{2}$ and $\delta = 0.7 < \delta_*(\frac{1}{2}) \approx 0.798$ [solid blue] and hence $\lambda_0 \approx 0.916 > \delta$, or $\delta = 0.95 > \delta_*(\frac{1}{2})$ [dashed red] and $\lambda_0 = 0.566 < \delta$. Along the thick [or thin] lines the front crests sit at $x = 0$ [or $x = \pi$] mod.$2\pi$, and are taller/sharper for smaller $\Re[B(\sigma)]$.

The third term of the HF expansion (6.3) is odd in $\sigma$, which follows from (6.1) once the integration constant $[= 0]$ is deduced from the solvability condition (6.2) at order $1/\omega$:

$$\lambda_1(\sigma) = \int_0^\sigma \sinh(\lambda_0 + \delta\cos(u))[1 - \nu\cosh(\lambda_0 + \delta\cos(u))]du. \qquad (6.6)$$

This can be used to improve the pole trajectory, via $\tanh(\frac{1}{2}B(\sigma)) \approx e^{-\delta\cos(\sigma) - \lambda_0 - \lambda_1(\sigma)/\omega}$; no extra analysis is needed here to handle timewise boundary-layers where $\tanh(\frac{1}{2}B(\sigma)) \approx 1$.

The HF analysis of $\hat{N}_{tot} = 1$ Burgers fronts yields $0 - \frac{1}{2}\sinh(2\hat{\lambda}_0)I_0(2\hat{\delta}) = 0$ instead of (6.4), whereby $\hat{\lambda}_0 = 0$ in any case, similar to the $|\delta| > \delta_c(\nu)$ or $\nu \geq 1$ DL-affected situations; and the integrand becomes $-\frac{1}{2}\hat{\nu}\sinh(2\hat{\delta}\cos(u))$ in the expression of $\hat{\lambda}_1(\sigma)$ that replaces (6.6).

The current length increment $V(\sigma)$ follows from (3.8) and (6.3). Up to $O(1)$ terms it reads:

$$V(\sigma) = \tfrac{1}{4}\delta^2\omega^2\sin^2(\sigma) + 2\nu\omega dB/d\sigma - 2\nu\omega\delta\sin(\sigma)\coth(\lambda(\sigma)) + O(1). \qquad (6.7)$$

As anticipated $V(t)$ is dominated by the base-front arclength $\frac{1}{4}\Phi(t)^2$, because $\Phi(t) \sim \omega$; only at orders $\leq O(\omega)$ does the DL effect modify $V(t)$ through $\lambda(\sigma)$ [compare to (4.2)]. Due to the rightmost $O(\omega)$ factor in (6.7) knowing $\lambda_0$ is not enough to get $V_{av}$ to $O(1)$ inclusive, $\lambda_1(\sigma)$ from (6.6) is also needed. Although viable via an integration by parts, the use of $\lambda_1(\sigma)$ can actually be bypassed if one multiplies (6.1) by $\coth(\lambda)$ before the $\sigma$-average is taken, to get $\omega[\delta \sin(\sigma) \coth(\lambda(\sigma))]_{av} = [\cosh(\lambda) - \nu \cosh^2(\lambda)]_{av}$. The latter quantity is $O(1)$ and can be evaluated from (6.3)(6.5) with at most $O(1/\omega)$ errors. This produces the leading-order result

$$V_{av} - \tfrac{1}{8}\delta^2\omega^2 - \mathcal{V}(\nu) = \nu^2[1 - I_0(2\delta)] \leq 0, \qquad |\delta| \leq \delta_c(\nu),$$
$$= \nu^2[1 + I_0(2\delta)] - 2\nu I_0(\delta), \quad |\delta| \geq \delta_c(\nu), \qquad (6.8)$$

in which the second line pertains to $\lambda_0 = 0$; see Fig.3. That $V_{av}$ transitions from DL-affected [$\lambda_0 \neq 0$] to Burgers like [$\lambda_0 = 0$] exactly at $|\delta| = \delta_c(\nu)$ originates from the special symmetry of (6.1) by $(\delta, \sigma) \leftrightarrow (-\delta, \sigma - \pi)$, see below (6.15). The Burgers fronts have $\hat{V}_{av} - \tfrac{1}{8}\hat{\delta}^2\hat{\omega}^2 = (2\hat{\nu})^2[I_0(2\hat{\delta}) - 1] \geq 0$ instead: again a CW type [30] law $\hat{V}_{av} \approx \hat{\delta}^2(\tfrac{1}{8}\hat{\omega}^2 + 4\hat{\nu}^2)$ when $|\hat{\delta}| \lesssim 1$, akin to the $|\delta| > \delta_c(\nu)$ expression of $V_{av}$ in the second line of (6.8), see Fig.3.

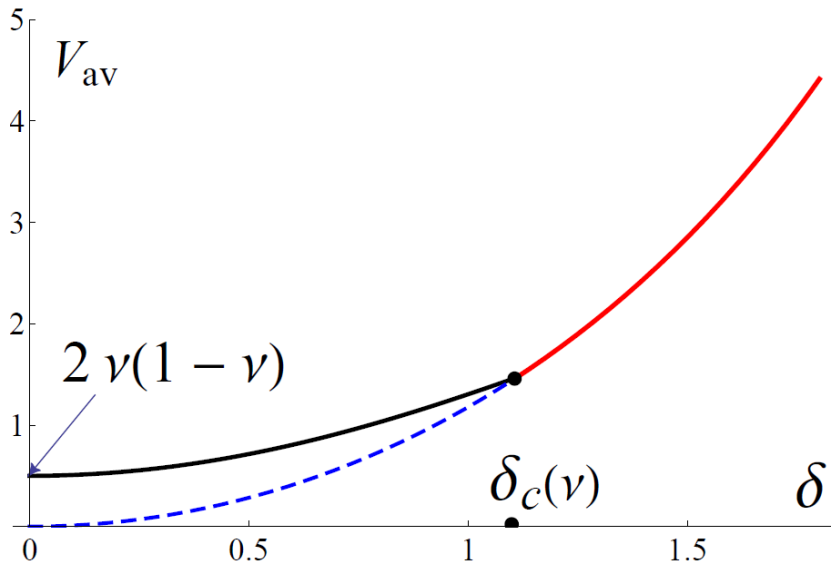

FIG. 3. (Color online) $V_{av}$ vs. $|\delta| = \max(\Phi(t)/\omega)$ from (6.8), for $\nu = \tfrac{1}{2}$ [$\delta_c \approx 1.109$] and $\omega = 3$ [solid lines]. The dashed curve pertains to the [then invalid] second line of (6.8) if plotted down to $|\delta| < \delta_c(\nu)$.

Time now is ripe to restore stretch effects, with $w(t) = w \sin(\omega t + \gamma)$ for some constant $w$

and $\gamma$ as in (5.1), and $w = -\omega d$ in the HF analysis. To this end one rewrites $\Phi(t) - w(t)$ as

$$\Phi(t) - w(t) = -\Delta\omega\sin(\sigma - \zeta) ,$$
$$\Delta^2 = \delta^2 - 2\delta d\cos(\gamma) + d^2, \quad \tan(\zeta) = \frac{d\sin(\gamma)}{\delta - d\cos(\gamma)} , \qquad (6.9)$$

and the previous results about $\lambda(\sigma)$ are adapted *via* the substitution $\delta\sin(\sigma) \to \Delta\sin(\sigma - \zeta)$. The series replacing (6.3) now has $\lambda_0 = \Re[\cosh^{-1}(I_0(\Delta)/\nu I_0(2\Delta))]$, which when plugged in (3.8) yields the two dominant orders of $V(t)$. But using $\lambda_1(\sigma)$ from the modified (6.6), $\lambda_1(\sigma) = \int_0^{\sigma-\zeta}\sinh(\lambda_0 + \Delta\cos(u))[1 - \nu\cosh(\lambda_0 + \Delta\cos(u))]du$, could not be avoided to determine all the $O(1)$ contributions to $V_{av}$. For example, $|\Delta| \leq \delta_c(\nu)$ and [the first line of ] (6.8) yield:

$$V(t) - \tfrac{1}{4}\delta^2\omega^2\sin^2(\sigma) = 2\nu\omega dB/d\sigma - 2\nu\omega\Delta\sin(\sigma - \zeta)\coth(\lambda_0) + O(1) , \qquad (6.10)$$

$$V_{av} - \tfrac{1}{8}\delta^2\omega^2 - \mathcal{V}(\nu) = \nu^2[1 - I_0(2\Delta)]$$
$$-2\nu d[\sin(\gamma + \zeta) \; \omega K'(\Delta) + \tfrac{1}{2}\cos(\gamma + \zeta)\nu K''(\Delta)] . \qquad (6.11)$$

Here $\omega K'(\Delta)$ and $K''(\Delta)$ result from the averaged $-\omega\sin(\sigma + \gamma)\tanh(\tfrac{1}{2}\lambda(\sigma))$ once the new $\lambda(\sigma) \approx \Delta\cos(\sigma - \zeta) + \lambda_0 + \lambda_1(\sigma)/\omega$ is made use of, and integration by parts is employed:

$$K'(\Delta) \equiv [\tanh[\tfrac{1}{2}\lambda_0 + \tfrac{1}{2}\Delta\cos(u)]\cos(u)]_{av} ,$$
$$\Delta\, K''(\Delta) \equiv \tfrac{2}{\nu}[\cosh(\lambda_0)I_0(\Delta) - 1] - 1 + 2\cosh(\lambda_0)I_0(\Delta) - \cosh(2\lambda_0)I_0(2\Delta) . \qquad (6.12)$$

Though out of analytical reach $K'(\Delta) = -K'(-\Delta)$ is accessible as a power series for $\Delta \to 0$ at fixed $\lambda_0$, $K'(\Delta) = \tfrac{1}{4}\Delta/\cosh^2(\tfrac{1}{2}\lambda_0) + O(\Delta^3)$, and can be tabulated. Better still, it is accessed accurately for all $|\Delta| \leq 3$ once rewritten as $\text{sech}(\lambda_0)[\mathcal{Y}(\cos(u), \Delta, \text{sech}(\lambda_0))]_{av}$, because :

$$\mathcal{Y}(c, \Delta, \Theta) \equiv \frac{c\sinh(\Delta c)}{(1 + \Theta\cosh(\Delta c))} \quad \Rightarrow$$
$$[\mathcal{Y}(\cos(u), \Delta, \Theta)]_{av} \approx \tfrac{1}{4}\mathcal{Y}(1, \Delta, \Theta) + \tfrac{1}{2}\mathcal{Y}(2^{-1/2}, \Delta, \Theta) . \qquad (6.13)$$

For $0 \leq \Theta \leq 1$, $|\Delta| \leq 3$ and $c^2 \leq 1$, $\mathcal{Y}(c, \Delta, \Theta)$ indeed is nicely fitted by $c^2(2c^2 - 1)\mathcal{Y}(1, \Delta, \Theta) + 4c^2(1 - c^2)\mathcal{Y}(2^{-1/2}, \Delta, \Theta)$; since adding a term $\propto c^2(1 - c^2)(2c^2 - 1)$ greatly improves the fit but keeps $[\mathcal{Y}(\cos(u), \Delta, \Theta)]_{av}$ as it is, $K'(\Delta) \approx \text{sech}(\lambda_0)[\tfrac{1}{4}\mathcal{Y}(1, \Delta, \text{sech}(\lambda_0)) + \tfrac{1}{2}\mathcal{Y}(2^{-1/2}, \Delta, \text{sech}(\lambda_0))]$ already is almost exact for all $|\Delta| \leq 3$ [at least]. Approximation (6.13) will be reused later.

Thanks to the relations $d\sin(\gamma) = \Delta\sin(\zeta)$ and $\delta - d\cos(\gamma) = \Delta\cos(\zeta)$ that led to (6.9), the stretch-induced contribution to $V_{av}$ also is $-2\nu[\omega\sin(\gamma)\delta K'(\Delta)+\tfrac{1}{2}(\cos(\gamma)\delta-d)\nu K''(\Delta)]d/\Delta$.
By contrast to (6.8) but in accord for $\omega^2 \gg 1$ with the linear analysis that gave (5.5), the averaged length increment $V_{av}$ in (6.11) now has a $O(\omega)$ piece, $-2\nu\omega\sin(\gamma)K'(\Delta)\delta d/\Delta$, the value and sign of which vary with the amplitude $w(t)$ of stretch and its phase $\gamma$ relative to the amplitude $\Phi(t)$ of the base-front oscillations, see Fig.4.

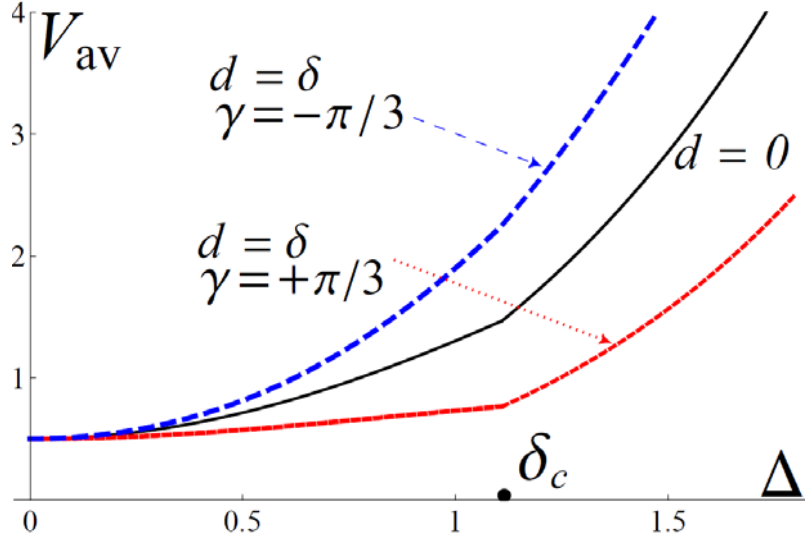

FIG. 4 (Color online). $V_{av}$ vs. combined intensity $\Delta$ from (6.11) with $\nu=\tfrac{1}{2}$ [$\delta_c\approx 1.109$] and $\omega=3$. $d=0$ [solid black]; $d=\delta$ and $\gamma=\tfrac{\pi}{3}$ [dotted red] or $\gamma=-\tfrac{\pi}{3}$ [dashed blue] that both give $[u(t,x)]_{av}=0$.

As promised in subsection V.A hints are now given on how such HF results can be generalized to forcing and stretch amplitudes different than the simple harmonic (5.1). Provided $\Phi(t)-w(t)$ can be written as $\Delta\omega\tfrac{d}{d\sigma}C(\sigma)$, where $C(\sigma)=C(\sigma+2\pi)$ has $[C(\sigma)]_{av}=0$ and is normalized to $[C(\sigma)^2]_{av}=\tfrac{1}{2}$, the series (6.3) becomes $\lambda(\sigma)=\Delta C(\sigma)+\lambda_0+\lambda_1(\sigma)/\omega+...$.

If $C(\sigma)$ has $2\pi$ as its minimum period and satisfies $C(\sigma-c)=-C(\sigma)$ for some $c=$ const., $c$ necessarily is $\pi(\mathrm{mod}\,2\pi)$ and $C(\sigma)$ only contains sines or/and cosines of $(2p+1)\sigma$ with $p\geq 0$ integer(s). The equation that generalizes (6.5) to this case requires each Bessel function $I_0(u)$ to be replaced with a new $\mathbf{I}_0(u)\equiv[e^{uC(\sigma)}]_{av}=\mathbf{I}_0(-u)$ that reads $1+\tfrac{1}{4}u^2+\tfrac{1}{24}u^4[C(\sigma)^4]_{av}$ +... at small enough arguments. By the same token, $\tfrac{d}{d\sigma}\lambda_1$ merely needs to be modified to $\sinh(\lambda_0+\Delta C(\sigma))[1-\nu\cosh(\lambda_0+\Delta C(\sigma))]$; the qualitative changes in pole trajectory, e.g., the number of jumps in $-\pi < \sigma \leq \pi$, depend on the specific shape of $C(\sigma)$. If a stretch $w(t) = -d\omega \, W(\sigma)$ is restored the expression of $V_{av}$ from (6.8) [with $|\Delta| \leq \delta_c(v)$] changes to:

$$V_{av} - \tfrac{1}{8}\delta^2\omega^2 - \mathcal{V}(v) = v^2[1 - \mathbf{I}_0(2\Delta)] - 2vd[\omega k' + k''] . \tag{6.14}$$

$k' \equiv [W(\sigma)\tanh(\tfrac{1}{2}\lambda_0 + \tfrac{1}{2}\Delta C(\sigma))]_{av}$, of order $\tfrac{1}{2}[W(\sigma)C(\sigma)]_{av}\Delta\operatorname{sech}^2(\tfrac{1}{2}\lambda_0)$ if $\Delta \ll 1$, and $k'' \equiv \tfrac{1}{2}[W(\sigma)\operatorname{sech}^2(\tfrac{1}{2}\lambda_0 + \tfrac{1}{2}\Delta C(\sigma))\lambda_1(\sigma)]_{av}$ must in general be tabulated, unless the symmetries of $W(\sigma)$ allow for an approximate $k'$ as in (6.12)(6.13) or/and an integration by parts to get $k''$.

If no constant $c$ such that $C(\sigma - c) = -C(\sigma)$ exists, (6.1) is not invariant by $(\Delta, C(\sigma)) \leftrightarrow (-\Delta, C(\sigma - \pi))$ any longer and $\mathbf{I}_0(-\Delta)$ generically differs from $\mathbf{I}_0(\Delta)$, modifying (6.4) to

$$e^{\lambda_0} \mathbf{I}_0(\Delta) - e^{-\lambda_0} \mathbf{I}_0(-\Delta) = \tfrac{1}{2}v[e^{2\lambda_0}\mathbf{I}_0(2\Delta) - e^{-2\lambda_0}\mathbf{I}_0(-2\Delta)] . \tag{6.15}$$

The former $\Delta \leftrightarrow -\Delta$ symmetry of (6.5) is broken, only mildly so if $\mathbf{I}_0(-\Delta)/\mathbf{I}_0(\Delta) - 1 \ll 1$ in which case $|\lambda_0|$ becomes small instead of $0$ as $|\Delta|$ crosses $\delta_c(v)$ from below; likewise, the Burgers analog of (6.15) [with 0 on the left] yields $\hat\lambda_0 = \tfrac{1}{4}\ln[\mathbf{I}_0(-2\hat\Delta)/\mathbf{I}_0(+2\hat\Delta)] \neq 0$. Going further would necessitate a separate analysis, an option that will not be pursued here because $\mathbf{I}_0(\Delta) = 1 + 0 + \tfrac{1}{4}\Delta^2 + \tfrac{1}{6}\Delta^3[C(\sigma)^3]_{av} + ...$ still is nearly quadratic if $\Delta$ is small, and stimuli with statistical properties invariant by $\Delta \leftrightarrow -\Delta$ [no skewness] seemed more relevant physically.

To close the subsection two ways to get $[u(t,x)]_{av} = 0$, besides selecting $\gamma = \pm\cos^{-1}(\Phi/2w)$ when (5.1) holds, are sketched. One may employ $\tfrac{1}{2}\Phi(t) - w(t)$ and $\Phi(t)$ that have opposite parities in some shifted time $\omega t - c$ : $u_2(t)$ in (3.7), $u(t,x)$ in (3.6) and $w(t,x)$ then vanish on time-average if $\Phi(t)$ and $w(t)$ do. Or $\tfrac{1}{2}\Phi(t) - w(t)$ is chosen as $\tfrac{d}{dt}m(\Phi(t))$ for some arbitrary function $m(.)$ : $u_1(t)$ and $u_2(t)$ in (3.7) become perfect $t$-derivatives of periodic functions if $\Phi(t)$ is, and all average to zero. The two ways can possibly be combined.

### B. Two-pair seesaw

The next nontrivial HF study corresponds to $N_{tot} = 2$, comprising $N = 1$ pair of poles with $z_1 = iB(\sigma)$, $z_{-1} = \overline{z}_1$ and $n = 1$ pair with $z_2 = \pi + ib(\sigma)$, $z_{-2} = \overline{z}_2$, This can be encoded in the auxiliary unknowns $(\lambda(\sigma), \mu(\sigma))$ defined by $\tanh(\tfrac{1}{2}B(\sigma)) = e^{-\lambda(\sigma)}$ and $\tanh(\tfrac{1}{2}b(\sigma)) = e^{-\mu(\sigma)}$;

$\lambda(\sigma) \geq 0$ [or $\mu(\sigma) \geq 0$] thus corresponds to $\Re(z_1) = 0 (\text{mod}.2\pi)$ [or $\Re(z_2) = \pi (\text{mod}.2\pi)$]. The pole equations (3.8) to be studied at $\omega \to \infty$, without stretch to begin with, translate to

$$\frac{d\lambda}{d\sigma} = \frac{\sinh(\lambda) - \frac{\nu}{2}\sinh(2\lambda)}{\omega} - \frac{2\nu \sinh(\lambda)\sinh(\mu)}{\omega \sinh(\lambda + \mu)} - \delta \sin(\sigma),$$
$$\frac{d\mu}{d\sigma} = \frac{\sinh(\mu) - \frac{\nu}{2}\sinh(2\mu)}{\omega} - \frac{2\nu \sinh(\lambda)\sinh(\mu)}{\omega \sinh(\lambda + \mu)} + \delta \sin(\sigma),$$
(6.16)

instead of (6.1). The above rightmost terms have opposite signs since both result from pole convection by the base-front slope $\Phi_x(t,z) = -\Phi(t)\sin(z)$ that indeed changes sign by $z \leftrightarrow z + \pi$. As a consequence (6.16) is invariant by $(\lambda, \delta) \leftrightarrow (\mu, -\delta)$ and, separately, by $(\sigma, \delta) \leftrightarrow (\sigma - \pi, -\delta)$ which results from $\sin(\sigma - \pi) = -\sin(\sigma)$. No surprise then that numerical integrations of (6.16) produced $B(\sigma)$ and $b(\sigma)$ curves executing 'seesaw' interlaced oscillations, about a common altitude when $\delta$ is moderate. This suggested the HF expansions:

$$\lambda(\sigma) = +\delta \cos(\sigma) + \lambda_0 + \lambda_1(\sigma)/\omega + \ldots,$$
$$\mu(\sigma) = -\delta \cos(\sigma) + \lambda_0 + \mu_1(\sigma)/\omega + \ldots.$$
(6.17)

Just like in subsection V.A one waits until the transients die and requires $\lambda(\sigma + 2\pi) = \lambda(\sigma)$, $\mu(\sigma) = \mu(\sigma + 2\pi)$, which imposes a pair of solvability conditions. Both give access to $\lambda_0$:

$$I_0(\delta)\frac{1}{\nu}\sinh(\lambda_0) + I_0(2\delta)[\frac{1}{\sinh(2\lambda_0)} - \frac{1}{2}\sinh(2\lambda_0)] = \coth(2\lambda_0) .$$
(6.18)

For $|\delta| = 0^+$ (6.18) retrieves the equilibriums $\lambda_\infty = 0$ or $\frac{1}{\nu} = \cosh(\lambda_\infty) + 1/\cosh(\lambda_\infty)$, i.e., $B_\infty = \infty = b_\infty$ for $\nu \geq \frac{1}{2}$ or $2B_\infty = \tanh^{-1}(2\nu) = 2b_\infty$ for $\nu < \frac{1}{2}$, as is expected of poles that belong to twin steady wrinkles sharing the same cell $-\pi < x \leq \pi$. By contrast to (6.4), (6.18) could only be solved numerically; sample curves $\lambda_0$ vs. $|\delta|$ are displayed in Fig.5, for various $\nu$ s. Below the first bisector of Fig.5 $|\delta|$ exceeds $\lambda_0$, in which case both $\lambda(\sigma)$ and $\mu(\sigma)$ change sign twice during the cycle; this will always occur if $\nu \geq \frac{2}{3}$. A single crest survives at $x = 0$ or $\pi$ (mod. $2\pi$) when $\lambda(\sigma)\mu(\sigma) < 0$, corresponding to $\Re[z_1(t)] = \Re[z_2(t)]$.

$\lambda_1(\sigma)$ and $\mu_1(\sigma)$ are still accessible by quadrature, yet integrations by parts again exempt one from invoking the solvability conditions at order $1/\omega$ to get the integration constants. For, as long as $w \equiv 0$ one can bypass $\lambda_1(\sigma)$ and $\mu_1(\sigma)$ to access all the $O(1)$ pieces of $V_{av}$: multiply

(6.16) by $\coth(\lambda(\sigma))$ or $\coth(\mu(\sigma))$ before the averages are taken. This yields:

$$V_{av} - \tfrac{1}{8}\omega^2\delta^2 - \mathcal{V}(2\nu) = -4\nu I_0(\delta)\cosh(\lambda_0) + 2\nu^2[\cosh(2\lambda_0)I_0(2\delta) + 3]. \qquad (6.19)$$

Weak forcing [$\delta \to 0$] leads to $V_{av} \to \mathcal{V}(2\nu)$ if $\nu \leq \tfrac{1}{2}$, where the latter argument $2\nu$ stems from the presence of twin unforced wrinkles of wavelengths $\pi$ in the same box $-\pi < x \leq \pi$; if $\nu > \tfrac{1}{2}$, one has $|\lambda_0(\delta,\nu)| \to 0$ and $V_{av} \to 0$ in the small-forcing limit, Figs. 5 and 6. Sample results from numerical integration of (6.16) are also shown in Fig. 6: when the unforced flat flame is unstable [$\nu < \tfrac{1}{2}$] the influences of DL-instability and curvature effects are patent… as far as $V_{av} - \tfrac{1}{8}\Phi^2$ is concerned. Yet $V_{av}$ itself is again dominated by the time-averaged increase in base-front arclength $\tfrac{1}{4}[\Phi(t)^2]_{av} = \tfrac{1}{8}\omega^2\delta^2$ and its parabolic growth, except for moderately forced DL-unstable fronts, see Fig.7.

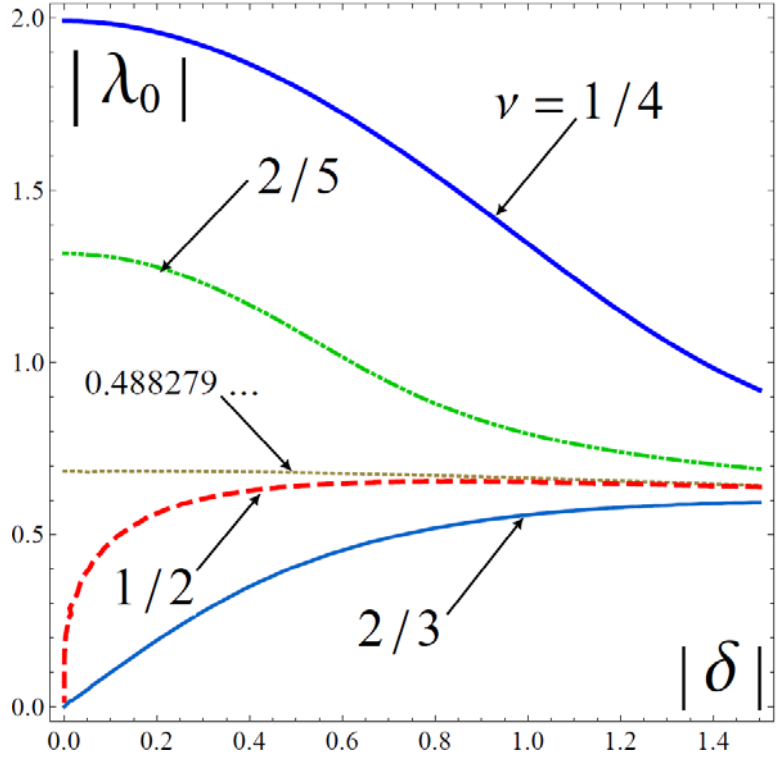

FIG. 5. (Color online). $|\lambda_0(\delta,\nu)|$ vs. $|\delta|$ from (6.18), for the indicated values of $\nu$. When $|\delta| > |\lambda_0|$, $\lambda(\sigma)$ and $\mu(\sigma)$ from (6.17) change sign twice per period, which induces pole jumps. $\lambda_0 \underset{\delta\to 0}{\to} 0$ only if $\nu \geq \tfrac{1}{2}$, and $|\tfrac{\partial}{\partial\delta}\lambda_0| \leq 1$ for $\nu \geq \tfrac{2}{3}$. Note how flat $\lambda_0(|\delta|<1,\nu)$ is when $\nu = \tfrac{1}{3}(7^{1/2} - \tfrac{1}{2})^{1/2} \approx 0.4882794$.

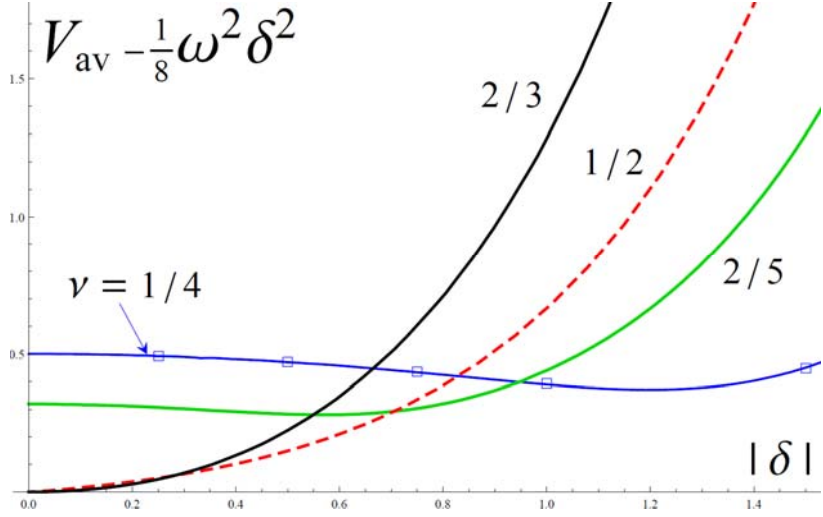

FIG.6. (Color online). Excess $V_{av} - \frac{1}{8}\omega^2\delta^2$ of $V_{av}$ over base-shape arclength, as is given by (6.19) for various $\nu$ s. The symbols (□) are results from numerical integrations of (6.16) with $\omega = 10$, $\nu = \frac{1}{4}$.

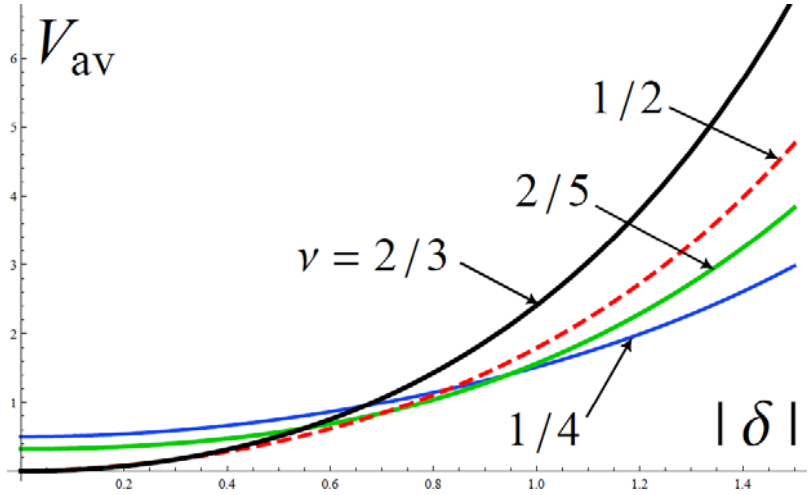

FIG.7. (Color online). $V_{av}$ vs. $|\delta|$, as given by (6.19) for $\omega = 3$ and the indicated $\nu$ s.

A nonzero stretch can be accounted for in the same way as in subsection VI B. Once $\Delta$ is substituted for $\delta$ in (6.18) and (6.19), the correction to (6.19) has the same structure as in the second line of (6.11): $K'(\Delta)$ is formally the same, accurately described as under (6.13) using the new $\lambda_0$; $\Delta K''(\Delta)$, accessible from $\lambda_1(\sigma)$ and $\mu_1(\sigma)$ via integrations by parts, now reads

$$\tfrac{2}{\nu}[\cosh(\lambda_0)I_0(\Delta) - 1] - 1 + 2\cosh(\lambda_0)I_0(\Delta) - \cosh(2\lambda_0)I_0(2\Delta) + 2[I_0(\Delta)\operatorname{sech}(\lambda_0) - 1] \underset{\Delta\to 0}{\sim} \Delta^2.$$

The Burgers version of (6.18), with carets and no $I_0(\hat\delta)/\hat\nu$-weighted term, is solved to give

$\sinh^2(\hat{\lambda}_0) = 2[1 + j - (1 + 3j)^{1/2}]/j$ where $j$ is a local shorthand for $I_0(2\hat{\delta})^2$. The averaged length increment ensues: $\hat{V}_{av} - \frac{1}{8}\hat{\omega}^2\hat{\delta}^2 = 2\hat{v}^2\{[2 + 3j - (1 + 3j)^{1/2}]^{1/2} - 1\}$. Once again its growth is parabolic at $\hat{\delta}^2 \ll 1$, $\hat{V}_{av} \approx (\frac{1}{8}\hat{\omega}^2 + 3\hat{v}^2)\hat{\delta}^2$, and steepens at higher amplitudes of $\hat{\Phi}(t)$. The influence of stretch is deduced in the usual way: carets are put, $\hat{\delta}$ is replaced by $\hat{\Delta}$ in $\hat{\lambda}_0$, $\hat{V}_{av} - \frac{1}{8}\hat{\omega}^2\hat{\delta}^2$ and $\hat{K}'(\hat{\Delta})$, and all the $I_0(\hat{\Delta})$-weighted pieces are deleted from $\hat{\Delta}\hat{K}''(\hat{\Delta})$.

### C. Twin piles of many

The pole equations (3.4) get increasingly difficult to solve analytically for $N_{tot} > 2$, even if restricted to $N$ pairs $z_k(t) = iB_k(t)$ or $n$ pairs $\pi + ib_\alpha(t)$, or both. Yet simplifications again occur with $N_{tot} \gg 1$ and wrinkle wavenumbers that ensure $\nu N_{tot} \sim \nu N_{opt}(\nu) \approx \frac{1}{2}$, as happened with the MS equation [18, 20]. Since such poles traverse the lines $\Re(z) = 0$ or $\pi$ (mod $2\pi$) along which they are smoothly distributed with $O(\nu)$ typical spacing, their populations are now viewed as two one-dimensional continua. Although a single pile of $z_k(t) = iB_k(t)$ poles [*i.e.*, $n = 0$] could be analyzed by similar – but actually bulkier – methods [Appendix C], only the case of equally populated piles of poles of either type [*i.e.*, $N = n$] is envisaged here. For clarity this subsection VI C will focus on stretch-free cases, $w(t) \equiv 0$, except at the end.

#### 1. Balances

Let $\rho(t, B) = \rho(t, -B)$ and $\varrho(t, b) = \varrho(t, -b)$ denote such densities that $\rho(t, B)dB \geq 0$ and $\varrho(t, b)db \geq 0$ are the numbers of $B$-poles or $b$-poles contained in $(B, B + dB)$ or $(b, b + db)$, respectively. As the poles are indestructible, move along lines and are unable to cross one another, Euler type of hydrodynamic descriptions of their dynamics begin with 1-D balances

$$\frac{\partial \rho}{\partial t} + \frac{\partial j}{\partial B} = 0 \;,\; \frac{\partial \varrho}{\partial t} + \frac{\partial q}{\partial b} = 0,$$
$$j \equiv -\omega\delta \sin(\omega t) \sinh(B) \rho(t, B) + J \;, \quad (6.20)$$
$$q \equiv +\omega\delta \sin(\omega t) \sinh(b) \varrho(t, b) + Q,$$

where $j(t, B)\,i/\rho(t, B)$ and $q(t, b)\,i/\varrho(t, b)$ stand for the Lagrangian pole speeds evaluated at $z = iB$ or $z = \pi + ib$. By (3.4), $-\omega\delta \sin(\omega t)\sinh(B)\rho(t,B)$ and $\omega\delta \sin(\omega t)\sinh(b)\varrho(t,b)$ result from convection caused by the base-front slope $\Phi_x(t, z) = \delta\omega \sin(\omega t)\sin(z)$, while $J(t, B)$ and $Q(t,b)$ account for the DL effect and the pole-to-pole interactions. The latter are handled in a continuous-medium manner, $\Sigma_j h(B_j) \to \int h(B')\rho(t,B')dB'$, with principal-value integrals to preclude self-interaction where needed [recall $j \neq k$ in (3.4)]. All in all (3.4) leads to:

$$J(t,B) = \rho(t,B)[\fint_{-B_{\max}(t)}^{+B_{\max}(t)} \frac{\nu\rho(t,B')dB'}{\tanh(\frac{1}{2}(B-B'))} + \int_{-b_{\max}(t)}^{+b_{\max}(t)} \frac{\nu\varrho(t,b')db'}{\coth(\frac{1}{2}(B-b'))} - \text{sgn}(B)],$$

$$Q(t,b) = \varrho(t,b)[\fint_{-b_{\max}(t)}^{+b_{\max}(t)} \frac{\nu\varrho(t,b')db'}{\tanh(\frac{1}{2}(b-b'))} + \int_{-B_{\max}(t)}^{+B_{\max}(t)} \frac{\nu\rho(t,B')dB'}{\coth(\frac{1}{2}(b-B'))} - \text{sgn}(b)]. \quad (6.21)$$

One may note the symmetry of (6.20)(6.21) by $(B, \rho(t,.), \delta) \leftrightarrow (b, \varrho(t,.), -\delta)$. The function $B_{\max}(t)$ [or $b_{\max}(t)$] corresponds to the largest $\tanh(\frac{1}{2}B)$ [or $\tanh(\frac{1}{2}b)$] where $\rho(t,B) \neq 0$ [or $\varrho(t,b) \neq 0$]; $B_{\max}(t)$ and $b_{\max}(t)$, unknown from the outset, must be found as part of the solutions to (6.20) by means of the normalizations $\int_0^{B_{\max}(t)} \rho(t,B)dB = N$ and $\int_0^{b_{\max}(t)} \varrho(t,b)db = n$, with $n = N$ in this subsection. The nonlinear and nonlocal system (6.20)(6.21) of partial differential equations with oscillating coefficients is now studied analytically for $\omega \to \infty$.

## 2. Pole densities

To this end one first remarks that setting $J(t,B) \equiv 0$ and $Q(t,b) \equiv 0$ in (6.20) would convert each balance to a linear first-order hyperbolic equation solvable analytically in terms of $\sigma \equiv \omega t$ and of a characteristic variable, $E \equiv \tanh(\frac{1}{2}B)e^{+\delta\cos(\sigma)}$ or $\varepsilon \equiv \tanh(\frac{1}{2}b)e^{-\delta\cos(\sigma)}$. With this hint as a guide, the balances (6.20) are rewritten as follows [mind the signs before the $\delta$ s]

$$\left.\frac{\partial R(\sigma,E)}{\partial \sigma}\right|_E = -\frac{1-E^2}{2\omega}\frac{\partial J(\sigma,E)}{\partial E},$$

$$\rho(t,B) = \frac{e^{+\delta\cos(\sigma)} - E^2 e^{-\delta\cos(\sigma)}}{1-E^2} R(\sigma,E), \quad (6.22)$$

$$\left.\frac{\partial r(\sigma,\varepsilon)}{\partial \sigma}\right|_\varepsilon = -\frac{1-\varepsilon^2}{2\omega}\frac{\partial Q(\sigma,\varepsilon)}{\partial \varepsilon},$$

$$\varrho(t,b) = \frac{e^{-\delta\cos(\sigma)} - \varepsilon^2 e^{+\delta\cos(\sigma)}}{1-\varepsilon^2} r(\sigma,\varepsilon), \quad (6.23)$$

where the partial derivatives $\partial/\partial\sigma$ must be taken at fixed $E$ or $\varepsilon$ as specified. By an abuse of notation $J(\sigma,E)$ denotes $J(t,B)$ once rewritten in terms of $\sigma$ and $E$; likewise, $Q(\sigma,\varepsilon)$ just is a rewrite of $Q(t,b)$. Omitting the partial flux $J(\sigma,E)$ of $B$-poles from (6.22) would yield a 'density' $R_0(E)(e^{\delta\cos(\sigma)} - E^2 e^{-\delta\cos(\sigma)})/(1-E^2)$ with $R_0(E)$ arbitrary; and similarly, up to a flipped sign of $\delta$ and another arbitrary $r_0(\varepsilon)$, by setting $Q(\sigma,\varepsilon)=0$ in (6.23). This is precisely what the aforementioned linear hyperbolic equations would give.

To determine $R(\sigma,E) = R(\sigma,-E)$ and $r(\sigma,\varepsilon) = r(\sigma,-\varepsilon)$ when $J(\sigma,E) = -J(\sigma,-E) \not\equiv 0$ and $Q(\sigma,\varepsilon) = -Q(\sigma,-\varepsilon) \not\equiv 0$, as to access $\rho(t,B)$ and $\varrho(t,b)$ via (6.22)(6.23), a key point to make is that integrations of the balances (6.22) or (6.23) at fixed $E$ or $\varepsilon$ over a full period of $R(\sigma,E) = R(\sigma+2\pi,E)$ or $r(\sigma,\varepsilon) = r(\sigma+2\pi,\varepsilon)$ yield a pair of solvability conditions:

$$[J(\sigma,E)]_{\text{av},E} = 0 = [Q(\sigma,\varepsilon)]_{\text{av},\varepsilon} . \tag{6.24}$$

By (6.22) $\frac{\partial}{\partial E}[J(\sigma,E)]_{\text{av},E}$ vanishes, hence $[J(\sigma,E)]_{\text{av},E} = -[J(\sigma,-E)]_{\text{av},E} \propto \text{sgn}(E)$; since no pole can cross the real axis one has $j(\sigma,0^\pm) = 0$ which implies $J(\sigma,0^\pm) = 0$, $[J(\sigma,0^\pm)]_{\text{av}} = 0$ and then $[J(\sigma,E)]_{\text{av},E} = 0$ whatever $\omega \neq 0$ is; $[Q(\sigma,\varepsilon)]_{\text{av},\varepsilon} = 0$ follows in the same manner.

Crucially, the right-hand sides of the balances (6.22)(6.23) also disappear in the HF limit. This motivated the asymptotic series for the auxiliary densities $R(\sigma,E)$ and $r(\sigma,\varepsilon)$:

$$\begin{aligned} R(\sigma,E) &= R_0(E) + R_1(\sigma,E)/\omega + ... , \\ r(\sigma,\varepsilon) &= R_0(\varepsilon) + r_1(\sigma,\varepsilon)/\omega + ... . \end{aligned} \tag{6.25}$$

Note that $R(\sigma,E)$ and $r(\sigma,\varepsilon)$ are anticipated to be described at leading order by the same $R_0(.)$, in analogy with having used the same $\lambda_0$ in the two-pair dynamics of subsection VI B, see (6.17); $r_0(.) \equiv R_0(.)$ could be deduced, but is simply postulated in (6.25) to save space.

Just like in VI.A-B finding the unknown $R(\sigma,E)$ and $r(\sigma,\varepsilon)$ requires (6.25) to be substituted into the solvability conditions: $J(t,B)$ and $Q(t,b)$ from (6.21) are re-expressed in terms of $(\sigma,E)$ or $(\sigma,\varepsilon)$, respectively; next (6.25) is plugged into $J(\sigma,E)$ and $Q(\sigma,\varepsilon)$; the results are finally averaged at fixed $E$ or $\varepsilon$ as prescribed in (6.24). At leading order for $\omega \to \infty$, (6.24) yields two copies of the same equation for $R_0(.)$, as anticipated in (6.25). A tedious rearrangement that uses the leading-order normalization condition(s) [see (6.28)] and $R_0(E) = R_0(-E)$ finally recasts the equation for $R_0(E)$ to : $R_0(E) \times \{\text{affine functional of } R_0(E)\} = 0$. The latter compact form thus has two possible outcomes.

- Either $R_0(E) \equiv 0$, over the range $E^2 \geq E_{max,0}^2$ belonging to $\tanh^2(\tfrac{1}{2}B) \geq \tanh^2(\tfrac{1}{2}B_{max,0}(\sigma))$ $\equiv \lim_{\omega \to \infty} \tanh^2(\tfrac{1}{2}B_{max}(\sigma))$ and to $\tanh^2(\tfrac{1}{2}b) \geq \tanh^2(\tfrac{1}{2}b_{max,0}(\sigma)) \equiv \lim_{\omega \to \infty} \tanh^2(\tfrac{1}{2}b_{max}(\sigma))$.

- Or $R_0(E) \not\equiv 0$ obeys a singular, yet *linear*, integral equation over $-E_{max,0} \leq E \leq E_{max,0}$:

$$\fint_{-E_{max,0}}^{E_{max,0}} \frac{2\nu R_0(E')dE'}{(E-E')} + E \int_{-E_{max,0}}^{E_{max,0}} \frac{2\nu R_0(E')dE'}{1-EE'} = g(E,2N) , \qquad (6.26)$$

$$g(E,2N) \equiv \frac{I_0(\delta)\,\text{sgn}(E)(1-E^2)^2 + 4\nu N(I_0(2\delta)-1)E(1+E^2)}{(1-E^2)^2 + (I_0(2\delta)-1)(1+E^4)} . \qquad (6.27)$$

$E_{max,0} = \lim_{\omega \to \infty} E_{max}(\sigma)$ was presumed constant to get (6.26), but it must in fact ensue from normalizing $\rho(t,B)$ and $\varrho(t,b)$. Both conditions fortunately coincide for $\omega \to \infty$ and yield

$$\int_0^{E_{max,0}} \frac{2\nu R_0(E)dE}{(1-E^2)} = \nu N , \qquad (6.28)$$

confirming that $E_{max,0}$ is $\sigma$-independent; the integral in (6.28) necessitates $E_{max,0} < 1$ to converge, or needs a fast enough decay of $R_0(E)$ towards $R_0(1) = 0$ if $E_{max,0}$ equals $1$.

From the definitions of $E$ and $\varepsilon$, the same constant $E_{max,0}$ determines $\tanh(\tfrac{1}{2}B_{max,0}(\sigma)) = E_{max,0}\, e^{-\delta \cos(\sigma)}$ and $\tanh(\tfrac{1}{2}b_{max,0}(\sigma)) = E_{max,0}\, e^{+\delta \cos(\sigma)}$. The leading order densities $\rho_0(t,B)$ and $\varrho_0(t,b)$ written in (6.22)(6.23) are in phase opposition, but time-wise shifts do not count in (6.24): both expansions in (6.25) consistently shared the same function $R_0(.)$. The constancy of $\tanh(\tfrac{1}{2}B_{max,0}(\sigma))\tanh(\tfrac{1}{2}b_{max,0}(\sigma)) = E_{max,0}^2$ may help one picture the 'seesaw' pile motions.

First, note that (6.26) departs by its right-hand side $g(E,2N)$ from what the MS case gave [18, 20], except at $\delta = 0$ when both share the $\text{sgn}[E]$ form characterizing a pure DL effect. Having no such $\text{sgn}[.]$ function the hatted, Burgers analogs of (6.21)-(6.27) involve no $I_0(\hat{\delta})$ either, nor $\mathbf{I}_0(\hat{\delta}) = \mathbf{I}_0(-\hat{\delta})$ if $\hat{\Phi}(\sigma) = -\hat{\Phi}(\sigma - \pi) \not\equiv -\hat{\omega}\hat{\delta}\sin(\sigma)$ as in subsection VIA: one just sets those to zero when adapting the DL-affected HF results to the Burgers case.

Second, (6.26) is not of Tricomi type yet is likely related to it, precisely because $g(E,2N)$ $= \text{sgn}[E]$ at $\delta = 0$ and one should then be back to the MS case up to wavelength halving. The new variables $\Theta \equiv 2E/(1+E^2)$ and $\chi(\Theta) \equiv R_0(E(\Theta))$ in effect transform (6.26)-(6.28) to:

$$\oint_{-\Theta_{max,}}^{\Theta_{max}} \frac{\nu\chi(\Theta')d\Theta'}{(\Theta-\Theta')} = G(\Theta, 2N), \quad -\Theta_{max} \leq \Theta \leq \Theta_{max},$$

$$G(\Theta, 2N) \equiv \frac{I_0(\delta)\Theta_*^2 \text{sgn}(\Theta)(1-\Theta^2)}{2I_0(2\delta)(\Theta_*^2-\Theta^2)} + \frac{2\nu N(\Theta_*^2-1)\Theta}{(\Theta_*^2-\Theta^2)}, \quad (6.29)$$

$$\int_0^{\Theta_{max}} \frac{\nu\chi(\Theta)d\Theta}{(1-\Theta^2)} = \nu N, \quad \Theta_{max} \equiv \frac{2E_{max,0}}{1+E_{max,0}^2},$$

where $1 < \Theta_* \equiv [2/(1+1/I_0(2\delta))]^{1/2} < \sqrt{2}$. Reducing (6.26) to the above Tricomi equation relates to wavelength halving, $\tanh^{-1}(\Theta) = 2\tanh^{-1}(E)$; and its right-hand side is simpler than $g(E, 2N)$ once $E = (1-(1-\Theta^2)^{1/2})/\Theta$ is plugged in (6.27) because $g(E, 2N)$ only depends on $E + 1/E = 2/\Theta$. Most importantly, the integral equation (6.29) is solvable by quadrature [33-34] for any presumed $0 < \Theta_{max} \leq 1$. Provided $R_0(E)$ [or $\chi(\Theta)$] does not vanish over any finite subinterval of $0 \leq E^2 < E_{max,0}^2$ [or of $0 \leq \Theta^2 < \Theta_{max}^2$] this yields [Appendix B]:

$$\nu\chi(\Theta) = \frac{I_0(\delta)\Theta_*^2}{\pi^2 I_0(2\delta)} \cosh^{-1}(\frac{\Theta_{max}}{|\Theta|})\frac{(1-\Theta^2)}{(\Theta_*^2-\Theta^2)} +$$

$$\frac{\Theta_*(\Theta_*^2-1)[2\nu N - \mathcal{N}(\Theta_{max}, \delta)]}{\pi\sqrt{\Theta_*^2-\Theta_{max}^2}} \frac{(\Theta_{max}^2-\Theta^2)^{1/2}}{(\Theta_*^2-\Theta^2)}, \quad (6.30)$$

$$\mathcal{N}(\Theta_{max}, \delta) \equiv \frac{I_0(\delta)\Theta_*}{\pi I_0(2\delta)} \sin^{-1}(\frac{\Theta_{max}}{\Theta_*}).$$

The resulting $R_0(E) = \chi(2E/(1+E^2))$ is bounded by $\text{const.}(E_{max,0}^2 - E^2)^{1/2}$ near the edges $E^2 = E_{max,0}^2$ of its support, and vanishes beyond; so do $\rho(t, B)$ and $\varrho(t, b)$. As checks, back-substitution of (6.30) into (6.29) and numerical evaluations of the integrals with the help of Mathematica [35] retrieved the sample $G(\Theta, N)$ and $2\nu N$ we input to $O(10^{-20})$ accuracy.

### 3. Slopes

With $R_0(E)$ now available the next step tackles the polar piece $\phi_x(t, x)$ of the front slope, starting with the continuous-medium version thereof deduced from (3.1):

$$\phi_x(t,x) = -\nu \int_{-B_{max}(t)}^{+B_{max}(t)} \cot(\tfrac{1}{2}(x-iB))\rho(t,B)\,dB + \nu \int_{-b_{max}(t)}^{+b_{max}(t)} \tan(\tfrac{1}{2}(x-ib))\varrho(t,b)\,db. \quad (6.31)$$

One next makes use of (6.22), (6.25) and (6.30). A lengthy rearrangement, based on $R_0(E) = R_0(-E)$ and similar to that invoked about (6.26), recasts the leading-order polar slope to

$$\phi_x(t,x) = -8\nu N \frac{\sinh(\delta\cos(\sigma))}{(1/\mathcal{T}+\mathcal{T})} - \Sigma(\mathcal{E})\frac{e^{\delta\cos(\sigma)}/\mathcal{T}+e^{-\delta\cos(\sigma)}\mathcal{T}}{(1/\mathcal{T}+\mathcal{T})},$$

$$\mathcal{T}(\sigma,x) \equiv \tan(\tfrac{1}{2}x)e^{\delta\cos(\sigma)}, \quad \mathcal{E}(\sigma,x) \equiv \frac{2}{1/\mathcal{T}-\mathcal{T}}, \quad \Sigma(\mathcal{E}) \equiv \int_{-\Theta_{max,0}}^{\Theta_{max,0}} \frac{2\nu\chi(\Theta)d\Theta}{(\mathcal{E}-i\Theta)},$$
(6.32)

where $\mathcal{E}(\sigma,x)$ also is $\tan[2\tan^{-1}(\tan(\tfrac{1}{2}x)e^{\delta\cos(\sigma)})]$. It 'simply remains' to plug (6.30) in the integral which defines $\Sigma(\mathcal{E})$ in (6.32) to complete the leading order polar slope $\phi_x(t,x)$. Analytical evaluation [Appendix D] produces $\Sigma(\mathcal{E}) = -\Sigma(-\mathcal{E})$ as

$$\Sigma(\mathcal{E}) = \frac{2I_0(\delta)\Theta_*^2}{\pi I_0(2\delta)}[\frac{(1+\mathcal{E}^2)}{(\Theta_*^2+\mathcal{E}^2)}\sinh^{-1}(\frac{\Theta_{max}}{\mathcal{E}}) - \frac{(\Theta_*^2-1)}{\Theta_*}\sin^{-1}(\frac{\Theta_{max}}{\Theta_*})\frac{\mathcal{E}}{(\Theta_*^2+\mathcal{E}^2)}] +$$
$$\frac{2\Theta_*(\Theta_*^2-1)[2\nu N - \mathcal{N}(\Theta_{max},\delta)]}{(\Theta_*^2-\Theta_{max}^2)^{1/2}}\frac{\mathcal{E}}{(\Theta_*^2+\mathcal{E}^2)}[(1+\frac{\Theta_{max}^2}{\mathcal{E}^2})^{1/2} - (1-\frac{\Theta_{max}^2}{\Theta_*^2})^{1/2}],$$
(6.33)

which fixes the analytical structure of the front slope $\varphi_x(t,x) = \omega\delta\sin(\sigma)\sin(x) + \phi_x(t,x) \ldots$ to *two* orders. For $\delta \to 0$, $\rho(t,B) \to R_0(\tanh(\tfrac{1}{2}B)) + \ldots$ by (6.22), $\Theta_* \to 1$ by its definition below (6.29), $\mathcal{E}(\sigma,x) \to \tan(x)$ and $\phi_x(t,x) \to -\frac{2}{\pi}\sinh^{-1}(\Theta_{max}\cot(x))$: the latter slope diverges at $x=0 \pmod{\pi}$ and coincides with that of long-waved MS flames comprising twin steady crests per cell [20]. The block weighted by $I_0(\delta)$ in (6.33) also results from the DL effect and still diverges at $\tan(\tfrac{1}{2}x)=0$ or $\infty$, where the crests oscillate in phase opposition [see Fig.8].

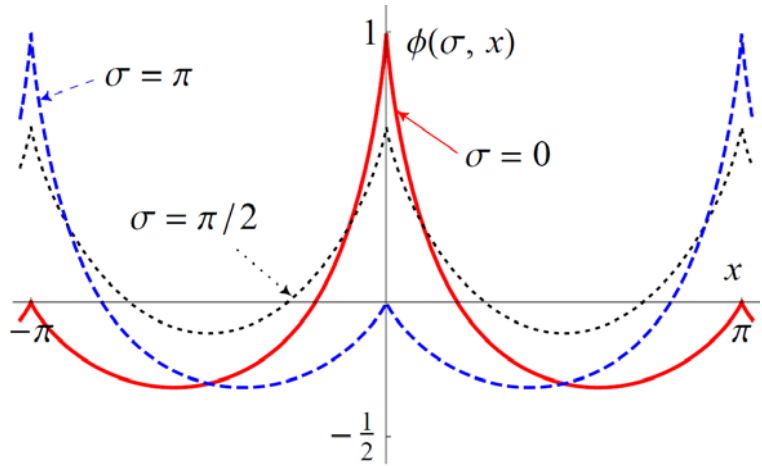

FIG.8. (Color online). Snapshots of *polar* shapes $\phi(\sigma,x)$ [some shifted for readability] deduced from (6.32) (6.33) and (6.34) by numerical integration over $x$, for $\delta=1$, $\nu N = \tfrac{1}{8}$ and $\sigma = 0$ [solid red], $\tfrac{\pi}{2}$ [dotted black] or $\pi$ [dashed blue]. Surprisingly, (6.31) integrated over $0 \le x \le \pi$, using (6.22)(6.23) and the normalization (6.28), simply yields $\phi(\sigma,0) - \phi(\sigma,\pi) = 8\nu N\delta\cos(\sigma)$; and the same, up to carets, for HF Burgers fronts.

As mentioned earlier it is simple to deduce Burgers analogs of the above HF results: one just puts the carets and discards everything with $I_0(\hat{\delta})$ as a factor. In particular $\hat{\mathcal{N}}(\hat{\Theta}_{\max},\hat{\delta}) \equiv 0$, and $\hat{\chi}(\Theta)$ reduces to its second line in the analog of (6.30); also, the whole first line drops out from the DL-free version of (6.33), rendering Burgers slopes discontinuous but bounded at the crests $[\hat{\mathcal{E}}(1+\hat{\Theta}_{\max}^2/\hat{\mathcal{E}}^2)^{1/2}/(\hat{\Theta}_*^2+\hat{\mathcal{E}}^2) \sim \text{sgn}(\tan(\tfrac{1}{2}x))$ for $x \to \mathbb{Z}\pi$ in the analog of (6.33)].

## *4. Normalizations*

To fully finalize the leading-order densities *via* $\chi(\Theta)$ and at last obtain the unsteady profiles of front slope it 'just remains' to determine $\Theta_{\max}$ from the normalization condition in (6.29), using (6.30). The needed integrals can again be done analytically [Appendix D], and the resulting condition that gives access to $\Theta_{\max}$ looks simple

$$[2\nu N - \mathcal{N}(\Theta_{\max},\delta)](1-\Theta_{\max}^2)^{1/2} = 0, \qquad (6.34)$$

yet is only superficially so because it offers two branches, $2\nu N = \mathcal{N}(\Theta_{\max},\delta)$ or $\Theta_{\max}=1$.

The branch $\Theta_{\max} \neq 1$ produces $\Theta_{\max} = \Theta_* \sin[2\pi\nu N I_0(2\delta)/\Theta_* I_0(\delta)]$; it resumes the expected MS result [15] in absence of forcing, *viz.* $\Theta_{\max} = \sin[\pi\nu N_{tot}]$ with $N_{tot} = 2N$ here. For $2\nu N < \nu N_{opt}(\nu) = \tfrac{1}{2}$ this ordinary branch is admissible at moderate intensities $\delta$ of forcing, which brings about drastic simplifications: the whole second lines of (6.30) and (6.33) disappear. Some concern might arise as $2\nu N < \tfrac{1}{2}$ crosses $2\nu N_* \equiv \mathcal{N}(\text{sech}(\delta),\delta)$ from below: due to $e^{-|\delta|} < E_{\max,0} < 1$, $\tanh(\tfrac{1}{2}B_{\max,0}(\sigma)) = E_{\max,0} e^{-\delta\cos(\sigma)}$ and $\tanh(\tfrac{1}{2}b_{\max,0}(\sigma)) = E_{\max,0} e^{+\delta\cos(\sigma)}$ momentarily get $>1$, whereby $B_{\max,0}(\sigma)$ or $b_{\max,0}(\sigma)$ [not both, for the latter two $\tanh(.)$s multiply to $E_{\max,0}^2$] then acquires an imaginary part of $-\pi$. Put in words, the leading edges of both pole-densities jump periodically, accompanied by such finite portions of the pole populations that $e^{-2|\delta|} < E^2 < E_{\max,0}^2$ or $e^{-2|\delta|} < \varepsilon^2 < E_{\max,0}^2$, and having $1 < \tanh^2(\tfrac{1}{2}B) < E_{\max,0}^2 e^{-2\delta\cos(\sigma)}$ or $1 < \tanh^2(\tfrac{1}{2}b) < E_{\max,0}^2 e^{+2\delta\cos(\sigma)}$ while they leak into the adjacent crest. The worrisome point is that, for leaking poles located in $dB'$ near $iB'$ [or $ib'$], $-B'$ [or $-b'$] is complex in the integrals (6.21) for the fluxes and no longer parameterizes the complex conjugate $\overline{iB'}$ [or $\overline{ib'}$] of $iB'$ [or $ib'$]. Fortunately $i(-B') = \overline{i(B'+2i\pi)}$ when $\Im[B'] = -\pi$, and the $2i\pi$-periodicity of $\tanh(\tfrac{1}{2}(B-B'))$ saves the applicability of the integrals over $B'$ once rewritten as $(\int_{-B_{\max}(t)}^{-\infty-i\pi} + \int_{-\infty}^{+\infty} + \int_{+\infty-i\pi}^{+B_{\max}(t)})dB'$ when $\tanh^2(\tfrac{1}{2}B_{\max,0}(\sigma)) > 1$; and similarly for the integrations over $b'$ when $b$-poles leak.

This apparent tension would not have appeared by choosing $\tanh(\frac{1}{2}B)$ and $\tanh(\frac{1}{2}b)$ as the independent variables instead of $B$ and $b$, and at any rate does not affect the fluxes $J(\sigma, E)$ and $Q(\sigma, \varepsilon)$ in (6.22)(6.23). Consequently the previous HF analysis still holds true when pole jumps/leakage occurs, at the price of having $\rho(t, B) < 0$ wherever $1 < \tanh^2(\frac{1}{2}B)$: this results from the definition of $\rho(t, B)$ as a local reciprocal spacing $1/(B_k - B_{k-1})$ that reverses its sign once both $B_k$ and $B_{k-1}$ have jumped, and the same for $\varrho(t, b)$. As convergence in (6.28) needs $E_{max,0}^2 \leq 1$ such variable-sign densities are correctly encoded in (6.22)(6.23), with one proviso: the viable root $\Theta_{max}$ of (6.34) must result in $R_0(E) = \chi(\Theta(E)) > 0$ all over $0 < E^2 < E_{max,0}^2$.

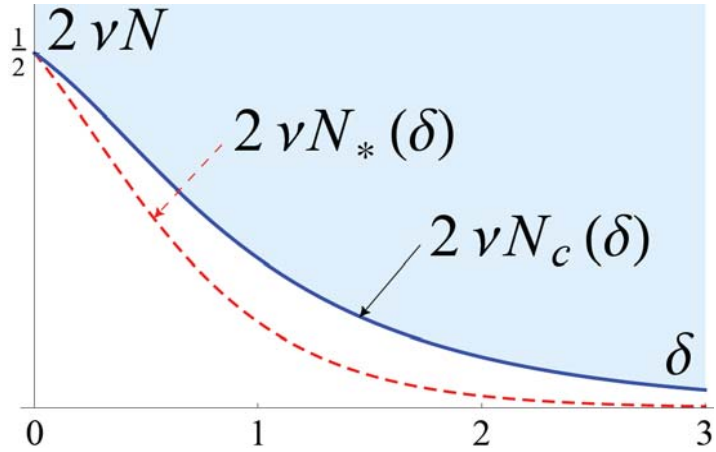

FIG. 9. (Color online). Phase diagram in $(\delta, 2N\nu)$ plane. Only above the dashed-red line $N = N_*(\delta)$ do pole jumps occur. Besides, $E_{max,0} = 1 = \Theta_{max}$ over the whole shaded area above the solid-blue line $N = N_c(\delta)$. For future reference, e.g., in Figs.10-13, one mentions that $2\nu N_*(1) \approx 0.1213$ and $2\nu N_c(1) \approx 0.21104$.

One can now complete the discussion of (6.34). Although $\Theta_{max} = 1 = E_{max,0}$ yields a genuine and numerically verified solution to the integral equation (6.29) whatever $\nu N$ and $\delta$ are, it is *spurious* below the curve $N = N_c(\delta)$ in Fig.9 defined by $2\nu N_c = \mathcal{N}(1, \delta)$, because it leads to $R_0(E) < 0$ at $E^2 \lesssim 1$: the ordinary solution $\Theta_{max} \leq 1$ to (6.34) is viable for $N \leq N_c(\delta)$ even if $N > N_*(\delta)$ and pole leakage occurs. But along and above the solid line $N = N_c(\delta)$ in Fig.9 the root $\Theta_{max} = 1$ to (6.34) takes over and then yields $R_0(E) \geq 0$ as required. In particular $\Theta_{max} = 1$ prevails for all $2\nu N \geq \frac{1}{2}$ as soon as forcing acts. Note that $R_0(E)$ develops bumps at $0 < 1 - E^2 = O(\delta^2)$ if $\delta \ll 1$ and $2\nu N > \frac{1}{2}$ [see second line of (6.30)], which reminds one of the surplus poles that escape to $\pm i\infty$ in absence of forcing. Those poles that leak have $E^2$ or $\varepsilon^2$ in $(e^{-2|\delta|}, E^2_{\max,0})$ and their number in either pile is $m = \int_{\exp(-|\delta|)}^{E_{\max,0}} 2R_0(E)dE/(1-E^2)$; $m/N \leq 1$ vanishes at $N = N_*(\delta)$, stays finite at $\delta \to 0$ if $2\nu N > \frac{1}{2}$, and goes to 1 for $|\delta| \to \infty$.

The snapshots in Fig.10 recapitulate the types of density profiles. Note that $\Theta_{\max} < 1$ implies $\tanh(\frac{1}{2}B_{\max,0}(\sigma))\tanh(\frac{1}{2}b_{\max,0}(\sigma)) = E^2_{\max,0} < 1$ and the existence of a gap between the $B$-pole and the $b$-pole populations [Fig.10, top row]; but $\coth(\frac{1}{2}B_{\max,0}(\sigma)) = \tanh(\frac{1}{2}b_{\max,0}(\sigma))$ if $\Theta_{\max} = 1 = E_{\max,0}$, implying $\Re[B_{\max,0}(\sigma)] \equiv \Re[b_{\max,0}(\sigma)]$ and gap suppression [Fig.10, bottom row].

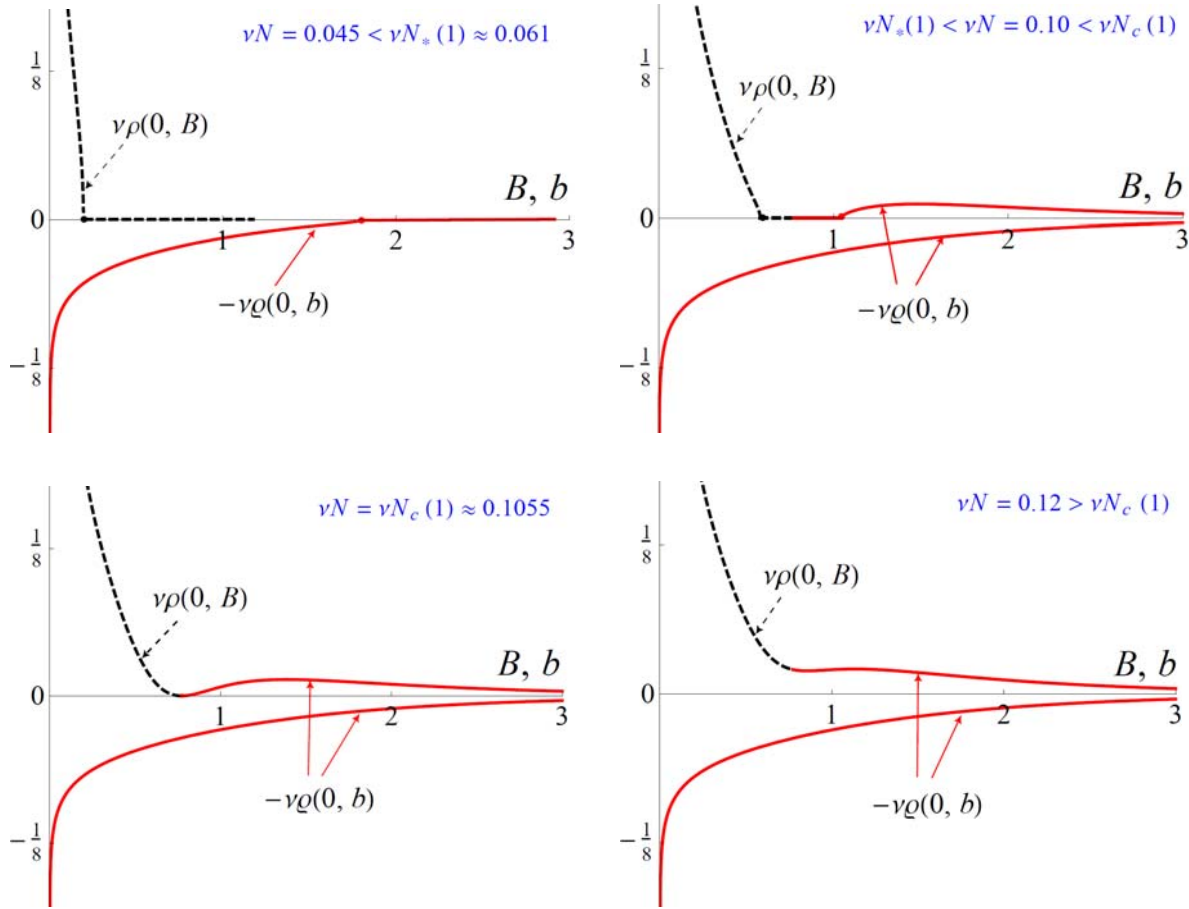

FIG.10. (Color online). Densities $+\nu\rho(\sigma,B)$ [dashed black] and $-\nu\varrho(\sigma,b)$ [solid red] vs. pole distance [$B$ or $b$] to real axis, for $\delta = 1$ and various $\nu N$. All curves correspond to $\sigma = 0 \pmod{2\pi}$, when the $iB$ [or $\pi + ib$] poles are the most pushed towards [or pulled away from] the real axis by forcing; half a period later the blue or red curves will have permuted shapes, because $\varrho(\sigma,b) = \rho(\sigma \pm \pi, b)$. The Burgers fronts have $\hat{N}_c(\hat{\delta}) = 0$ and $\hat{\Theta}_{\max} = 1$ in the analogs of (6.30)(6.33), and lead to gapless densities for any $\hat{\nu}\hat{N} > 0$; also, $\hat{\nu}\hat{\rho}(\sigma,0) = \hat{\nu}\hat{N}\exp(\hat{\delta}\cos(\sigma))[1 - 1/I_0(2\hat{\delta})]^{1/2}/(2^{1/2}\pi) = \hat{\nu}\hat{\varrho}(\sigma + \pi, 0)$ are bounded.

The Burgers fronts have $\hat{\mathcal{N}}(\hat{\Theta}_{max}, \hat{\delta}) \equiv 0$, and $\hat{\Theta}_{max} = 1$ always holds: at any one time some of their poles leak. The formula found for the leaking fraction $\hat{m}/\hat{N}$ is comparatively simple, and reads $1 - \frac{2}{\pi}\tan^{-1}[\sin(\hat{\alpha})/\sinh(|\hat{\delta}|)]$ where $2\hat{\alpha} \equiv \cos^{-1}(1/I_0(2\hat{\delta}))$; for $|\hat{\delta}| \to 0$ this reduces to $1 - \frac{1}{\pi}\cos^{-1}[\frac{1}{3}] \approx 0.608$, which numerical integrations of the Burgers version of the discrete pole dynamics (3.4) confirmed [*e.g.*, $\hat{N} = 10$, $\hat{v} = \frac{1}{20}$, $\hat{\omega} = 10$ and $\hat{\delta} = \frac{1}{10}$ gave $\hat{m} = 6$].

## *5. Average arclength*

To complete subsection VI.C one must proceed to the arclength increment $V(t)$, in principle accessible from the continuous version of (3.8). Due to difficulties in explicitly evaluating the large-$N$ version of the needed $\Sigma_{k=1}^{N_{tot}} \Im[dz_k/dt]$ from the previously computed densities, we could not determine $V(t)$ analytically. $V_{av} = [V(t)]_{av}$ seemed simpler to access because time derivatives average to zero, yet that happened to be anything but trivial. Taking the average of (3.8), when $\Phi(t) = -\omega\delta\sin(\sigma)$ and $w(t) \equiv 0$, indeed boils down to computing the large-$N$ version $S + s$ of $-\omega\delta[\sin(\sigma)2\nu\Sigma_{k=1}^N(\cosh(B_k) - \cosh(b_k))]_{av}$ in the HF limit $\omega \to \infty$:

$$V_{av} - \tfrac{1}{8}\delta^2\omega^2 - \mathcal{V}(2\nu N) = (S + s),$$
$$S = -\frac{\omega\delta}{2\pi}\int_{-\pi}^{+\pi}\sin(\sigma)d\sigma\int_0^{+B_{max}(\sigma)}2\nu\rho(\sigma, B)(\cosh(B) - 1)dB, \qquad (6.35)$$
$$s = +\frac{\omega\delta}{2\pi}\int_{-\pi}^{+\pi}\sin(\sigma)d\sigma\int_0^{+b_{max}(\sigma)}2\nu\varrho(t, b)(\cosh(b) - 1)\,db.$$

As expected, (6.35) is symmetric by $(S, \delta, \rho(\sigma, B)) \leftrightarrow (s, -\delta, \varrho(t, b))$. Even though $\omega \to \infty$, $S + s$ is $O(1)$ since the leading-order densities depend on $\sigma$ *via* $\cos(\sigma)$, see (6.22)(6.23): as in subsections VI.A and VI.B one needs to take the next-to-leading-order densities, hence the still unknown $R_1(\sigma, E)$ and $r_1(\sigma, \varepsilon)$ featured in (6.25), into account. These are in principle accessible from (6.22)(6.23). For example $R_1(\sigma, E) = -\tfrac{1}{2}(1 - E^2)\frac{\partial}{\partial E}\int_0^\sigma J_0(\sigma', E)d\sigma' + C_1(E)$, where the integration 'constant' $C_1(E)$ can be shown from (6.24) to vanish and the leading-order flux $J_0(\sigma, E)$ is expressible via (6.21) in terms of now known material; and similarly for $r_1(\sigma, \varepsilon)$. However, because $J_0(\sigma, E)$ and $Q_0(\sigma, \varepsilon)$ are quadratic integral functionals of the auxiliary density $R_0(E) = \chi(2E/(1 + E^2))$ provided by (6.30), *e.g.*,

$$J_0(\sigma, E) = [-\text{sgn}(E) + \fint_{-E_{\max,0}}^{E_{\max,0}} \frac{(e^{\delta\cos(\sigma)} - e^{-\delta\cos(\sigma)}E'^2)}{1-E'^2} \frac{2\nu R_0(E')dE'}{E-E'} +$$
$$E\int_{-E_{\max,0}}^{+E_{\max,0}} \frac{(e^{-\delta\cos(\sigma)} - e^{\delta\cos(\sigma)}E'^2)}{1-E'^2} \frac{2\nu R_0(E')dE'}{1-EE'}]\frac{R_0(E)(e^{\delta\cos(\sigma)} - e^{-\delta\cos(\sigma)}E^2)}{1-E^2}, \quad (6.36)$$

the resulting $R_1(\sigma, E)$ and $r_1(\sigma, \varepsilon)$ are extremely bulky and not revealing, except about one fact: $R_1(\sigma, E) \underset{E\to 0}{\sim} 1/|E|$ and $r_1(\sigma, \varepsilon) \underset{\varepsilon\to 0}{\sim} 1/|\varepsilon|$, which results from $R_0(E) \underset{E\to 0}{\sim} \ln(1/|E|)$. This is why the subtractions $\cosh(.) \to \cosh(.) - 1$ were effected in the inner integrals defining $S$ and $s$ in (6.35), which ensures their convergence by eliminating formally infinite integrals proportional to $\pm \omega\, \delta N \nu \sin(\sigma)$: these vanish on $\sigma$-average, and actually do not affect $S+s$. More importantly, just like in previous subsections the intricate $R_1(\sigma, E)$ and $r_1(\sigma, \varepsilon)$ can be bypassed. Once $E$ is used instead of $B$ in the expressions for $S$ all the integrals in (6.35) acquire the constant bounds $\pm\pi, 0$ or $E_{\max,0}$ at leading order in $\omega$, hence the integrations over $E$ and $\sigma$ may be swapped if needed. Combining integration over $\sigma$ by parts to make $\frac{\partial}{\partial \sigma}R_1(\sigma, E)$ appear, use of the differential equation for $R_1(\sigma, E)$, integration swapping, and integration over $E$ by parts, one deduces an alternate leading-order expression of $S$, *viz.*:

$$S = 4\int_0^{+E_{\max,0}} [\frac{\nu J_0(\sigma, E) E\, e^{-2\delta\cos(\sigma)}}{(1-E^2 e^{-2\delta\cos(\sigma)})}]_{\text{av}}\, dE. \quad (6.37)$$

A similar one, obtained by the substitutions $(\delta, E, J_0(\sigma, E)) \to (-\delta, \varepsilon, Q_0(\sigma, \varepsilon))$, gives $s = S$.

The nice point as to (6.37) is that the denominator of $J_0(\sigma, E)/(1 - E^2 e^{-2\delta\cos(\sigma)})$ goes out if (6.36) holds, rendering the above $\sigma$-average analytically accessible. Through a cumbersome rearrangement which relies on the evenness of $R_0(E)$ and exploits Eqs. (6.26)(6.28) it obeys to get rid of the integrations over $E'$ featured in (6.36), one transforms (6.37) into

$$s = S = \int_0^{E_{\max,0}} \frac{4\nu R_0(E)}{(1-E^2)^2} h(E, \delta) E dE, \quad (6.38)$$
$$h(E, \delta) \equiv [1 - E^2 I_0(2\delta)]g(E, 2N) + 4\nu N[I_0(2\delta) - 1]E - I_0(\delta)(1-E^2),$$

where $g(E, 2N)$ is the same known function as in (6.27). Before coming to the final forms of $S+s$ and of the averaged increment $V_{\text{av}}$ in front length through (6.35), a few remarks are due:

(i) Both crests contribute equally to $V_{\text{av}}$, as expected from their motions in phase opposition.

(ii) Because $R_0(E)$ reaches zero for $E^2 \to E_{max,0}^2$ at least as quickly as $(E_{max,0}^2 - E^2)^{1/2}$, and $g(1,2N) \equiv 4\nu N$, (6.38) holds even inside the shaded region of Fig.9 where $\Theta_{max} = 1 = E_{max,0}$.

(iii) Due to $g(E, 2N) \underset{\delta \to 0}{\to} \text{sgn}(E)$ and $I_0(u) - 1 \underset{u \to 0}{\sim} u^2$, $V_{av} - \tfrac{1}{8}\delta^2 \omega^2 - \mathcal{V}(2\nu N) = S + s$ again is quadratic in the [small] reduced intensity $\delta$ of forcing if $R_0(E) \underset{\delta \to 0}{\to} \cosh^{-1}(E_{max,0}/|E|)/\nu\pi^2$ *uniformly* in $E$; this happens when $2\nu N < \nu N_{opt}(\nu) = \tfrac{1}{2}$ and produces $V_{av} - \mathcal{V}(2\nu N) \underset{\delta \to 0}{\sim} +\delta^2$ with $\mathcal{V}(2\nu N) < \tfrac{1}{2}$. But the peaks that $R_0(E)/(1-E^2)$ acquires near the edges $E^2 = 1$ of its support if $2\nu N > \tfrac{1}{2}$ make $V_{av}$ behave somewhat differently at weak forcing, $V_{av} - \tfrac{1}{8}\delta^2\omega^2 = \tfrac{1}{2} + |\delta|(2N\nu - \tfrac{1}{2})\sqrt{2}/\pi + ...$, compatibly with the value $V_\infty = \tfrac{1}{2}$ belonging to overpopulated, unforced MS flames. The aforementioned peaks at $E \approx \pm 1$ 'store' for $\delta \to 0$ the weakly bound surplus poles that would drift to $\pm i\infty$ when $\delta = 0$.

Such behaviors can be checked using the explicit form of (6.38), obtained by switching to $\Theta = 2E/(1+E^2)$ as the integration variable in (6.38) and doing the integral analytically with $R_0(E) = \chi(\Theta)$ from (6.30) [Appendix D].

Along the ordinary branch of (6.34), where $2\nu N = \mathcal{N}(\Theta_{max}, \delta)$ and $\Theta_{max} \leq 1$, one gets

$$V_{av} - \tfrac{1}{8}\omega^2\delta^2 - \mathcal{V}(2\nu N) \equiv S + s = -\frac{I_0(\delta)(\Theta_*^2 - 1)4\nu N}{\pi \Theta_*} \sin^{-1}(\frac{\Theta_{max}}{\Theta_*}) \;, \qquad (6.39)$$

which also is $(2\nu N)^2(1 - I_0(2\delta)) \leq 0$ here and can be compared with the first line of (6.8).

If $2\nu N > 2\nu N_c(\delta) \equiv \mathcal{N}(1,\delta)$ in Fig.9, the root $\Theta_{max} = 1$ of (6.34) takes over and yields:

$$S + s = -\frac{I_0(\delta)(\Theta_*^2 - 1)4\nu N}{\pi \Theta_*}\sin^{-1}(\frac{1}{\Theta_*}) +$$
$$[2\nu N - \mathcal{N}(1,\delta)][2\nu N(I_0(2\delta) + 1) - 2I_0(\delta)\frac{(\Theta_*^2 - 1)^{1/2} + \Theta_*^2 \sin^{-1}(1/\Theta_*)}{\pi\Theta_*}] \;, \qquad (6.40)$$

which for $2\nu N \geq \tfrac{1}{2}$ goes to $2[2\nu N - \tfrac{1}{2}]^2$ at weak forcing and then gives $V_{av} = \tfrac{1}{2}$. Along the border $N = N_c(\delta)$ in Fig.9 where $2\nu N = \mathcal{N}(1,\delta)$ and $\Theta_{max} = 1$ is a double root of (6.34), the second line of (6.40) vanishes, which matches (6.39) as it must. Sample curves giving $V_{av} = \tfrac{1}{8}\omega^2\delta^2 + 4\nu N(1 - 2\nu N) + S + s + o(1)$ *vs.* $|\delta|$ from (6.39)(6.40) are plotted in Fig.11.

How $V_{av}$ and $\Theta_{max}$ vary with $2\nu N$ at fixed $\omega$ and $\delta$ is illustrated in Fig.12.

Equation (6.40) simplifies a lot for Burgers fronts [add carets and set $I_0(\hat{\delta}) = 0 = \hat{\mathcal{N}}(1,\hat{\delta})$], to give $\hat{S} + \hat{s} = (2\hat{\nu}\hat{N})^2(I_0(2\hat{\delta}) + 1)$. Then $\hat{V}_{av} - \frac{1}{8}\hat{\omega}^2\hat{\delta}^2 = 4\hat{\nu}\hat{N}(0 - 2\hat{\nu}\hat{N}) + \hat{S} + \hat{s}$ shrinks to

$$\hat{V}_{av} - \tfrac{1}{8}\hat{\omega}^2\hat{\delta}^2 = (2\hat{\nu}\hat{N})^2(I_0(2\hat{\delta}) - 1), \qquad (6.41)$$

again a CW-type, parabolic law $\hat{V}_{av} \approx \hat{\delta}^2[\tfrac{1}{8}\hat{\omega}^2 + (2\hat{\nu}\hat{N})^2]$ at small intensities $\hat{\delta}$ of forcing.

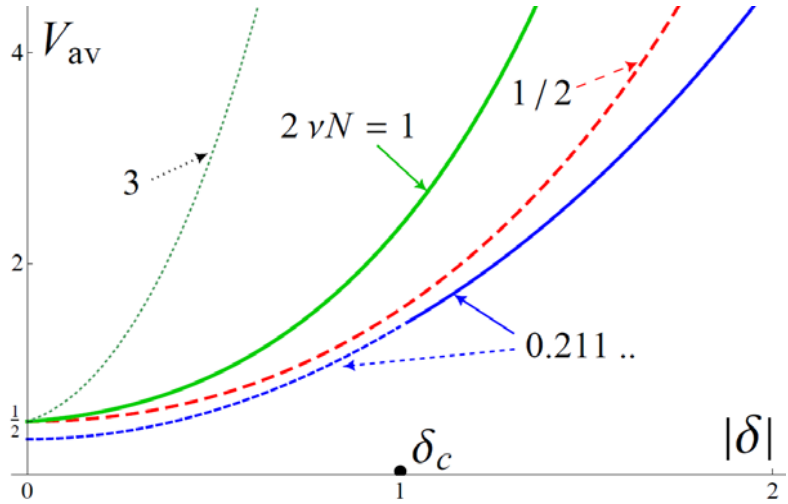

FIG.11. (Color online). $V_{av}$ vs. $|\delta|$ for $\omega = 3$: with $2\nu N = \mathcal{N}(1,1) \approx 0.21104$ [blue lines] for which $\Theta_{max} < 1$ if $|\delta| < \delta_c = 1$ [dotted] and $\Theta_{max} = 1$ otherwise [solid], and $V_{av} \approx \tfrac{1}{3}$ at $\delta = 0$; or with $2\nu N = \tfrac{1}{2}$ [dashed red], $1$ [solid green] and $3$ [dotted black] that have $V_{av} = \tfrac{1}{2}$ at $\delta = 0$ and $\Theta_{max} = 1$ for $\delta \neq 0$.

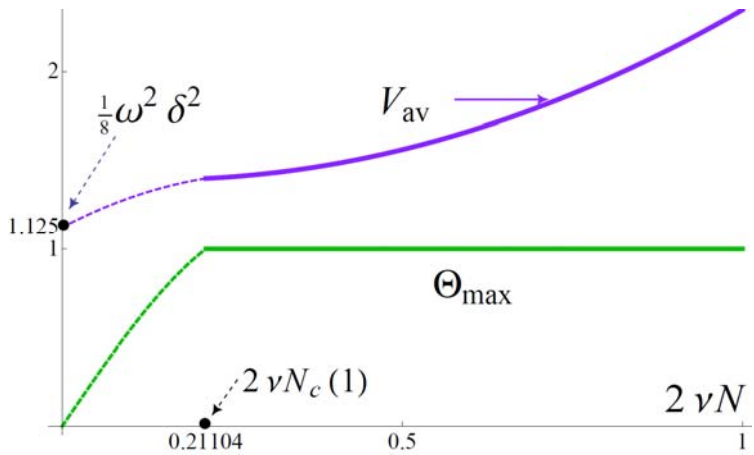

FIG.12. (Color online). $V_{av} = \tfrac{1}{8}\omega^2\delta^2 + 4\nu N(1 - 2\nu N) + S + s$ [violet lines] vs. $2\nu N$, for $\omega = 3$ and $\delta = 1$ according to (6.39)(6.40), and associated root $\Theta_{max}$ of (6.34) [green]. The $N < N_c(1)$ curves [dashed] belong to the unshaded area in Fig.9; for $N > N_c(1)$, $\Theta_{max} = 1$ and $V_{av}$ is a quadratic polynomial of $2\nu N$.

As mentioned earlier, changing the base amplitude $\Phi(t)$ from $-\omega\delta\sin(\sigma)$ to $\delta\omega\frac{d}{d\sigma}C(\sigma)$, with $C(\sigma-\pi) = -C(\sigma)$, $[C(\sigma)]_{av} = 0$ and $[C(\sigma)^2]_{av} = \frac{1}{2}$, just requires to replace each Bessel function $I_0(u)$ by $\mathbf{I}_0(u) \equiv [e^{uC(\sigma)}]_{av} = \mathbf{I}_0(-u)$. No attempt was made to extend subsection VI.C to $\mathbf{I}_0(u) \neq \mathbf{I}_0(-u)$; if $\mathbf{I}_0(-u)/\mathbf{I}_0(u) - 1 << 1$ the transition to $\Theta_{max} = 1$ *at the line* $N = N_c(\delta)$ in Fig.9 is presumably replaced by a softer one, in analogy with (6.15) in Subsection VI.A.

### *6. Stretch*

Once a stretch intensity $w(t) = -d\omega\sin(\sigma+\gamma)$ is restored nothing is changed formally in $R_0(E)$, $E_{max,0}$ and polar slope, up to the substitutions $(\delta, \sigma) \to (\Delta, \sigma-\zeta)$ where the effective amplitude $\Delta$ and the phase $\zeta$ are defined as in Eq. (6.9) of subsection VI A; similarly about adapting the caption to Fig.10 and the grouping $S + s$ in $V_{av} - \frac{1}{8}\omega^2\delta^2 - \mathcal{V}(2\nu N)$.

Yet the full-fledged, stretch affected $V_{av}$ from (3.8) demands significantly more,

$$V_{av} - \tfrac{1}{8}\omega^2\delta^2 - \mathcal{V}(2\nu N) = (S+s) + (S'+s'), \tag{6.42}$$

where $S'+s'$ stems from the large-$N$ limit of $-2d\omega[\sin(\sigma+\gamma)\Sigma_{k=1}^{N}\nu(e^{-B_k(\sigma)} - e^{-b_k(\sigma)})]_{av}$:

$$\begin{aligned}S' &= -\frac{\omega d}{2\pi}\int_{-\pi}^{+\pi}\sin(\sigma+\gamma)d\sigma\int_0^{B_{max}(\sigma)}2\nu\rho(\sigma,B)(e^{-B}-1)dB\ , \\ s' &= +\frac{\omega d}{2\pi}\int_{-\pi}^{+\pi}\sin(\sigma+\gamma)\, d\sigma\int_0^{b_{max}(\sigma)}2\nu\varrho(\sigma,b)(e^{-b}-1)db\ . \end{aligned} \tag{6.43}$$

Since the leading orders $\rho_0(\sigma,B)$ or $\varrho_0(\sigma,b)$ depend on $\sigma$ through $\cos(\sigma-\zeta)$ only, $S'$ [or $s'$] is $\sim \omega$: a *two*-term HF expansion of $S'+s'$ is needed, as in Eq. (6.11) of subsection VI.A.

Because $\rho_0(\sigma,B)$ and $\varrho_0(\sigma,b)$ share the same $R_0(.)$, and a change $\Delta \to -\Delta$ can be compensated by a global time-shift $\sigma \to \sigma - \pi$ that does not affect $\sigma$-averaging, the leading orders $S_0'$ and $s_0'$ of $S'$ and $s'$ coincide:

$$S_0' = s_0' = 4\omega d\nu\sin(\gamma+\zeta)\int_0^{E_{max,0}}\frac{R_0(E)dE}{1-E^2}[\frac{\cos(u)Ee^{-\Delta\cos(u)}}{1+Ee^{-\Delta\cos(u)}}]_{av} . \tag{6.44}$$

Here we used $u \equiv \sigma - \zeta$ and $E = \tanh(\tfrac{1}{2}B)e^{\Delta\cos(u)}$ to get $e^{-B} = (1 - Ee^{-\Delta\cos(u)})/(1+Ee^{-\Delta\cos(u)})$; the zero average of a $u$-derivative was also dropped. The above double integral over $E$ and $u$ could not be exactly done, even in the simpler form that uses its being odd in $\Delta$ and switches to $\Theta = 2E/(1+E^2)$ as the new variable of outer integration : $S_0'/2\omega d \sin(\gamma+\zeta) = -\int_0^{\Theta_{\max}} \nu\chi(\Theta)\Theta[\mathcal{Y}(\cos(u),\Delta,\Theta)]_{av} d\Theta/(1-\Theta^2)$, $\mathcal{Y}(c,\Delta,\Theta) \equiv c\sinh(\Delta c)/(1+\Theta\cosh(\Delta c))$. Yet approximation (6.13) can provides one with a one-variable integral, very accurate if $|\Delta| \leq 7$,

$$S_0' + s_0' \approx -\omega d \sin(\gamma+\zeta) \int_0^{\Theta_{\max}} \frac{\nu\chi(\Theta)\Theta}{(1-\Theta^2)}[\mathcal{Y}(1,\Delta,\Theta) + 2\mathcal{Y}(2^{-1/2},\Delta,\Theta)]d\Theta , \qquad (6.45)$$

with $\nu\chi(\Theta)$ and $\Theta_{\max} \leq 1$ from (6.30)(6.34). As occurred in (6.12) $S_0' + s_0'$ can have any sign and is of order $\omega d\Delta \sin(\gamma+\zeta) \equiv \omega d\delta \sin(\gamma)$ at small intensity $\Delta$ of combined stretch. Though possibly doable [see hints in Appendix D] the integral in $S_0' + s_0'$ was evaluated numerically.

The leading-order Burgers analog $\hat{S}_0' + \hat{s}_0'$ of $\hat{V}_{av} - \frac{1}{8}\hat{\omega}^2\hat{\delta}^2 - (2\hat{\nu}\hat{N})^2(I_0(2\hat{\Delta})-1) = \hat{S}' + \hat{s}'$ has $\hat{\Theta}_{\max} = 1$ in (6.45) and no $I_0(\hat{\Delta})$- weighted grouping in the function $\hat{\chi}(\hat{\Theta})$ defined as in (6.30).

As for the $O(1)$ contributions $S_1'$ and $s_1'$ to $S' + s'$, $B_{\max}(\sigma) - B_{\max,0}(\sigma) \ll 1$ and $b_{\max}(\sigma) - b_{\max,0}(\sigma) \ll 1$ may be omitted from (6.43) because the leading-order pole densities vanish at $B_{\max,0}(\sigma)$ or $b_{\max,0}(\sigma)$. Fortunately one can again bypass the need to plug the bulky expressions of $R_1(E,\sigma)$ and $r_1(\varepsilon,\sigma)$ into the next-to-leading order of (6.43), rewritten in terms of $u = \sigma - \zeta$ and $E = \tanh(\frac{1}{2}B)e^{\Delta\cos(u)}$ or $\varepsilon = \tanh(\frac{1}{2}b)e^{-\Delta\cos(u)}$. Through parity arguments, integrations by parts over $u$, integration swapping, integration by parts over $E$, and by invoking $J_0(\sigma,E_{\max,0}) = 0 = \lim_{E\to 0}(EJ_0(\sigma,E))$, one derives an alternate expression of $S_1'$ [analogous to (6.37)] that only makes use of leading-order quantities:

$$S_1' = \frac{4d\cos(\zeta+\gamma)}{\Delta} \int_0^{E_{\max,0}} \frac{dE}{(1+E)} [\nu J_0(\sigma,E) \frac{(1-e^{-\Delta\cos(u)})}{(1+Ee^{-\Delta\cos(u)})}]_{av} . \qquad (6.46)$$

And $s_1'$, also in principle accessible in terms of previously determined quantities, ensues *via* substitutions $\Delta \to -\Delta$, $E \to \varepsilon$, $J_0(\sigma,E) \to Q_0(\sigma,\varepsilon)$ in (6.46). The remaining difficulty as to effectively evaluate $S_1' + s_1'$ is that the pole fluxes $J_0(\sigma,E)$ and $Q_0(\sigma,\varepsilon)$ still are quadratic functionals of the auxiliary density $R_0(E)$, see (6.36) with $\delta \to \Delta$. Luckily enough, $s_1' = S_1'$. Furthermore, as the denominator of $J_0(\sigma,E)/(1+Ee^{-\Delta\cos(u)})$ can again be cleared when (6.36) holds, (6.46) is reducible to a one-variable integral upon use of the normalization (6.28) and of equation (6.26) which $R_0(E)$ obeys. Switching to $\Theta$ as the integration variable yields

$$S_1' + s_1' = 4d\cos(\zeta+\gamma)\int_0^{\Theta_{max}} \frac{\nu\chi(\Theta)}{(1-\Theta^2)}\kappa(\Theta,\Delta)d\Theta, \quad \Delta\,\kappa(\Theta,\Delta) \equiv$$
$$2\nu N\frac{(1-I_0(2\Delta))\Theta}{(\Theta+1)} + [\frac{(1-I_0(2\Delta))\Theta}{2(\Theta+1)} + I_0(2\Delta) - I_0(\Delta)]\,G(\Theta,2N) + 1 - I_0(\Delta),$$
(6.47)

where $G(\Theta, 2N)$ is the same known function as in (6.29), up to the substitution $\delta \to \Delta$. The latter result shows that $S_1' \propto 2d\Delta\cos(\zeta+\gamma) \equiv 2d[\delta\cos(\gamma)-d]$ at $|\Delta|<<1$. Though plausibly doable exactly when $\chi(\Theta)$ is taken from (6.30), (6.45) and (6.47) are easier to tabulate.

Besides carets the Burgers version of (6.47) again needs to suppress all the $I_0(\hat{\Delta})$- weighted pieces from $\hat{\chi}(\hat{\Theta})$; the final grouping $1 - I_0(\hat{\Delta}) = I_0(0) - I_0(\hat{\Delta})$, stemming from the sgn[.] functions in the pole fluxes (6.21), must also be deleted from $\kappa(\hat{\Theta}, \hat{\Delta})$ in the analog of (6.47).

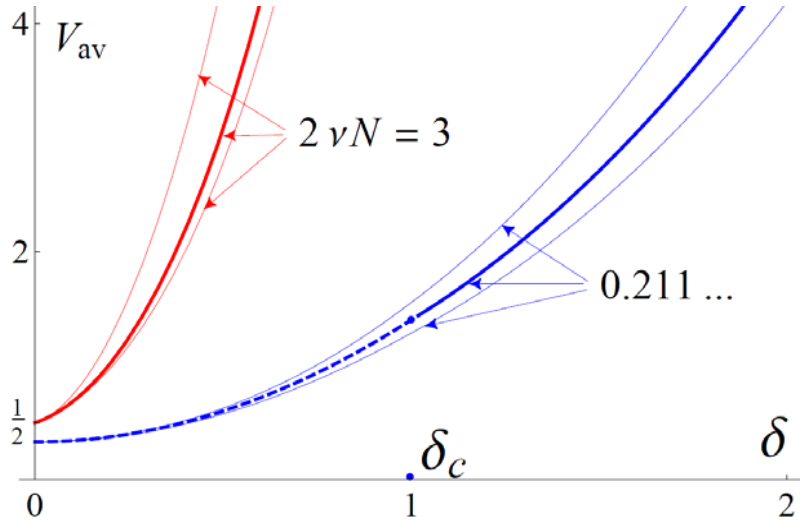

FIG.13. (Color online) $V_{av}$ vs. $\delta$ for $\omega = 3$ and $2\nu N = 3$ [red lines] or $2\nu N_c(1) \approx 0.21104$ [in blue]. In each triplet of curves the thickest one is stretch-free, $d = 0$; that below [or above] it accounts for the influence of a stretch $w(\sigma,x) = -\omega d\sin(\sigma+\gamma)\sin(x)$ of amplitude $d = \delta$ and phase $\gamma = +\frac{\pi}{3}$ [ or $\gamma = -\frac{\pi}{3}$ ].

In sum, equations (6.45)(6.47) along with (6.38)-(6.40) determine how the *time-averaged* increment $V_{av} = \frac{1}{8}\omega^2\delta^2 + \mathcal{V}(2\nu N) + (S+s) + (S_0'+s_0'+S_1'+s_1') + o(1)$ in front arc-length responds to HF variations of forcing and stretch functions whose dependence on time and space are specified in (3.3)(3.6)(3.7) and (5.1)(6.9), in the case that twin crests with $N = n \sim 1/\nu$ $\gg 1$ pairs of poles interact. This *three*-term HF expansion of $V_{av}$ fully accounts for the non-linear coupling with the Darrieus-Landau instability term and the non-trivial pole dynamics illustrated in Fig.10. Yet the sample curves $V_{av}$ *vs.* $\delta$ in Fig.13 stubbornly exhibit a CW-type parabolic growth, except for an offset $\mathcal{V}(2\nu N)$ known from free propagations and a nonzero initial slope of $(2\nu N - 1/2)2^{1/2}/\pi$ if [and only if] $2\nu N \geq 1/2$.

## 7. Intensity of outer flow

Though useful in analyses the reduced amplitudes $\delta$ or $\Delta$ are not particularly telling. More intuitive norms of the forcing and stretch intensities can be defined as $u_{rms}^2 \equiv [<u^2(t,x)>]_{av}$ and $w_{rms}^2 \equiv [<w^2(t,x)>]_{av}$, in which spatial and time-wise averages may be swapped. These could serve to quantify the imposed flow-field modulations $(w_\infty, u_\infty)$ ahead of the flame.

Because $u(t,x) = u_\infty \, a(\mathcal{A})/u_L \Omega(\mathcal{A})^2$ and $w(t,x) = w_\infty / u_L \Omega(\mathcal{A})$ are scaled in different ways with respect to the Atwood number $\mathcal{A}$ [Appendix A] the simple grouping $(u_{rms}^2 + w_{rms}^2)^{1/2} = (a/\Omega^2 u_L)[u_{\infty, rms}^2 + w_{\infty, rms}^2 \Omega^2/a^2]^{1/2}$ mainly measures $u_\infty$ if $\mathcal{A} \ll 1$ [$\Omega(\mathcal{A}) \to 0$, $a(\mathcal{A}) \to 1$] and needs $\mathcal{A} = O(1)$ to account for $w_\infty$. Recall that the maximum arclength increment of free flames, belonging to $V_{av} = \max(V_\infty) = \frac{1}{2}$, is $\frac{1}{2}\Omega^2/a^2$ [37]. Besides, using the DL growth rate $\Omega(\mathcal{A})$ [2, 3] and $\Omega(\mathcal{A})/a(\mathcal{A})$ from [12] [which gives fair agreement with experiments on free flames] one gets $\Omega = a$ for $\mathcal{A} \approx 0.755$ [fresh-to-burnt density ratio $E \approx 7.16$, $a \approx 1.61$].

So, whereas $(u_{rms}^2 + w_{rms}^2 \, a^2/\Omega^2)^{1/2}$ would possibly be better suited [but $\mathcal{A}$-dependent] in general, $(u_{rms}^2 + w_{rms}^2)^{1/2}$ is selected here as a fiducial global measure of $(u_\infty, w_\infty)$. When (3.6) (3.7) and (5.1)(6.9) hold this is

$$u' \equiv (u_{rms}^2 + w_{rms}^2)^{1/2} = \tfrac{1}{2}\omega |\delta| [\omega^2 + \tfrac{3}{64}\omega^2 \delta^2 + (2\nu N_{tot} - 1 + \nu)^2 + \alpha]^{1/2},$$
$$\alpha = 0 \text{ if } d = 0, \quad \alpha = 1 + 2\nu N_{tot}(1 - \nu \mp \omega\sqrt{3}) \text{ if } d = \delta, \; \gamma = \pm\tfrac{\pi}{3}.$$
(6.48)

The results in Figs 3 and 13 are re-plotted in terms of $u'/(\tfrac{1}{2}\omega^2)$ in Figs 14 and 15. Besides the change of scale encoded in $\tfrac{1}{2}\omega^2$, the main trend is a milder growth of $V_{av}$ than when plotted *vs.* $\delta$, be it when $d = 0$ or with $d = \delta$ and $\gamma = \pm\tfrac{\pi}{3}$, because $u'/(\tfrac{1}{2}\omega^2)$ is $|\delta|[1 + \tfrac{3}{64}\delta^2]^{1/2}$ for $\omega \to \infty$. Similar curve distortions are expected in the Low-Frequency analyses to come.

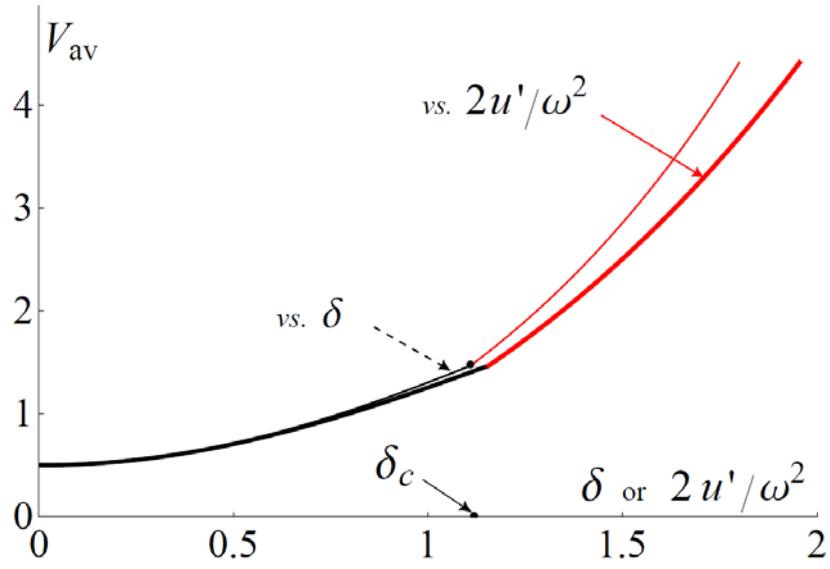

FIG.14. (Color online) $V_{av}$, here with $\omega = 3$, $d = 0$, $N = 1$, $\nu = \tfrac{1}{2}$, plotted vs. $\delta$ as in Fig.3 [thin lines, in black for $\delta \le \delta_c(\tfrac{1}{2}) \approx 1.109$ and red otherwise] or vs. $u'/(\tfrac{1}{2}\omega^2)$ from (6.48) [thicker lines].

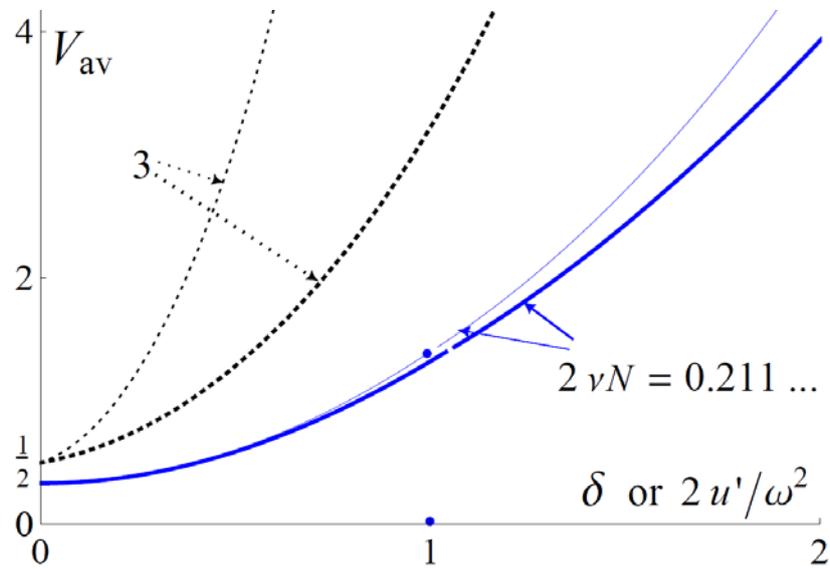

FIG.15. (Color online) $V_{av}$, here with $\omega = 3$, $d = 0$ and $2\nu N = 2\nu N_c(1) \approx 0.21104$ [solid blue lines] or $2\nu N = 3$ [dotted black], plotted vs. $\delta$ as in Fig.13 [thin lines] or vs. $u'/(\tfrac{1}{2}\omega^2)$ [thicker lines].

# VII. LOW FREQUENCIES

The other extreme $\omega \to 0$ of low frequencies [LF] is now envisaged. To explore the fully nonlinear dynamics $\Phi(\sigma) = \Phi \sin(\sigma)$ and $w(\sigma) = w \sin(\sigma + \gamma)$, with $\Phi = -\omega \delta$ and $w = -\omega d$, are now viewed $O(1)$. So is $\Phi(t) - w(t) = -\omega \Delta \sin(\sigma - \zeta)$, whose combined amplitude $\Delta$ and phase $\gamma$ are defined as in (6.9); and the same for the hatted, Burgers analogs thereof.

## A. One pair of poles

To take up the LF limit it is now simpler to begin with the Burgers case. The dynamics of the upper pole $i\hat{B}(\sigma)$ is governed by the analog of (5.3) where $\tanh(\tfrac{1}{2}\hat{B}(\sigma)) = e^{-\hat{\lambda}(\sigma)} > 0$

$$\frac{d\hat{\lambda}}{d\sigma} = \frac{-\hat{v}\sinh(\hat{\lambda})\cosh(\hat{\lambda}) - \hat{\omega}\hat{\Delta}\sin(\sigma - \hat{\zeta})}{\hat{\omega}}, \quad \sigma \equiv \hat{\omega} t. \tag{7.1}$$

Slow motions with $\tfrac{d}{d\sigma}\hat{\lambda} = O(1)$ clearly require for $\hat{\omega} \to 0$ that the numerator on the right of (7.1) be $O(\hat{\omega})$. The so-called slow manifold of (7.1) is here a curve, defined by:

$$\hat{\omega}\hat{\Delta}\sin(\sigma - \hat{\zeta}) \underset{\hat{\omega} \to 0}{=} -\tfrac{1}{2} v \sinh(2\hat{\lambda}). \tag{7.2}$$

Should the initial conditions give (7.1) a $O(1)$ numerator, $\tfrac{d}{d\sigma}\hat{\lambda} \sim 1/\hat{\omega}$ would make the $\hat{\lambda}(\sigma)$ trajectory rush at nearly constant $\sigma$ towards (7.2), along which $\tfrac{d}{d\hat{\lambda}}[-\tfrac{1}{2}v\sinh(2\hat{\lambda})] < 0$. As the slow evolution along (7.2) makes $\hat{\lambda}(\sigma)$ change sign at $\sigma - \zeta = \mathbb{Z}\pi$, the crest spends half its lifetime at $x = 0 (\mathrm{mod}\, 2\pi)$ and is shifted by $\pi$ during the other half. From the Burgers version of (3.8) the leading-order current increase $\hat{V}(\sigma)$ in front arclength ensues,

$$\hat{V}(\sigma) \underset{\hat{\omega} \to 0}{=} 2\hat{v}(0-\hat{v}) + \tfrac{1}{4}\hat{\omega}^2 \hat{\delta}^2 \sin^2(\sigma - \hat{\zeta}) + \\ -2\hat{v}\hat{\omega}\hat{\Delta}\sin(\sigma - \hat{\zeta})\coth[\hat{\lambda}(\sigma)] - 2\hat{v}\hat{\omega}\hat{d}\sin(\sigma + \hat{\gamma})\tanh(\tfrac{1}{2}\hat{\lambda}(\sigma)), \tag{7.3}$$

in which a time derivative featured in (3.8) could be omitted. Once time-averaged, using that $\lambda(\sigma)$ is an odd function of $u = \sigma - \zeta$, (7.3) produces

$$\hat{V}_{\mathrm{av}}/\hat{v}^2 - \tfrac{1}{8}(\hat{\omega}\hat{\delta}/\hat{v})^2 = \tfrac{2}{\pi}E[-(2\hat{\omega}\hat{\Delta}/\hat{v})^2] - 1 + 2(\hat{\omega}\hat{d}/\hat{v})\cos(\hat{\gamma} + \hat{\zeta})F[2\hat{\omega}\hat{\Delta}/\hat{v}] \tag{7.4}$$

where $F[\varpi] \equiv [\sin(u)\tanh(\tfrac{1}{4}\sinh^{-1}(\varpi \sin(u)))]_{\mathrm{av}}$ and $E[.]$ is the complete elliptic integral of 2nd kind. This simple case points to a difficulty of LF analyses: even once the pole dynamics is available, taking time-averages generates integrals that cannot be done analytically; here the easy-to-tabulate $F[\varpi]$ lies fairly close to $\frac{1}{8}\sinh^{-1}[\varpi/(1+(\frac{1}{40}\varpi)^2)]$ for $|\varpi| \leq 20$.

Interestingly, the phase $\hat{\gamma}$ between stretch $\hat{w}(\sigma)$ and $\hat{\Phi}(\sigma)$ enters $\hat{V}_{av}$ through $\cos(\hat{\gamma}+\hat{\zeta}) = (\hat{\delta}\cos(\hat{\gamma})-\hat{d})/\hat{\Delta}$ which is even in $\hat{\gamma}$, whereas the leading-order HF result mainly involved $\sin(\hat{\gamma}+\hat{\zeta}) = \sin(\hat{\gamma})\hat{\delta}/\hat{\Delta}$ that changes sign with $\hat{\gamma}$, see (6.11). As is indicated by the linear analyses prior to (5.5), this results from the pole motions and $\hat{\Phi}(\sigma)-\hat{w}(\sigma)$ being here in phase, whereas they were in quadrature of phase at high frequencies. The curves $\hat{V}_{av}$ vs. $\hat{\omega}\hat{\delta}$ drawn in Fig.16 for $\hat{d}=0$, or with $\hat{d}=\hat{\delta}$ and $\hat{\gamma}=\pm\frac{\pi}{3}$, illustrate the point.

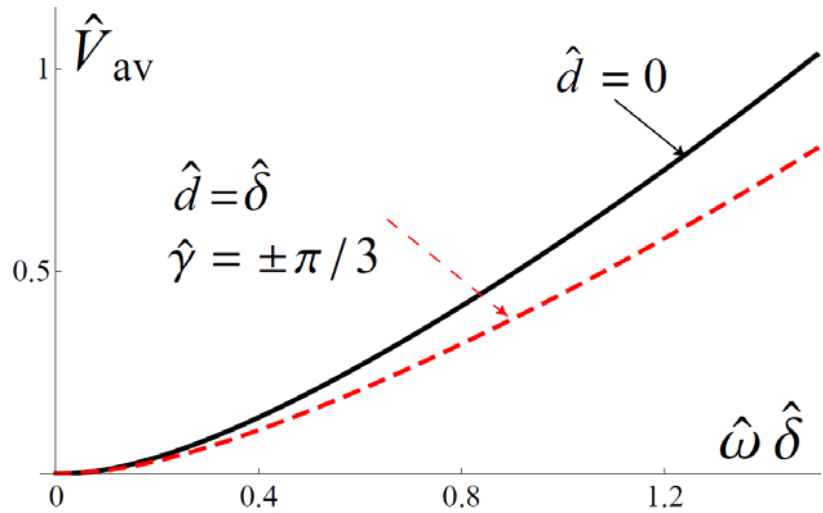

FIG.16 (Color online). Time-averaged length increment $\hat{V}_{av}$ of $\hat{N}=1$ quasi-steady Burgers fronts vs. $\hat{\omega}\hat{\delta}$, for $\hat{v}=\frac{1}{2}$ and $\hat{d}=0$ [solid black], or stretch affected with $\hat{d}=\hat{\delta}$ and $\gamma=\pm\frac{\pi}{3}$ [dashed red].

Accounting for the DL instability effect removes the carets and merely modifies (7.1) to

$$\frac{d\lambda}{d\sigma} = \frac{\sinh(\lambda)-v\sinh(\lambda)\cosh(\lambda)-\omega\Delta\sin(\sigma-\zeta)}{\omega}. \tag{7.5}$$

If $v<1$ the innocuous-looking change significantly affects the slow curve

$$\omega\Delta\sin(\sigma-\zeta) \underset{\omega\to 0}{=} \sinh(\lambda)-v\sinh(\lambda)\cosh(\lambda), \tag{7.6}$$

because its right-hand side becomes a non-monotonic function of $\lambda$, see Fig.17.

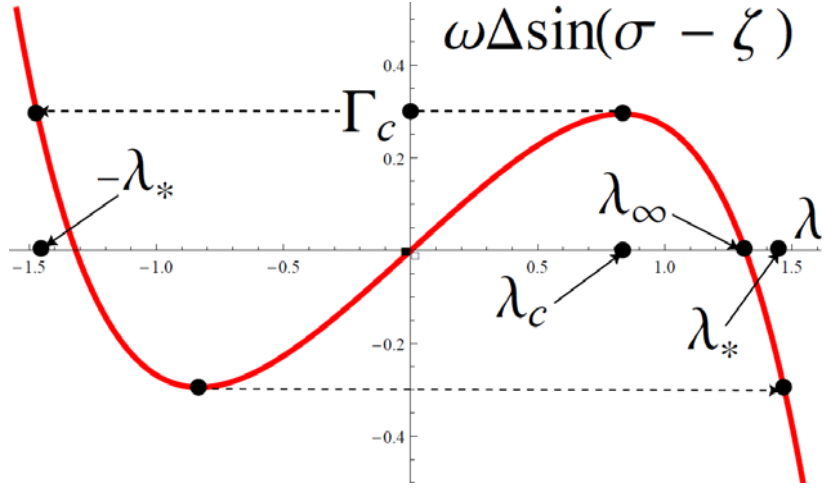

FIG.17 (color online) Slow manifold (7.6) of (7.5) for $\nu = \frac{1}{2}$ [solid red line], and the fast motions [dashed]

As long as $|\omega\Delta| < \Gamma_c \equiv \sinh(\lambda_c) - \frac{1}{2}\nu\sinh(2\lambda_c)$ with $\lambda_c \equiv \cosh^{-1}((1+\sqrt{1+8\nu^2})/4\nu)$, e.g., $\Gamma_c = \frac{1}{4}(6\sqrt{3}-9)^{1/2} \approx 0.295$ and $\lambda_c = \cosh^{-1}(\frac{1}{2}+\frac{\sqrt{3}}{2}) \approx 0.8314$ for $\nu = \frac{1}{2}$, the slow motion consists of a constant-sign $\lambda(\sigma)$ that smoothly oscillates about $\pm\lambda_\infty = \pm\cosh^{-1}(1/\nu)$: this pertains to a crest wobbling at a fixed abscissa, $x = 0$ if $\lambda > 0$ or $x = \pi$ if $\lambda < 0$, in the central front cell.

When $|\omega\Delta| > \Gamma_c$ though, $\lambda(\sigma)$ undergoes relaxation oscillations whose fast motions at $\sigma \approx$ const. drive $\lambda(\sigma)$ towards the parts of the slow curve where $\frac{\partial}{\partial\lambda}[\sinh(\lambda) - \nu\frac{1}{2}\sinh(2\lambda)] < 0$; in doing so $\lambda(\sigma)$ changes sign, implying jumps by $\pm\pi$ in pole and crest locations. The near-discontinuous trajectories make the time-derivative from (3.8), and $V(\sigma)$ itself, akin to Dirac deltas, yet with weights of alternating signs so these disappear on time-average: the leading-order $V_{av}$ only depends on the slower motions. As the stable slow-curve arcs are traversed twice per period when jumps are absent, $\lambda(\sigma)$ is an even function of $\sigma - \zeta - \frac{\pi}{2}$, just like $\sin(\sigma - \zeta)$; but the jumps break the symmetry and endow the $\lambda(\sigma)$ trajectories in Fig.17 with an overall counterclockwise rotation.

For $|\omega\Delta| < \Gamma_c$, $V_{av}$ is ultimately found to be given by:

$$V_{av} - \tfrac{1}{8}(\omega\delta)^2 - 2\nu(1-\nu) = -2\nu J'(\omega\Delta) + 2\nu\cos(\gamma+\zeta)J''(\omega\Delta)d/\Delta , \qquad (7.7)$$

$$J'(\omega\Delta) = \int_{\lambda_+(\omega\Delta)}^{\lambda_-(\omega\Delta)} \frac{\mathfrak{r}(\lambda,\omega\Delta,\nu)d\lambda}{\pi\,\sinh^2(\lambda)}, \quad J''(\omega\Delta) = \tfrac{1}{2}\int_{\lambda_+(\omega\Delta)}^{\lambda_-(\omega\Delta)} \frac{\mathfrak{r}(\lambda,\omega\Delta,\nu)d\lambda}{\pi\,\cosh^2(\tfrac{1}{2}\lambda)},$$
$$\mathfrak{r}(\lambda,\omega\Delta,\nu) \equiv \sqrt{(\omega\Delta)^2 - [\sinh(\lambda) - \tfrac{\nu}{2}\sin(2\lambda)]^2} \qquad (7.8)$$

with $\lambda_+(\omega\Delta) < \lambda_-(\omega\Delta)$ the positive roots of $\sinh(\lambda_\pm) - \frac{1}{2}\nu\sinh(2\lambda_\pm) = \pm\omega|\Delta|$. To get (7.7) (7.8), $[\cos(\sigma-\zeta)\tanh(\frac{1}{2}\lambda)]_{av} = 0$ was used and the non-zero contributions to $V_{av}$ were accessed from the slow motion over $-\frac{\pi}{2} \leq \sigma - \zeta \leq \frac{\pi}{2}$. One indeed notes that $[\sin(u)\coth(\lambda)]_{av} \propto [\cosh(\lambda) - \nu\cosh^2(\lambda)]_{av}$ and $[\sin(u)\tanh(\frac{1}{2}\lambda)]_{av} \propto [\sinh^2(\frac{1}{2}\lambda)(1-\nu\cosh(\lambda))]_{av}$ with $u = \sigma - \zeta$, which result from (7.5) *via* trig identities; since $\lambda$ is an odd function of $\sin(u)$, each of the latter two even functions of $\lambda$ has the same average over $(0, 2\pi)$, $(-\frac{1}{2}\pi, \frac{1}{2}\pi)$ or $(\frac{1}{2}\pi, \frac{3}{2}\pi)$, just like $\sin^2(u)$. One next trades $u = \sigma - \zeta$ for $\lambda$ as the integration variable and performs an integration by parts, taking advantage that $\mathfrak{r}(\lambda_\pm, \omega\Delta, \nu) = 0$. For $|\omega\Delta| \to 0$ one deduces $\lambda_\pm(\omega\Delta) \approx \lambda_\infty \mp \omega|\Delta|/\omega_\infty$ with $\lambda_\infty = \cosh^{-1}(1/\nu)$ and $\omega_\infty = 1/\nu - \nu$ as in subsection V, and $\int_{\lambda_+}^{\lambda_-}\mathfrak{r}(\lambda,\omega\Delta,\nu)d\lambda \to \frac{\pi}{2}(\omega\Delta)^2/\omega_\infty$: as it should, the $O(\omega^2\Delta^2)$ right-hand side of (7.7) resumes the LF limit of (5.5) and, in particular, $V_{av} \to 2\nu(1-\nu)$ for $\omega\Delta \to 0$. Being out of analytical reach for general $|\omega\Delta| < \Gamma_c$ the above integrals must be evaluated numerically.

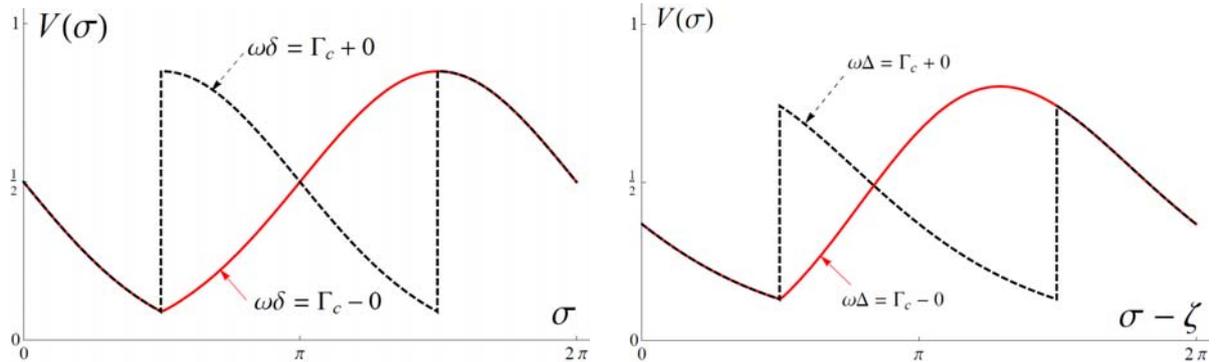

FIG.18 (color online) Front-length increment $V(\sigma)$ for $\nu = \frac{1}{2}$. *Left panel*: no-stretch situation [$d = 0$], with $\omega\delta = 0.2949$ [solid red] slightly below $\Gamma_c(\frac{1}{2}) = \frac{1}{4}(6\sqrt{3}-9)^{1/2} \approx 0.29499$, or slightly above it, $\omega\delta = 0.2950$ [dashed black]; the time-averages are $V_{av} \approx 0.494$ or $V_{av} \approx 0.495$, respectively. *Right panel*: a stretched case with $d = \delta$, $\gamma = +\frac{\pi}{3}$ and $\omega\Delta = 0.2949$ [solid red] or $0.2950$ [dashed black]; the resulting $V_{av}$ are $\approx 0.487$ and $\approx 0.396$, respectively [$\gamma = -\frac{\pi}{3}$ would instead give $V_{av} \approx 0.487$ or $\approx 0.580$].

Noise added to $\Phi(\sigma) - w(\sigma)$ will induce erratic jumps in crest location if $|\omega\Delta| \approx \Gamma_c$, for *two* trajectories are allowed in the critical case $|\omega\Delta| = \Gamma_c$ and only one has jumps. Albeit quite different in shapes the $V(\sigma)$s share the same average if stretch is omitted, *i.e.*, when $d = 0$ and $\Delta = \delta$ [Fig.18, left]: both $V_{av}$ follows (7.8) with $\lambda_+(\omega\delta) = \lambda_c(\nu)$ and $\lambda_-(\omega\delta)$ equal to the

then sole root $\lambda_*(\nu) > \lambda_\infty$ of $\sinh(\lambda_*) - \frac{1}{2}\nu\sinh(2\lambda_*) = -\Gamma_c$, see Fig.17. If $d \neq 0$ though, the parity of $\lambda(\sigma)$ in $\sigma - \zeta - \frac{\pi}{2}$ and its anti-symmetry in $\sigma - \zeta$ are destroyed whenever jumps occur, rendering $V_{av}$ discontinuous at $|\omega\Delta| = \Gamma_c$, see Figs. 18 and 20.

If $\omega|\Delta| > \Gamma_c$ the jumps slightly complicate the access to $V_{av}$ but the $\sigma$-average can still be done by changing the variable to $\lambda(\sigma)$ piecewise, along each stable branch of (7.6) where $\frac{d}{d\lambda}[\omega\Delta\sin(\sigma - \zeta)] < 0$. For example the contribution $J'(\omega\Delta)$ to (7.7) acquires the new form

$$\pi J'(\omega\Delta) = \{\int_{\lambda_c(\nu)}^{\lambda_*(\nu)} + 2\int_{\lambda_*(\nu))}^{\lambda_-(\omega\Delta)}\}\frac{\mathfrak{r}(\lambda,\omega\Delta,\nu)d\lambda}{\sinh^2(\lambda)} - \sqrt{(\omega\Delta)^2 - \Gamma_c^2}[\coth(\lambda_*) + \coth(\lambda_c)] \ , \qquad (7.9)$$

and that of $J''(\omega\Delta)$ presents the same structure provided $-\coth(\lambda)$ and $1/\sinh^2(\lambda)$ are replaced with $\tanh(\frac{1}{2}\lambda)$ and $\frac{1}{2}/\cosh^2(\frac{1}{2}\lambda)$, respectively. Note that (7.9)(7.8) do coincide if $|\omega\Delta| = \Gamma_c$, in which case $\lambda_-(\omega\Delta) = \lambda_*(\nu)$. The main novelty is that $V_{av}$ acquires yet another contribution $2\nu\sin(\gamma + \zeta)J'''(\omega\Delta)d/\Delta$ because the jumps of $\lambda(\sigma)$ break its symmetry in $\sigma - \zeta = \frac{\pi}{2}$, implying $[\cos(\sigma - \zeta)\tanh(\frac{1}{2}\lambda(\sigma))]_{av} \neq 0$. As the expression of $J'''(\omega\Delta)$ is even less telling than (7.9), and in any case must be evaluated numerically, it is omitted here.

When $\nu \geq 1$, corresponding to DL-stable fronts, the slow manifold of Fig.17 resumes a decreasing odd function of $\lambda$ akin to (7.2). As a result $\lambda(\sigma)$ slowly oscillates evenly between $\lambda_-(\omega\Delta)$ and $-\lambda_-(\omega\Delta)$, with $\lambda_-(\omega\Delta) > 0$ being the now unique root of $\sinh(\lambda) - \frac{1}{2}\nu\sinh(2\lambda) = -|\omega\Delta|$; since the parity of $\lambda(\sigma)$ in $\sigma - \zeta - \frac{\pi}{2}$ is restored, $[\cos(\sigma - \zeta)\tanh(\frac{1}{2}\lambda)]_{av} = 0$, and $V_{av}$ does not contain a $\propto \sin(\gamma + \zeta)$ contribution any longer. However, because $\lambda(\sigma)$ is now driven to zero at a slow pace, twice in a cycle, the integral $J'(\omega\Delta)$ again needs be modified to avoid the divergence that $1/\sinh^2(\lambda)$ would otherwise induce in (7.8). A suitable subtraction in the integration-by-parts step towards $[\sin(\sigma - \zeta)\coth(\lambda)]_{av}$ then produces:

$$\pi J'(\omega\Delta) = \int_{-\lambda_-(\omega\Delta)}^{+\lambda_-(\omega\Delta)} \frac{(\mathfrak{r}(\lambda,\omega\Delta,\nu) - |\omega\Delta|)d\lambda}{\sinh^2(\lambda)} - 2|\omega\Delta|\coth(\lambda_-(\omega\Delta)) \ . \qquad (7.10)$$

Expression (7.8) of $J''(\omega\Delta)$ still holds true formally, with $-\lambda_-(\omega\Delta)$ substituted for $\lambda_+(\omega\Delta)$. For $|\omega\Delta| \to 0$, $\lambda_- \approx |\omega\Delta|/(\nu - 1)$ whereby the integral in (7.10) reduces to $(\nu - 1)(2 - \pi)$ and $-2\nu J'(\omega\Delta)$ goes to $-2\nu(1-\nu)$; because $J''(\omega\Delta)$ vanishes at $\omega\Delta = 0$, $V_{av}$ also does and is of the Clavin-Williams type [30, 31] for $\omega\delta \ll 1$ in all $\nu > 1$ situations, see Fig.19.

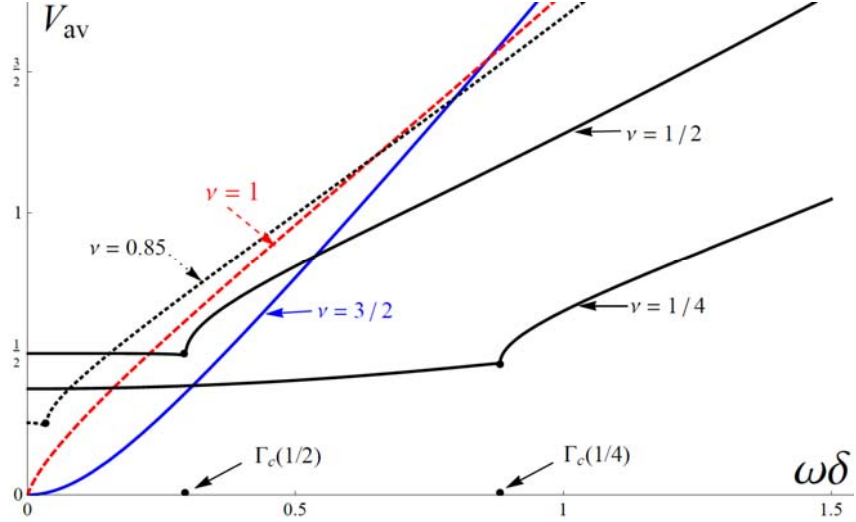

FIG.19 (color online) Time-averaged length increment $V_{av}$ of a one-pair LF flame, *vs.* amplitude $\omega\delta$ of base-shape oscillations, for $d = 0$ and selected values of $\nu$. For $\nu \to 1^-$, $\Gamma_c(\nu) \approx [\tfrac{1}{3}(1-\nu^2)]^{3/2}$ fades and the $V_{av}$ *vs.* $\omega\delta$ curve acquires an infinite initial slope. Yet all $\nu > 1$ cases lead to $V_{av} \sim (\omega\delta)^2$ near the origin.

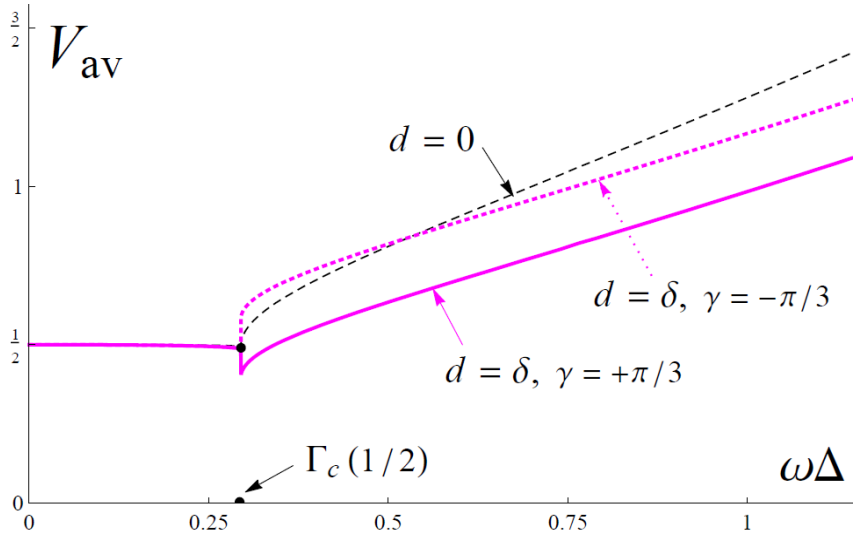

FIG. 20 (color online) Time-averaged length increment $V_{av}$ *vs.* amplitude $\omega\delta$ of base front oscillations, for $\nu = \tfrac{1}{2}$. The thick magenta lines account for stretch with $d = \delta$ and $\gamma = +\tfrac{\pi}{3}$ [solid] or $-\tfrac{\pi}{3}$ [dotted] in (3.3); the $d = 0$ curve belonging to stretch-less fronts as in Fig.19 is recalled for comparison [thin black dashed]. The variation of $V_{av}$ with $\gamma$ at fixed $\nu < 1$ and $\omega\Delta > \Gamma_c(\nu)$ generically scales as $\sin(\gamma+\zeta)$ [$= \sin(2\gamma)$ here].

To complete Sub-Section VII.A the stretch affected Fig.20 is re-plotted against the outer-flow intensity $u'$ from (6.48), which with $N=1$, $\nu=\frac{1}{2}$, $d=\delta$ and phases $\gamma=\pm\frac{\pi}{3}$ reads $u'=\frac{1}{2}\omega\delta[\frac{3}{64}\omega^2\delta^2+\frac{7}{4}]^{1/2}$ for $\omega\to 0$.

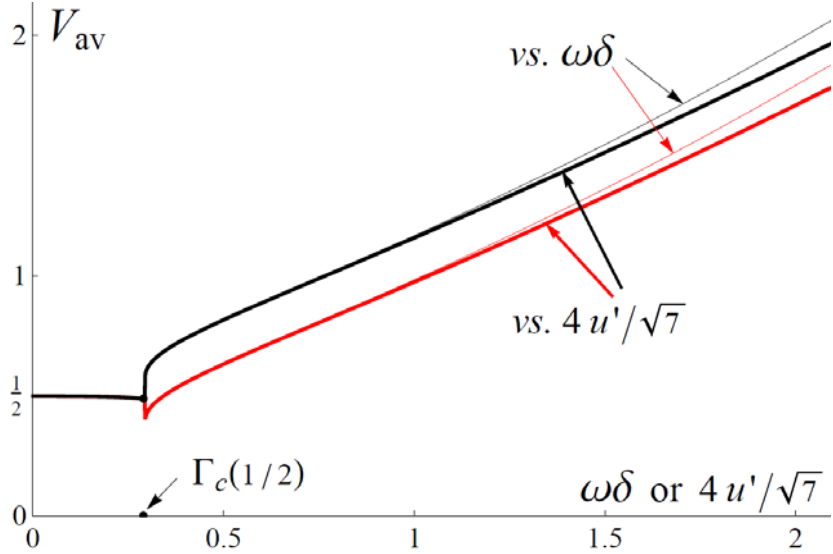

FIG.21 (color online) Time-averaged length increment $V_{av}$ for $\nu=\frac{1}{2}$ and $d=\delta$, plotted *vs.* the amplitude $\omega\delta$ of base front oscillations [thin curves] or *vs.* $4u'/\sqrt{7}=\omega\delta\,[1+\frac{3}{112}\omega^2\delta^2]^{1/2}$, see (6.48). The upper [or lower] doublet of thin/thick curves corresponds to a phase of $\gamma=-\frac{\pi}{3}$ [or $\gamma=+\frac{\pi}{3}$]. Curve distortion brought about by re-plotting *vs.* $u'$ is stronger in the $d=0$ situation, for which $4u'=\omega\delta\,[1+\frac{3}{16}\omega^2\delta^2]^{1/2}$.

One may finally acknowledge that accessing $V_{av}$ via (7.7)-(7.10) actually is more intricate than numerical evaluations of $[\omega\Delta\sin(u)\coth(\lambda(u))]_{av}$ and $[\omega d\sin(u+\gamma+\zeta)\tanh(\frac{1}{2}\lambda(u))]_{av}$, with $\lambda(u)$ iteratively deduced from (7.6) at any current $u=\sigma-\zeta$. Moreover, the analyses and effective evaluations of the integrals required by the simplest LF dynamics starkly contrast with the fact that numerical routines [*e.g.*, in [35]] supply $\lambda(u)$ and $V_{av}$ in no time at all, even if employed with a tiny $\omega$. All this will help one shorten next Subsection.

### B. Two pairs at Low Frequency

The starting point adopted here is a version of (6.16) for the functions $\lambda(\sigma)$ and $\mu(\sigma)$ that encode the dynamics of $z_1=iB=2i\tanh^{-1}(e^{-\lambda})$, $z_2-\pi=ib=2i\tanh^{-1}(e^{-\mu})$ and conjugates. Specifically, (6.16) is rewritten in terms of $p\equiv\frac{1}{2}(\lambda+\mu)$ and $m=\frac{1}{2}(\lambda-\mu)$:

$$\omega \frac{dp}{d\sigma} = \sinh(p)\cosh(m) - \tfrac{\nu}{2}\sinh(2p)\cosh(2m) - \nu\frac{\cosh(2p)-\cosh(2m)}{\sinh(2p)},$$

$$\omega \frac{dm}{d\sigma} = \sinh(m)\cosh(p) - \tfrac{\nu}{2}\sinh(2m)\cosh(2p) - \omega\delta\sin(\sigma) \ .$$

(7.11)

Dropping both left-hand sides in the low-frequency limit $\omega \to 0$ leads to

$$\sinh(p)\cosh(m) - \tfrac{\nu}{2}\sinh(2p)\cosh(2m) - \nu\frac{\cosh(2p)-\cosh(2m)}{\sinh(2p)} = 0, \qquad (7.12)$$

$$\omega\delta\sin(\sigma) = \sinh(m)\cosh(p) - \tfrac{\nu}{2}\sinh(2m)\cosh(2p) \ . \qquad (7.13)$$

To convert (7.13) to a slow curve, (7.12) could be solved for $\cosh(m)$, but $m(p;\nu)$ is multi-valued and has turning points; so, it proved more convenient to express $p$ as a function of $m$.

Pole repulsion [18] precludes pole-crossing, which forbids $\lambda + \mu = 0$ [i.e., $z_1 = z_2 \pmod{2\pi}$] unless $\lambda = 0 = \mu$ [i.e., both $z_1$ and $z_2$ at $i\infty$] : $p$ stays nonnegative if it starts so. Then (7.12) provides a single $p(|m|;\nu) \geq 0$ for any $-\infty < m < \infty$, rendering the right-hand side of (7.13) expressible numerically in terms of $m$ and $\nu$. Given the near-identity of (7.12) and (6.18) up to the correspondence $[p, \cosh(m), \cosh(2m)] \leftrightarrow [\lambda_0, I_0(\delta), I_0(2\delta)]$, the curves $p(|m|;\nu)$ vs. $m$ resembles those in Fig.5, especially at small $m$ since $\cosh(u) \approx I_0(u\sqrt{2})$ for $2u^2 \leq 1$.

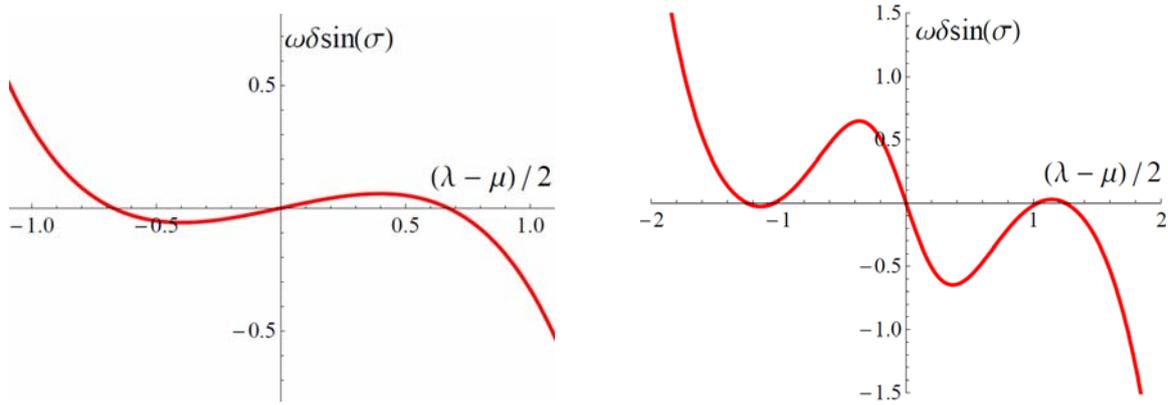

FIG.22. (Color online). Slow curves $\omega\delta\sin(\sigma)$ vs. $m = \tfrac{1}{2}(\lambda - \mu)$ pertaining to (7.11) for $\omega \to 0$, with $\nu = 0.488279$ [left] or $\nu = \tfrac{1}{4}$. Fast motions to the left [right] drive $m(\sigma)$ at $\sigma \approx \text{const.}$ towards the decreasing parts of each slow curve if $\omega\delta\sin(\sigma)$ is above [below] it. For $\nu \geq 1$ [DL-stable fronts] the slow curves monotonically decrease as in Subsection VII.A; and the same for Burgers fronts [as in (7.2)].

For all $\nu \geq 1$ the resulting slow curve $\omega\delta\sin(\sigma) = f(m;\nu) = -f(-m;\nu)$ has $f(0;\nu) = 0$ and $\tfrac{\partial}{\partial m}f(m;\nu) \leq 0$. In such instances $\lambda(\sigma)\mu(\sigma)$ is negative throughout, meaning that the poles belong to the same crest; the crest spends half its lifetime [$m < 0$] at $x = 0$ and is shifted by $\pi$ during the other half, as expected of DL-stable fronts enslaved by forcing. The analog of (7.12) for Burgers fronts gives $\sinh^2(2\hat{p}) = 2[1 + \cosh^2(2\hat{m}) - (1 + 3\cosh^2(2\hat{m}))^{1/2}]/\cosh^2(2\hat{m})$ whatever $\hat{v} \geq 0$ is, indeed leading to exactly the same trends.

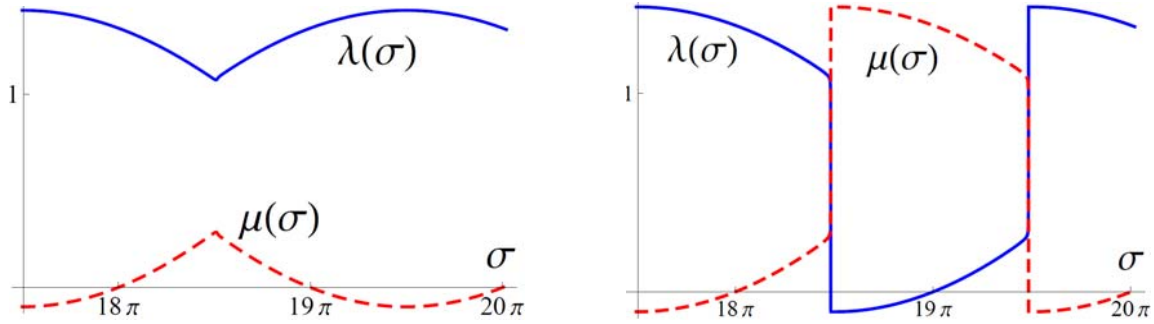

FIG.23. (Color online). Periodic evolutions of $\lambda = p + m$ and $\mu = p - m$ with $\sigma = \omega t$ as described by (7.12)(7.13) in the quasi-steady limit $\omega \to 0^+$, for $\nu = 0.488279$ and $|\omega\delta| = 0.05868$ [left] or $|\omega\delta| = 0.05869$ [right]. Though not obvious at first sight $V_{av}$ is continuous at the transition between the two cases: the above left or right trajectories indeed are either identical, or related by $\lambda \leftrightarrow \mu$ and $\sigma - 19\pi \leftrightarrow 19\pi - \sigma$.

If $1 \geq \nu \geq \tfrac{1}{2}$ the slow curves get S-shaped, as $f(m;\nu)$ also vanishes at $m = \pm\tfrac{1}{2}\cosh^{-1}(\tfrac{1}{\nu})$; the latter unforced solutions have only one pair of poles at finite distance [$B = B_\infty = \tanh^{-1}(\nu)$ for $m > 0$] the other one being rejected to $\pm i\infty$ [i.e., $\mu = 0$]. A forcing with a small $|\omega\delta|$ [below about 0.025845430 for $\nu = 0.8$] then makes $\mu(\sigma)$ oscillate about zero in phase opposition with the larger $\lambda(\sigma)$. Increasing $|\omega\delta|$ [above $\approx 0.025845433$ for $\nu = 0.8$] triggers relaxation oscillations of similar type as in Fig.23, right panel. Either $\lambda(\sigma)$ or $\mu(\sigma)$ smoothly changes sign during each half-period, and both later permute jumpwise: the poles cooperate for a while to form single a crest at $x = 0$ or $\pi$ [mod($2\pi$)] that subsequently splits.

The slow curves still are S-shaped if $\tfrac{1}{2} \geq \nu \geq \tfrac{1}{3}\sqrt{2} \approx 0.4714$, in particular for $\nu = 0.488279$ which yields a nearly constant $p$, see Fig.5. As long as $|\omega\delta|$ is small, $m(\sigma)$ and $p(\sigma)$ *may now feebly oscillate about the steady unforced values* [19] pertaining to four poles in a single pile; such histories have a very small basin of attraction though. Most often, or if $|\omega\delta|$ is beyond a $\nu$-dependent threshold [$\approx 0.058685$ for $\nu = 0.488279$], once during each cycle $m$ jumps from positive to negative [this permutes the values of $\lambda$ and $\mu$ and in some sense re-verses the arrow of time, see Fig23], and back: $\lambda\mu$ changes sign twice per temporal period, again leading to crests that periodically split and then merge at $x = 0$ or $x = \pi$ [$\mod(2\pi)$].

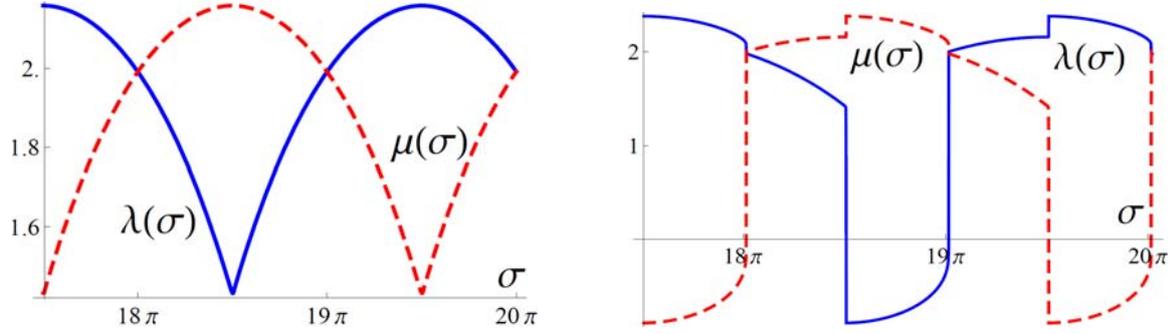

FIG.24 (Color online). $\lambda(\sigma)$, $\mu(\sigma)$ for $\nu = \frac{1}{4}$ and $|\omega\delta| = 0.64927$ [left] or $|\omega\delta| = 0.64928$ [right].

When $\nu < \frac{1}{3}\sqrt{2}$ the slow curve develops central secondary bumps whose height exceeds the lateral ones for $\nu$ below about $\frac{4}{9}$, see Fig.22. Up to 4 jumps of $\lambda(\sigma)$ and $\mu(\sigma)$ may then take place in a period, $\lambda\mu$ changes sign twice during each half-cycle and the crests flicker. Yet fine-tuned initial conditions also yield fronts whose single crest weakly breathes [ $\lambda\mu < 0$, see Fig.22, left] if $|\omega\delta|$ is small [below $\approx 0.0261$ for $\nu = \frac{1}{4}$], except if $\nu = \frac{1}{3}$: $m = \pm\frac{1}{2}\cosh^{-1}(3) \approx \pm 0.8814$ are double roots of $f(m, \frac{1}{3}) = 0$ shared by the unforced fronts with two or four poles [19, 29], and these equilibriums are semi-stable saddle-nodes that cannot resist $|\omega\delta| \neq 0$.

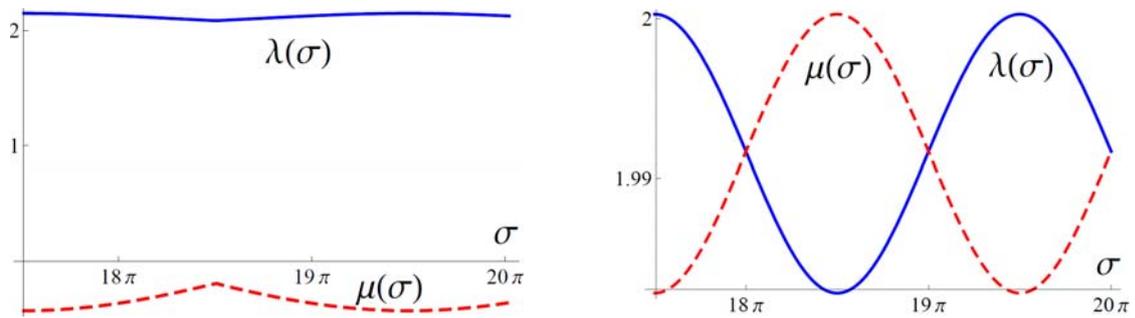

FIG.25 (Color online). Same as in Fig.24, with $\nu = \frac{1}{4}$ but using $|\omega\delta| = 0.0261107$ (left) or $0.0261108$ (right) and other initial conditions. Note the different vertical scale in rightmost panel.

In principle (7.13) enables one to plot the slow evolutions of $\tanh(\frac{1}{2}B(\sigma)) = e^{-p(\sigma)-m(\sigma)}$ and $\tanh(\frac{1}{2}b(\sigma)) = e^{-p(\sigma)+m(\sigma)}$; yet this soon gets tedious when jumps exist, especially different types thereof. It is way faster to resort to a numerical integration of (7.11): figures 23-25 were in fact obtained using the NDSolve routine [35] with $\omega = 10^{-4}$. The ensuing front length increment $V(\sigma)$ follows from (3.8). A numerical integration parallel to that of (7.11) yields $V_{av}$, leading to Fig.26: provided $v < \tfrac{1}{2}$, the excess of $V_{av}$ over the base-front contribution $\tfrac{1}{8}\omega^2\delta^2$ is remarkably flat before pole jumps set in and endow $V_{av}$ with an upward discontinuity.

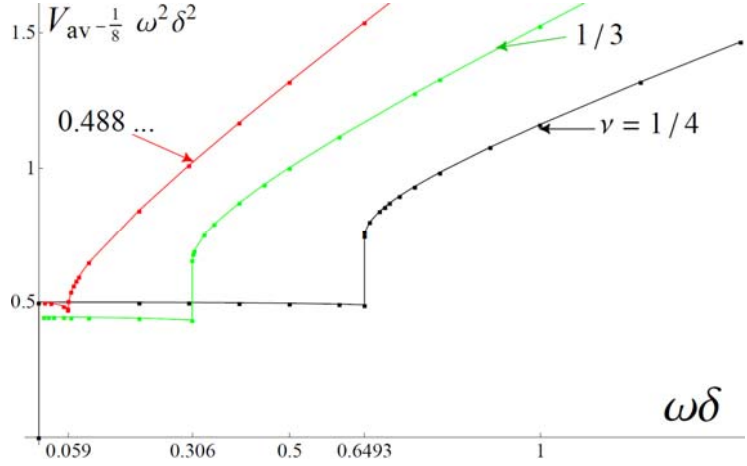

FIG.26 (Color online). Excess of $V_{av}$ over averaged base-shape arclength $\tfrac{1}{8}\omega^2\delta^2$ for $\omega \to 0^+$ and various $v < \tfrac{1}{2}$. The symbols are numerical results from (7.11)(3.8) with $\omega = 10^{-4}$, and the lines are guides to the eye.

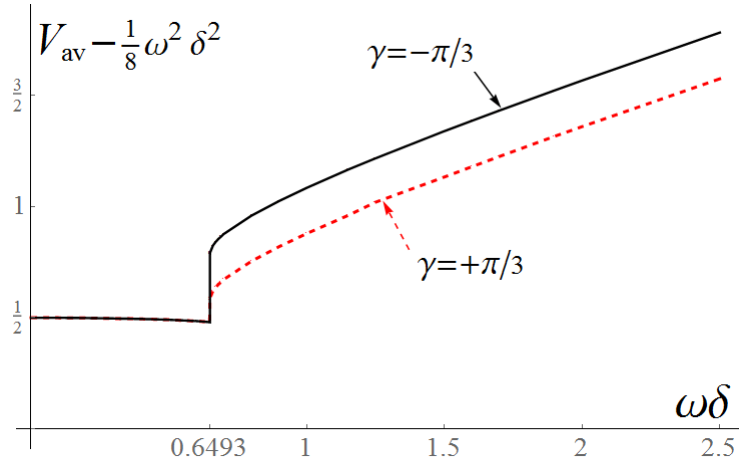

FIG.27 Excess of the time-averaged arclength increment $V_{av}$ over that, $\tfrac{1}{8}\omega^2\delta^2$, of the base front, for a low-frequency flame with $2N = 4$ poles per cell, $v = \tfrac{1}{4}$ and stretch [$d = \delta$, $\gamma = \pm \tfrac{\pi}{3}$]. Switching to $u' = \tfrac{\sqrt{29}}{8}\omega\delta[1 + \tfrac{3}{116}\omega^2\delta^2]^{1/2}$ from (6.48) as the abscissa would nearly be just a change of scale if $\omega\delta \leq 2.5$.

Stretch is incorporated in the usual way, by substituting $\omega\Delta\sin(\sigma - \zeta)$ for $\omega\delta\sin(\sigma)$ in the slow curve (7.13). Yet the numerical evaluation of $2vw(t)[e^{-B} - e^{-b}] \equiv -2v\omega d\sin(\sigma + \gamma) \times$ $[\tanh(\tfrac{1}{2}\lambda) - \tanh(\tfrac{1}{2}\mu)]$ is again necessary in practice to access the stretch-affected $V(\sigma)$ from (3.8), and its average. Sample results indicate that, if $\gamma = \pm \cos^{-1}(\delta/2d)$ to ensure $[u(t,x)]_{av} = 0$, stretch notably affects $V_{av}$ only if jumps of $\lambda(\sigma)$ exist, see Fig.27: when they do the influence of $\text{sgn}(\gamma)$ is one-sided, contrary to the $N = 1$ situation of depicted in Fig.20.

The above approach can be transposed to other $\Phi(t)$ and $w(t)$. For example if $-\Phi(\sigma) = \omega \Sigma_{\ell=1}^{\ell_{max}} \delta_\ell \sin(\ell\sigma + \sigma_\ell)$ and $w(\sigma) = -\omega \Sigma_{\ell=1}^{\ell_{max}} d_\ell \sin(\ell\sigma + \sigma_\ell + \gamma_\ell)$, $u_2(t)$ in (3.7) has $[u_2(t)]_{av} \sim \Sigma_{\ell=1}^{\ell_{max}} \delta_\ell(\delta_\ell - 2d_\ell \cos(\gamma_\ell))$, which vanishes if $\cos(\gamma_\ell) = \delta_\ell/2d_\ell$ for each $\delta_\ell \neq 0$ and leads to $[u(t,x)]_{av} = 0$. Provided $\ell_{max}\omega \ll 1$ the above 2-pair LF analysis is adapted by substituting $w(t) - \Phi(t) = \omega \Sigma_{\ell=1}^{\ell_{max}} \Delta_\ell \sin(\ell\sigma + \sigma_\ell - \zeta_\ell)$ for $\omega\delta\sin(\sigma)$ in (7.13), with $(\Delta_\ell, \zeta_\ell)$ defined in terms of $(\delta_\ell, d_\ell, \gamma_\ell)$ as in (6.9), but keeping the right-hand side of (7.13) as it is. The evolutions of $m(\sigma)$, $p(\sigma)$ and then of $\lambda(\sigma)$, $\mu(\sigma)$ and the pole locations can still be discussed graphically; yet final evaluations of $V_{av}$ require numerical integrations, especially when $\delta_\ell \sim \delta \sim d_\ell$ and $1 \ll \ell_{max} \ll \omega^{-1}$ so that $\Phi(t)$ and $w(t)$ mimic 'turbulent' bursts.

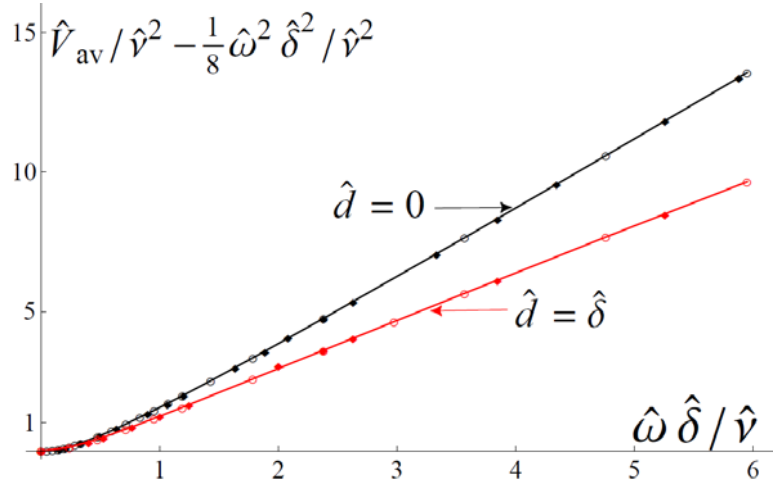

FIG.28 (Color online) Scaled excess of $\hat{V}_{av}$ over the averaged base-front length $\tfrac{1}{8}\hat{\omega}^2\hat{\delta}^2$ vs. $\hat{\omega}\hat{\Delta}/\hat{v}$, as deduced numerically from the Burgers analog of (7.11) with $\hat{\omega} = 10^{-4}$ and $\hat{d} = 0$ [upper, black] or $\hat{d} = \hat{\delta}$ and $\hat{\gamma} = \pm\tfrac{\pi}{3}$ [lower, red]. The solid diamonds [or empty circles] were obtained by varying $\hat{\omega}\hat{\delta}$ [or $\hat{v}$].

Burgers fronts will have a simpler LF response: they are flat if unforced and $\hat{V}_{av} = 0$ for $\hat{\omega}\hat{\Delta} = 0$. Moreover, $(\hat{V}_{av} - \tfrac{1}{8}\hat{\omega}^2\hat{\delta}^2)/\hat{v}^2$ only depends on $\hat{\omega}\hat{\Delta}/\hat{v}$, $\hat{d}/\hat{\delta}$, $\hat{N}_{tot}$ for $\hat{\omega} \to 0$ [the HF limit involved $\hat{\Delta}$, $\hat{\omega}\hat{d}/\hat{\delta}$, $\hat{N}_{tot}$]. Such a scaling law existed when $\hat{N}_{tot} = 1$ [see (7.4) with $\hat{d}/\hat{\delta}=0$, or $\hat{d}/\hat{\delta}=1$ and $\hat{\gamma}=\pm\frac{\pi}{3}$ ] and still does in the present two-pair situation, see Fig.28. As noted earlier the analog of (7.12) gives a $\hat{v}$-independent $\hat{p}=\hat{p}(|\hat{m}|)$; the slow curve replacing (7.13), $(\hat{\omega}\hat{\Delta}/\hat{v})\sin(\sigma-\hat{\zeta})=0-\frac{1}{2}\sinh(2\hat{m})\cosh(2\hat{p})$, relates $\hat{m}$ [then $\hat{p}$, $\hat{\lambda}$, $\hat{\mu}$,... ] to $(\hat{\omega}\hat{\Delta}/\hat{v})\sin(\sigma-\hat{\zeta})$ and the averaged (3.8) provides the scaling law in question. Setting $\hat{\varphi}=\hat{v}\tilde{\varphi}$ in the Burgers equation (2.4) actually allows $\hat{v}$ to be lumped in $\hat{u}(t,x)/\hat{v}^2$ and $\hat{w}(t,x)/\hat{v}$, both of which only involve $\hat{N}_{tot}$, $\hat{\Phi}(t)/\hat{v}$ and $\hat{w}(t)/\hat{v}$ when $\hat{\omega}\to 0$; besides, $\hat{V}_{av} \equiv [\langle\frac{1}{2}\hat{\varphi}_x^2\rangle]_{av} = \hat{v}^2[\langle\frac{1}{2}\tilde{\varphi}_x^2\rangle]_{av}$. Such scaling laws will persist for $\hat{N}_{tot} \sim 1/\hat{v} \gg 1$.

The situation when $N_{tot} > 2$ grows is far less clear for DL-affected fronts, as the sequence of saddle-node bifurcations encountered as $\omega\delta$ varies will get increasingly intricate. This is due to the quick proliferation [29] of steady unforced shapes allowed by $N_{tot} \sim N_{opt}(\nu) \approx \frac{1}{2\nu}$, and of $V_{av}$ vs. $\omega\delta$ branches that start from $0, \mathcal{V}(\nu), \mathcal{V}(2\nu)...$ at $\omega\delta = 0$. Slow curves with so many zeros also have many extrema, each of which can trigger the jump of a pair of poles when slowly reached and crossed in the appropriate direction; as hardly attainable zeros and semi-stable saddle-nodes are intertwined and get densely packed for $\nu \to 0^+$, it is not obvious from the outset whether $N_{tot} \sim 1/\nu \gg 1$ fronts are amenable to continuous-medium LF analyses.

### C. Single-pile dynamics.

#### 1. Burgers front

As a warm-up consider a unique pile of $\hat{N} \sim 1/\hat{v} \gg 1$ Burgers poles of $B$-type, corresponding to crests located at $x=0(\mathrm{mod}\,2\pi)$. A balance for the pole density $\hat{\rho}(\sigma, B)$, akin to (6.20) or, better still, its integrated version $\hat{\omega}\int_0^B \frac{\partial}{\partial\sigma}\hat{\rho}(\sigma, B')dB' + \hat{j}(t,B) = 0$ that accounts for $\hat{j}(t,0) = 0$, reduces to a zero-flux condition $\hat{j}(t,B) = 0$ at leading order in the limit $\hat{\omega} \to 0$. Using (6.21) with $\hat{\varrho}(t,b) \equiv 0$ [since the current pile is alone] leads to an integral equation

$$\fint_{-\hat{T}_{max}(\sigma)}^{\hat{T}_{max}(\sigma)} \frac{2\hat{v}\hat{\rho}(\sigma, T')dT'}{(T-T')} = -\hat{\omega}\hat{\delta}\sin(\sigma)\frac{2T}{1-T^2}, \qquad (7.14)$$

in which $\hat{\rho}(\sigma, T)$ is a shorthand for $\hat{\rho}(\sigma, B)$ once it is rewritten in terms of $T \equiv \tanh(\frac{1}{2}B)$.

This Tricomi equation already encountered in [25] is solved [Appendix B] to give

$$\hat{v}\hat{\rho}(\sigma, T) = -\hat{\omega}\hat{\delta}\sin(\sigma)\frac{(\hat{T}_{\max}^2 - T^2)^{1/2}}{\pi(1-\hat{T}_{\max}^2)^{1/2}(1-T^2)}, \tag{7.15}$$

where $\hat{T}_{\max} = \hat{T}_{\max}(\sigma)$ is fixed by the density normalization $\int_0^{\hat{T}_{\max}} 2\hat{\rho}(\sigma,T)dT/(1-T^2) = \hat{N}$:

$$-\hat{\omega}\hat{\delta}\sin(\sigma)\frac{\hat{T}_{\max}^2}{2(1-\hat{T}_{\max}^2)} = \hat{N}\hat{v} . \tag{7.16}$$

The inequalities $\hat{N}\hat{v} > 0$ and $0 < \hat{T}_{\max}(\sigma) \leq 1$ impose a restriction: (7.16) is valid only for $\hat{\delta}\sin(\sigma) \leq 0$. When $\hat{\delta}\sin(\sigma) \to 0^-$, $\hat{T}_{\max}(\sigma) \to 1^-$ and $\hat{\rho}(\sigma,T)$ gets bumps at $T^2 - 1 \sim \hat{T}_{\max}^2 - 1$ $\sim \hat{\omega}\hat{\delta}\sin(\sigma) \ll 1$, thereby signaling that the pile of poles is to jump to $z = \pi + ib \pmod{2\pi}$ *as a whole* because there is no mechanism to keep it at $z = iB \pmod{2\pi}$; once this has occurred the poles [now of the $b$-type] have enough time $1 \ll t \ll 1/\hat{\omega}$ to spread there. The integral equation for $\hat{\varrho}(\sigma,\tau)$, $\tau \equiv \tanh(\frac{1}{2}b)$, is obtained from (7.14) on substituting $(\tau, \hat{\varrho}(\sigma,\tau), \hat{\tau}_{\max}, \sin(\sigma))$ for $(T, \rho(\sigma,T), T_{\max}, -\sin(\sigma))$; and the same in (7.15)(7.16), to get slow solutions $\hat{\varrho}(\sigma,\tau)$ and $\hat{\tau}_{\max}(\sigma)$ that now hold for $\hat{\delta}\sin(\sigma) \geq 0$ only. Overall, the maximum pole altitude at either $x$-wise location is $2\tanh^{-1}[(1+|\hat{\omega}\hat{\delta}\sin(\sigma)|/2\hat{N}\hat{v})^{-1/2}]$ and blows up if $\sin(\sigma) = 0$; as indicated by (7.15)(7.16), $d\hat{B} \sim d\hat{T}/(1-\hat{T}^2)$ and $d\hat{b} \sim d\hat{\tau}/(1-\hat{\tau}^2)$, all the poles are at infinity when $\sigma = 0 \pmod{\pi}$, consistently with the Burgers front being then flat.

Restoring stretch is again achieved by replacing $(\sigma, \hat{\delta})$ with $(\sigma - \hat{\zeta}, \hat{\Delta})$, where $\hat{\zeta}$ and $\hat{\Delta}$ are defined analogously to (6.9). With stretch included, and for $\hat{\Delta}\sin(\sigma - \hat{\zeta}) \leq 0$, the *total* front slope $\hat{\varphi}_x(\sigma,x) = \hat{\omega}\hat{\delta}\sin(\sigma)\sin(x) + \hat{\phi}_x(\sigma,x)$ reads [25]

$$\hat{\varphi}_x(\sigma,x) = \hat{\omega}\hat{\delta}\sin(\sigma)\sin(x) + \hat{\omega}\hat{\Delta}\sin(\sigma - \hat{\zeta})\left[\frac{(1+\hat{T}_{\max}^2\cot^2(\frac{1}{2}x))^{1/2}}{(1-\hat{T}_{\max}^2)^{1/2}} - 1\right]\sin(x) , \tag{7.17}$$

where $\hat{\delta}\sin(\sigma) - \hat{\Delta}\sin(\sigma - \hat{\zeta})$ also is $\hat{d}\sin(\sigma + \hat{\gamma})$ by the Burgers analog of (6.9); whenever $\hat{\Delta}\sin(\sigma - \hat{\zeta}) \geq 0$ one replaces $(x, \hat{\Delta}, \hat{T}_{\max})$ by $(x - \pi, -\hat{\Delta}, \hat{\tau}_{\max})$ in (7.17). The shape $\hat{\varphi}(\sigma, x)$ is comprised of a continuous reunion of parabola-like arches whose pointed up crests spend half their lives [$\hat{\Delta}\sin(\sigma - \hat{\zeta}) < 0$] centered on $x = 0$, the other half on $x = \pi$; the polar component of this Burgers front slope vanishes when $\sin(\sigma - \hat{\zeta}) = 0$, but $\hat{\varphi}_x(\sigma,x)$ and $\hat{V}(\sigma)$ do not.

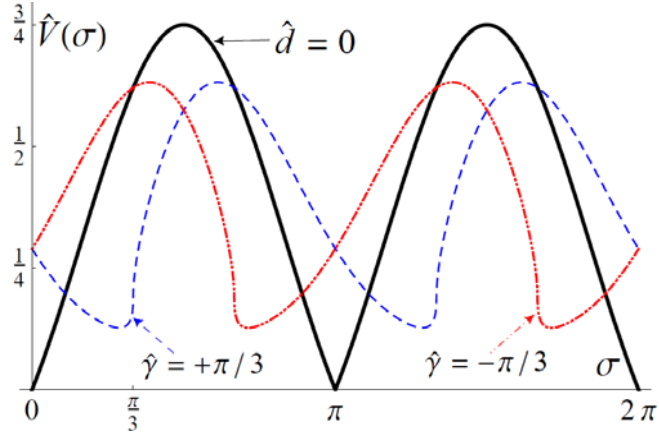

FIG.29 (Color online). Length increment $\hat V(\sigma)$ of LF Burgers fronts for $\hat N \hat v = \tfrac{1}{4}$ and $\hat\omega\hat\delta = 1$. Stretch-free case: $\hat d = 0$ [solid black]. With stretch: $\hat d = \hat\delta$ and $\hat\gamma = +\tfrac{\pi}{3}$ [dashed blue] or $\hat\gamma = -\tfrac{\pi}{3}$ [dot-dashed red].

The arclength increment $\hat V(\sigma)$ is accessible from the analog of (3.8) by plugging $\cosh(B) = (1+T^2)/(1-T^2)$ and $e^{-B} = (1-T)/(1+T)$ in the integrals $\int_0^{\hat T_{max}} 2\hat\rho(\sigma,T)\cosh(B)dT/(1-T^2)$ and $\int_0^{\hat T_{max}} 2\hat\rho(\sigma,T)e^{-B}dT/(1-T^2)$, with $\hat\rho(\sigma,T)$ taken from (7.15). Or (7.17) is injected in $\hat V(\sigma) = \pi^{-1}\int_0^\pi \tfrac{1}{2}\hat\varphi_x(\sigma,x)^2 dx$ and $x$- wise integrations are performed [Appendix D]. This yields

$$\hat V(\sigma) - \tfrac{1}{4}\hat\omega^2\hat\delta^2 \sin^2(\sigma) \underset{\hat\omega\to 0}{=} 2\hat v\hat N \,|\,\hat\omega\hat\Delta\sin(u)| + \hat\omega^2\hat\Delta\hat d \sin(u)\sin(u+\hat\zeta+\hat\gamma)\, U(\hat\theta(u))\;,$$
$$U(\theta) \equiv \frac{\theta(1-2\theta^2)(1-\theta^2)^{1/2} + \theta^2(1-\tfrac{1}{2}\theta^2)\pi - \sin^{-1}(\theta)}{\pi(1-\theta^2)^2}\;, \qquad (7.18)$$

where $u$ is a shorthand for $\sigma-\hat\zeta$; $\hat\theta(u) \equiv (1+|\hat\omega\hat\Delta\sin(u)|/2\hat N\hat v)^{-1/2} = \hat\theta(-u)$ resumes $\hat T_{max}(\sigma)$ when $\hat\Delta\sin(u) < 0$, see (7.16), and is $\hat\tau_{max}(\sigma)$ otherwise. Once time-averaged (7.18) produces $\hat V_{av}$ in a form that can be recast as a scaling law

$$\frac{\hat V_{av}}{(\hat v\hat N)^2} - \frac{1}{8}\frac{\hat\omega^2\hat\delta^2}{(\hat v\hat N)^2} = \frac{4}{\pi}\frac{|\hat\omega\hat\Delta|}{\hat v\hat N} + \frac{\hat\omega\hat\Delta\;\hat\omega\hat d}{(\hat v\hat N)^2}\cos(\hat\zeta+\hat\gamma)[\sin^2(u)\,U(\hat\theta(u))]_{av}\;, \qquad (7.19)$$

where the latter average needs to be tabulated *vs.* $\hat\varpi = |\hat\omega\hat\Delta|/\hat N\hat v$. This will not be made more explicit: similar to Fig.28, $\hat d = \hat\delta$ and $\hat\gamma = \pm\tfrac{\pi}{3}$ [hence $\hat\Delta\hat\delta\cos(\hat\zeta+\hat\gamma) = -\tfrac{1}{2}\hat\delta^2$] just lower the stretch-free $\hat d = 0$ result [here $\hat V_{av} = \tfrac{1}{8}\hat\omega^2\hat\delta^2 + \tfrac{2}{\pi}2\hat v\hat N\,|\hat\omega\hat\delta|$] by a nearly uniform fraction, 22 % [or 24 %] for $\hat v\hat N = \tfrac{1}{4}$ [or $\tfrac{1}{2}$], despite their having rather different $\hat V(\sigma)$ histories, see Fig.29.

Before proceeding to the DL mechanism, a lesson from the above analysis must be stressed: In the LF limit, otherwise *quasi-steady evolutions of the pile contents may need to be complemented by faster transients that look discontinuous* when observed at the $\sigma = \omega t = O(1)$ scale.

## 2. DL-unstable front

Besides removing the carets, restoring the DL instability augments equation (7.14) to

$$\fint_{-T_{\max}(\sigma)}^{T_{\max}(\sigma)} \frac{2\nu\rho(\sigma,T')dT'}{(T-T')} = \text{sgn}(T) - \omega\Delta\sin(\sigma-\zeta)\frac{2T}{1-T^2}, \quad (7.20)$$

whose solution follows [25] from (7.15) and the MS case [20] by linearity. Specifically,

$$\nu\rho(\sigma,T) = \frac{1}{\pi^2}\cosh^{-1}(\frac{T_{\max}}{|T|}) - \omega\Delta\sin(\sigma-\zeta)\frac{(T_{\max}^2 - T^2)^{1/2}}{\pi(1-T_{\max}^2)^{1/2}(1-T^2)} \quad (7.21)$$

is the sum of a MS-type piece and of a Burgers-like one. Likewise $T_{\max}$ obeys

$$\frac{1}{\pi}\sin^{-1}(T_{\max}) - \omega\Delta\sin(\sigma-\zeta)\frac{T_{\max}^2}{2(1-T_{\max}^2)} = N\nu, \quad (7.22)$$

which if solved for $\omega\Delta\sin(\sigma-\zeta)$ plays the role of a slow curve; yet it is here restricted to $\omega\Delta\sin(\sigma-\zeta) \leq (1-T_{\max}^2)^{3/2}/\pi T_{\max}$ to ensure $\rho(\sigma,T) \geq 0$ at $T^2 \lesssim 1$, and to a single-pile case.

Depending on the number $2N$ of poles, two situations are encountered.

- $N\nu < \frac{1}{2}$.

Equation (7.22) is valid up to $\omega\Delta\sin(\sigma-\zeta) = 0$, where it gives $T_{\max}$ the MS front [20] value $T_\infty = \sin(\pi N\nu) < 1$, and still holds true for some $\omega\Delta\sin(\sigma-\zeta) > 0$. The admissible values of $\omega\Delta\sin(\sigma-\zeta)$ are bounded above at a $\nu N$-dependent $\Gamma_\star > 0$ by the appearance of a double root $T_{\max} = T_\star$ of the normalization (7.22); importantly, this occurs when the inequality below (7.22) becomes an equality and $\rho(\sigma,T)$ starts being depleted at $T^2 = 1^-$. The link between $N\nu$ and $\Gamma_\star$ is available in a parametric form as $\pi\Gamma_\star = (1-T_\star^2)^{3/2}/T_\star$ and $\pi N\nu = \sin^{-1}(T_\star) - \frac{1}{2}T_\star(1-T_\star^2)^{1/2}$, which implies $1-T_\star \sim (\frac{1}{2}-N\nu)^2$ and $\Gamma_\star \sim (\frac{1}{2}-N\nu)^3$ for $2N\nu \to 1^-$ whereas $T_\star \sim N\nu$ and $\Gamma_\star \sim 1/N\nu \gg 1$ if $N\nu \to 0^+$: as in [36] the DL instability allows small quasi-steady crests to resist strong combined stretch, *i.e.*, geometrical stretch from $-\Phi_x(t,z)$ and/or actual one from $w(t,z)$.

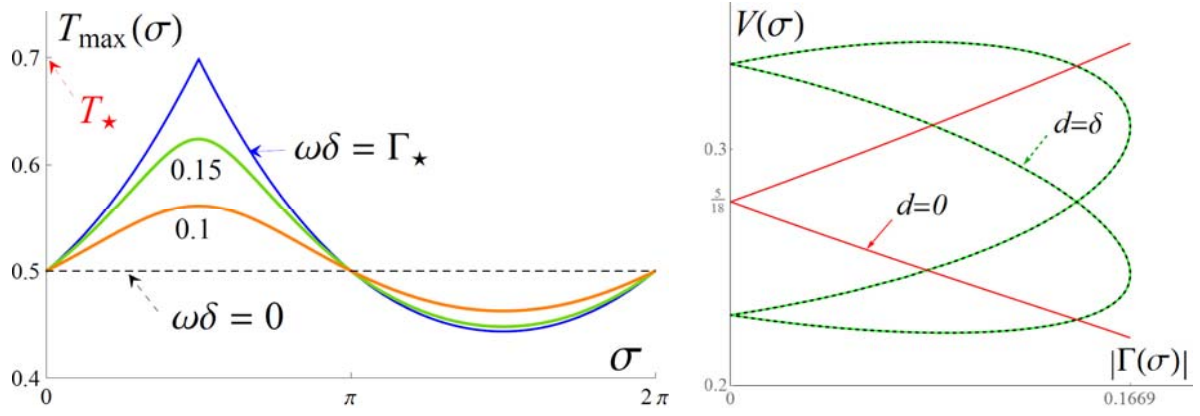

FIG.30 (Color online). (Left) $T_{\max}(\sigma)$ computed from (7.20) with $\nu N = \frac{1}{6}$, $\omega d = 0$ and $\omega \delta$ from 0 to $\Gamma_\star(\frac{1}{6}) \approx 0.1669$. (Right) Lissajous-type curves $V(\sigma)$ vs. $|\Gamma(\sigma)| \equiv |\omega \Delta \sin(\sigma - \zeta)|$ for $\omega \Delta = \Gamma_\star(\frac{1}{6})$ and $d = 0$ [red solid] or $d = \delta$, $\gamma = \pm \frac{\pi}{3}$ [green, or black dashed]; $\mathcal{V}(\nu N) = 2\nu N(1 - \nu N)$ is $\frac{5}{18}$ here. The red curve resembles Fig.7 of [25]: for $\omega \to 0$ and $\omega d = 0$, $|\Gamma(\sigma)| \ll 1$ indeed is $\propto \langle u(t,x)^2 \rangle^{1/2} \ll 1$.

Provided $\omega \Delta \sin(\sigma - \zeta)$ stays below $\Gamma_\star > 0$ the above one-pile analysis adequately describes a single crest [per cell] that slowly breathes periodically at $x = 0 \pmod{2\pi}$. The polar slope $\phi_x(\sigma, x)$ and the arc-length increment $V(\sigma)$, obtained in the same way as (7.17)(7.18), read

$$\phi_x(\sigma, x) = \frac{-2}{\pi} \sinh^{-1}\left(\frac{T_{\max}}{\tan(\frac{1}{2}x)}\right) + \omega \Delta \sin(\sigma - \zeta) \left[\frac{(1 + T_{\max}^2 \cot^2(\frac{1}{2}x))^{1/2}}{(1 - T_{\max}^2)^{1/2}} - 1\right] \sin(x), \qquad (7.23)$$

$$V(\sigma) - \mathcal{V}(\nu N) - \tfrac{1}{4}\omega^2 \delta^2 \sin^2(\sigma) = -2\omega \Delta \sin(\sigma - \zeta)[\Lambda_1(T_{\max}) - \omega \Delta \sin(\sigma - \zeta)\Lambda_3(T_{\max})] \\ - 2\omega d \sin(\sigma + \gamma)[\Lambda_2(T_{\max}) - \omega \Delta \sin(\sigma - \zeta)\Lambda_4(T_{\max})], \qquad (7.24)$$

$$\Lambda_1(T_{\max}) = \frac{T_{\max}}{\pi(1 - T_{\max}^2)^{1/2}}, \quad \Lambda_2(T_{\max}) = \frac{2 T_{\max} \cos^{-1}(T_{\max})}{\pi^2 (1 - T_{\max}^2)^{1/2}}, \quad \Lambda_3(T_{\max}) = \frac{T_{\max}^2 (2 - T_{\max}^2)}{4(1 - T_{\max}^2)^2},$$

$$\Lambda_4(T_{\max}) = \frac{2 T_{\max}(1 - 2 T_{\max}^2)\sqrt{1 - T_{\max}^2} + T_{\max}^2(2 - T_{\max}^2)\pi - 2\sin^{-1}(T_{\max})}{4\pi(1 - T_{\max}^2)^2}. \qquad (7.25)$$

The functions $\Lambda_j(T_{\max}(\sigma))$ stem from $\rho(\sigma, T)$-weighted integrals that crop up once (7.21) is injected in the one-pile, small-$\omega$ version of (3.8); they were checked using Mathematica [35].

Given $0 < \nu N < \frac{1}{2}$ and $0 < \omega \Delta \leq \Gamma_\star(\nu N)$, normalization (7.22) is used to numerically access $0 < T_{\max}(\sigma) \leq T_\star(\nu N)$ [see Fig.30] which when plugged in (7.24) provides $V(\sigma)$. This single-pile configuration ceases to be viable whenever $\omega \Delta \sin(\sigma - \zeta) > \Gamma_\star > 0$. Although rooted in nonlinearity $\Gamma_\star$ is pretty small unless $N\nu$ is below about $\frac{1}{5}$: $\Gamma_\star(\frac{1}{6}) \approx 0.1669$, $\Gamma_\star(\frac{1}{5}) \approx 0.1095$,

$\Gamma_\star(\frac{1}{4}) \approx 0.0564$, $\Gamma_\star(\frac{1}{3}) \approx 0.0147$ and $\Gamma_\star = 0$ for $N\nu \geq \frac{1}{2}$ [see below]. Then $V(\sigma) - V_{av}$ nearly is a pure sine if $\omega\Delta < \Gamma_\star$, and $V_{av}$ is desperately close to $\mathcal{V}(\nu N) = 2\nu N(1-\nu N)$, see Fig.30.

- $N\nu \geq \frac{1}{2}$.

$N\nu - \frac{1}{\pi}\sin^{-1}(T_{max})$ stays nonnegative and (7.22) is qualitatively the same as in Burgers cases: a unique solution $0 < T_{max}(\sigma) \leq 1$ if $\omega\Delta\sin(\sigma-\zeta) \leq 0$, and none otherwise. And $T_{max}(\sigma) \to 1^-$ when $\omega\Delta\sin(\sigma-\zeta) \to 0^-$, making (7.22) acquire the same form as (7.16) except for $N\nu - \frac{1}{2}$ instead of $\hat{N}\hat{\nu}$. Given that only the Burgers-like piece of $\rho(\sigma,T)/(1-T^2)$ builds up peaks of total content $N - N_{opt}(\nu)$ pairs at $T^2 \lesssim 1$, and that the DL instability can help small crests survive strong stretch, will the pile initially located at $x=0$ jump as a whole to $x=\pi$ when $\omega\Delta\sin(\sigma-\zeta)$ goes positive [option (*i*)] or will it do this only partly [option (*ii*)] ?

Option (*i*) looks viable, whereupon the same substitution $\Delta\sin(\sigma-\zeta) \to |\Delta\sin(\sigma-\zeta)|$ as for Burgers fronts would extend the validity of (7.20)-(7.24), yet this calls for words of caution. First, even if option (*i*) is the correct one at leading order in a $N \sim 1/\nu \gg 1$ analysis, it is still compatible with $n = O(1) \ll N$ pairs of discrete poles staying at $x=0$ for $\Delta\sin(\sigma-\zeta) > 0$. This is suggested by the results of [36] where the [then geometrical] stretch intensity had to exceed $1/4\nu \gg 1$ to dislodge $n=1$ pair of poles from a front trough; the threshold stretch decreased as $1/n\nu$ for $n>1$, similar to $\Gamma_\star \sim 1/N\nu$ deduced from (7.22) for $N\nu \ll 1$. Second, crest resistance to [geometrical-] stretch is the very reason why bi-coalesced, steady solutions to the MS equation (1.1) exist for *any* large numbers $(N,n)$ of pairs in the piles, provided $N + n \leq N_{opt}(\nu) = \frac{1}{2\nu} + O(1)$: being fixed by initial conditions $n/N$ is not accessible to steady analyses, and the arbitrariness about $n/N$ might survive if forcing and/or stretch are weak enough. Even if option (*i*) is a viable possibility, it unlikely is the only one.

Option (*ii*), where only some of the $N$ pairs initially at $x=0$ jump as $\sin(\sigma-\zeta)$ changes sign, also is conceivable; but how then to fix the filling fractions $N(\sigma)/N_{tot}$ and $n(\sigma)/N_{tot}$ ?

At any rate, except for Burgers fronts [$\hat{N}.\hat{n} \equiv 0$ over the $\sigma = O(1)$ scale] or DL-unstable ones with $N\nu < \frac{1}{2}$ and $|\omega\Delta|$ below the often small $\Gamma_\star$, envisaging only a single quasi-steady pile is too restrictive a framework: *the LF evolution of bi-coalesced fronts needs to be addressed*.

## D. Bi-coalesced fronts

### *1. Densities*

Thus, now consider two piles pertaining to crests centered on $x = 0$ or $x = \pi$, along with the associated densities $\rho(\sigma, T) \geq 0$ and $\varrho(\sigma, \tau) \geq 0$ and the corresponding current pole contents $2N(\sigma)$ and $2n(\sigma)$, respectively. In the LF limit the [integrated] balances (6.20) reduce to a pair of zero-flux conditions, $j(\sigma, T) = 0 = q(\sigma, \tau)$.

Provided none of the densities identically vanishes, those lead to coupled integral equations,

$$\fint_{-T_{\max}}^{T_{\max}} \frac{2\nu\rho(\sigma, T')dT'}{(T-T')} + T \int_{-\tau_{\max}}^{\tau_{\max}} \frac{2\nu\varrho(\sigma, \tau')d\tau'}{(1-T\tau')} = \text{sgn}(T) - \omega\Delta\sin(\sigma - \zeta)\frac{2T}{1-T^2}, \qquad (7.26)$$

$$\fint_{-\tau_{\max}}^{\tau_{\max}} \frac{2\nu\varrho(\sigma, \tau')d\tau'}{(\tau-\tau')} + \tau \int_{-T_{\max}}^{T_{\max}} \frac{2\nu\rho(\sigma, T')dT'}{(1-\tau T')} = \text{sgn}(\tau) + \omega\Delta\sin(\sigma - \zeta)\frac{2\tau}{1-\tau^2},$$

valid for $0 < T^2 \leq T_{\max}^2 \leq 1$ or $0 < \tau^2 \leq \tau_{\max}^2 \leq 1$, respectively, and constrained by $\rho(\sigma, T) \geq 0$, $\varrho(\sigma, \tau) \geq 0$. The unknown $T_{\max} = T_{\max}(\sigma)$ and $\tau_{\max} = \tau_{\max}(\sigma)$ must obey the normalizations:

$$\int_0^{T_{\max}} \frac{2\rho(\sigma, T)dT}{1-T^2} = N(\sigma), \qquad \int_0^{\tau_{\max}} \frac{2\varrho(\sigma, \tau)d\tau}{1-\tau^2} = n(\sigma). \qquad (7.27)$$

These are themselves linked by a global pole-number budget in which $N_{tot}$ is specified

$$N(\sigma) + n(\sigma) = N_{tot}, \qquad (7.28)$$

but $N(\sigma)$ and $n(\sigma)$ have to be found as part of the solution. However, *the three conditions in (7.27)(7.28) do not suffice to determine all four $T_{\max}(\sigma), \tau_{\max}(\sigma), N(\sigma)$ and $n(\sigma)$*; except if $N$ and $n$ may stay constant and are arbitrary in a range, which did happen for $\Delta = 0$ [20].

Deferring this subtle point to latter entries of subsection VII.D, it is convenient to first solve (7.26) without specifying $T_{\max}(\sigma)$ and $\tau_{\max}(\sigma)$. This is eased by the linearity of (7.26) for any presumed $0 < T_{\max} \leq 1$ and $0 < \tau_{\max} \leq 1$, which allows $\rho(\sigma, T)$ and $\varrho(\sigma, \tau)$ to be split as

$$\rho(\sigma, T) = \rho_{MS}(\sigma, T) + \rho_{Burg}(\sigma, T), \qquad \varrho(\sigma, \tau) = \varrho_{MS}(\sigma, \tau) + \varrho_{Burg}(\sigma, \tau). \qquad (7.29)$$

The MS-type components $\rho_{MS}(\sigma, T)$ and $\varrho_{MS}(\sigma, \tau)$ of (7.29) belong to $\Delta\sin(\sigma - \zeta) = 0$ in (7.26). They were already computed in [20] and can be rewritten as

$$\rho_{MS}(\sigma,T) = F_{MS}(\sigma,\xi)\big|_{\xi=T/T_{\max}} \ , \quad \varrho_{MS}(\sigma,\tau) = F_{MS}(\sigma,\xi)\big|_{\xi=\tau/\tau_{\max}} ,$$
$$\nu F_{MS}(\sigma,\xi) \equiv \frac{1}{\pi^2}\cosh^{-1}\!\left(\frac{(1+\eta\xi^2)}{(1+\eta)|\xi|}\right) \ , \quad 0\le \eta\equiv T_{\max}\tau_{\max}\le 1 , \tag{7.30}$$

hence $-1\le \xi \le 1$ spans either support. That both MS-type densities derive from the same function $F_{MS}(\sigma,\xi)$ evaluated at different $\xi$ s, and the way Ref. [20] obtained (7.30), gave hints on how to access the Burgers-type pieces of (7.29). Making the Ansatz

$$\rho_{Burg}(\sigma,T) = -\omega\Delta\sin(\sigma-\zeta)\tfrac{1}{2}[F_{-}(\sigma,\xi)+F_{+}(\sigma,\xi)]\big|_{\xi=T/T_{\max}} ,$$
$$\varrho_{Burg}(\sigma,\tau) = +\omega\Delta\sin(\sigma-\zeta)\tfrac{1}{2}[F_{-}(\sigma,\xi)-F_{+}(\sigma,\xi)]\big|_{\xi=\tau/\tau_{\max}} , \tag{7.31}$$

in the sgn[.]-free version of (7.26), and using that $\int_{-1}^{+1} f(\xi')d\xi'/(1-\eta\xi\xi')$ is even in $\xi$ when $f(\xi)$ is, indeed lead to *decoupled* equations that share the same $\eta = T_{\max}\tau_{\max}$ as in (7.30):

$$\fint_{-1}^{+1}\frac{\nu F_{\pm}(\sigma,\xi')d\xi'}{(\xi-\xi')} \pm \eta\xi\int_{-1}^{+1}\frac{\nu F_{\pm}(\sigma,\xi')d\xi'}{(1\mp\eta\xi\xi')} = \frac{\xi T_{\max}}{1-\xi^2 T_{\max}^2} \mp \frac{\xi\tau_{\max}}{1-\xi^2\tau_{\max}^2} . \tag{7.32}$$

One next introduces $-1\le \Theta_{\pm}\equiv (1\pm\eta)\xi/(1\pm\eta\xi^2)\le 1$ and $F_{\pm}(\sigma,\xi) = \chi_{\pm}(\sigma,\Theta_{\pm})$. This converts (7.32) to a pair of Tricomi equations [compare with (6.26)(6.29)] whose right-hand sides have the same structure as in (7.14), up to rescaling. Resolution [Appendix B] then yields:

$$\nu\chi_{\pm}(\sigma,\Theta_{\pm}) = \frac{(T_{\max}\mp\tau_{\max})(1\pm T_{\max}\tau_{\max})}{\pi(T_{\max}\pm\tau_{\max})^2(1-1/\vartheta_{\pm}^2)^{1/2}}\frac{(1-\Theta_{\pm}^2)^{1/2}}{(\vartheta_{\pm}^2-\Theta_{\pm}^2)} ,$$
$$\vartheta_{\pm}\equiv (1\pm T_{\max}\tau_{\max})/|T_{\max}\pm\tau_{\max}|>1 . \tag{7.33}$$

$\rho_{Burg}(\sigma,T)$ and $\varrho_{Burg}(\sigma,\tau)$ follow by substituting $F_{\pm}(\sigma,\xi) = \chi_{\pm}(\sigma,\Theta_{\pm}(\xi))$ in (7.31). Jointly with (7.29)(7.30) this ultimately yields:

$$\nu\rho(\sigma,T) = \frac{1}{\pi^2}\cosh^{-1}\!\left(\frac{(T_{\max}+\tau_{\max}T^2)}{(1+T_{\max}\tau_{\max})|T|}\right) - \frac{\Gamma(\sigma)}{\pi(1-T_{\max}^2)^{1/2}}\frac{(T_{\max}^2-T^2)^{1/2}}{(1-T^2)}\frac{(1-\tau_{\max}^2 T^2)^{1/2}}{(1-\tau_{\max}^2)^{1/2}} ,$$
$$\nu\varrho(\sigma,\tau) = \frac{1}{\pi^2}\cosh^{-1}\!\left(\frac{(\tau_{\max}+T_{\max}\tau^2)}{(1+\tau_{\max}T_{\max})|\tau|}\right) + \frac{\Gamma(\sigma)}{\pi(1-\tau_{\max}^2)^{1/2}}\frac{(\tau_{\max}^2-\tau^2)^{1/2}}{(1-\tau^2)}\frac{(1-T_{\max}^2\tau^2)^{1/2}}{(1-T_{\max}^2)^{1/2}} , \tag{7.34}$$

$$\Gamma(\sigma)\equiv \omega\Delta\sin(\sigma-\zeta) .$$

One may incidentally note that $\rho_{Burg}(\sigma,T)$ and $\varrho_{Burg}(\sigma,\tau)$ cannot be simultaneously positive in (7.34): accordingly, *no bi-coalesced quasi-steady Burgers front may exist here* because $\hat{\rho}(\sigma,T)$ or $\hat{\varrho}(\sigma,\tau)$ would be negative. Entry VII C.1, where $\hat{\rho}(\sigma,T)\hat{\varrho}(\sigma,\tau) \equiv 0$, will suffice.

If $\tau_{\max} \to 0$, $\nu\rho(\sigma,T)$ in (7.34) resumes (7.21), $\nu\varrho(\sigma,\tau) \to \pi^{-2}\cosh^{-1}(\tau_{\max}/|\tau|)$ over $0 \leq \tau^2 \leq \tau_{\max}^2$ and $\pi\nu n \approx \tau_{\max}$ by (7.27). Next, assuming that $(1-T_{\max}^2)(1-\tau_{\max}^2)$ remains $O(1)$ and setting $\sin(\sigma-\zeta)=0$ in (7.34) retrieves the densities of *one* of the steady, under-populated, bi-coalesced MS fronts, each of which is known [20] to exist only when $\nu N_{tot} < \nu N_{opt}(\nu) = \frac{1}{2}$. Conversely, $\nu N_{tot} \geq \frac{1}{2}$ *implies that* $(1-T_{\max}^2)(1-\tau_{\max}^2)$ *will vanish at* $\sigma-\zeta = 0(\mod \pi)$.

Consider such a $\nu N_{tot} > \frac{1}{2}$ case as $\sigma-\zeta \to \mathbb{Z}\pi$, say with $1-T_{\max}^2 \to 0$ but $1-\tau_{\max}^2 = O(1)$. For compatibility of (7.34) with (7.27)(7.28) and $\nu N_{tot} < \infty$, $1-T_{\max}^2$ may not be much less than $\omega\Delta\sin(\sigma-\zeta)$. The opposite may not hold either: this would make the pole densities resume their MS-type components but $\nu\rho_{MS}(\sigma,T)$ and $\nu\varrho_{MS}(\sigma,\tau)$ jointly contribute only $\frac{1}{2}$ to $\nu N_{tot}$ if $T_{\max} = 1$, whatever $\tau_{\max}$ is [Appendix E]. The only viable compromise has $1-T_{\max}^2 \sim \omega\Delta\sin(\sigma-\zeta)$, hence $\nu\varrho(\sigma,\tau)$ still resumes $\nu\varrho_{MS}(\sigma,\tau)$ but $\nu\rho(\sigma,T) \to \nu\rho_{MS}(\sigma,T)$ only at fixed $T^2 < 1$. To store $2N_{tot} - 1/\nu$ 'missing' poles $2\nu\rho_{Burg}(\sigma,T)/(1-T^2)$ has to grow Dirac-like peaks near $T_{\max}^2 \to 1$, of weights $\nu N_{tot} - \frac{1}{2}$ and of *same sign* as $-\omega\Delta\sin(\sigma-\zeta)$. Thus, if $\nu N_{tot} > \frac{1}{2}$ and $\tau_{\max} < 1$, $T_{\max}^2$ will reach $1^-$ when and only when $\omega\Delta\sin(\sigma-\zeta) \to 0^-$; this also implies $\nu n(\sigma) \to \int_0^{\tau_{\max}} 2\nu\varrho_{MS}(\sigma,\tau)d\tau/(1-\tau^2) \leq \frac{1}{4} = \frac{1}{2}\nu N_{opt}(\nu)$ [Appendix E] and $\nu N(\sigma) > \frac{1}{4}$.

When $\nu N_{tot} > \frac{1}{2}$ and $T_{\max} < 1$, $\tau_{\max}(\sigma) = 1^-$ is likewise attained iff $\omega\Delta\sin(\sigma-\zeta) \to 0^+$, then implying $\nu N(\sigma) \leq \frac{1}{4}$ and $\nu n(\sigma) > \frac{1}{4}$.

The net result is that $\nu N_{tot} - \frac{1}{2}$ pairs of surplus poles from the pile at $x=0$ [or $x=\pi$] jump by $+\pi$ [or $-\pi$] when $\omega\Delta\sin(\sigma-\zeta)$ crosses $0$ from below [or above]; and the pile contents jump from below $\frac{1}{2}\nu N_{opt}(\nu)$ pairs to above it, and *vice versa*. As in the Burgers case in SubSec.VII. C1, this occurs 'instantly' if observed on the $t=1/\omega$ scale; in a shorter while $\sigma-\zeta-\mathbb{Z}\pi = o(1)$ after their jump, commuted poles have ample time for $\omega \to 0$ to settle and reach a new quasi-equilibrium before the slow dynamics takes over.

The above relative signs and inequalities survive if $2\nu N_{tot} = 1^+$, some in a weaker sense [*e.g.*,

for $\tau_{max}^2 \to 1^-$ the previous $\nu n(\sigma) > \tfrac{1}{4}$ gets $\nu n(\sigma) \geq \tfrac{1}{4}$]. Why jumps still exist for $2\nu N_{tot} = 1$ in absence of surplus poles, and what happens after $\sigma - \zeta \approx \mathbb{Z}\pi$, will be tackled later.

## 2. Slope.

The large-$N_{tot}$ expression (6.31) of the polar slope $\phi_x(\sigma, x)$ is rewritten as

$$\phi_x(\sigma, x) = -\int_{-T_{max}}^{T_{max}} \frac{2\nu\rho(\sigma,T)dT}{\tan(\tfrac{1}{2}x) + iT} + \tan(\tfrac{1}{2}x) \int_{-\tau_{max}}^{\tau_{max}} \frac{2\nu\varrho(\sigma,\tau)d\tau}{1 - i\tau\tan(\tfrac{1}{2}x)} , \qquad (7.35)$$

by the tangent-of-a-sum formula and the parities in $T$ or $\tau$. Doing the latter integrals yields the total slope $\varphi_x(\sigma, x) = \omega\delta\sin(\sigma)\sin(x) + \phi_x(\sigma, x)$ of quasi-steady bi-coalesced fronts as

$$\begin{aligned}\varphi_x(\sigma, x) &= \omega\delta\sin(\sigma)\sin(x) - \frac{2}{\pi}\sinh^{-1}[\frac{(T_{max}\cot(\tfrac{1}{2}x) - \tau_{max}\tan(\tfrac{1}{2}x))}{(1 + T_{max}\tau_{max})}] + \\ &\quad \omega\Delta\sin(\sigma - \zeta)\left[\frac{(1 + T_{max}^2\cot^2(\tfrac{1}{2}x))^{1/2}(1 + \tau_{max}^2\tan^2(\tfrac{1}{2}x))^{1/2}}{(1 - T_{max}^2)^{1/2}(1 - \tau_{max}^2)^{1/2}} - 1\right]\sin(x) ,\end{aligned} \qquad (7.36)$$

in which $\omega\delta\sin(\sigma) - \omega\Delta\sin(\sigma - \zeta) = \omega d\sin(\sigma + \gamma) = -w(t)$ [by (6.9)] vanishes if the stretch intensity does. Recall that $\omega\Delta\sin(\sigma - \zeta) = w(t) - \Phi(t)$ measures the combined stretch, whereby (7.36) can be transposed to any quasi-steady $w(t)$ and $\Phi(t)$ through obvious substitutions.

The slope profile (7.36) could give direct numerical access to $V(\sigma) \equiv \pi^{-1}\int_0^\pi \tfrac{1}{2}\varphi_x^2 dx$ … if the current support bounds $T_{max}(\sigma)$ and $\tau_{max}(\sigma)$ were available; the onus is now on them.

## 3. Constraints and Normalizations

Before proceeding to evaluations of $T_{max}(\sigma)$ and $\tau_{max}(\sigma)$ proper, one must check if $\rho(\sigma, T)$ and $\varrho(\sigma, \tau)$ in (7.34) are nonnegative, which is more likely violated by remote poles: these are indeed endangered if the combined stretch $w(\sigma) - \Phi(\sigma)$ tends to expel them. To see when or whether this occurs one defines two scalars $\mathbb{K}$ and $\Bbbk$ that encode the near-end density signs:

$$\frac{\rho(\sigma,T)}{(T_{max}^2 - T^2)^{1/2}} \underset{T^2 \to T_{max}^2}{=} \mathbb{K}(T_{max}, \tau_{max}, \Gamma(\sigma)), \qquad \frac{\varrho(\sigma,\tau)}{(\tau_{max}^2 - \tau^2)^{1/2}} \underset{\tau^2 \to \tau_{max}^2}{=} \Bbbk(T_{max}, \tau_{max}, \Gamma(\sigma)) , \qquad (7.37)$$

in which $\Gamma(\sigma) \equiv w(\sigma) - \Phi(\sigma)$ is the abridged notation for the combined stretch, e.g., $\Gamma(\sigma) = \omega\Delta\sin(\sigma - \zeta)$ in the case of (7.34). The $(T_{max}, \tau_{max}, +\Gamma) \leftrightarrow (\tau_{max}, T_{max}, -\Gamma)$ symmetry of (7.34) of course extends to $\mathbb{K}$ and $\Bbbk$.

Due to the analytical structure of $\rho(\sigma,T)$ and $\varrho(\sigma,\tau)$ in (7.34) it so happens that

$$\begin{aligned}
\rho(\sigma,T)\big|_{0\leq T^2\leq 1} \geq 0 &\quad\Leftrightarrow\quad \mathbb{K}(T_{\max},\tau_{\max},\Gamma(\sigma)) \geq 0 \quad\Leftrightarrow\quad +\Gamma(\sigma) \leq h(T_{\max},\tau_{\max}), \\
\varrho(\sigma,\tau)\big|_{0\leq \tau^2\leq 1} \geq 0 &\quad\Leftrightarrow\quad \Bbbk(T_{\max},\tau_{\max},\Gamma(\sigma)) \geq 0 \quad\Leftrightarrow\quad -\Gamma(\sigma) \leq h(\tau_{\max},T_{\max}),
\end{aligned} \tag{7.38}$$

$$h(T_{\max},\tau_{\max}) \equiv \frac{(1-\tau_{\max}^2)^{1/2}}{(1+\tau_{\max}T_{\max})} \frac{(1-T_{\max}^2)^{3/2}}{\pi T_{\max}} \geq 0 \ . \tag{7.39}$$

One next notes that *only* if $\Gamma(\sigma)$ is positive [or negative] can $\mathbb{K} > 0$ [or $\Bbbk > 0$] be violated. And in case $\mathbb{K} = 0$ [or $\Bbbk = 0$], the other constraint $\Bbbk > 0$ [or $\mathbb{K} > 0$] is met; as explained later on, this means that only one pile at a time can be losing its content. One may also remark that $\tau_{\max} = 0$ in (7.39) makes the first line of (7.38) resume the inequality quoted below (7.22) in SubSec. VII C about the single-crest dynamics.

When specialized to (7.34) the normalizations conditions (7.27) acquire the structure

$$\begin{aligned}
M(T_{\max},\tau_{\max}) - \Gamma(\sigma)L(T_{\max},\tau_{\max}) - \nu N &= 0, \\
M(\tau_{\max},T_{\max}) + \Gamma(\sigma)L(\tau_{\max},T_{\max}) - \nu n &= 0,
\end{aligned} \tag{7.40}$$

which is manifestly symmetric by $(T_{\max},\tau_{\max},+\Gamma,N,n) \leftrightarrow (\tau_{\max},T_{\max},-\Gamma,n,N)$, and where:

$$M(a,b) \equiv \frac{2}{\pi^2} a \int_0^1 \cosh^{-1}\left[\frac{(1/\xi + ab\xi)}{(1+ab)}\right] \frac{1}{(1-a^2\xi^2)} d\xi \ ,$$

$$L(a,b) \equiv \frac{2}{\pi(1-a^2)^{1/2}(1-b^2)^{1/2}} a^2 \int_0^1 \frac{(1-\xi^2)^{1/2}(1-a^2b^2\xi^2)^{1/2}}{(1-a^2\xi^2)^2} d\xi \ . \tag{7.41}$$

The terms $\propto \Gamma(\sigma)$ of (7.40) stem from the Burgers-type components in (7.34); $L(T_{\max},\tau_{\max})$ is reducible to elliptic integrals [Appendix E, §3], with $L(T_{\max},0) = T_{\max}^2/2(1-T_{\max}^2)$ as a limit. The DL-induced $M(T_{\max},\tau_{\max})$ needs numerical quadrature, except $M(T_{\max},0) = \frac{1}{\pi}\sin^{-1}(T_{\max})$ [single-crested cell] and $M(T_{\max},T_{\max}) = \frac{1}{\pi}\tan^{-1}(T_{\max})$ [twin crests]. Equations (E.8) and (E.11) (E.13) of Appendix E provide more convenient expressions to calculate $M(a,b)$ and $L(a,b)$.

Given $\nu N$ and $\nu n$ adding up to a prescribed $\nu N_{tot}$, (7.40) constitute coupled nonlinear equations in two unknowns $T_{\max}(\sigma)$ and $\tau_{\max}(\sigma)$; this involves $\sigma$ only as a parameter hidden in $\Gamma$, so the resolution of (7.40) does not mandatorily conform to chronological order.

*4. To leak, or not to leak…*

Besides a need for numerical integrations to get $M(T_{max}, \tau_{max})$ the main difficulty of (7.40) is that it may fail to have solutions in $[0,1]^2$ that obey (7.38). A similar problem was already encountered at entry 2 of Subsection VII C. In order for $N = N_{tot}$ to yield a viable solution with a single crest at $x = 0$, $\Gamma(\sigma)$ had to be lower than a threshold defined as $\Gamma_\star(\nu N_{tot}) = 2\max_{0 \leq T_{max} \leq 1}[(\frac{1}{\pi}\sin^{-1}(T_{max}) - \nu N_{tot})(1/T_{max}^2 - 1)]$, so $\Gamma_\star(\nu N_{tot}) = 0$ if $\nu N_{tot} \geq \frac{1}{2}$ and $\Gamma_\star(\nu N_{tot}) > 0$ otherwise; this largest $\Gamma(\sigma)$ compatible with (7.40) and $\tau_{max} = 0 = n$, met at $T_\star(\nu N_{tot})$, satisfies $\mathbb{K}(T_\star, 0, \Gamma_\star) = 0$, i.e., $T_{max} = T_\star$ is a double root of $M(T_{max}, 0) - \Gamma_\star L(T_{max}, 0) = \nu N_{tot}$. Double roots and 'instant' pole jumps were also encountered at semi-stable maxima [saddle-nodes] of slow curves in the LF dynamics of one-pair or two-pair fronts [SubSecs. VII A, B].

Here (7.40) has double roots exactly when $\mathbb{K}\Bbbk = 0$ [see Appendix E], precisely when *one* of the pole densities [not both, see above] starts flattening at the ends of its support, which signals nascent leakage. As stressed earlier $N + n = N_{tot}$ and the normalizations in (7.40) do not suffice to get $T_{max}$, $\tau_{max}$ *and* $N$ when all vary. To encode the now-continuous shift with $N(\sigma)$ of saddle-nodes where pole jumps occur, it is surmised that

$$\mathbb{K}(T_{max}, \tau_{max}, \Gamma) = 0 \quad or \quad \Bbbk(T_{max}, \tau_{max}, \Gamma) = 0 \qquad (7.42)$$

*is the missing information needed to compute $N(\sigma)$ and $n(\sigma)$ when one of the piles is losing its poles*; the other pile collects them since $\nu N + \nu n = \nu N_{tot}$ is conserved. Pile leakage is then controlled by those remotest poles still present in the pile but on the verge of leaving it, hence by (7.40) having a double root $T_{max}(\sigma)$ [if $\mathbb{K} = 0$] *or* $\tau_{max}(\sigma)$ [if $\Bbbk = 0$]. This is qualitatively analogous to pouring water by quasi-steadily tilting a partially filled vessel at an angle $\Gamma(\sigma)$ to the vertical: leakage starts as the liquid surface gets tangent to the vessel brim, and the tangency condition akin to (7.42) holds while $\Gamma$ further increases and overflowing continues.

It is also known from Appendix E that the normalizations (7.40) supplemented with (7.42) [and $\nu n = \nu N_{tot} - \nu N$] constitute a solvable $3 \times 3$ system for $T_{max}(\sigma)$, $\tau_{max}(\sigma)$ *and* $\nu N(\sigma)$; all of them depend on $\sigma$ through $\Gamma(\sigma)$ only, and on the prescribed $\nu N_{tot}$.

A first question arises then: when does pole leakage stop, given that in most physically interesting cases $\Gamma(\sigma)$ will oscillate, *e.g.*, $\Gamma(\sigma) = \omega \Delta \sin(\sigma - \zeta)$ ? For definiteness consider (7.40) supplemented by $\mathbb{K}=0$ in a situation where the crest at $x=0$ is currently losing its poles, and near the time $\sigma_m$ when $\Gamma(\sigma)$ approaches a maximum $\Gamma_m$, e.g., near $\sigma_m - \zeta = \frac{\pi}{2}$ if $\Gamma(\sigma) = \omega\Delta\sin(\sigma - \zeta)$. Because it would only depend on $\Gamma(\sigma)$ if (7.40) and $\mathbb{K}=0$ were to hold, $N(\sigma)$ would increase for $\sigma \gtrsim \sigma_m$: that is not viable though, because $\mathbb{K}=0$ implies $\Bbbk>0$, hence the pile at $x=\pi$ is not yet losing its poles so $N(\sigma) = N_{tot} - n(\sigma)$ cannot increase.

The way out is thus: the normalizations (7.40) being irremovable, $\mathbb{K}=0$ *must cease to hold, and N and n must stop varying, immediately after* $\Gamma(\sigma)$ *has reached its maximum.* For $\sigma \gtrsim \sigma_m$, $N$ and $n$ keep the values determined by $\Gamma_m$ they acquired at $\sigma = \sigma_m$, leaving (7.40) alone to provide $T_{\max}(\sigma)$ and $\tau_{\max}(\sigma)$. Pole leakage from the pile at $x = \pi$ stops similarly, just *after* $\Gamma(\sigma)$ has reached a minimum. By distinguishing between $\sigma \leq \sigma_m$ and $\sigma > \sigma_m$ this mechanism identifies the forward time direction as the preferred one, see Fig.31. In the hydraulic analogy evoked above, the liquid ceases to leak when the vessel angle of tilt starts decreasing; from then on the vessel content keeps the value it got when outflow stopped.

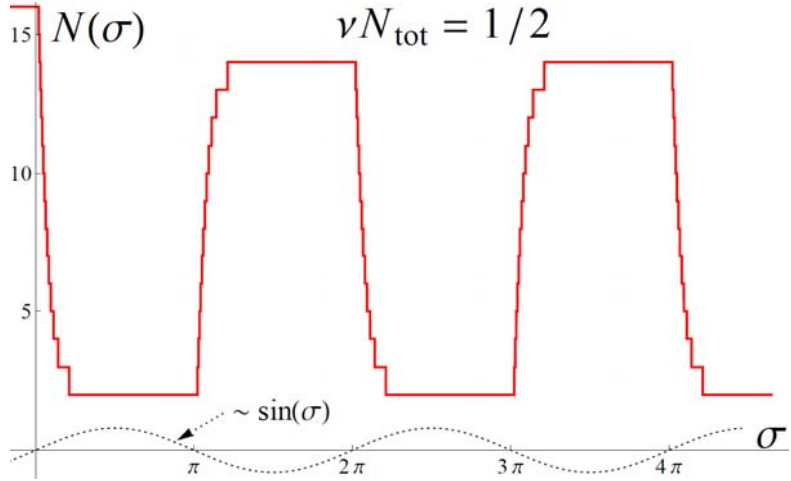

FIG.31 (Color online). Number $N(\sigma)$ of poles in the pile sat at $x = 0 (\mathrm{mod}\, 2\pi)$, as computed numerically from the pole equations (3.4) with $\Gamma(\sigma) = \omega\delta\sin(\sigma)$, $\omega\delta = 1$, $\nu N_{tot} = \frac{1}{2}$, $\omega = \frac{1}{40}$, $N_{tot} = 16$. Note the upper or lower plateaus of equal durations, the looking alike steep transitions between them, the absence of $\sigma \leftrightarrow -\sigma$ symmetry, and the one-shot approach to the limit cycle despite non-MS initial pole locations.

A second question is: when does pole leakage start? If it does, this occurs when $\mathbb{K}=0$ is first encountered while $\Gamma(\sigma) > 0$ is increasing towards $\Gamma_m$, say at $\sigma = \sigma_c$, or when $\Bbbk = 0$ is met while $\Gamma(\sigma) < 0$ still decreases. As $\Gamma(\sigma)$ is to cross such points, the Jacobian associated with (7.40) *at fixed N and* $\sigma$ also fades [Appendix E, §1]; provided $\frac{d}{d\sigma}\Gamma \neq 0$, $\frac{d}{d\sigma}T_{\max}$ or $\frac{d}{d\sigma}\tau_{\max}$ diverges at $\sigma = \sigma_c - 0$, signaling that one of the piles is to undergo pole leakage. Yet this assumes a smooth enough dynamics prior to $\sigma = \sigma_c$ to allow $\mathbb{K}\Bbbk \to 0$ to be reached in a limit $\sigma \to \sigma_c - 0$. This is what happens if $\nu N_{tot} < \frac{1}{2}$; when $\nu N_{tot} > \frac{1}{2}$ discontinuous pile contents are encountered, and $\nu N_{tot} = \frac{1}{2}$ sort of combines both trends.

If $\nu N_{tot} < \frac{1}{2}$ and $|\Gamma|$ is small enough, there is also the possibility that no leakage at all occurs.

### *5. Communicating vessels*

Although believed sufficient to compute $N(\sigma)$, $T_{\max}(\sigma)$ and $\tau_{\max}(\sigma)$ unambiguously in the general case, the above rules soon get tedious to exploit in practice if $\Gamma(\sigma)$ exhibits several extrema and many zeros per period. So, only $\Gamma(\sigma) = \omega\Delta \sin(\sigma - \zeta)$ is to be envisaged here. The symmetry of (7.40)(7.42) by $(T_{\max}, \tau_{\max}, +\Gamma, N, n) \leftrightarrow (\tau_{\max}, T_{\max}, -\Gamma, n, N)$, along with the sign flip of $\sin(\sigma - \zeta)$ by $\sigma \leftrightarrow \sigma + \pi$, imply that knowing the half-period $0 < \sigma - \zeta < \pi$ enables one to reconstruct the whole pile-content history if it is $2\pi$-periodic. The $\nu N_{tot} \geq \frac{1}{2} = \nu N_{opt}(\nu)$ cases are to be considered first, $\nu N_{tot} < \frac{1}{2}$ with its several variants will come next.

- $\nu N_{tot} \geq \frac{1}{2}$

As stressed a few lines after (7.34), $\nu N_{tot} \geq \frac{1}{2}$ implies $(1 - T_{\max})(1 - \tau_{\max}) = 0$ when $\Gamma(\sigma) = 0$, which one may locate at $\sigma - \zeta = 0$, hence $h(T_{\max}, \tau_{\max}) = 0$ by (7.39); to prevent $\mathbb{K} \geq 0$ from being violated as soon as $\Gamma(\sigma) > 0$, the pile sat at $x = 0$ immediately starts to lose it content at $\sigma_c - \zeta = 0^+$. As is detailed in Appendix E, Eqs. (7.39)(7.40), if recast in the form

$$M(T_{\max}, \tau_{\max}) + M(\tau_{\max}, T_{\max}) - h(T_{\max}, \tau_{\max})[L(T_{\max}, \tau_{\max}) - L(\tau_{\max}, T_{\max})] = \nu N_{tot} ,$$
$$\Gamma = h(T_{\max}, \tau_{\max}) ,$$
(7.43)

determine $T_{\max}$ and $\tau_{\max}$ *once for all* as $\nu N_{tot}$-dependent functions of $\Gamma > 0$. The first line of (7.40) next provides $\nu N(\sigma) = \mathfrak{N}(\Gamma(\sigma), \nu N_{tot})$, where $\mathfrak{N}(\Gamma, \nu N_{tot})$ does not depend on the detailed evolutions of $\Gamma(\sigma)$ and decreases with $\Gamma$ from $\mathfrak{N}(0^+, \nu N_{tot}) = \frac{1}{4}$ to $\mathfrak{N}(\infty, \nu N_{tot}) = 0$.

Leakage will continue until $\sigma_m - \zeta = \frac{\pi}{2}$, at which point $\nu N(\sigma)$ reaches its minimum $\nu N_m \equiv \mathfrak{N}(\Gamma_m, \nu N_{tot})$, $\Gamma_m = \omega\Delta$ here. By (7.42) $\nu N(\sigma)$ keeps the same $\nu N_m$ for $\frac{\pi}{2} \leq \sigma - \zeta < \pi$, over which interval $\Gamma(\sigma) > 0$ and the pile at $x = \pi$ does not leak yet because $\Bbbk > 0$, by (7.38).

Periodicity and the aforementioned symmetries determine what will happen over $\pi \leq \sigma - \zeta < 2\pi$; this is illustrated in Figs. 32 and 33.

Importantly, $\nu N = \nu N_{tot} - \mathfrak{N}(\Gamma_m, \nu N_{tot})$ [reached at $\sigma - \zeta = 2\pi^-$] and $\nu N = \frac{1}{4} - 0$ [obtained as $\sigma - \zeta \to 0^+$] do differ: as expected [see VII D §1] $\nu N(\sigma)$ is *discontinuous* at $\sigma - \zeta = \pm 0$ (mod $\pi$). It follows from $\mathfrak{N}(0^+, \nu N_{tot}) = \frac{1}{4}$ that $[\nu N]_{0-}^{0+} = \frac{1}{4} - \nu N_{tot} + \mathfrak{N}(\Gamma_m, \nu N_{tot}) \leq \frac{1}{2} - \nu N_{tot}$, hence the discontinuity of $\nu N(\sigma)$ disappears *if and only if* $\nu N_{tot} \to \frac{1}{2}^+$ and $\Gamma_m = \omega\Delta \to 0$. As stressed earlier jumps represent $\omega = 0^+$ idealizations of fast transients, here the pile-to-pile pole transfers, that actually take some time if $\omega$ is small but finite, see Fig.31. For $\nu N_{tot} > \frac{1}{2}$ their persistence as $\Gamma_m \to 0$ relates to the absence of a steady pole population when $\Gamma(\sigma)$ is cut off, since $2N_{tot} - 1/\nu$ poles must then escape to $\pm i\infty$: sgn($\Gamma$) tells them where to land.

Note that $\Gamma_m \to 0$ implies $\nu N(\sigma) \to \frac{1}{4} - 0$ [or $\to \nu N_{tot} - \frac{1}{4}$] over $0 < \sigma - \zeta < \pi$ [or over $\pi < \sigma - \zeta < \pi$]. Furthermore $\mathfrak{N}(\Gamma \gg 1, \nu N_{tot}) \approx 1/2\pi^2\Gamma$ tells that strong combined stretch yields $N(\sigma)n(\sigma) \approx N_{tot}/2\pi^2 |\Gamma(\sigma)|$: similar to the Burgers case, but never quite zero. Here, $\nu N(\sigma)$ is sandwiched between two square waves with extreme values $(\frac{1}{4}, \nu N_{tot} - \frac{1}{4})$ or $(0, \nu N_{tot})$, and is always bounded by one oscillating between $\nu N(\sigma_m)$ and $\nu N_{tot} - \nu N(\sigma_m)$.

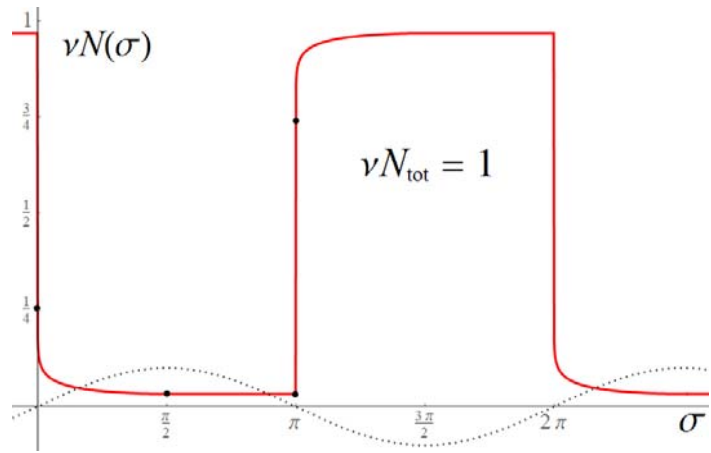

FIG.32 (Color online) Evolution of $\nu N(\sigma)$ for a LF bi-coalesced front [solid red line] with $\nu N_{tot} = 1, \Gamma(\sigma) \equiv \omega\delta \sin(\sigma)$, $\omega\delta = 0.75$. Equation (E.7) gives $\nu N(\sigma)$ over $0 < \sigma \leq \frac{\pi}{2} (\mathrm{mod}\, 2\pi)$ while $\nu N(\sigma) = \nu N(\frac{\pi}{2})$ for $\frac{\pi}{2} \leq \sigma < \pi (\mathrm{mod}\, 2\pi)$; $\nu N(\sigma)$ and $\nu n(\sigma) = \nu N_{tot} - \nu N(\sigma)$ exchange their roles during the second half-periods $\pi \leq \sigma < 2\pi (\mathrm{mod}\, 2\pi)$. The dotted sine curve representing $\Gamma(\sigma)$ sets the global phase.

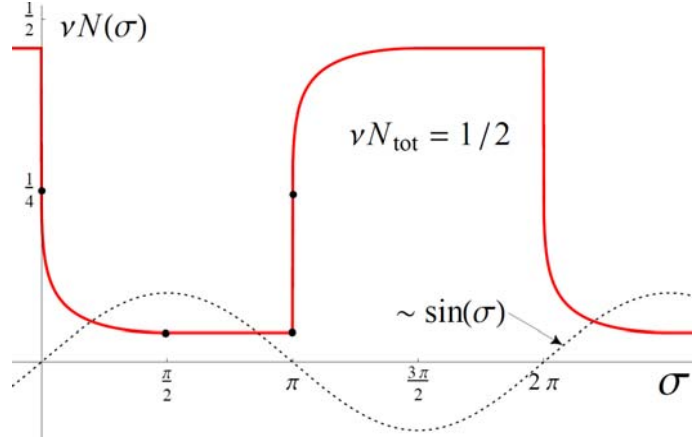

FIG.33 (Color online) Same as in Fig.32, for $\nu N_{tot} = \frac{1}{2}$; note the milder variations of $\nu N(\sigma)$ compared to Fig.32. The condition $\mathbb{K} = 0$ [or $\Bbbk = 0$] again holds only over $0 < \sigma \leq \frac{\pi}{2}$ [or $\pi < \sigma \leq \frac{3\pi}{2}$] $(\mathrm{mod}\, 2\pi)$.

- $\nu N_{tot} < \frac{1}{2}$

It is still convenient to first tabulate $T_{\max}(\Gamma)$, $\tau_{\max}(\Gamma)$ and $\nu N = \mathfrak{N}(\Gamma, \nu N_{tot})$ for a leaking pile located at $x = 0$, again by using (7.43).

The first novelty as compared to $\nu N_{tot} > \frac{1}{2}$ is that in order for (7.43) to have solutions, and for some leakage to occur, $\Gamma$ must *exceed* $\Gamma_\star(\nu N_{tot})$ met apropos one-crested cells [SubSec. VII C, entry 2]. $\Gamma \to \Gamma_\star(\nu N_{tot}) + 0$ in effect leads to $\tau_{\max}(\Gamma) \to 0^+$ and $\mathfrak{N}(\Gamma, \nu N_{tot}) \to \nu N_{tot}$ which pertain to an empty pile at $x = \pi$; in this limit $T_{\max}(\Gamma) \to T_\star(\nu N_{tot}) < 1$ from below.

The second new point is this: if $\Gamma = 0$ the normalizations (7.40) have roots $(T_{\max}, \tau_{\max})$ for *any* $\nu N$ and $\nu n$ such that $\nu N + \nu n = \nu N_{tot} < \frac{1}{2}$, and they will continue to do so with the same constant $\nu N$ and $\nu n$ as long as $\Gamma$ is 'sufficiently' small. The front dynamics may thus start at $\sigma - \zeta = 0 (\mathrm{mod}\, 2\pi)$ with *some* MS shape [20]. To determine which MS solution and $\nu N$ are picked one anticipates that $\nu N(\sigma)$ is continuous here, and periodicity is invoked next: when leakage just begins, at time $\sigma = \sigma_c$ and current $\Gamma_c = \Gamma(\sigma_c)$, $\nu N(\sigma)$ from (7.43) must coincide with $\nu N_{tot} - \mathfrak{N}(\Gamma_m, \nu N_{tot})$, reached at $\sigma - \zeta = 2\pi$. This translates to:

$$\nu N_{tot} - \mathfrak{N}(\Gamma_m, \nu N_{tot}) = \mathfrak{N}(\Gamma_c, \nu N_{tot}) \ . \tag{7.44}$$

With $\mathfrak{N}(\Gamma, \nu N_{tot})$ monotonously decreasing from $\mathfrak{N}(\Gamma_\star, \nu N_{tot}) = \nu N_{tot}$ to $\mathfrak{N}(\infty, \nu N_{tot}) = 0$ [see Appendix E, Fig. E2], (7.44) determines a unique $\Gamma_c(\Gamma_m, \nu N_{tot}) \geq \Gamma_\star(\nu N_{tot})$ that yields the value $\nu N_{tot} - \mathfrak{N}(\Gamma_m, \nu N_{tot}) = \nu n(\sigma_m)$ of $\nu N(\sigma)$ all over $0 \leq \sigma - \zeta \leq \sigma_c - \zeta$. Strong combined stretch implies $\mathfrak{N}(\Gamma, \nu N_{tot}) \approx 1/2\pi^2 \Gamma \ll 1$ when (7.42) holds, in which case $\Gamma_c(\Gamma_m, \nu N_{tot})$ tends to $\Gamma_\star(\nu N_{tot})$ from above, by (7.44).

While $\Gamma_m \geq \Gamma_\star(\nu N_{tot})$ is a prerequisite for leakage, it is not sufficient. To ensure that $\Gamma(\sigma)$ reaches $\Gamma_c(\Gamma_m, \nu N_{tot})$ *sooner* than $\sigma_m - \zeta$, $\Gamma_m$ will have to exceed *a minimum amplitude* $\Gamma_{\min}(\nu N_{tot}) \geq \Gamma_\star(\nu N_{tot})$ defined by $\Gamma_c(\Gamma_{\min}, \nu N_{tot}) = \Gamma_{\min}$. By equation (7.44) this requires

$$\mathfrak{N}(\Gamma_{\min}(\nu N_{tot}), \nu N_{tot}) = \tfrac{1}{2}\nu N_{tot} , \qquad (7.45)$$

which selects $\Gamma_{\min}(\nu N_{tot})$ from the already tabulated function $\mathfrak{N}(\Gamma, \nu N_{tot})$. Interestingly, $\Gamma_{\min}(\nu N_{tot})$ also is the value of $|\Gamma(\sigma)|$ whenever $\nu N(\sigma)$ and $\nu n(\sigma) = \nu N_{tot} - \nu N(\sigma)$ cross.

If the condition $\Gamma_m \geq \Gamma_{\min}(\nu N_{tot})$ is met, $\sigma_c$ itself ensues from $\Gamma(\sigma_c) = \Gamma_c$, e.g., $\sigma_c - \zeta = \sin^{-1}[\Gamma_c(\omega\Delta, \nu N_{tot})/\omega\Delta] \leq \tfrac{\pi}{2}$ in the present situation. The pile at $x = \pi$ starts losing poles ½ period later. Figure 34 shows the $\nu N(\sigma)$ history belonging to $\nu N_{tot} = \tfrac{1}{4}$ and $\Gamma_m = \omega\Delta = 0.75$.

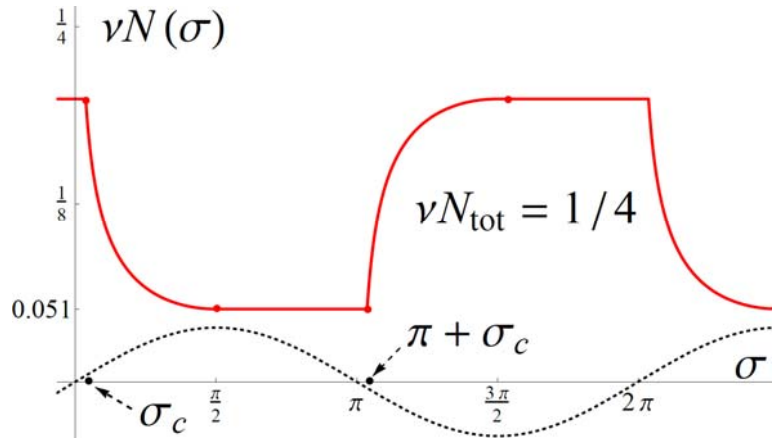

FIG.34 (Color online) $\nu N(\sigma)$ *vs.* $\sigma$ for $\nu N_{tot} = \tfrac{1}{4}$, $\Gamma_m = \omega\delta = \tfrac{3}{4}$ whereby $\mathfrak{N}(\Gamma_m, \tfrac{1}{4}) \approx 0.0510$, $\sigma_c \approx 0.1149$. The oscillation of $\nu N(\sigma)$ about $\tfrac{1}{2}\nu N_{tot}$ fades if $\Gamma_m \to \Gamma_{\min}(\tfrac{1}{4}) \approx 0.1969$ from above.

Appendix E illustrates how $\nu N_{tot} = \tfrac{1}{2} - 0$ makes $\mathfrak{N}(\Gamma \ll 1, \nu N_{tot})$ 'instantly' drop from $\mathfrak{N}(\Gamma_\star, \nu N_{tot}) = \nu N_{tot}$ to about $\tfrac{1}{4}$, whilst $\tau_{\max}(\Gamma)$ springs out of $0$ to cross $T_{\max}(\Gamma)$ at $T_{\max} = \tan(\tfrac{1}{2}\pi\nu N_{tot}) = 1^-$ [Figs. E2, E3]. The extra sensitivity is due to $M(1, \tau_{\max}) + M(\tau_{\max}, 1) \equiv \tfrac{1}{2}$, an identity valid for *all* $0 \leq \tau_{\max} \leq 1$ [Appendix E, §3], whereby infinitesimal $\Gamma$s induce dramatic rearrangements of near-optimally filled, near-MS pole populations. Moreover $\Gamma_\star(\nu N_{tot}) \sim$ $(\frac{1}{2} - \nu N_{tot})^3$ [where $\mathfrak{N}(\Gamma_\star, \nu N_{tot}) = \nu N_{tot} \approx \frac{1}{2}$] vanishes faster than $\Gamma_{min}(\nu N_{tot}) \sim (\frac{1}{2} - \nu N_{tot})^2$ [where $\mathfrak{N}(\Gamma_{min}, \nu N_{tot}) = \frac{1}{2} \nu N_{tot} \approx \frac{1}{4}$]. $\mathfrak{N}(\Gamma, \nu N_{tot})$ can resume a decay below $\frac{1}{4}$ similar to what $\nu N_{tot} = \frac{1}{2}$ gave, but only beyond $\Gamma \sim \Gamma_{min}(\nu N_{tot}) \gg \Gamma_\star(\nu N_{tot})$. Put in words, *the continuous $\nu N(\sigma)$ from* (7.43) *steepens for* $\nu N_{tot} \to \frac{1}{2} - 0$, *which builds up the discontinuity* from $\frac{1}{2}$ to $\frac{1}{4}$ found at $\nu N_{tot} = \frac{1}{2}$; this is due to the MS patterns being infinitely degenerate when $\nu N_{tot} = \frac{1}{2}$.

If $\nu N_{tot} < \frac{1}{2}$ is fixed and $\Gamma_m \to \Gamma_{min}(\nu N_{tot}) + 0$, pole leakage lasts only for a short while prior to $\sigma_m - \zeta$, $\Gamma_c(\Gamma_m, \nu N_{tot}) \to \Gamma_m$ and (7.44) indicates that $\mathfrak{N}(\Gamma_c, \nu N_{tot}) \approx \mathfrak{N}(\Gamma_m, \nu N_{tot})$ both approach $\frac{1}{2} \nu N_{tot}$: $\nu N(\sigma)$ and $\nu n(\sigma)$ then feebly oscillate in phase opposition around $\frac{1}{2} \nu N_{tot}$. Because $\Gamma_{min}(\nu N_{tot}) \sim (\frac{1}{2} - \nu N_{tot})^2 \to 0$ when $\nu N_{tot} - \frac{1}{2} \to 0^-$, this matches the vanishingly small oscillations of $\nu N(\sigma)$ and $\nu n(\sigma)$ which $\Gamma \to 0$ *and* $\nu N_{tot} - \frac{1}{2} \to 0^+$ produce.

LF *oscillating bi-coalesced fronts involving pole leakage cease to exists* if $\Gamma_m < \Gamma_{min}(\nu N_{tot})$. In such circumstances the normalizations in (7.40) suffice to get $T_{max}(\sigma)$ and $\tau_{max}(\sigma)$ for a *continuum* of pairs $(N, n)$ obeying $\nu N + \nu n = \nu N_{tot} < \frac{1}{2}$; yet this is restricted by the fact that all durably oscillating fronts have $\min(N,n)/\max(N,n)$ not smaller than a $\Gamma_m$-dependent bound $\Omega(\Gamma_m, \nu N_{tot})$ if $\Gamma_\star(\nu N_{tot}) \leq \Gamma_m \leq \Gamma_{min}(\nu N_{tot})$. Since pole leakage provisionaly involves $\mathbb{K}\Bbbk = 0$ when pruning away overpopulated piles, the bound relates to the tabulated solution $\nu N = \mathfrak{N}(\Gamma, \nu N_{tot}) = \nu N_{tot} - \nu n$ to (7.40)(7.42). To guarantee that the more loaded pile will not be chopped off any further, the allowed $n$ and $N$ must satisfy:

$$\frac{\min(N,n)}{\max(N,n)} \geq \Omega(\Gamma_m, \nu N_{tot}) = \frac{\nu N_{tot}}{\mathfrak{N}(\Gamma_m, \nu N_{tot})} - 1 . \qquad (7.46)$$

One verifies that $\Omega(\Gamma_\star(\nu N_{tot}), \nu N_{tot}) = 0$ [even one-crested fronts are then allowed], whereas $\Omega(\Gamma_{min}(\nu N_{tot}), \nu N_{tot}) = 1$ implies $N = n$ [the continuum of solutions has shrunk to a point]. Because $\max(N,n) + \min(N,n) = N_{tot}$, (7.46) also reads as $\max(N,n) \leq \mathfrak{N}(\Gamma_m, \nu N_{tot})$, where the definition (7.45) of $\Gamma_{min}(\nu N_{tot})$ implies $\frac{1}{2} \nu N_{tot} \leq \mathfrak{N}(\Gamma_m, \nu N_{tot}) \leq \nu N_{tot}$ if $\Gamma_\star(\nu N_{tot}) \leq \Gamma_m \leq \Gamma_{min}(\nu N_{tot})$, see Fig. (E.2) of Appendix E. The otherwise arbitrary $\max(N,n)$ must then lie in the domain $[\frac{1}{2} \nu N_{tot}, \mathfrak{N}(\Gamma_m, \nu N_{tot})]$, which extends to $[\frac{1}{2} \nu N_{tot}, \nu N_{tot}]$ for $\Gamma_m \to \Gamma_\star(\nu N_{tot})^+$.

*Arbitrary* $N \geq 0$ and $n \geq 0$ obeying $\nu N + \nu n = \nu N_{tot} < \tfrac{1}{2}$ get allowed if $0 \leq \Gamma_m \leq \Gamma_\star(\nu N_{tot})$; they also produce crests breathing is phase opposition at $x=0$ and/or $\pi \pmod{2\pi}$. The cases $\Gamma(\sigma) \equiv 0$ are $N_{tot} \sim 1/\nu \gg 1$ analogs of Fig.1: the polar slope (7.36) reduces to *one* of the steady MS profiles with constant $T_{\max}$ and $\tau_{\max}$, while the base slope $\Phi_x(\sigma, x)$ still fluctuates.

In sum, only a bi-coalesced front is viable in the $\omega \to 0$ limit if $\Gamma_m \neq 0$ and $\nu N_{tot} \geq \tfrac{1}{2}$; the evolution of the pile contents is then only piecewise continuous, and includes plateaus. If $\nu N_{tot} < \tfrac{1}{2}$ a two-pile LF front exists as long as $\Gamma_m$ exceeds the threshold $\Gamma_{\min}(\nu N_{tot})$, with the pile contents evolving continuously in a definite way, and plateaus still present. But if $0 \leq \Gamma_m \leq \Gamma_{\min}(\nu N_{tot})$ the pile contents stay constant and *arbitrary in a range*; the minimum durable pile-content ratio depends on $\Gamma_m \geq \Gamma_\star(\nu N_{tot})$ in a known way. Lastly, *any* $N \geq 0$ *and* $n \geq 0$ adding up to $\nu N_{tot} < \tfrac{1}{2}$ are allowed if $0 \leq \Gamma_m \leq \Gamma_\star(\nu N_{tot})$. Whether all this influences the flame shape, the arclength increment $V(\sigma)$ and its time-average $V_{av}$ is examined next.

### 6. Shapes & Speeds

- Simple Solutions.

The aforementioned LF fronts with $w(\sigma) = \Phi(\sigma)$ but $w(\sigma) \not\equiv 0$ arguably are the simplest nontrivial ones. For them $T_{\max}$ and $\tau_{\max}$ are the $\sigma$-independent roots of $M(T_{\max}, \tau_{\max}) = \nu N$ and $M(\tau_{\max}, T_{\max}) = \nu n$, where the *arbitrary* $\nu N$ and $\nu n$ add to $\nu N_{tot} < \tfrac{1}{2}$. Here $\Gamma \equiv 0$, so the front slope reduces to the first line of (7.36). From the LF form of (3.8) one deduces

$$V(\sigma) = \tfrac{1}{4} w(\sigma)^2 + \mathcal{V}(\nu N_{tot}) + 2w(\sigma)[\int_0^{T_{\max}} \frac{2\nu \rho_{MS}(T)dT}{(1+T)^2} - \int_0^{\tau_{\max}} \frac{2\nu \varrho_{MS}(\tau)d\tau}{(1+\tau)^2}] \ . \tag{7.47}$$

This is quite simple if $n = \tfrac{1}{2} N_{tot}$, as $T_{\max} = \tau_{\max} = \tan(\tfrac{1}{2}\pi\nu N_{tot})$ and the above square bracket vanishes. Next, $n = 0$ leads to $\tau_{\max} = 0$ and $T_{\max} = \sin(\pi\nu N_{tot})$, a special case of (7.24)(7.25): $V(\sigma) = \mathcal{V}(\nu N_{tot}) + \tfrac{1}{4} w(\sigma)^2 + 2w(\sigma)\tan(\pi\nu N_{tot})(1 - 2\nu N_{tot})/\pi$. Such $N_{tot} \sim 1/\nu \gg 1$ versions of (4.2) share the same $V_{av} = \mathcal{V}(\nu N_{tot}) + \tfrac{1}{4}[w(\sigma)^2]_{av}$, as do all $\Gamma \equiv 0$ solutions that only differ by the above two constant integrals in $V(\sigma)$. As noted earlier, however, these are principally useful as computational checks, because the forcing function they share for $(\omega, \nu) \to 0$, $u(t,x) \approx -w(t)\cos(x) + \tfrac{1}{4} w(t)^2 \cos(2x)$ from (3.6)(3.7), has $[u(t,x)]_{av} \neq 0$ even if $[w(t)]_{av} = 0$.

- $\nu N_{tot} \geq \frac{1}{2}$

These are the next-to-simplest nontrivial cases, where one of the pole piles starts loosing its content as soon as $\Gamma(\sigma)$ changes sign, *e.g.*, at $\sigma - \zeta = +0 (\mathrm{mod}\,\pi)$ if $\Gamma(\sigma) = \Gamma_m \sin(\sigma - \zeta)$, $\Gamma_m = \omega \Delta$, see Figs. 32 and 33. Over the first quarter-period $0 < \sigma - \zeta \leq \frac{\pi}{2}$, $\nu N(\sigma) \leq \frac{1}{4}$ decays according to the universal curves $\mathfrak{N}(\Gamma, \nu N_{tot})$ tabulated in (E.4c) or (E.5c) for $\nu N_{tot} = 1$ or $\frac{1}{2}$, respectively, with associated $T_{\max}(\Gamma(\sigma))$ and $\tau_{\max}(\Gamma(\sigma))$ given in (E.4a,b) or (E.5a,b). Pole leakage stops once $\Gamma(\sigma)$ has just reached its maximum $\Gamma_m = \omega \Delta$ and $\nu N(\sigma)$ remains at $\mathfrak{N}(\Gamma_m, \nu N_{tot})$ all over $\frac{\pi}{2} \leq \sigma - \zeta < \pi$, see Figs. 32 & 33; the associated $(T_{\max}(\sigma), \tau_{\max}(\sigma))$ are found by solving (7.40) by means of Mathematica's FindRoot [35]. The next half-period $\pi < \sigma - \zeta < 2\pi$ exploits the $(\sigma, \Gamma, N, n, T_{\max}, \tau_{\max}) \leftrightarrow (\sigma + \pi, -\Gamma, n, N, \tau_{\max}, T_{\max})$ symmetry.

For $\nu N_{tot} = 1$, this produced the sample $T_{\max}(\sigma)$ and $\tau_{\max}(\sigma)$ histories displayed in Fig. 35.

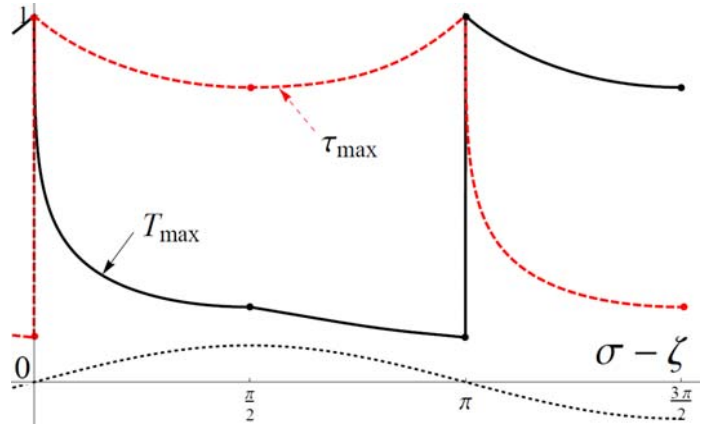

FIG.35 (Color online) $T_{\max}(\sigma)$ [solid black] and $\tau_{\max}(\sigma)$ [dashed red] *vs.* $\sigma - \zeta$ computed from (7.43) [for $0 \leq \sigma - \zeta \leq \frac{\pi}{2}$] or (7.40) [over $\frac{\pi}{2} < \sigma - \zeta < \pi$] with $\nu N_{tot} = 1$, and extended by symmetries & periodicity. The dotted line is proportional to $\Gamma(\sigma) = \omega \Delta \sin(\sigma - \zeta)$, with $\omega \Delta = 0.75$ as in Fig.32. Both $T_{\max}$ and $\tau_{\max}$ equal 1 at $\sigma - \zeta = 0^+ (\mathrm{mod}\,\pi)$, and elsewhere decrease as $\omega \Delta$ grows. Besides, $\tau_{\max} \to 1^-$ for $\omega \Delta \to 0^+$.

With $T_{\max}(\sigma)$ and $\tau_{\max}(\sigma)$ available, the current front slope $\varphi_x(\sigma, x)$ and its polar component $\phi_x(\sigma, x)$ are fully determined from formula (7.36). The resulting polar shape $\phi(\sigma, x)$, here a numerical antiderivative of $\phi_x(\sigma, x)$, is sampled at a few selected instants in Fig.36, which illustrates how the front shape jumps [only $\phi(\sigma, x)$ does] across $\sigma - \zeta = 0 (\mathrm{mod}\,\pi)$; one may also note that the polar shape $\phi(\sigma, x)$ is comprised of twin crests when $\sigma - \zeta = 0^+ (\mathrm{mod}\,\pi)$.

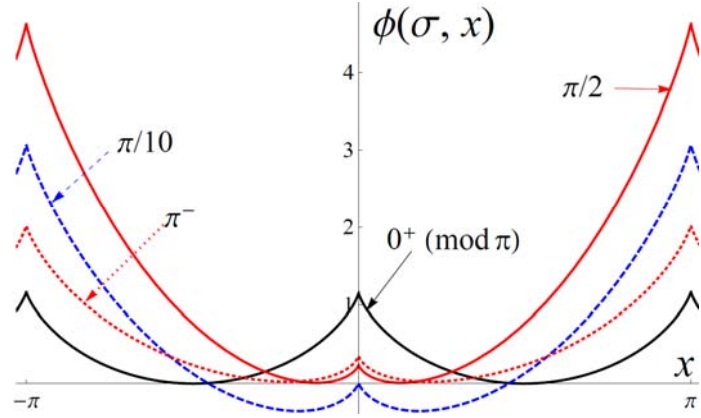

FIG.36 (Color online) Snapshots of the *polar* contribution to the front shape, deduced by numerical integration of $\phi_x(\sigma, x)$ from (7.36), for $\nu N_{tot} = 1, \omega\Delta = 0.75$ as in Fig.34, at the selected times $\sigma - \zeta = +0$ [solid black], $\frac{\pi}{10}$ [dashed blue], $\frac{\pi}{2}$ [red] and $\pi - 0$ [dotted magenta]. Over $\pi < \sigma - \zeta < 2\pi$ the crests centered on $x = 0$ or $\pi$ exchange their roles: $\phi(\sigma, x) = \phi(\sigma - \pi, x - \pi)$, just like $\Phi(\sigma, x) = -\omega\delta\cos(\sigma)\cos(x)$. The total front shape has $|\varphi(\sigma, 0) - \varphi(\sigma, \pi)| = 2\omega\delta$ when $\sigma - \zeta = 0^+ (\mathrm{mod}\,\pi)$.

The same $T_{\max}(\sigma)$ and $\tau_{\max}(\sigma)$ give access to the $\pi$-periodic current front-length increment $V(\sigma)$, via analytical evaluation of the quasi-steady version of (3.8) in terms of pole densities (7.34) [see Appendix E], or from its definition $V(\sigma) = \frac{1}{\pi}\int_0^\pi \frac{1}{2}(\varphi_x(\sigma, x))^2 dx$ and the total slope profile (7.36) via a numerical quadrature; for all its flexibility the latter option was chosen. A subsequent numerical time-average yields $V_{av} = \frac{1}{\pi}\int_0^\pi V(\sigma)d\sigma$ for $\nu N_{tot} = 1$, without stretch [$d = 0$] or with it [$d = \delta, \gamma = \pm\frac{\pi}{3}$].

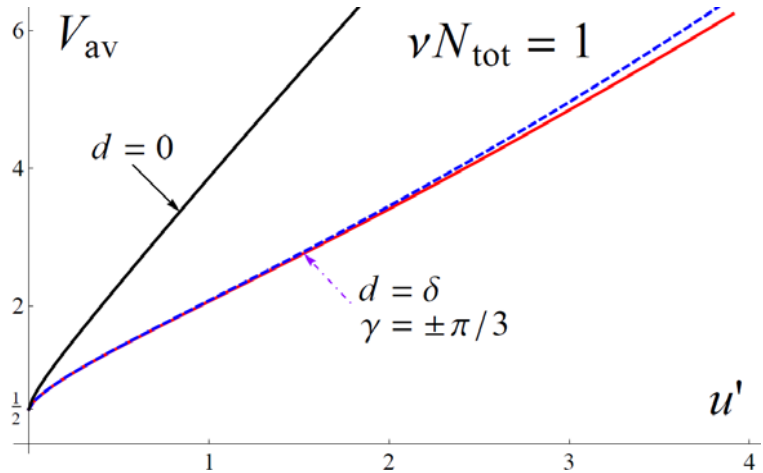

FIG.37 (Color online). Length increment $V_{av}$ vs. flow intensity $u'$ for $\nu N_{tot} = 1$ with $d = 0$ [solid black], or with $d = \delta$ and $\gamma = +\frac{\pi}{3}$ [red] or $\gamma = -\frac{\pi}{3}$ [dashed blue]. The three curves have $V_{av} - \frac{1}{2} \sim (u')^{2/3}$ initially.

When plotted against the maximum combined stretch $0 < \Gamma_m = \omega\delta \leq 3.5$, the stretch affected $V_{av} - \tfrac{1}{2}$ curves nearly are in a constant ratio of 0.84 to the $d = 0$ result, and begin as $\sim (\omega\delta)^{2/3}$ just like static fronts [25]; all next transition to nearly straight, concave-up parabolic growths. The very same responses look rather different, however, when $V_{av}$ is plotted *vs.* the global outer-flow intensity $u'$ defined in (6.48) and specialized here to $(\omega, \nu) \to 0$, namely:

$$u' = \tfrac{1}{2}\omega\delta[(2\nu N_{tot} - 1)^2 + \tfrac{3}{64}(\omega\delta)^2]^{1/2} \quad \text{if } d = 0,$$
$$= \tfrac{1}{2}\omega\delta[(2\nu N_{tot} - 1)^2 + 2\nu N_{tot} + 1 + \tfrac{3}{64}(\omega\delta)^2]^{1/2} \quad \text{if } d = \delta. \quad (7.48)$$

The initial $u'/(\omega\delta)$ ratio indeed significantly varies with $d$, and the ultimate $u' \sim (\omega\delta)^2$ growth encoded in (7.48) tends to straighten the $V_{av}$ *vs.* $u'$ curves, see Fig.37. Similar to the HF results in Figs.14 & 15, the latter effect can be traced back to the quadratic dependence on $\Phi(t)$ of the second spatial overtone $u_2(t)\cos(2x)$ in the forcing function $u(t,x)$, see (3.7).

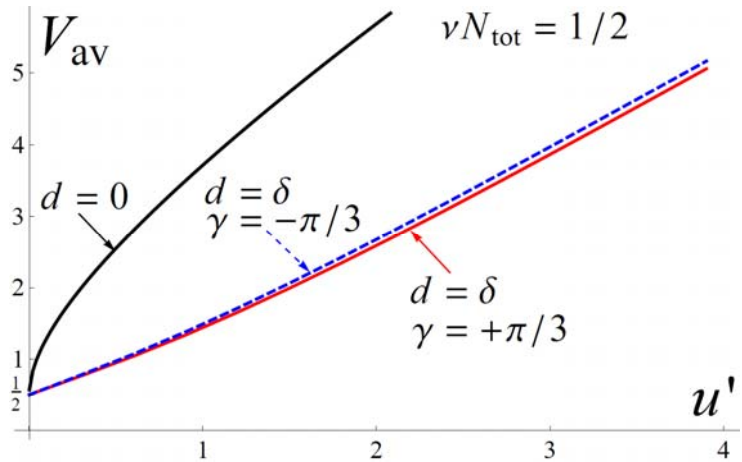

FIG.38 (Color online). Same as in Fig.37, for $\nu N_{tot} = \tfrac{1}{2}$. The $d = \delta$ curves now have $V_{av} - \tfrac{1}{2} \sim u'$ initially, whereas $d = 0$ leads to $V_{av} - \tfrac{1}{2} \sim (u')^{1/2}$ as a consequence of (7.48).

The above steps carry over to $\nu N_{tot} = \tfrac{1}{2}$, to produce qualitatively similar results up to minor variants. For example, $1 - T_{max}(\sigma)$ and $1 - \tau_{max}(\sigma)$ now have a square-root behavior at $\sigma - \zeta \to 0^+ (\text{mod } \pi)$ [stemming from the slower decay of the "universal" $\nu N(\Gamma) = \mathfrak{N}(\Gamma, \tfrac{1}{2})$ in Fig. E.1 of Appendix E], as does $1 - \tau_{max}(\sigma)$ at $\sigma - \zeta \to 0^- (\text{mod } \pi)$. Also, the maximum peak-to-trough amplitude in the counterpart of Fig.36 is reduced by a factor of about 0.75, whereas the polar shape at $\sigma - \zeta = 0^+ (\text{mod } \pi)$ is unchanged. Being still parabolic and in almost fixed ratios to the $d=0$ result, all $V_{av} - \frac{1}{2}$ response curve now are initially linear in $\omega\delta$. The main difference with $\nu N_{tot} = 1$ manifests itself when $V_{av}$ is replotted vs. $u'$: $d = \delta$ produces $V_{av} - \frac{1}{2} \sim u'$ if small, but $d = 0$ gives $V_{av} - \frac{1}{2} \sim (u')^{1/2}$. As with static forcing [25] this faster rise follows from (7.48), as does the nearly linear final growth of $V_{av}$ seen in Fig. 38 at larger $u$'s.

- $\nu N_{tot} < \frac{1}{2}$

These configurations illustrated below for $\nu N_{tot} = \frac{1}{4}$ definitely are the most intricate ones, as the pole dynamics then depends qualitatively on the location of $\Gamma_m = \omega\Delta$ relative to special values: $\Gamma_\star(\frac{1}{4}) = 2\max[(\frac{1}{\pi}\sin^{-1}(T) - \frac{1}{4})(1/T^2 - 1)] \approx 0.0564$ [where the continuum of solutions starts to shrink], $\Gamma_{\min}(\frac{1}{4}) \approx 0.1965$ [onset of pole transfers, defined in (7.45)] and $\Gamma_c(\Gamma_m, \frac{1}{4})$ [value of $\Gamma(\sigma)$ when the pile located at $x = 0 (\bmod 2\pi)$ starts losing its content, see (7.44)].

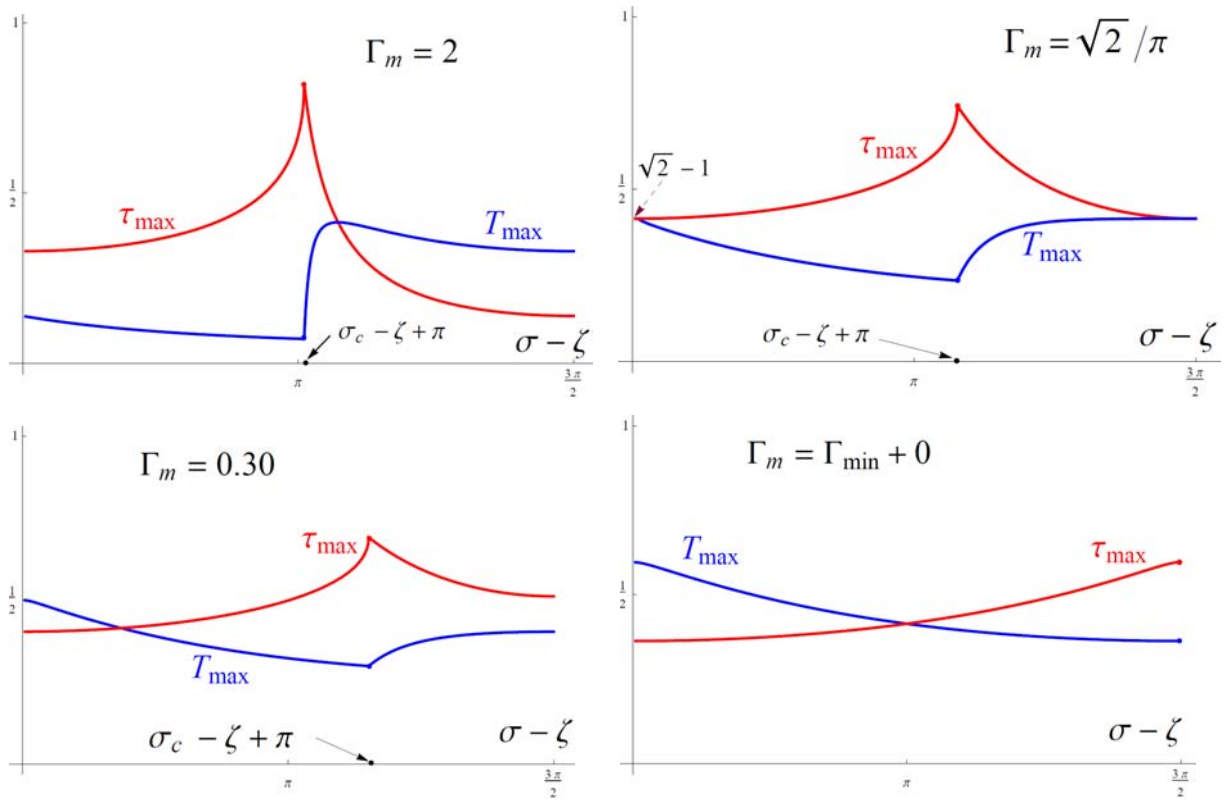

FIG.39 (Color online). Half-periods $\frac{\pi}{2} \leq \sigma - \zeta \leq \frac{3\pi}{2}$ of $T_{\max}(\sigma)$ and $\tau_{\max}(\sigma)$ for $\nu N_{tot} = \frac{1}{4}$ and the indicated values of $\Gamma_m = \omega\Delta$. If $\Gamma_m > \Gamma_{\min}(\frac{1}{4}) \approx 0.1965$, the pile of poles sat at $x = \pi (\bmod 2\pi)$ starts losing its content at $\sigma - \zeta = \sigma_c - \zeta > \pi$; no leakage occurs if $\frac{\pi}{2} \leq \sigma - \zeta < \sigma_c - \zeta$. Over the next half-period, $T_{\max}(\sigma)$ and $\tau_{\max}(\sigma)$ exchange their roles.

The sample $T_{\max}(\sigma)$ and $\tau_{\max}(\sigma)$ histories displayed in Fig. 39 for $\Gamma_m \geq \Gamma_{\min}(\tfrac{1}{4})$ begin their $2\pi$-periodic cycle at $\sigma - \zeta = \tfrac{\pi}{2}$, when $\Gamma(\sigma)$ reaches its maximum $\Gamma_m$ and hence poles stop escaping from their pile at $x = 0 \pmod{2\pi}$. Over the subinterval $\tfrac{\pi}{2} \leq \sigma - \zeta < \pi + \sigma_c - \zeta$, with $\sigma_c - \zeta = \sin^{-1}(\Gamma_c(\Gamma_m, \tfrac{1}{4}) / \Gamma_{\min}(\tfrac{1}{4}))$ ranging from $+0$ to $\tfrac{\pi}{2}$ as $\Gamma_m$ goes from $\infty$ to $\Gamma_{\min}(\tfrac{1}{4})$, no leakage at all takes place and $\nu N$ keeps the value $\nu N_m = \mathfrak{N}(\Gamma_m, \tfrac{1}{2})$ [tabulated in Appendix E] it got at $\sigma - \zeta = \tfrac{\pi}{2}$; the larger $\Gamma_m$, the smaller $\nu N_m$ is. Until $\sigma - \zeta = \pi + \sigma_c - \zeta$, $T_{\max}(\sigma)$ and $\tau_{\max}(\sigma)$ were again computed from (7.40)(7.41) using Mathematica's FindRoot. The rest of the half-period employs the swapped tabulated functions $\tau_{\max}(-\Gamma)$ and $T_{\max}(-\Gamma)$ from (E.4.a,b), and the next half-period ensues from its predecessor by exchanging the two trajectories.

As is implied by (7.40)(7.41), along with $M(a,a) = \tfrac{1}{\pi}\tan^{-1}(a)$ and $\nu N + \nu n = \tfrac{1}{4}$, $T_{\max}(\sigma)$ and $\tau_{\max}(\sigma)$ take on the value $\tan(\tfrac{\pi}{8}) = \sqrt{2} - 1 \approx 0.414$ whenever they cross, e.g. at $\sigma - \zeta = \tfrac{\pi}{2}$ $(\bmod\,\pi)$ if $\Gamma_m = \sqrt{2}/\pi \approx 0.450$. Also, both histories get a smooth derivative as $\Gamma_m$ reaches $\Gamma_{\min}(\tfrac{1}{4})$ and pole leakage completely disappears, and look more and more sinusoidal if $\Gamma_m$ is further reduced. As was mentioned earlier, $\Gamma_\star(\tfrac{1}{4}) < \Gamma_m < \Gamma_{\min}(\tfrac{1}{4})$ allows for a continuum of solutions for which $\nu N$ stays constant and may have any value between $\tfrac{1}{8}$ and $\mathfrak{N}(\Gamma_m, \tfrac{1}{4}) < \tfrac{1}{4}$.

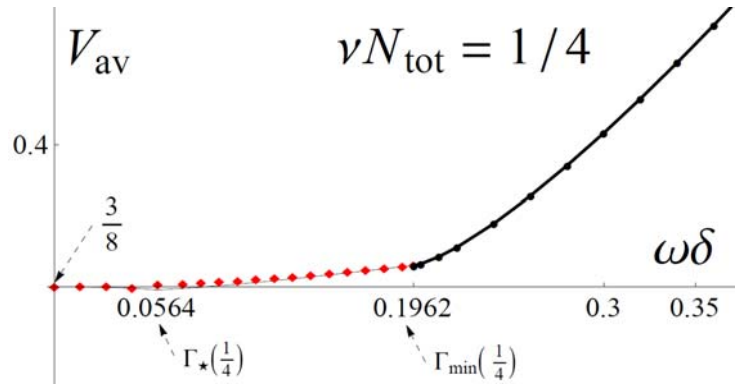

FIG.40 (Color online). Close-up view of the curve $V_{av}$ vs. $\Gamma_m = \omega\delta$ belonging to $\nu N_{tot} = \tfrac{1}{4}, d = 0$. The square symbols [or thin black line] pertain to the upper [or lower] bound of the range of allowed values of $\nu N$ for $0 \leq \omega\delta < \Gamma_{\min}(\tfrac{1}{4})$, viz.: $[\tfrac{1}{8}, \mathfrak{N}(\Gamma_m, \tfrac{1}{2})]$ if $\Gamma_\star(\tfrac{1}{4}) \leq \omega\delta < \Gamma_{\min}(\tfrac{1}{4})$, or $[\tfrac{1}{8}, \tfrac{1}{4}]$ if $0 \leq \omega\delta \leq \Gamma_\star(\tfrac{1}{4})$.

All the resulting values of $V_{av}$, be it when $d = 0$ or with $d = \delta$ and $\gamma = \pm\tfrac{\pi}{3}$, reach the MS $\mathcal{V}(\nu N_{tot}) = 2\nu N_{tot}(1 - \nu N_{tot}) = \tfrac{3}{8}$ as the combined-stretch intensity $\Gamma_m$ $[= \omega\delta$ in either case] goes to zero, yet in practice reach $V_{av} \approx \frac{3}{8} = 0.375$ for all $\Gamma_m \leq \Gamma_{min}(\frac{1}{4})$; Fig.40 illustrates this. A two-fold reason accounts for this state of affairs. First, $\Gamma_{min}(\frac{1}{4})$ is small and $\Gamma_\star(\frac{1}{4})$ even more so. Second, taking $\nu N = \frac{1}{8}$ yields twin crests that feebly oscillate in phase opposition so that their influences on $V_{av}$ nearly compensate one another; besides, choosing $\nu N = \mathfrak{N}(\Gamma_m, \frac{1}{4})$ and a smaller and smaller $\Gamma_m$ rapidly leads to a depleted crest at $x = \pi \pmod{2\pi}$, while the crest at $x = 0 \pmod{2\pi}$ oscillates too weakly to noticeably affect $V_{av}$. The larger choice extends to $\nu N = \frac{1}{4}$ [hence $\nu n = 0$] if $\Gamma_m \leq \Gamma_\star(\frac{1}{4})$, which brings one back to SubSec C, §2 or Fig.30.

Beyond $\Gamma_{min}(\frac{1}{4})$, all $V_{av}(\Gamma_m) - V_{av}(\Gamma_{min}(\frac{1}{4}))$ are parabolic functions of $\Gamma_m = \omega\delta$, roughly proportional to the $d = 0$ result. Once replotted in terms of $u'$ though, these lead to different curves, again due to the $\nu N_{tot}$-dependent correspondence $u' \leftrightarrow \omega\Delta$ encoded in (7.48).

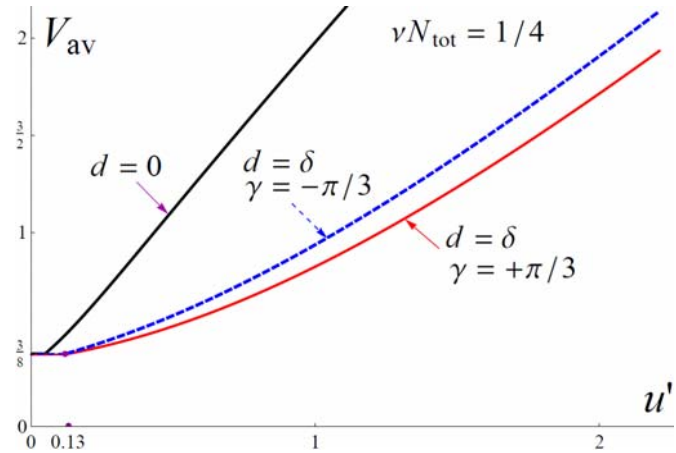

FIG.41 (Color online). $V_{av}$ vs. $u'$ response curves for $\nu N_{tot} = \frac{1}{4}$, in a stretch-free configuration $d = 0$ [solid black] or when stretch is included: $d = \delta$ and $\gamma = +\frac{\pi}{3}$ [lower, red] or $\gamma = -\frac{\pi}{3}$ [dashed blue]. The plateaux at small $u'$ have $V_{av} \approx \frac{3}{8} = 0.375$ and pertain to $0 \leq \omega\delta < \Gamma_{min}(\frac{1}{4}) \approx 0.196$, as in Fig.40.

• Fixed-shape forcing & stretch.

In most of the situations retained so far, three main configurations were focused on: (i) pure shear-flow [$d = 0$, hence no stretch, $w(\sigma, x) = 0$], kept here as a benchmark case and for comparison with the static wrinkled flames of reference [25], even though it yields a nonzero $[u(\sigma, x)]_{av} = -\frac{1}{4}[\Phi(\sigma)^2]_{av} \cos(2x)$, see (3.6)(3.7); (ii) harmonically oscillating base-shape and stretch amplitudes with $d = \delta$, and a phase $\gamma = \pm\frac{\pi}{3}$ chosen to guarantee $[u(\sigma, x)]_{av} = 0$ since $[\frac{1}{2}\Phi(\sigma)(w(\sigma) - \frac{1}{2}\Phi(\sigma))]_{av} = \frac{1}{4}\omega\delta(\omega d\cos(\gamma) - \frac{1}{2}\omega\delta)$ then vanishes.

Selecting $\gamma = 0$ and $d = \frac{1}{2}\delta$ is an alternate means to achieve $[u(\sigma, x)]_{av} = 0$. More generally $\Phi(\sigma) = 2w(\sigma)$ makes (3.3) and (3.6)(3.7) simplify to

$$w(\sigma, x) = \tfrac{1}{2}\Phi(\sigma)\sin(x) ,$$
$$u(\sigma, x) = [\omega \tfrac{d}{d\sigma}\Phi(\sigma) + (\nu N_{tot} - 1 + \nu)\Phi(\sigma)]\cos(x) , \qquad (7.49)$$

hence both stimuli acquire *intensity-independent* spatial dependences. Accordingly $u'$ now is

$$u' = \omega\Delta[\omega^2 + (\nu N_{tot} - 1 + \nu)^2 + \tfrac{1}{4}]^{1/2}, \quad \omega\Delta = \tfrac{1}{2}\omega\delta. \qquad (7.50)$$

Once specialized to long, LF wrinkles and to $\Phi(\sigma) = -\omega\delta\sin(\sigma)$, (7.49) becomes $w(\sigma, x) = -\tfrac{1}{2}\omega\delta\sin(\sigma)\sin(x)$, $u(\sigma, x) = -\omega\delta(\nu N_{tot} - 1)\sin(\sigma)\cos(x)$, and adapting the procedures that gave Figs. 35-41 ultimately produces the curves displayed in Fig. 42.

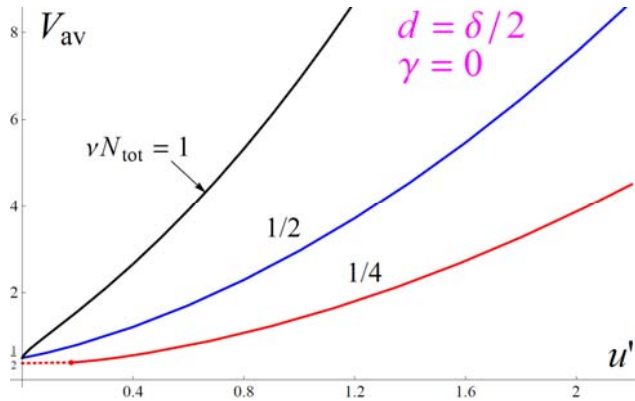

FIG. 42 (Color online) $V_{av}$ vs. $u'$ for quasi-steady bi-coalesced fronts with $d = \tfrac{1}{2}\delta$ and several $\nu N_{tot}$.

The now *constant*, $\nu N_{tot}$-dependent ratio $u'/\omega\Delta$ from (7.50) is responsible for the responses being different and for their *parabola*-like shapes. No curve straightening is expected beyond the $u$'s accessible to the fits from Appendix E. Indeed, $\omega\Delta \gg 1$ implies $\min(\nu N) \sim 1/2\pi^2\omega\Delta$ while $\omega\delta \to 2\omega\Delta$ reinforces the base-shape contribution to $\varphi_x(\sigma, x)$ in (7.36), rendering the front Burgers-like; but $\hat{\gamma} = 0$ and $\hat{d} = \tfrac{1}{2}\hat{\delta}$ in (7.19) *also* give $\hat{V}_{av} \sim \hat{\omega}^2\hat{\Delta}^2$ if $\hat{u}' \sim \hat{\omega}\hat{\Delta} \gg 1$.

The price to pay for stimuli of fixed spatial shapes seems to be a final $V_{av} \sim u'^2$ growth. Yet one may recall that, though $\nu N_{tot}$ in (7.49)(7.50) must be $\sigma$-independent for overall consis-tency of the present approach, its variation with $\omega$ or/and $\delta$ is allowed. Similar to the static case [25] one could assume $\nu N_{tot} = \frac{1}{2} + \frac{1}{8} a\, \omega\delta$, $a = const. > 0$, without spoiling the spatial dependences in (7.49), to yield $u' = \omega\Delta[(\frac{1}{4} a\, \omega\Delta - \frac{1}{2})^2 + \frac{1}{4}]^{1/2}$ in the small-$\nu$ and small-$\omega$ version of (7.50). Such stimuli would tend to pure unsteady shear-flows, $u(\sigma, x) \gg w(\sigma, x)$, at large $\omega\delta$, then resulting in $u' \sim (\omega\Delta)^2 \sqrt{a} \gg 1$. It is unclear at present whether the new way $u'$ and $\omega\Delta$ are linked can convert a $V_{av} \sim u'^2$ law to $V_{av} \sim u'$, because $V_{av}$ also grew with $\nu N_{tot}$ in Fig. 42; and the strategy used thus far employed tabulations available only for $\nu N_{tot} = 1, \frac{1}{2}, \frac{1}{4}$. The Burgers case gives hints though, if an affine growth of $\hat{\nu}\hat{N}$ with $\hat{\omega}\hat{\delta}$ is allowed: the grouping $\hat{\varpi} = \hat{\omega}\hat{\Delta} / \hat{N}\hat{\nu}$ in (7.19) then asymptotes a positive constant, $\hat{\gamma} = 0$ and $\hat{d} = \frac{1}{2}\hat{\delta}$ still result in $\hat{V}_{av} \sim (\hat{\omega}\hat{\Delta})^2$, but now this means $\hat{V}_{av} \sim \hat{u}'$ as guessed. Just like in the static case [25] positing that $2\nu N_{tot} = const. + a\frac{1}{4}\omega\delta$ is expected to end up with $V_{av} \sim u'$ if $a > 0$, up to $\sim \omega\Delta \sim u'^{1/2}$ corrections induced by the DL term in (7.36) and the cross-product it generates in $(\varphi_x)^2$ and hence in $V_{av} \equiv [\langle \frac{1}{2}(\varphi_x)^2 \rangle]_{av}$; $a < 0$ and a large enough $\omega\Delta$ will eventually render the front poleless, at which point $V_{av} = [\langle \frac{1}{2}(\Phi_x)^2 \rangle]_{av} = \frac{1}{2}(\omega\Delta)^2$ as in Sec.IV.

Another aspect of (7.49) is that the outer flow $(u(t, x), w(t, x))$ derives from a stream function $\Psi(\sigma, x, y) = \sin(x)[\Psi_0(\sigma) + \Psi_1(\sigma)\cos(y)]$. More precisely, in $w(\sigma, x) = +\Psi_y(\sigma, x, W(\sigma))$ and $u(\sigma, x) = -\Psi_x(\sigma, x, W(\sigma))$ the derivatives are evaluated at some $y = W(\sigma)$ one may choose, whilst $\Psi_0(\sigma)$ and $\Psi_1(\sigma)$ are adjusted to (7.49). This $\Psi(\sigma, x, y)$ encodes a space-periodic, fluctuating superposition of a shear-flow and stacked line-arrays of incompressible 'square' vortices; randomness is even allowed via $\Phi(\sigma)$, which might be of some use.

## VIII. SUMMARY and DISCUSSION

A previous study [25] of steady wrinkled flames subjected to static shear-flows concluded that "… *the real theoretical problem to address next concerns unsteadily forced flames…*".
To meet the challenge this essay added a time-dependent *forcing* term $u(t, x)$ to the right-hand side of a Michelson-Sivashinsky [MS] equation for front-shape evolutions. Besides the classical MS ingredients – Darrieus-Landau wrinkling instability, geometric nonlinearity and curvature effect – and the space-periodic shear-flow encoded in $u(t,x)$, the new model also included an imposed inhomogeneous flow $w(t,x)$ parallel to the mean front location, which brings about wrinkle *stretch* or compression in the $x$ direction.

Building on [25] the front shape is split as $\varphi(t,x) = \Phi(t,x) + \phi(t,x)$, with a prescribed base shape $\Phi(t,x)$ and a partner $\phi(t,x)$ to determine. The slope $\phi_x(t,x)$ is next posited to admit a decomposition in terms of $N_{tot} = N + n$ pairs of complex-conjugate poles: $N$ imaginary pairs belong to the front crests centred on $x = 0 (\mod 2\pi)$, while $n$ other pairs sharing $\pi (\mod 2\pi)$ as their real part constitute the other crests. The total front shape $\varphi(t,x) = \varphi(t,-x)$ satisfies Neumann conditions at $x = 0(\mod \pi)$ and $2\pi$-periodicity in $x$ if $\Phi(t,x)$ does.

Overall consistency with the existence of a pole-decomposition actually imposes the forms $\Phi(t,x) = \Phi(t)\cos(x)$ and $w(t,x) = w(t)\sin(x)$, with $\Phi(t)$ and $w(t)$ being taken at will. The forcing $u(t,x)$ must read $u_1(t)\cos(x) + u_2(t)\cos(2x)$ in this semi-inverse approach. The $u_j(t)$ depend on the chosen $\Phi(t)$, $w(t)$, $N_{tot}$ and on the selected neutral-to-actual wrinkle wavelength ratio $\nu > 0$; for future reference note that $u_1(t)$ involves $\Phi(t)$ and $w(t)$ linearly, while $u_2(t) = \frac{1}{2}\Phi(t)(w(t) - \frac{1}{2}\Phi(t))$ generically is quadratic in the amplitudes. Most of the analyses developed here assume $\Phi(t) = \Phi \sin(\omega t)$ and $w(t) = w\sin(\omega t + \gamma)$ where the constant $(\Phi, w)$ and the frequency $\omega$ may be chosen; enforcing that $u(t,x)$ has zero time-average $[u(t,x)]_{av}$, to imitate flame propagations across stacked line-arrays of eddies, next requires the relative phase $\gamma$ to satisfy $\cos(\gamma) = \Phi/2w$.

This converts flame dynamics to coupled nonlinear ordinary differential equations [ODEs] for $2N_{tot}$ complex-pole trajectories, in which the combined-stretch amplitude $w(t) - \Phi(t)$ is an input. If solved such ODEs yield $\varphi(t,x)$, the wrinkling induced front arc-length increment $V(t)$ and its time-average $V_{av}$. A Burgers model, free of Darrieus-Landau [DL] instability, is also amenable to this approach. This program could be worked out analytically only in the context of High-Frequency [HF], $\omega \to \infty$, or Low-Frequency [LF], $\omega \to 0$, analyses.

In the HF limit the amplitudes $\Phi = -\omega\delta$ and $w = -\omega d$ are viewed $O(\omega)$ to account for non-linearity; so is the maximum amplitude $\omega\Delta$ of combined stretch $w(t) - \Phi(t)$. The pole dynamics is then easily accessible at leading order. That, however, is not accurate enough to get the space-average $V(t)$ of $\frac{1}{2}(\varphi_x)^2$ to $O(1)$ inclusive. The needed corrections to $V_{av}$ were computed analytically for $N=1$ and when $N=1=n$. Explicit formulae were also obtained for long and large wrinkles comprised of 'twin' dense piles of poles, $N=n \sim 1/\nu \gg 1$, at the expense of analytically solving an integral equation relating to their unsteady densities.

The variations of $V_{av}$ with combined-stretch amplitude $\Delta$, its geometrical component alone [$d=0$] or accompanied by $w(t,x)$ [$d=\delta$, $\gamma = \pm \pi/3$], have the overall shape of a concave-up *parabola*. The main difference with that derived for passive fronts [30], or in the Burgers model, is an overall *offset* caused by the DL instability if $\nu < 1$ and available beforehand: $2\nu N_{tot}(1-\nu N_{tot})$ for $N_{tot} = N+n$ less than the bound $N_{opt}(\nu) \equiv \lfloor \frac{1}{2\nu} + \frac{1}{2} \rfloor$ allowed in steady MS fronts, or $2\nu N_{opt}(1-\nu N_{opt})$. If $N_{tot} > N_{opt}(\nu) \gg 1$ surplus poles exist and $\frac{d}{d\Delta} V_{av}|_{\Delta=0} > 0$ was found. Replotting $V_{av}$ in terms of a measure $u'$ of the outer flow [$\sim (w(t,x), u(t,x))$] ahead of the front slightly straightens the HF response curves, since $u' \sim \Delta^2$ at [quite] large $\Delta$: as said earlier, the second spatial harmonics of $u(t,x)$ generically is quadratic, here in $\Delta$.

But figures 3-15 were purposely restricted to moderate $\Delta$s, despite Bessel-function-related corrections whose fast growth can compete with base-slope contributions. The $\Delta \gg 1$ limit of HF combined stretch indeed is singular: repulsion from conjugates catches up with combined stretch if the poles get too close to the real axis, *i.e.*, if the crests become too sharp. This can be seen in the $N=1$ dynamics in (6.1). Instead of ruled by $\delta \frac{d}{d\sigma} \cos(\sigma)$, $\frac{d}{d\sigma} \lambda$ gets affected by the 'smaller' $\nu \sinh(2\lambda)/2\omega$ while $\lambda(\sigma) \approx \delta \cos(\sigma)$ is still $O(1)$ to its maximum [say, if $|\sigma| \leq (2/\delta)^{1/2}$], provided $\delta > \delta_c(\nu)$ exceeds a $\delta_{sup}(\omega)$ defined by $\nu \exp(2\delta_{sup})/4\omega = (2\delta_{sup})^{1/2}$: though unbounded for $\omega \to \infty$, $\delta_{sup}(\omega)$ is only $\approx 2.5$ for $\nu = \frac{1}{2}$ and $\omega = 10$, and doubling $\omega$ pushes it to a mere $\approx 3$. Over such restricted ranges even Bessel functions look parabolic.

The LF dynamics, where $\Phi = -\omega\delta$, $w = -\omega d$ and $\omega\Delta$ are now $O(1)$, is richer because the DL instability competes at leading order and the period $2\pi/\omega \to \infty$ leaves ample time for the poles to rearrange. This results in multiple solutions and relaxation phenomena corresponding to pole jumps from one crest to its neighbor, caused by a combined stretch intense enough to overcome hydrodynamic instability. In the Burgers version all poles jump at once as soon as the combined stretch changes sign, as no DL instability opposes it: they all reside in the same pile at low frequencies, hence at any one time form a single crest at $x=0$ *or* $\pi$ (mod $2\pi$).

The $V_{av}$ vs. $\omega\Delta$ curves obtained for $N_{tot} = 1$ exhibit the mentioned above offset, and stay 'flattish' until pole jumps set in; this holds when the geometric stretch is alone [$d = 0$] or is accompanied by $w(t,x)$ [$d = \delta$, $\gamma = \pm\pi/3$]. If $\omega\Delta$ reaches known wavelength-dependent thresholds, $V_{av}$ abruptly bifurcates to a curve that ultimately becomes 'weakly parabolic'. In channels so narrow that no unstable wrinkle fits in [$\nu > 1$], the offset disappears and $V_{av}$ is initially quadratic in $\omega\Delta$ and $u'$, to ultimately grow as above. The $u' \sim \Delta^2$ dependence at large $\omega\Delta$ s will render all $V_{av}$ vs. $u'$ curves slightly sub-linear, yet only for $u' \gg 1$. The trends are similar for $N_{tot} = 2$, though the bifurcations at $\nu \lesssim 1$ are way more intricate.

As for long and high quasi-steady wrinkles with $N \sim n \sim 1/\nu \gg 1$, accessing the front slope requires analytical resolution of coupled integral equations for unequal pole densities. The pile contents $N(t)$ and $n(t)$ need auxiliary numerical work, as do $V(t)$ and $V_{av}$. Most often $N(t)$ is only piecewise constant, and evolves in a way that crucially depends on $\omega\Delta$ and the $N_{tot}/N_{opt}(\nu)$ ratio; this noticeably complicate the analyses. Still, as long as the geometric stretch from the base slope acts alone [$d = 0$] or with an hydrodynamic stretch of equal weight [$d = \delta$, $\gamma = \pm\pi/3$], the resulting $V_{av}$ vs. $\omega\Delta$ curves again grossly resemble offset parabolas. If $N_{tot} < N_{opt}(\nu)$ they include a flat initial portion though, shown to correspond to a continuum of no-jump solutions; and $V_{av} - \frac{1}{2}$ starts as $\propto (\omega\Delta)^{2/3}$ [or $\omega\Delta$] for $N_{tot} > N_{opt}(\nu)$ [or $N_{tot} = N_{opt}(\nu)$]. Switching to $u'$ eventually changes the asymptotic laws from parabolic to linear, since the configurations at hand imply $u' \sim (\omega\Delta)^2$ in the large-$\Delta$ limit.

But taking $2d = \delta$ and $\gamma = 0$ to model stimuli $(u(t,x), w(t,x))$ with $\omega\Delta$- independent spatial shapes, kills the 2$^{nd}$ spatial harmonics of $u(t,x)$ and results in $u' \sim \omega\Delta$, producing $V_{av} \sim u'^2$ as the final trend; that the growth of $V_{av}$ depends on the spatial spectral content of $u(t,x)$ and $w(t,x)$ is not much of a surprise, yet this resurgence of $V_{av} \sim u'^2$ as in the CW passive models [30] somewhat is. Similar to [25] though, allowing the time-independent pole number $2N_{tot}$ to vary from run to run with $\omega\Delta$ will modify the late behavior, to yield a $V_{av} \sim u'$ final growth while keeping $u(t,x) \sim \cos(x)$ and $w(t,x) \sim \sin(x)$ irrespective of the imposed outer flow intensity; $\nu N_{tot} \sim \text{const.} + \omega\Delta$ in fact results in a final $V_{av} \sim u'$ even if $d = \delta$ and $\gamma = \pm\frac{\pi}{3}$.

In view of so simple $V_{av}$ graphs or front shapes, the machinery needed to get them might seem disproportionate. This is built in the present approach though, since $\varphi(t,x)$ little feels remote poles, however fancy their exact motions are. The slightest front distortions, not to mention noticeable changes in $V_{av}$, conversely require spectacular pole rearrangements. After all, even tiny Burgers wrinkles $\sim f(t)\cos(x)$ also involve pole jumps, from $\pm i\infty$ to $\pi \pm i\infty$ (mod $2\pi$) depending on sgn$[f(t)]$, and crests that correlatively spring from $x = 0$ to $x = \pi$ (mod.$2\pi$). In a sense, jump-free front motions at low $\omega\Delta$ correspond to a MS dynamics modulated by combined stretch, whereas larger amplitudes of forcing/stretch yield a DL-affected Burgers type of dynamics. Pole jumps must appear if the combined stretch is intense.

When it comes to dense piles their densities, the front slope and its length happen to be concisely parametrized by two remotest-pole attributes, $T_{\max}(t)$ and $\tau_{\max}(t)$ [*e.g.*, see (7.40)]; the ransom to pay is the difficulty to determine those and their sudden/wild changes. It would be of interest to develop approximate methods that are less sensitive to the remotest poles. An approach yet to exploit is that steady or quasi-steady pole locations result from minimizing an electrostatic-like energy [20, 38], and the determination of $T_{\max}(t)$ and $\tau_{\max}(t)$ can possibly be solved as one of optimization with constraints. It could be even more rewarding to set up a variational principle for the 'polar' slope $\phi_x(t,x)$ itself: in view of the obtained front shapes, piecewise linear slopes could serve as test profiles. Though seemingly only academic now that the $\omega \to 0$ version of (2.1)(2.2) is solved, the point will get crucial when more general models than MS-based are envisaged. Those built on the Zhdanov-Trubnikov equation [39], which has an additional nonlinearity $\propto [\mathcal{I}(\varphi)]^2$ and is still amenable to pole decompositions [40], could admittedly benefit of the Resolvent method [41] to bypass some of the lengthy and more pedestrian calculations developed here. At any rate, this will also require to determine $T_{\max}(t)$, $\tau_{\max}(t)$ and the filling fraction $N(t)/N_{tot}$. The functional forms of some building blocks that enter the pole densities or slopes found for LF or HF bi-coalesced fronts could complementarily provide test profiles or interpolating functions parametrized by only two numbers $T_{\max}(t)$ and $\tau_{\max}(t)$, for use in other non-MS situations or/and if $\omega = O(1)$.

The base-front amplitude $\Phi(t)$ was assumed here to vanish on average, and the combined stretch $w(t) - \Phi(t)$ also enjoyed this property. Envisaging $\Phi(t) = -D - \omega\delta \sin(\omega t)$, with $D = const.$, could also be of interest if $2\nu N_{tot}$ is chosen to comply with $[u_1(t)]_{av} = 0$. Choosing $\Phi(t) = -D(\omega't) - \omega\delta\sin(\omega t)$ with $\omega \gg 1$ and $\omega' \leq O(1)$ is even more interesting: $N = 1$ already leads to a nontrivial 'slower dynamics which takes place on the $t = O(1)$ time scale. This indeed produces $\frac{d}{dt}\lambda_0 = I_0(\Delta)\sinh(\lambda_0) - \frac{\nu}{2}I_0(2\Delta)\sinh(2\lambda_0) - D(\omega't)$ instead of (6.4), and hysteresis can occur if $(\omega't)$- dependent, zero-mean $\Delta$ or $D$ are used. Should those evolve on a $t \sim 1/\omega' \gg 1$ scale, one would end up with a 'slow curve' resembling (7.6) up to Bessel functions or kin that result from averages over the fastest oscillations. Whole pole piles with $\omega \gg 1$ combined with $D = D(\omega't)$ seem intractable analytically if $\omega' = O(1)$. But combining $\omega \gg 1$ and $\omega' \ll 1$ possibly is within reach, via coupled integral equations resembling (7.6), yet with different right-hand sides 'dressed' by the fast oscillations; this could help one study fronts subject to slowly modulated high-frequency 'turbulent' bursts.

The present text focused on spatially 1-dimensional front shapes, but its results can be readily adapted [20] to the class of 2D shapes $\varphi(t,\mathbf{x}) = \varphi(t,x) + \varphi'(t,x')$, $\mathbf{x} \equiv (x,x')$, provided $w(t,x)$ be replaced by the 2-dimensional field $\mathbf{w}(t,\mathbf{x}) = [w(t)\sin(x), w'(t)\sin(x')]$; then $u(t,\mathbf{x})$ has a 'separated' structure $u(t,x) + u'(t,x')$ since $\mathbf{w}.\nabla\varphi$ and $|\nabla\varphi|^2$ do. Envisaging *uncoupled*, independent histories – different ones but with identical average properties – along $x$ and $x'$ is the best the current approach can do to mimic 2D flame surfaces, analytically or numerically. Such sums can also model flames in Hele-Shaw cells [42, 43] where the wrinkle wavelength in the $x$ direction parallel to the plates vastly exceeds the gap between them: one could in effect superimpose solutions with $N_{tot}$ pairs of poles allotted to $\varphi(t,x)$ and $N'_{tot}$ pairs to $\varphi'(t,x')$. Different tools to handle the two components of $\varphi(t,\mathbf{x})$ can even be envisaged. At any rate, this will amount to adding the contributions to $V_{av}$ from both directions.

Periodic boundary conditions were not addressed here: they make the problem much tougher than Neumann's, as the pole phases along $x$ become continuous degrees of freedom replacing the frozen pole locations $x = 0 (\mod \pi)$. In particular global $x$-wise translation invariance is restored, which brings about a new zero-mode and possibly allows for travelling wrinkles. The HF analysis of a single pair of poles already constitutes a non-trivial exercise if periodic conditions are employed: a few numerical attempts suggest that the two poles can oscillate about $x = \frac{\pi}{2}(\mod \pi)$. Dense piles are even more difficult to handle: far off the real axis the poles will condense along oscillating *curved* lines whose shape(s) must be found besides the pole densities along them. Even the number of such curved piles is then an unknown, actually.

A fully numerical approach to solve the pole equations (3.4) was not the main scope in the present, analytically oriented, text. But this clearly is an option that deserves to be further developed. As such coupled ODEs lead to a computational burden that roughly grows as $(c_1 + c_2/\omega)N_{tot}^2$, $c_{1,2} = const.$, a fair compromise seems to lie about $N_{tot} = 10$ which can describe six types of situations [from single-pile fronts to twin crests] at moderate cost. This is also portable enough to handle various $w(t)$ or $\Phi(t)$, and any $\omega$, *e.g.* to mimic turbulent burst [see VII B]; quasi-periodic stimuli involving several frequencies or even stochastic ones could possibly be afforded to get statistics on $V_{av}$ via long and repeated runs.

One may finally recall a property of the pole-dynamics approach adopted here, namely: their number $2N_{tot}$ is a constant of motion, rendering the front evolutions weakly sensitive to noise. By the same token though, it is uneasy, even numerically, to mimic the influence of a weak random noise, *e.g.*, by sprinkling new poles [44, 45]. It is even more delicate to remove some of them, except possibly at imaginary infinity, without bringing about an unwanted forcing that would be correlated with the front distortions. How to cleanly cope with a time dependent, genuinely dynamical, number of poles is one of the theoretical/numerical gaps yet to fill. Incorporating additive random noise – random in time *and* in space – is another one.

-----------------

## ACKNOWLEDGEMENTS

We thank the *Agence Nationale pour la Recherche* [ANR] for support under grant ANR-14-CE05-0006. One of us [GJ] thanks H. El-Rabii [CNRS and Poitiers Univ.] for discussions, help about Mathematica and the figures, and a careful checking of so many integrals.

## APPENDIX A: Front shape kinematics

Let $(w, u_L + u)$ denote the fresh-gas velocities along the Cartesian $(X, Y)$ axes, just ahead of the front of dimensioned equation $Y = F(T, X)$ at time $T$. An assumed-linear law $u_n / u_L = 1 - \mathcal{L}\mathcal{C}$ for the normal ($u_n$) to flat-flame ($u_L$) local burning-speed ratio, with $\mathcal{C}$ denoting the front curvature and $\mathcal{L}$ a Markstein [5] length, leads to an evolution equation for $F(T, X)$:

$$(u_L + u - F_T - wF_X)/(1 + F_X^2)^{1/2} = u_L(1 - \mathcal{L}F_{XX}/(1 + F_X^2)^{3/2}). \tag{A.1}$$

The left-hand side of (A.1) indeed defines $u_n$ while the right-hand side is $u_L(1 - \mathcal{L}\mathcal{C})$. A small-slope expansion, valid if $F_X^2 \ll 1$ and acceptable up to $F_X^2 \approx 1$, of the reciprocal direction cosine $1/\cos(\alpha) = (1 + F_X^2)^{1/2}$ between front and $Y$ direction simplifies (A.1) to:

$$u - F_T - wF_X = \tfrac{1}{2} u_L F_X^2 - u_L \mathcal{L} F_{XX} + ... \ . \tag{A.2}$$

We next assume that $(w, u)$ is comprised of $o(u_L)$ contributions from the DL mechanism, and of imposed ones $\mathbf{u}_\infty = (w_\infty, u_\infty)$ of comparable (hence small) magnitude [1]

$$\begin{aligned} u &= u_\infty(T, X) + u_L \mathcal{A}\, \mathcal{H}[F_X](T, X) + ..., \\ w &= w_\infty(T, X) + u_L \mathcal{A}\, F_X(T, X) + ... \quad , \end{aligned} \tag{A.3}$$

from which smaller cross terms were omitted; the Hilbert transform $\mathcal{H}[.]$ in (A.3) is such that $\mathcal{H}[iKe^{iKX}] = |K|e^{iKX}$, and linear; $\mathcal{A} \ll 1$ is the Atwood number. Then (A.2) becomes

$$F_T + w_\infty F_X + \tfrac{1}{2} u_L a(\mathcal{A})(F_X)^2 - u_L \Omega(\mathcal{A})(F_{XX}/K_n + \mathcal{H}[F_X]) = u_\infty + ... \ . \tag{A.4}$$

Here $a(\mathcal{A}) = 1 + 2\mathcal{A}$, $\Omega(\mathcal{A}) = \mathcal{A}$, $K_n = \mathcal{A}/\mathcal{L}$, yet other expressions would also do below.

If $\mathbf{u}_\infty \equiv 0$ (A.4) resumes the MS equation (1.1) on using the variables $\varphi = KF\, a(\mathcal{A})/\Omega(\mathcal{A})$, $t = u_L T K\, \Omega(\mathcal{A})$, $x = KX$ and $\nu = K/K_n$, where $K = 2\pi/L$ denotes the actual wavenumber of the front wrinkles, assumed to possess $L$-periodicity along the $X$ direction.

Accounting for $\mathbf{u}_\infty \neq 0$ yields (2.1) where $w(t, x)$ or $u(t, x)$ denotes $w_\infty(T, X)/u_L \Omega(\mathcal{A})$ or $u_\infty(T, X) a(\mathcal{A})/u_L \Omega(\mathcal{A})^2$, respectively, with $\langle u(t, x) \rangle = 0$ assumed: any originally non-zero $\langle u_\infty \rangle \equiv \int_{-L/2}^{+L/2} u_\infty(T, X) dX / L$ can indeed be lumped in with $-F_T$ without affecting the front *shape* or $\varphi_x(t,x)$ in (2.1), and its arc-length $L_e \equiv \int_{-L/2}^{+L/2}(1+F_X^2)^{1/2}dX$. Next, $L_e/L - 1 \propto V(t) \equiv \int_{-\pi}^{+\pi}\frac{1}{2}\varphi_x^2 dx/2\pi$ in the small-slope approximation, hence the effective burning speed $u_e(T) \equiv \int_{-L/2}^{+L/2} u_n (1+F_X^2)^{1/2}dX/L = u_L L_e/L$ satisfies $u_e/u_L - 1 = V(t)(\Omega(\mathcal{A})/a(\mathcal{A}))^2 \sim V(t)\mathcal{A}^2$.

## APPENDIX B: Tricomi equation and kin.

Let $f(\xi) = f(-\xi)$ be a solution to the singular, yet linear, Tricomi integral equation

$$\fint_{-\xi_{\max}}^{\xi_{\max}} \frac{f(\xi')d\xi'}{\xi - \xi'} = g(\xi), \quad g(-\xi) = -g(\xi), \quad -\xi_{\max} < \xi < \xi_{\max}, \tag{B.1}$$

that would be bounded at $\xi = \pm\xi_{\max}$, where $\xi_{\max} > 0$ is given and $\fint(.)d\xi'$ is a principal part.

If hole-free [*i.e.*, nonzero over any finite subrange of $-\xi_{\max} < \xi < \xi_{\max}$] $f(\xi)$ reads [33, 34]

$$f(\xi) = \frac{\pi^{-2}}{\sqrt{\xi_{\max}^2 - \xi^2}}[C - \fint_{-\xi_{\max}}^{\xi_{\max}} \frac{g(\xi')\sqrt{\xi_{\max}^2 - \xi'^2}d\xi'}{\xi - \xi'}], \tag{B.2}$$

$$C = \fint_{-\xi_{\max}}^{\xi_{\max}} \frac{g(\xi')\sqrt{\xi_{\max}^2 - \xi'^2}d\xi'}{\xi_{\max} - \xi'}. \tag{B.3}$$

Its piece $\propto C = $ const. satisfies the $g(\xi) \equiv 0$ version of (B.1), and (B.3) is chosen to ensure $f(\xi_{\max}) < \infty$; it in fact yields $f(\pm\xi_{\max}) = 0$ if $g(\xi)(\xi_{\max}^2 - \xi^2)^{1/2} = o(1)$ as $\xi \to \pm\xi_{\max}$. For example $g(\xi) = \xi$ yields Wigner's semicircle $f(\xi) = \pi^{-1}(\xi_{\max}^2 - \xi^2)^{1/2}$ [33]. It also holds that

$$g(\xi) = \mathrm{sgn}(\xi) \Rightarrow f(\xi) = \pi^{-2}\cosh^{-1}(\xi_{\max}/|\xi|), \tag{B.4}$$

$$g(\xi) = \frac{\xi}{\Xi^2 - \xi^2}, \quad \Xi > \xi_{\max} \Rightarrow f(\xi) = \frac{\pi^{-1}\Xi}{(\Xi^2 - \xi_{\max}^2)^{1/2}}\frac{(\xi_{\max}^2 - \xi^2)^{1/2}}{(\Xi^2 - \xi^2)}, \tag{B.5}$$

[see Appendix D how to do the needed integrals]; this is used to solve (6.29)(7.20) and (C.2).

Now set $\xi = ku/(1+cu^2)$, $\frac{d}{du}\xi = k(1-cu^2)/(1+cu^2)^2$, for constant $c$ and $k > 0$. Using

$$\frac{k(1-cu'^2)}{(1+cu'^2)^2}[\frac{1}{ku/(1+cu^2) - ku'/(1+cu'^2)}] \equiv \frac{1}{u-u'} + \frac{cu}{1-cuu'} + \frac{2cu'}{1+cu'^2}, \tag{B.6}$$

and the fact that $F(u) \equiv f(ku/(1+cu^2))$ is even in $u$, transforms (B.1) to

$$\fint_{-u_{\max}}^{u_{\max}} \frac{F(u')du'}{u-u'} + cu\int_{-u_{\max}}^{u_{\max}} \frac{F(u')du'}{1-cuu'} = G(u), \quad G(u) \equiv g(\frac{ku}{1+cu^2}) = -G(-u), \tag{B.7}$$

where $u_{max}$ meets $\xi_{max} = ku_{max}/(1+cu_{max}^2)$. The hole-free solution to (B.7) that is bounded at $u = \pm u_{max}$ thus follows from (B.2)(B.3); this is used to solve (6.26)(7.32) and (E.1). Note that the $\xi \leftrightarrow u$ link also is $\tan^{-1}(2i\xi\sqrt{c}/k) = 2\tan^{-1}(iu\sqrt{c})$, a change of length scale in disguise. The same kind of trick is invoked to analytically evaluate the front-slope profile from (7.35).

## APPENDIX C: Single pile at High Frequency

The HF analysis of a single pile with density $\rho(\sigma, B)$ parallels that of Subsection VI.C. It involves a single balance $\partial \rho/\partial t + \partial j/\partial B = 0$, where $j \equiv -\omega\delta\sin(\omega t)\sinh(B)\rho(t,B) + J(t,B)$,

$J(t,B) = \rho(t,B)[\fint_{-B_{max}(t)}^{+B_{max}(t)} \frac{\nu\rho(t,B')dB'}{\tanh(\frac{1}{2}(B-B'))} - \text{sgn}(B)]$ are the pole fluxes. This is transformed to

$$\left.\frac{\partial R(\sigma, E)}{\partial \sigma}\right|_E + \frac{1-E^2}{2\omega}\frac{\partial J(\sigma, E)}{\partial E} = 0 ,$$
$$\rho(t,B) = \frac{e^{\delta\cos(\sigma)} - E^2 e^{-\delta\cos(\sigma)}}{1-E^2} R(\sigma, E), \quad E \equiv \tanh(\tfrac{1}{2}B)e^{\delta\cos(\sigma)}, \quad \sigma = \omega t ,$$
(C.1)

where $R(\sigma, E) = R_0(E) + R_1(\sigma, E)/\omega + ....$ is an auxiliary density. As a result of the solvability condition $[J(\sigma, E)]_{av,E} = 0$ [compare with (6.24)] the leading order $R_0(E)$ vanishes if $E^2$ exceeds a presumed-constant $E_{max,0}^2$ to be found, and for $-E_{max,0} \leq E \leq E_{max,0}$ it satisfies

$$\fint_{-E_{max,0}}^{E_{max,0}} \frac{2\nu R_0(E')dE'}{(E-E')} = g(E,N) ,$$
$$g(E,N) \equiv \frac{I_0(\delta)\text{sgn}(E)(1-E^2)^2 + 2\nu N(I_0(2\delta)-1)E(1+E^2)}{(1-E^2)^2 + (I_0(2\delta)-1)(1+E^4)}.$$
(C.2)

Contrary to (6.26) (C.2) already has the Tricomi form, and its right-hand side just differs from (6.27) by $2N \to N$. The constant $0 < E_{max,0} \leq 1$ in (C.2) follows from density normalization

$$\int_0^{E_{max,0}} \frac{2\nu R_0(E)dE}{(1-E^2)} = \nu N .$$
(C.3)

By invoking Eqs. (B.2)(B.3) of Appendix B, and hints from Appendix D to do the needed integrals, the hole-free solution $R_0(E)$ to (C.2) that has $R_0(\pm E_{max,0}) < \infty$ evaluates to

$$R_0(E) = R_{MS}(E) + R_{Burg}(E) =$$

$$\frac{I_0(\delta)}{\nu\pi^2 I_0(2\delta)} \frac{(1-E^2)^2 \cosh^{-1}(E_{max,0}/|E|)}{(E^4 - 2E^2\cos(2\alpha)+1)} - \frac{2I_0(\delta)\tan(\alpha)}{\nu\pi^2 I_0(2\delta)} \Im[\frac{\sin^{-1}(E_{max,0}e^{-i\alpha})}{\sqrt{1-E_{max,0}^2 e^{-2i\alpha}}} \frac{e^{i\alpha}\sqrt{E_{max,0}^2 - E^2}}{(E^2 - e^{2i\alpha})}]$$

$$+ \frac{2\nu N \sin(\alpha)}{\nu\pi} \Im[\frac{e^{i\alpha}}{\sqrt{1-E_{max,0}^2 e^{-2i\alpha}}} \frac{\sqrt{E_{max,0}^2 - E^2}}{(E^2 - e^{2i\alpha})}] \qquad (C.4)$$

Such involved expressions stem from the Tricomi equation (C.2) still having the polynomial $(1-E^2)^2 + (I_0(2\delta)-1)(1+E^4)$ as a denominator, as opposed to (6.29); its four complex roots $E = \pm e^{\pm i\alpha}$, $0 < \alpha = \frac{1}{2}\cos^{-1}(1/I_0(2\delta)) < \frac{\pi}{4}$, weigh down the integrals needed in (B.2)(B.3).

The polar component $\phi_x(\sigma,x) \equiv \varphi_x(\sigma,x) - \omega\delta\sin(x)$ of the corresponding front slope reads

$$\phi_x(\sigma,x) = -4\nu N \frac{\sinh(\delta\cos(\sigma))}{(1/\mathcal{T} + \mathcal{T})} - \Sigma(\mathcal{E}) \frac{e^{\delta\cos(\sigma)}/\mathcal{T} + e^{-\delta\cos(\sigma)}\mathcal{T}}{(1/\mathcal{T} + \mathcal{T})},$$

$$\mathcal{T}(\sigma,x) \equiv \tan(\tfrac{1}{2}x)e^{\delta\cos(\sigma)}, \quad \mathcal{E}(\sigma,x) = \mathcal{T}, \quad \Sigma(\mathcal{E}) \equiv \int_{-E_{max,0}}^{E_{max,0}} \frac{2\nu R_0(E)dE}{(\mathcal{E}-iE)}, \qquad (C.5)$$

which formally mirrors (6.32) up to $2N \to N$ and a simpler definition of $\mathcal{E}(\sigma,x)$. Doing the integral featured in (C.5) [see Appendix D for hints] produces $\Sigma(\mathcal{E}) = -\Sigma(-\mathcal{E})$ as:

$$\Sigma(\mathcal{E}) = \Sigma_{DL}(\mathcal{E}) + \Sigma_{Burg}(\mathcal{E}) =$$

$$\frac{2I_0(\delta)}{\pi I_0(2\delta)} \{ \frac{(1+\mathcal{E}^2)^2 \sinh^{-1}(E_{max,0}/\mathcal{E})}{[\mathcal{E}^4 + 2\cos(2\alpha)\mathcal{E}^2 + 1]} + 2\tan(\alpha)\mathcal{E}\sqrt{1 + \frac{E_{max,0}^2}{\mathcal{E}^2}} \Im[\frac{\sin^{-1}(E_{max,0}e^{-i\alpha})e^{i\alpha}}{(\mathcal{E}^2 + e^{2i\alpha})\sqrt{1-E_{max,0}^2 e^{-2i\alpha}}}]\}$$

$$+ 4\nu N \sin(\alpha)\Im[\frac{\mathcal{E}e^{i\alpha}}{(\mathcal{E}^2 + e^{2i\alpha})} - \mathcal{E}\sqrt{1 + \frac{E_{max,0}^2}{\mathcal{E}^2}} \frac{e^{i\alpha}}{(\mathcal{E}^2 + e^{2i\alpha})\sqrt{1-E_{max,0}^2 e^{-2i\alpha}}}] \; . \qquad (C.6)$$

The normalization (C.3) finally determines $E_{max,0}$ through

$$\nu N = \frac{I_0(\delta)\sec(\alpha)}{\pi I_0(2\delta)} \frac{\Re[\sin^{-1}(E_{max,0}e^{-i\alpha})/\sqrt{1-E_{max,0}^2 e^{-2i\alpha}}]}{\Re[1/\sqrt{1-E_{max,0}^2 e^{-2i\alpha}}]} \quad \text{or} \quad E_{max,0} = 1, \qquad (C.7a, b)$$

which just like (6.34) offers two possibilities. The ordinary branch in (C.7a) has $\nu N \approx E_{max,0} I_0(\delta)/\pi I_0(2\delta)$ for $\nu N \ll 1$, resumes the MS result $\nu N = \frac{1}{\pi}\sin^{-1}(E_{max,0}) \leq \frac{1}{2}$ [20] when $\delta = 0 = \alpha$ and prevails for $\nu N \leq \nu N_c(\delta) \equiv \mathcal{N}(\delta,1)$, where $\mathcal{N}(\delta, E_{max,0})$ is the right-hand side of (C.7a); see Fig.C1, to compare with Fig.9. In particular $E_{max,0} < 1$ holds if $\nu N_c(\delta) >$

$\nu N \geq \nu N_*(\delta) \equiv \mathcal{N}(\delta, e^{-|\delta|})$: for $\delta \cos(\sigma) < 0$, $e^{-|\delta|} \leq E_{max,0} < 1$ provisionally yields $\tanh[\frac{1}{2}B] > 1$ for those poles with labels $e^{-|\delta|} < E \leq E_{max,0}$.

Only if $\nu N \geq \nu N_c(\delta)$ may the root $E_{max,0} = 1$ take over in (C.7), for it then yields a non-negative $R_0(E)$ over its support; in particular $E_{max,0} = 1$ for all $\nu N \geq \nu N_{opt}(\nu) = \frac{1}{2}$, if $\delta \neq 0$.

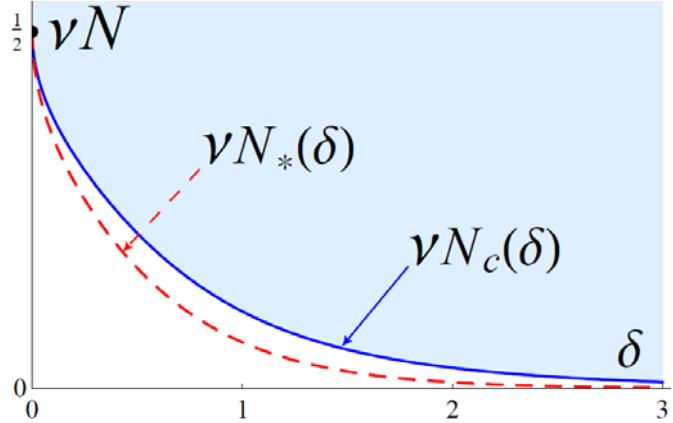

FIG.C1 (Color online) Phase diagram for a single pile of poles at high frequency. $E_{max,0} = 1$ prevails along and above the solid-blue line, and $0 < E_{max,0} \leq e^{-|\delta|}$ along and below the dashed-redline, see (C.7); thus $e^{-|\delta|} < E_{max,0} < 1$ in between. Besides, all $\nu N \geq 1/2$ lead to $E_{max,0} = 1$ whatever $\delta \neq 0$ is. Compare with Fig.9.

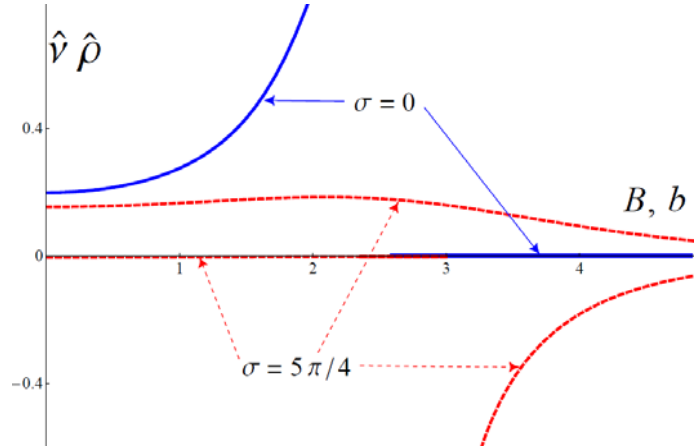

FIG.C2 (Color online): Pole density for single-pile HF Burgers fronts, vs. distance above real axis. The solid-blue curve has $\hat{\delta} = 0.15$, $\sigma = 0$: it diverges at $B = 2\tanh^{-1}(\exp(-\hat{\delta})) \approx 2.59$ and vanishes beyond. The red-dashed curves have $\hat{\delta} = 0.15$, $\sigma = 5\pi/4$: the upper one, corresponding to poles still along the imaginary axis [$z = iB$], vanishes only at $B = \infty$; the lower one, negative since it corresponds to poles that jumped to $z = \pi + ib$, diverges at $b = 2\tanh^{-1}(\exp(\hat{\delta}\cos(\frac{5\pi}{4}))) \approx 2.94$ and vanishes below, as well as at $b = \infty$.

For Burgers fronts all symbols but $(\sigma, E, x)$ get carets, and one deletes everything weighted by $I_0(\hat{\delta})$: only the last line survives in the analogs of (C.4) or (C.6), $\hat{E}_{max,0} = 1$ always holds and some poles provisionally leak to $z = \pi + ib \pmod{2\pi}$ as soon as $\hat{\delta}\cos(\sigma) < 0$, see Fig.C2.

The Darrieus-Landau affected component $\propto R_{MS}(E)/(1-E^2)$ of $\rho(\sigma, B)$ from (C.1)(C.4) always vanishes at $E = E_{max,0} \leq 1$ [*i.e.*, at $B = 2\tanh^{-1}(e^{-\delta\cos(\sigma)}E_{max,0})$], but its Burgers-type partner $\propto R_{Burg}(E)/(1-E^2)$ does so only if $E_{max,0} < 1$ and diverges if $E_{max,0} = 1$. This effect, best shown in the Burgers case where $\hat{R}_{MS}(E) \equiv 0$, see Fig. C2, does *not* occur with twin piles : $\Theta \equiv 2E/(1+E^2)$ implies $(1-\Theta^2)^{1/2} \sim (1-E^2)$ for $E^2 \to 1^-$ in (6.30) and its Burgers analog.

The time-averaged arclength increment $V_{av} = (2\pi)^{-1}\int_{-\pi}^{+\pi} V(\sigma)d\sigma$ can be accessed in the same way from (3.8) as in §5 of Subsection VI.C. In the stretch-free cases this ultimately produces

$$V_{av} - \tfrac{1}{8}\omega^2\delta^2 - \mathcal{V}(\nu N) = 2\int_0^{E_{max,0}} \frac{4\nu R_0(E)}{(1-E^2)^2} h(E,\delta) E\, dE,\qquad\text{(C.8)}$$
$$h(E,\delta) \equiv [1 - I_0(2\delta)E^2]g(E,N) + 2\nu N[I_0(2\delta) - 1]E - I_0(\delta)(1-E^2).$$

This differs from (6.38) by $2N \to N$ and $R_0(E)$ [(C.4) in lieu of $\chi(2E/(1+E^2))$ and (6.30)].

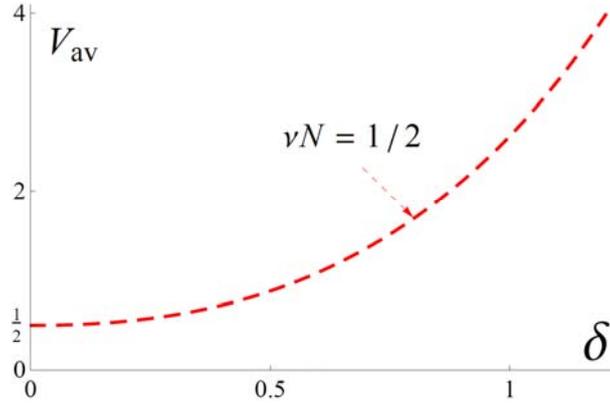

FIG.C3 (Color online): $V_{av}$ vs. $\delta$ from (C.8), for $\omega = 3$, $\nu N = \tfrac{1}{2}$.

Since $g(1,N) = 2\nu N$, and $R_0(E) \sim (E_{max,0}^2 - E^2)^{1/2}$ near the ends of its support, the integral in (C.8) converges even if $E_{max,0} = 1$; that also holds for the Burgers analog of (C.8), yet with $2\hat{\nu}\hat{N}(0 - \hat{\nu}\hat{N})$ instead of $\mathcal{V}(\nu N) = 2\nu N(1-\nu N)$ and with simpler $\hat{g}(E,\hat{N})$ and $\hat{R}_0(E)$ where all $I_0(\hat{\delta})$-weighted pieces inherited from (C.2)(C.4) have been deleted.

A comparison of Fig.C3 with Fig.11 reveals that the evolutions of $V_{av}$ vs. $\delta$ deduced from (C.8) have the same type of parabolic-with-offset curve as in twin-pile cases.

Single-pile fronts were not studied further at High Frequency, as numerical integrations of the pole equations (3.4) showed them to slowly evolve [$t >> 1/\omega$] to twin-pile configurations.

### APPENDIX D: A few integrals

Detailed determinations of all the integrals needed in this text being far too lengthy, only hints to calculate a few pivotal examples – sorted according to their origin – are given below.

• Solutions to Tricomi equations.

$$J_1 \equiv \int_0^{\xi_{max}} \frac{\sqrt{\xi_{max}^2 - \xi'^2}}{(\xi^2 - \xi'^2)} \frac{b^2 \xi' d\xi'}{(\xi'^2 - b^2)} = \frac{b^2}{(\xi^2 - b^2)} \times \qquad (D.1)$$
$$[b\sqrt{1 - \xi_{max}^2/b^2} \sin^{-1}(\xi_{max}/b) - \sqrt{\xi_{max}^2 - \xi^2} \cosh^{-1}(\xi_{max}/|\xi|)].$$

where $\pm b$ do not belong to the real segment $[-\xi_{max}, \xi_{max}]$ but $\xi$ does. The above integral is elementary since anti-derivatives are available once $u = \xi^2/\xi_{max}^2$ and partial fraction decomposition are used. Using (D.1) with $b = \infty$ in (B1)(B.2) yields (B.4) ; with $b^2 = e^{\pm 2i\alpha}$ (D.1) is used in Appendix C to get (C.4).

Another useful integral, in which the rational function $Q(\xi)$ is $\leq O(1/\xi)$ at infinity and has its poles $\xi_q$ outside the real interval $-\xi_{max} \leq \xi \leq \xi_{max}$, reads :

$$J_2 \equiv \int_{-\xi_{max}}^{\xi_{max}} \{\frac{\sqrt{\xi_{max}^2 - \xi'^2} Q(\xi') d\xi'}{(\xi - \xi')}\} dE' = \frac{i}{2} \oint \frac{\sqrt{1 - \xi_{max}^2/\zeta^2} Q(\zeta) \zeta d\zeta}{(\xi - \zeta)} \qquad (D.2)$$
$$= \pi [Q(\zeta)\zeta]_{\zeta = \infty} + \pi [\Sigma_{\xi_q} res(\xi_q)] \ .$$

In the second form of $J_2$ the counter-clockwise integration contour closely wraps the cut $-\xi_{max} < \zeta < \xi_{max}$ and the singularity at $\zeta = \xi$; the latter does not count, since $\sqrt{1 - \xi_{max}^2/\zeta^2}$ changes sign across the cut. Deforming the contour to $\zeta = Le^{i\beta}$, $L \to \infty$, picks contributions from the residues $res(\xi_q)$ of the integrand at the $\xi_q$'s and at $\infty$; this completes (D.2). For example, $Q(\xi) = \xi(1 + \xi^2)/[(\xi^2 - a^2)(\xi^2 - a^{-2})]$ is needed in (C.2) and (B.2) with $a = e^{i\alpha}$; it yields $J_2 = \pi + \pi\mathrm{cosec}(\alpha)\Im[a^3\sqrt{1-\xi_{max}^2/a^2}/(E^2-a^2)]$ and, all in all, (C.4).

- Normalizations for HF densities.

$$J_3 \equiv \int_0^{E_{max}} \frac{\cosh^{-1}(E_{max}/E)dE}{(b^2-E^2)} = \frac{\pi}{2b}\sin^{-1}(\frac{E_{max}}{b}), \quad b \notin [-E_{max}, E_{max}]. \tag{D.3}$$

$E = E_{max}/\cosh(u)$ converts $J_3$ to $\frac{1}{2}E_{max}\int_{-\infty}^{+\infty} u\sinh(u)du/(b^2\cosh^2(u)-E_{max}^2)$, which also is $\frac{1}{2}E_{max}\frac{\partial}{\partial\lambda}K\big|_{\lambda\to 0}$ where $K \equiv \int_{-\infty}^{+\infty}\sinh(u)e^{\lambda u}du/(b^2\cosh^2(u)-E_{max}^2)$. The hyperbolic functions – but not $e^{\lambda u}$ – being anti-periodic by $u \to u+i\pi$ the latter integral is evaluated by integration along the contour in complex $u$- plane that is comprised of the real axis [run from $-\infty$ to $\infty$] and of a parallel to it [traversed from $i\pi+\infty$ to $i\pi-\infty$]; $K$ is next specialized to $\lambda \to 0$. Acting on (D.3) with $\frac{1}{2b}\partial/\partial b$ will make $(b^2-E^2)$ transform to $(b^2-E^2)^2$ in $J_3$, whereby partial fraction decompositions give access to those normalization conditions that involve $\cosh^{-1}(E_{max}/E)$ multiplied by rational functions of $E^2$. The expansion of $b^2 J_3$ for $b^2 \to \infty$ yields $\int_0^{E_{max}}\cosh^{-1}(E_{max}/E)E^{2m}dE$ for all integer $m \geq 0$ [e.g., $\frac{\pi}{2}E_{max}$ if $m=0$].

$$J_4 \equiv \int_{-E_{max}}^{E_{max}} \frac{\sqrt{E_{max}^2-E^2}dE}{(b^2-E^2)} = \pi - \pi\sqrt{1-E_{max}^2/b^2}, \quad b \notin [-E_{max}, E_{max}]. \tag{D.4}$$

The contour method sketched below (D.2) adapts to $J_4$, yet with one variant : as they now lie outside the segment $[-E_{max}, E_{max}]$ the integrand singularities at $E = \pm b$ now count *via* their residues. The result (D.4) was made use of to get (6.30)(7.33). The above remarks as to the action of $\partial/\partial(b^2)$ on $J_3$ apply here. One can check, *e.g.* numerically [35], that (D.3)(D.4) hold if $b^2 < 0$ or complex, which was used in the slopes (6.33)(7.23)(7.36) ; and in (C.5)(C.6)

- Several integrals specially designed for bi-coalesced LF fronts are detailed at the end of Appendix E. They are all related to the three complete elliptic integrals [46] defined by:

$$K(a^2) \equiv \int_0^1 \frac{d\xi}{\sqrt{(1-\xi^2)(1-a^2\xi^2)}}, \quad E(a^2) \equiv \int_0^1 \frac{\sqrt{1-a^2\xi^2}}{\sqrt{1-\xi^2}}d\xi,$$
$$\Pi(a^2|b^2) \equiv \int_0^1 \frac{d\xi}{(1-b^2\xi^2)\sqrt{(1-\xi^2)(1-a^2\xi^2)}}, \quad 0 \leq a \leq b \leq 1. \tag{D.5}$$

# APPENDIX E: Normalizations and arc-length for LF bi-coalesced fronts.

## 1. *Jacobians*

Let $\lim_{T \to \pm T_{\max}} \rho(\sigma,T)/(T_{\max}^2 - T^2)^{1/2}$ and $\lim_{\tau \to \pm \tau_{\max}} \varrho(\sigma,\tau)/(\tau_{\max}^2 - \tau^2)^{1/2}$ be $\mathbb{K}(T_{\max},\tau_{\max},\Gamma(\sigma))$ and $\Bbbk(T_{\max},\tau_{\max},\Gamma(\sigma))$, respectively, with $\Gamma(\sigma) \equiv w(\sigma) - \Phi(\sigma)$ the combined stretch intensity [*e.g.*, $\omega\Delta\sin(\sigma-\zeta)$]; $\mathbb{K}$ and $\Bbbk$ locally encode the density signs and may not be negative.

The $2\times 2$ Jacobian matrix $\bar{\bar{J}}$ associated at fixed $(\sigma, N, n)$ with the normalizations (7.27) has $J_{11} = \frac{\partial}{\partial T_{\max}} \int_0^{T_{\max}} 2\nu\rho \, dT/(1-T^2)$ and $J_{21} = \frac{\partial}{\partial T_{\max}} \int_0^{\tau_{\max}} 2\nu\varrho \, d\tau/(1-\tau^2)$ in its first column; and similarly for $J_{12}$ and $J_{22}$, now in terms of $\frac{\partial}{\partial \tau_{\max}}$. Since $\rho(\sigma,T)$ vanishes at $T = T_{\max}$, $J_{11}$ also is $\int_0^{T_{\max}} 2\nu \frac{\partial \rho}{\partial T_{\max}} dT/(1-T^2)$; the other three matrix elements can be processed likewise. Differentiating (7.26) with respect to $T_{\max}$ at fixed $(\sigma, T, \tau)$ shows that the functions $\frac{\partial \rho}{\partial T_{\max}}$ and $\frac{\partial \varrho}{\partial T_{\max}}$ involved in $J_{11}$ and $J_{21}$ obey the pair of *homogeneous* integral equations

$$\begin{aligned}
&\fint_{-T_{\max}}^{T_{\max}} \frac{\partial \rho(\sigma,T')}{\partial T_{\max}} \frac{dT'}{(T-T')} + T \int_{-\tau_{\max}}^{\tau_{\max}} \frac{\partial \varrho(\sigma,\tau')}{\partial T_{\max}} \frac{d\tau'}{(1-T\tau')} = 0, \\
&\fint_{-\tau_{\max}}^{\tau_{\max}} \frac{\partial \varrho(\sigma,\tau')}{\partial T_{\max}} \frac{d\tau'}{(\tau-\tau')} + \tau \int_{-T_{\max}}^{T_{\max}} \frac{\partial \rho(\sigma,T')}{\partial T_{\max}} \frac{dT'}{(1-\tau T')} = 0,
\end{aligned} \quad (E.1)$$

with boundary conditions, $(T_{\max}^2 - T^2)^{1/2} \frac{\partial \rho}{\partial T_{\max}} = \mathbb{K}(T_{\max},\tau_{\max},\Gamma(\sigma))T_{\max}$ at $T^2 \to T_{\max}^2$ and $\frac{\partial \varrho}{\partial T_{\max}} = 0$ at $\tau^2 \to \tau_{\max}^2$, deduced from the near-end forms of $\rho(\sigma,T)$ and $\varrho(\sigma,\tau)$. Thus $\frac{\partial \rho}{\partial T_{\max}}$, $\frac{\partial \varrho}{\partial T_{\max}}$, $J_{11}$ and $J_{21}$ are proportional to $\mathbb{K}(T_{\max},\tau_{\max},\Gamma(\sigma))$; in the same way $\frac{\partial \rho}{\partial \tau_{\max}}$, $\frac{\partial \varrho}{\partial \tau_{\max}}$, $J_{12}$ and $J_{22}$ are $\propto \Bbbk(T_{\max},\tau_{\max},\Gamma(\sigma))$. The Jacobian $\det(\bar{\bar{J}}) = J_{11}J_{22} - J_{12}J_{21}$ of (7.27) therefore satisfies

$$\det(\bar{\bar{J}}) \propto \mathbb{K}(T_{\max},\tau_{\max},\Gamma(\sigma)).\Bbbk(T_{\max},\tau_{\max},\Gamma(\sigma)). \quad (E.2)$$

This can be checked by solving (E.1) exactly, thanks to the same trick as in (7.31)(7.32) and by combining solutions to homogeneous Tricomi equations, see (B.1)(B.3); or from (7.34). It is therefore anything but coincidental that the normalization conditions get singular exactly as *one* pole density starts flattening at the ends of its support: the remotest poles, at $T^2 = T_{\max}^2$ if $\mathbb{K}(T_{\max},\tau_{\max},\Gamma(\sigma)) = 0$ *or* $\tau^2 = \tau_{\max}^2$ if $\Bbbk(T_{\max},\tau_{\max},\Gamma(\sigma)) = 0$, are to leave; or are leaving.

Now consider (7.27)(7.28) rewritten as $F(T_{\max}, \tau_{\max}, \Gamma(\sigma)) - N = 0$, $f(T_{\max}, \tau_{\max}, \Gamma(\sigma)) + N = N_{tot}$, together with $\mathbb{K}(T_{\max}, \tau_{\max}, \Gamma(\sigma)) = 0$. This enlarged system can be solved for $T_{\max}$, $\tau_{\max}$ and $N$. The new $3 \times 3$ Jacobian matrix indeed has $J_{11} = 0 = J_{21}$ [since $\mathbb{K} = 0$], $J_{13} = -1$, $J_{23} = +1$, $J_{31} = \frac{\partial \mathbb{K}}{\partial T_{\max}}$, $J_{32} = \frac{\partial \mathbb{K}}{\partial \tau_{\max}}$, $J_{33} = 0$, hence its determinant is $(J_{12} + J_{22}) \frac{\partial \mathbb{K}}{\partial T_{\max}} \neq 0$.

Since $F(T_{\max}, \tau_{\max}, \Gamma(\sigma))$, $f(T_{\max}, \tau_{\max}, \Gamma(\sigma))$, $\mathbb{K}(T_{\max}, \tau_{\max}, \Gamma(\sigma))$ involve $\sigma$ only via $\Gamma(\sigma)$, see (7.34), $(T_{\max}, \tau_{\max}, N)$ and $n(\sigma) = N_{tot} - N(\sigma)$ also do *whatever the evolution of $\Gamma(\sigma)$ is*. So their time derivative vanishes when $\frac{d}{d\sigma} \Gamma = 0$, around which time all get quadratic [*e.g.*, near $\sigma - \zeta = \frac{\pi}{2} (\mod 2\pi)$] …. provided $\mathbb{K}(T_{\max}, \tau_{\max}, \Gamma(\sigma)) = 0$ holds.

Similar conclusions hold when using (7.27)(7.28) supplemented by $\mathbb{k}(T_{\max}, \tau_{\max}, \Gamma(\sigma)) = 0$, which needs $\Gamma(\sigma) \leq 0$. One just swaps $T_{\max}$ and $\tau_{\max}$, and replaces $(N, \Gamma(\sigma))$ by $(n, -\Gamma(\sigma))$.

## 2. *Tabulating one-pile leakage*

When augmented by $\mathbb{K}(\Gamma(\sigma), T_{\max}, \tau_{\max}) = 0$, which requires $\Gamma(\sigma) \geq 0$, the normalizations (7.27)(7.28) used to compute $N(\sigma)$ alongside $T_{\max}(\sigma)$ and $\tau_{\max}(\sigma)$ can be processed to give

$$M(T_{\max}, \tau_{\max}) - \Gamma\, L(T_{\max}, \tau_{\max}) + M(\tau_{\max}, T_{\max}) + \Gamma\, L(\tau_{\max}, T_{\max}) = \nu N_{tot},$$

$$\Gamma = h(T_{\max}, \tau_{\max}), \qquad h(T_{\max}, \tau_{\max}) \equiv \frac{(1 - \tau_{\max}^2)^{1/2} (1 - T_{\max}^2)^{3/2}}{\pi T_{\max}(1 + T_{\max} \tau_{\max})}, \qquad \text{(E.3)}$$

$$\nu N(\sigma) = M(T_{\max}, \tau_{\max}) - \Gamma\, L(T_{\max}, \tau_{\max}),$$

with two functions $M(T_{\max}, \tau_{\max})$ and $L(T_{\max}, \tau_{\max})$ specified in terms of integrals later on in Appendix E. The second line of (E.3) encodes $\mathbb{K}(T_{\max}, \tau_{\max}, \Gamma) = 0$, by (7.38); the first one is the sum of the two normalizations in (7.27), by (7.28); the third line gives access to $\nu N(\sigma)$.

For a given $\nu N_{tot}$, $T_{\max}$, $\tau_{\max}$ and $\nu N$ are accessible *once for all*, as they only depend on $\Gamma$.

Firstly, $\Gamma = h(T_{\max}, \tau_{\max})$ is solved analytically for $\tau_{\max}$, which is plugged into the first line of (E.3). $T_{\max}$ is next determined iteratively *vs.* $\Gamma$ using the `FindRoot` Mathematica routine [35]. $T_{\max}$, $\tau_{\max}$ and $\nu N$ are finally fitted to yield the sample approximants given below, all valid to sub-percent relative accuracy over $0 \leq |\Gamma| \leq 6$.

$\boxed{\nu N_{tot} = 1}$
$$T_{\max}(\Gamma) \approx 1 - \Gamma^{1/3} / (A_0 + A_1\Gamma^{1/3} + A_2\Gamma^{2/3} + A_3\Gamma) ,$$
$$\tau_{\max}(\Gamma) \approx [1 - \Gamma / (a_0 + a_1\Gamma^{2/3} + a_2\Gamma^{4/3} + a_3\Gamma^2)]^{1/2}$$
(E.4a,b)

$A_0 = +0.592762$, $A_1 = +0.456921$,
$A_2 = +0.180919$, $A_3 = -0.019589$,
$a_0 = +0.993710$, $a_1 = +1.104548$,
$a_2 = +0.327290$, $a_3 = -0.014650$.

$$\nu N(\Gamma) = \mathfrak{N}(\Gamma,1) \approx 1/(4 + b_{1/3}\Gamma^{1/3} + b_1\Gamma + b_{5/3}\Gamma^{5/3} + b_2\Gamma^2)$$
$b_{1/3} = +16.0307$, $b_1 = +14.377$, $b_{5/3} = +1.82169$, $b_2 = -0.50944$
(E.4c)

It can be analytically shown from (E.3) that the large-$\Gamma$ limit produces $T_{\max} \approx 1/\pi\Gamma$, $\nu N \approx 1/2\pi^2\Gamma$, $\tau_{\max}^2 \approx 2\nu N_{tot}/\Gamma$ for any $\nu N_{tot}$: the pile with density $\rho(\sigma,T)$ in (7.34) loses most of its poles via 'tunneling' to next pile; the Burgers piece then dominates $\varrho(\sigma,\tau)$ in (7.34). The converse process takes place for $-\Gamma \gg 1$, when (7.27)(7.28) and $\Bbbk = 0$ act in conjunction.

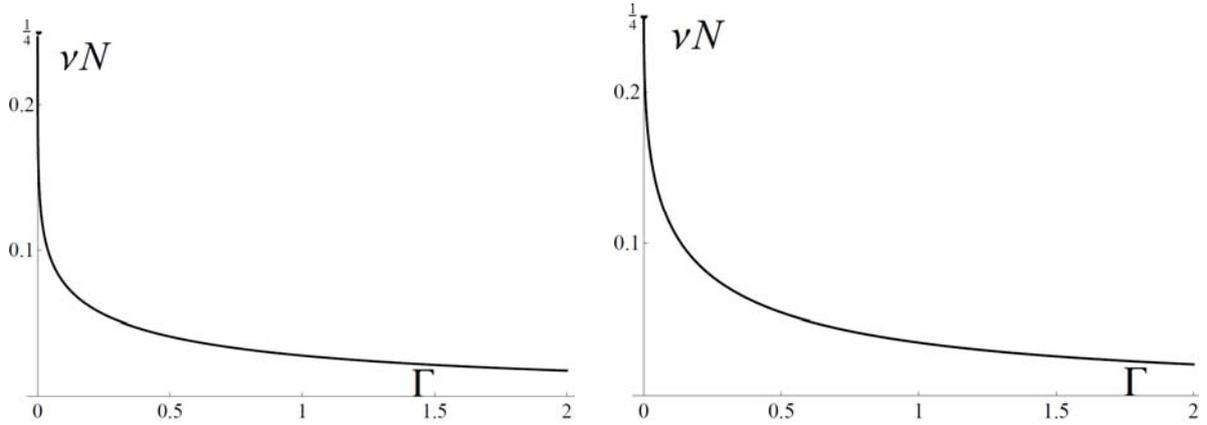

FIG. E1 (Color online) $\nu N$ vs. current combined stretch intensity $\Gamma$ as is computed from (E.4)(E.5), for $\nu N_{tot} = 1$ [left panel] or $\nu N_{tot} = \frac{1}{2}$ [right panel]. Only the steepest parts of the curves are explored if $\Gamma(\sigma) \ll 1$. In either case $\nu n = \nu N_{tot} - \nu N$ grows above $\frac{1}{4}$, to slowly asymptote $\nu N_{tot}$ at $\Gamma(\sigma) = \infty$.

Given the symmetries of $\sin(\sigma - \zeta)$, (E.4)(E.5) suffice to construct whole periodic $\nu N(\sigma)$s as in Fig.32, by substituting $(\tau_{\max}, T_{\max}, n, -\Gamma)$ for $(T_{\max}, \tau_{\max}, N, \Gamma)$ over $\pi < \sigma - \zeta \leq 2\pi$.

$\boxed{\nu N_{tot} = \tfrac{1}{2}}$
$$T_{\max}(\Gamma) \approx 1 - \Gamma^{1/2} / (A_0 + A_1\Gamma^{1/2} + A_2\Gamma^{2/2} + A_3\Gamma^{3/2}) ,$$
$$\tau_{\max}(\Gamma) \approx [1 - \Gamma^{1/2} / (a_0 + a_1\Gamma^{1/2} + a_2\Gamma^{2/2} + a_3\Gamma^{3/2})]^{1/2},$$
(E.5a,b)

$$A_0 = +0.604377, \quad A_1 = +0.505024,$$
$$A_2 = +0.186807, \quad A_3 = -0.026139,$$
$$a_0 = +0.909305, \quad a_1 = +0.501145,$$
$$a_2 = +0.147330, \quad a_3 = -0.017418.$$

$$\nu N(\Gamma) = \mathfrak{N}(\Gamma, \tfrac{1}{2}) \approx 1/[4 + b_{1/2}\Gamma^{1/2} + b_1\Gamma + b_{3/2}\Gamma^{3/2} + b_2\Gamma^2 + b_{3/2}\Gamma^{5/2} + b_3\Gamma^3] \quad (E.5c)$$
$$b_{1/2} = 14.039, \quad b_1 = +1.68731, \quad b_{3/2} = +14.0838,$$
$$b_2 = -6.6047, \quad b_{5/2} = +1.69595, \quad b_3 = -0.18112.$$

Note that $1 - \tau_{max}$, $1 - T_{max}$ and $\tfrac{1}{4} - \nu N$ now start as $O(\Gamma^{1/2})$, which implies a milder initial decay of $N$ than for all $\nu N_{tot} > \tfrac{1}{2}$, see right panel in Fig. E1 and Fig.33 in the main text.

$\boxed{\nu N_{tot} < \tfrac{1}{2}}$ Excluding $\Gamma$ from the first lines of (E.3) defines a curve in $(T_{max}, \tau_{max})$-plane that still makes sense for $\nu N_{tot} < \tfrac{1}{2}$. Along it $\tau_{max}$ initially grows with $T_{max}$ as $(2\pi \nu N_{tot} T_{max})^{1/2}$ and equals $T_{max}$ at $T_{max} = \tan(\tfrac{\pi}{2}\nu N_{tot})$; after having reached a maximum $\tau_{max}$ resumes 0 at $T_\star(\nu N_{tot}) < 1$ already encountered about the LF single-crested fronts, see entry VII.C.2.

Only when $\Gamma > 0$ lies above the value $\Gamma_\star(\nu N_{tot}) \equiv 2\max_{0<T<1}[(\tfrac{1}{\pi}\sin^{-1}(T) - \nu N_{tot})(1/T^2 - 1)]$, reached at $T_{max} = T_\star(\nu N_{tot})$, can the tabulations be extended to $\nu N_{tot} < \tfrac{1}{2}$. Piles that recurrently exchange poles require $\Gamma \geq \Gamma_{min}(\Gamma_m, \nu N_{tot}) \geq \Gamma_\star(\nu N_{tot})$, where $\Gamma_m = \max(\Gamma(\sigma))$ and $N(\Gamma_{min}) = \tfrac{1}{2}N_{tot}$. The range $0 < \Gamma < \Gamma_\star(\nu N_{tot})$ pertains to a continuum of non-leaking piles.

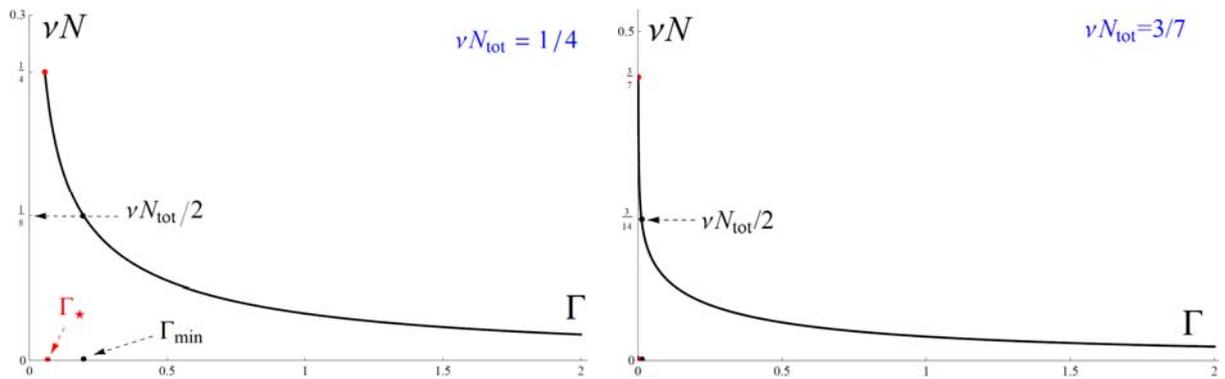

FIG.E2 (Color online). [Left panel] $\nu N = \mathfrak{N}(\Gamma, \tfrac{1}{4})$ deduced from (E.3), for which $\Gamma_\star(\tfrac{1}{4}) \approx 0.0564$ and $\Gamma_{min}(\tfrac{1}{4}) \approx 0.1968$. [Right panel] $\nu N = \mathfrak{N}(\Gamma, \tfrac{3}{7})$: the continuous drop from $\nu N_{tot} = \tfrac{3}{7} \approx 0.4286$ to $\tfrac{3}{14}$ is perceptible and $\Gamma_{min}(\tfrac{3}{7}) \approx 0.01230$ barely detectable visually, but $\Gamma_\star(\tfrac{3}{7}) \approx 0.0011$ is too small to see.

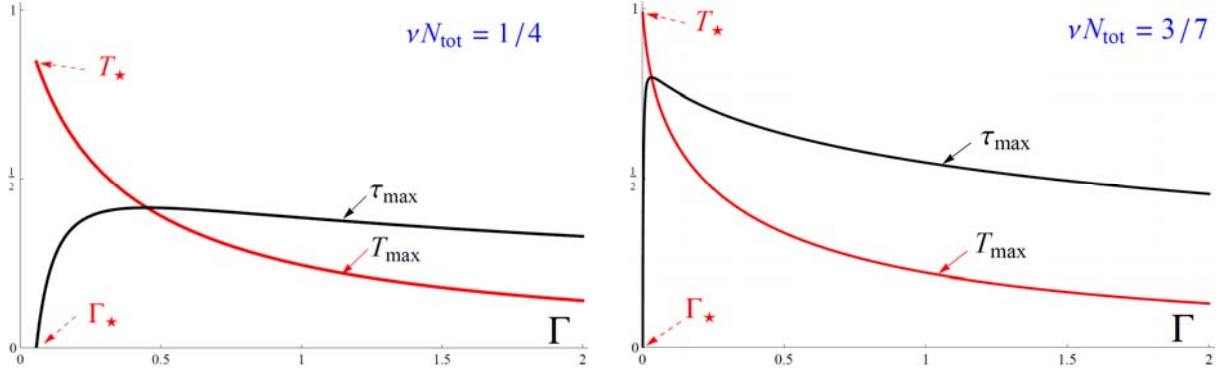

FIG.E3 (Color online). Variations of $T_{max}$ and $\tau_{max}$ vs. combined stretch intensity $\Gamma$ when the pile at $x=0$ is leaking [$\mathbb{K}=0$]. (Left) $\nu N_{tot}=\frac{1}{4}$, $T_\star(\frac{1}{4})\approx 0.8471$. (Right) $\nu N_{tot}=\frac{3}{7}$, $T_\star(\frac{3}{7})\approx 0.9887$. At crossing $T_{max}=\tau_{max}=\tan(\frac{\pi}{2}\nu N_{tot})\leq 1$, by (E.3) and $M(a,a)=\frac{1}{\pi}\tan^{-1}(a)$; (E.3) also tells that at this point $\Gamma = 2\cos(\pi\nu N_{tot})\cot(\pi\nu N_{tot})/\pi$, $\sim(1-2\nu N_{tot})^2$ if small, and $\mathfrak{N}(\Gamma,\nu N_{tot})$ is accessible via (E.3)(E.11). If $\mathbb{k}=0$ instead, $T_{max}$ and $\tau_{max}$ exchange their roles; so do $(\nu N,\Gamma)$ and $(\nu n,-\Gamma)$ in Figs. E2, E3.

The next fits, where $\Gamma_\star = \Gamma_\star(\frac{1}{4})$ and $T_\star = T_\star(\frac{1}{4})$, are accurate to $\leq 0.05\%$ for $\Gamma_\star \leq \Gamma \leq 3$:

$$T_{max}(\Gamma) \approx T_\star / [1 + A_{1/2}(\sqrt{\Gamma+\Gamma_\star} - \sqrt{2\Gamma_\star}) + \sum_{s=1}^{s=7} A_s(\Gamma - \Gamma_\star)^s]$$
$$\tau_{max}(\Gamma) \approx (\Gamma - \Gamma_\star) / [a_0 + a_{1/2}(\sqrt{\Gamma+\Gamma_\star} - \sqrt{2\Gamma_\star}) + \sum_{s=1}^{s=7} a_s(\Gamma - \Gamma_\star)^s]$$
(E.6a,b)

$A_{1/2} = +0.286226$, $A_1 = +2.431046$, $A_2 = -0.264506$, $A_3 = +0.468379$,
$A_4 = -0.344184$, $A_5 = +0.135401$, $A_6 = -0.027671$, $A_7 = +0.002305$.
$a_0 = +0.133071$, $a_{1/2} = -0.746095$, $a_1 = +2.650403$, $a_2 = +0.305520$,
$a_3 = +0.155099$, $a_4 = -0.118135$, $a_5 = +0.043615$, $a_6 = -0.008430$,
$a_7 = +0.000672$.

$$\nu N(\Gamma) = \mathfrak{N}(\Gamma,\tfrac{1}{4}) \approx 1/[4 + b_{1/2}(\sqrt{\Gamma+\Gamma_\star} - \sqrt{2\Gamma_\star}) + \sum_{s=1}^{s=7} b_s(\Gamma - \Gamma_\star)^s]$$ (E.6c)

$b_{1/2} = +42.49779$, $b_1 = -26.30069$, $b_2 = +33.69945$, $b_3 = -27.03354$,
$b_4 = +15.07545$, $b_5 = -5.18945$, $b_6 = +0.98316$, $b_7 = -0.07808$.

These give exact values at $\Gamma = \Gamma_\star$, yet constitute a bare minimum, for $\nu N_{tot} = \frac{1}{4}$ *only*. More clever fits should incorporate the $\Gamma \gg 1$ behaviors and hold for any $\nu N_{tot} < \frac{1}{2}$, while also reproducing the singular limit $\nu N_{tot} \to \frac{1}{2} - 0$. These remain to be constructed.

The approximants (E.4)-(E.6) provide what is missing in (7.34)(7.36) to complete the pole-density and front-slope profiles, and to access $V(\sigma)$, but solely during those fractions of the period $0 \leq \sigma - \zeta < 2\pi$ allowing (7.27)(7.28) *and* $\mathbb{K}\Bbbk = 0$ to be used. If $N$ and $n$ get $\sigma$-independent, the only route to $T_{max}(\sigma)$ and $\tau_{max}(\sigma)$ is to solve (7.27)(7.28) in the form

$$M(T_{max}, \tau_{max}) - \Gamma(\sigma) L(T_{max}, \tau_{max}) = \nu N,$$
$$M(\tau_{max}, T_{max}) + \Gamma(\sigma) L(\tau_{max}, T_{max}) = \nu n, \quad (= \nu N_{tot} - \nu N), \tag{E.7}$$

in which $\sigma$ again is a parameter. $T_{max}$ and $\tau_{max}$ from (E.7) still need not be found in chronological order, but they now depend on $\Gamma$, $\nu N_{tot}$ and $\nu N(\Gamma_m)$. This renders fits impractical.

### 3. *Elliptic integrals, and a recalcitrant one.*

$M(T_{max}, \tau_{max})$ and $L(T_{max}, \tau_{max})$ involved in (E.3)(E.7) are themselves defined as:

$$M(T_{max}, \tau_{max}) = \frac{2\mathbf{M}(T_{max}, T_{max}\tau_{max})}{\pi^2}, \qquad \mathbf{M}(a, \eta) \equiv a \int_0^1 \cosh^{-1}\left(\frac{(1+\eta\xi^2)}{(1+\eta)\xi}\right) \frac{d\xi}{1-a^2\xi^2}$$

$$L(T_{max}, \tau_{max}) = \frac{2\mathbf{L}(T_{max}, T_{max}\tau_{max})}{\pi(1-T_{max}^2)^{1/2}(1-\tau_{max}^2)^{1/2}}, \quad \mathbf{L}(a, \eta) \equiv a^2 \int_0^1 \frac{\sqrt{(1-\xi^2)(1-\eta^2\xi^2)}}{(1-a^2\xi^2)^2} d\xi \tag{E.8a,b}$$

- By decomposing $(1-\xi^2)(1-\eta^2\xi^2)/(1-a^2\xi^2)^2$ in simple fractions one can show that

$$\mathbf{L}(a,\eta) = \frac{\eta^2}{a^2} K(\eta^2) + (1 + \eta^2 - 2\frac{\eta^2}{a^2}) \Pi(a^2 | \eta^2) +$$
$$(1 - \frac{1}{a^2})(a^2 - \eta^2)[a^2 \frac{\partial}{\partial a^2} \Pi(a^2 | \eta^2) + \Pi(a^2 | \eta^2)] \tag{E.10}$$

where, using the notation of [46], $K(\eta^2)$ and $\Pi(a^2 | \eta^2)$ are the complete elliptic integrals of 1st and 3rd kinds, respectively, see (D.5). Moreover $a^2 \frac{\partial}{\partial a^2} \Pi(a^2 | \eta^2)$ itself is available [46] as $[(\eta^2 - a^2)K(\eta^2) + a^2 E(\eta^2) + (a^4 - \eta^2) \Pi(a^2 | \eta^2)] / 2(a^2 - 1)(\eta^2 - a^2)$, where $E(\eta^2)$ is the complete elliptic integral of 2nd kind, and can be substituted in (E.10). The resulting abstruse formula for $\mathbf{L}(a,\eta)$ one might fear actually shrinks to

$$\mathbf{L}(a,\eta) = \frac{1}{2}[(1 + \frac{\eta^2}{a^2}) K(\eta^2) - E(\eta^2) + (a^2 - \frac{\eta^2}{a^2}) \Pi(a^2 | \eta^2)]. \tag{E.11}$$

It noticeably accelerates the evaluation of $L(T_{max}, \tau_{max})$, up to 50-fold if $T_{max}$ nears 1.

- $\mathbf{M}(a,\eta)$ cannot be done exactly, except in special cases: $\mathbf{M}(a,0) = \frac{\pi}{2}\sin^{-1}(a)$, $\mathbf{M}(a, a^2) = \frac{\pi}{2}\tan^{-1}(a)$. One may also mention the identity $\mathbf{M}(1,\eta) + \mathbf{M}(\eta,\eta) = \frac{\pi^2}{4}$, valid whatever $0 \leq \eta \leq 1$ is and obtained by substituting $\xi = (1+\eta)u/(1+\eta u^2)$ in $\mathbf{M}(1,0)$: if $T_{\max} = 1$ the MS-type pole density components in (7.34) contribute a mere total of $\frac{1}{2}$ to $\nu N + \nu n$, for *any* $0 \leq \tau_{\max} \leq 1$; this also is what (7.28) gives for *all* bi-coalesced MS flames when $\nu N_{tot} = \frac{1}{2}$. By $\frac{d}{d\tau_{\max}} M(1,\tau_{\max}) = \frac{-2}{\pi^2(1+\tau_{\max})} K(\tau_{\max}^2)$, deduced by swapping $\frac{d}{d\tau_{\max}}$ and $\int_0^1 (.)d\xi$ in (E.8a), this tells that $M(\tau_{\max},1)$ grows from $M(0,1) = 0$ to $M(1,1) = \frac{1}{4}$: if $\varrho(\sigma,\tau) \to \varrho_{MS}(\sigma,\tau)$ as $T_{\max}^2 - 1 \propto \Gamma(\sigma) \to 0^-$ while $1 - \tau_{\max}^2 = O(1)$, one must have $\nu n(\sigma) \to M(\tau_{\max},1) < \frac{1}{4}$, see entry VII D.1.

Though not reducible to elliptic functions, $\mathbf{M}(a,\eta)$ is related to them through

$$\frac{\partial}{\partial a} \mathbf{M}(a,\eta) \equiv \frac{\eta}{a^2} K(\eta^2) + (1 - \frac{\eta}{a^2}) \Pi(a^2 \mid \eta^2) \ ,$$
$$\frac{\partial}{\partial \eta} \mathbf{M}(a,\eta) \equiv -[\frac{1}{a} K(\eta^2) + (a - \frac{1}{a}) \Pi(a^2 \mid \eta^2)] \frac{1}{(1+\eta)} ,$$
(E.12a, b)

to consider with the initial values $\mathbf{M}(0,\eta) = 0$ and $\mathbf{M}(a,0) = \frac{\pi}{2} \sin^{-1}(a)$. This did not avert all numerical integrations, it just helped accelerate them. From (E.12b) it is readily deduced that $\frac{\partial}{\partial b}[\mathbf{M}(a,ab)] = -[K(a^2b^2) + (a^2 - 1)\Pi(a^2, a^2b^2]/(1+ab)$, whereby there holds:

$$\mathbf{M}(a,ab) = \frac{\pi}{2} \sin^{-1}(a) - \int_0^b [K(a^2z^2) + (a^2 - 1)\Pi(a^2, a^2z^2] \frac{dz}{(1+az)} \ . \quad \text{(E.13)}$$

Numerical evaluations of $M(T_{\max}, \tau_{\max})$ from (E.13) are significantly quicker than via (E.8a), the more so if $T_{\max}$ approaches 1. Once Taylor expanded and integrated term-wise (E.12b) also provides $\mathbf{M}(a,\eta)$ as a convergent power series in $\eta < 1$, the first terms of which read $\frac{\pi}{2} \sin^{-1}(a) - \frac{\pi}{2} [1/a - (1/a^2 - 1)^{1/2}] \eta (1 - \frac{1}{2}\eta) + ...$ and can serve as checks.

• Accessing $V(\sigma)$ by means of the quasi-steady continuous version of (3.8) requires to know $\int_0^{T_{\max}} 2\nu\rho(\sigma,T)\cosh(B)dT/(1-T^2) = \int_0^{T_{\max}} 2\nu\rho(\sigma,T)(1+T^2)dT/(1-T^2)^2$; by (7.27) this also reads $2\int_0^{T_{\max}} 2\nu\rho(\sigma,T)dT/(1-T^2)^2 - \nu N$. The poles $\pi + ib$ contribute similarly, up to a flipped sign since $\cos(\pi + ib) = -\cosh(b)$. The resulting arclength increment $V(\sigma)$ takes the form

$$V(\sigma) - \tfrac{1}{2}\Gamma^2 - \mathcal{V}(\nu N_{tot}) = 2(\nu N - \nu n)\Gamma$$
$$-4[P(T_{\max}, \tau_{\max}) - P(\tau_{\max}, T_{\max})]\Gamma + 4[(Q(T_{\max}, \tau_{\max}) + Q(\tau_{\max}, T_{\max})]\Gamma^2 \ .$$
(E.14)

$P(T_{max}, \tau_{max}) - P(\tau_{max}, T_{max})$ crops up once $\nu \rho_{MS}(\sigma, T)$ and $\varrho_{MS}(\sigma, \tau)$ from (7.34) are made use of in $V(\sigma)$. The identity $2a/(1-a^2\xi^2)^2 \equiv \frac{\partial}{\partial a} a^2/(1-a^2\xi^2)$ and (E.8a) jointly lead to

$$P(T_{max}, \tau_{max}) = \frac{1}{\pi^2} \mathbf{P}(T_{max}, \tau_{max} T_{max}), \quad \mathbf{P}(a,\eta) \equiv 2\int_0^1 \cosh^{-1}\left(\frac{(1+\eta\xi^2)}{(1+\eta)\xi}\right) \frac{a\,d\xi}{(1-a^2\xi^2)^2},$$

$$\mathbf{P}(a,\eta) \equiv \mathbf{M}(a,\eta) + a\frac{\partial}{\partial a}\mathbf{M}(a,\eta).$$ 
(E.15)

$Q(T_{max}, \tau_{max}) + Q(\tau_{max}, T_{max})$, stemming from the Burgers-type contributions in (7.34), can be simplified thanks to $2a^2/(1-a^2\xi^2)^3 \equiv \frac{\partial}{\partial a^2} a^4/(1-a^2\xi^2)^2$ and (E.8b), which implies:

$$Q(T_{max}, \tau_{max}) = \frac{\mathbf{Q}(T_{max}, \tau_{max} T_{max})}{\pi(1-T_{max}^2)^{1/2}(1-\tau_{max}^2)^{1/2}}, \quad \mathbf{Q}(a,\eta) \equiv 2\int_0^1 \frac{\sqrt{(1-\xi^2)(1-\eta^2\xi^2)}}{(1-a^2\xi^2)^3} a^2 d\xi$$

$$\mathbf{Q}(a,\eta) \equiv \mathbf{L}(a,\eta) + a^2 \frac{\partial}{\partial(a^2)}\mathbf{L}(a,\eta).$$
(E.16)

As shown earlier $\frac{\partial}{\partial a}\mathbf{M}(a,\eta)$, $\mathbf{L}(a,\eta)$, $a^2\frac{\partial}{\partial a^2}\mathbf{L}(a,\eta)$, hence also $\mathbf{Q}(a,\eta)$, are expressible in terms of elliptic integrals; but $\mathbf{M}(a,\eta)$ and $\mathbf{P}(a,\eta)$ need numerical evaluation. So $V(\sigma)$ in (E.14) apparently also does, even if $T_{max}(\sigma)$ and $\tau_{max}(\sigma)$ are known. Still, the resisting $P(T_{max}, \tau_{max}) - P(\tau_{max}, T_{max})$ one needs to finalize (E.14) is also involved when $\nu N - \nu n$ is extracted from the normalizations (E.7), and can be excluded.

In absence of stretch $V(\sigma)$ can then be recast as

$$V(t) - \tfrac{1}{4}\Phi(t)^2 - \mathcal{V}(\nu N_{tot}) =$$
$$+4\Gamma^2[Q(T_{max}, \tau_{max}) + Q(\tau_{max}, T_{max})] - 2\Gamma^2[L(T_{max}, \tau_{max}) + L(\tau_{max}, T_{max})]$$
$$-\Gamma\frac{4T_{max}}{\pi^2}\frac{\partial \mathbf{M}(a,\eta)}{\partial a}\bigg|_{a=T_{max}, \eta=\tau_{max}T_{max}} + \Gamma\frac{4\tau_{max}}{\pi^2}\frac{\partial \mathbf{M}(a,\eta)}{\partial a}\bigg|_{a=\tau_{max}, \eta=\tau_{max}T_{max}},$$
(E.17)

with $Q(.,.), L(.,.)$ and $\frac{\partial}{\partial a}\mathbf{M}(a,\eta)$ accessible from (E.16), (E.11) and (E.12a), respectively.

Stretch adds still another assortment of elliptic integrals to (E.17). So, though $V(\sigma)$ is reducible to known functions, an easier route plugs the front-slope profile $\varphi_x(\sigma, x)$ from (7.35) into the definition $V(\sigma) \equiv \pi^{-1}\int_0^\pi \tfrac{1}{2}(\varphi_x)^2 dx$ and resorts to numerics, *e.g.*, NIntegrate [35]. As for $V_{av} = \pi^{-1}\int_0^\pi V(\sigma)d\sigma$ Simpson's rule [47], possibly used piecewise, can be well faster.